\documentstyle[eps,12pt]{report}
 
\newcommand{\etal}{{ $et\,al.$\ }}
\newcommand{\simgt}{\raisebox{-0.7ex}{\mbox{$\stackrel{\textstyle >}{\sim}$}}}
\newcommand{\simlt}{\raisebox{-0.7ex}{\mbox{$\stackrel{\textstyle <}{\sim}$}}}
\def\onehalfspace{\baselineskip=19pt}

\begin{document}
\onehalfspace
 
\begin{titlepage}
  \vspace*{\fill}
  \begin{center} {\huge\bf
               Obscuration of Quasars by  \\ \vspace{2mm}
        Dust and the Reddening Mechanism in Parkes-Quasars\\ \vspace{5mm}}
  \end{center}
 
  \vspace{1.0cm}
  \begin{center} {\large\bf
                by \\ \vspace{5mm}
                Franco John Masci \\ \vspace{5mm}}
  \end{center}
  \vspace{1.5cm}
 
  \begin{center} {\large\rm
                A thesis submitted in total fulfillment of the requirements\\
                for the degree of Doctor of Philosophy\\}
  \end{center}
 
\vspace{1.0cm}
 
\begin{center}
{\large\rm September 1997}
\end{center}
 
  \vspace{2.0cm}
  \begin{center} {\large\rm School of Physics \\
                             The University of Melbourne\\}
  \end{center}
  \vspace*{\fill}
\end{titlepage}
\begin{titlepage}
 
\vspace*{50mm}
\begin{center}
{\large\it  I dedicate this thesis to my mother,\\
whose love and support has made everything possible\\}
\end{center}
 
\end{titlepage}
\begin{titlepage}
  \leftline {\huge\bf Declaration}
\vspace{25mm}
 
\noindent
I hereby declare that my thesis entitled `Obscuration of Quasars by Dust and
the Reddening Mechanism in Parkes-Quasars' is the result of my own work
and contains nothing which is the outcome of work done in
collaboration, with the exception of certain observations.
Where work by others
has been used, appropriate acknowledgement and references are
given.
This thesis is not substantially the same, as a whole or in part, as anything
I have submitted for a degree or diploma or any other qualification
at any other University.
 
\vspace{20mm}
 
\noindent
This thesis is less than 100,000 words in length, exclusive of figures,
tables, appendices and bibliographies.
 
\vspace{30mm}
 
{\hspace{90mm} Franco J. Masci}
 
\vspace*{\fill}
\end{titlepage}
\newpage
\pagenumbering{roman}
 
\section*{\huge\bf{Acknowledgements}}
\addcontentsline{toc}{section}{\bf Acknowledgements}
 
\vspace{25mm}
 
\noindent
What a pleasure it is to wake up on this beautiful Spring Melbourne
day with a finished thesis, and the thoughts of all those people
who have made it possible.
 
Firstly, I thank my official supervisor, Rachel Webster and my unofficial
supervisor, Paul Francis for their enthusiasm and guidance throughout the
last four years. I feel I have been very lucky to have had two
supervisors both of whom have been generous with their time and ideas.
I owe an enormous debt of gratitude to Rachel Webster who has taught me
the benefits of a careful and rigorous approach to research.
I thank her for shortening my sentences and dealing with all those
administrative
issues in the University.
I extend my warmest thanks to her for everything she has done for me.
 
Paul Francis' contribution to this thesis has been incalculable. He
displayed enormous patience in teaching me the techniques of
observing and data reduction with {\it IRAF}.
I also thank him for his seemingly boundless knowledge of statistics.
I wish to say {\it grazie molto}.
 
This work has benefited enormously from many conversations, long and
short, both with numerous visitors to the
department and at various conferences and meetings.
I wish to extend a special thanks to Michael Drinkwater, Dick Hunstead,
Bev Wills, Anne Kinney, Loredana Bassani, Bill Priedhorsky, Bruce Peterson,
Kurt Liffman and Vic Kowalenko.

I am also indebted to the graduate students of `Room 359' (the `Astro Room')
for discussions and suggestions which only rarely were concerned
with the field of astrophysics.
Thanks in particular for all the chess games and all the help with my
computer-related traumas and
spelling dilemmas.
I wish you all a prosperous future.
A warm thanks to our school librarian Kamala Lekamge for her patience and
gratitude.
I specially thank Maurizio Toscano and Mathew Britton for lifting my
spirits with the finest of Melbourne's cuisine.
 
A special thanks to various members of the Melbourne University Mountaineering
Club for rejuvenating my academic life with numerous
expeditions, far and
close, and for
sharing my appreciation of
the natural environment.
 
I would particularly like to thank my mother for her tremendous love and
support throughout my entire scholastic life.
Given the hardships faced following the passing away of my father in 1977,
my academic position would not of been possible without her enormous help.
I thank her wholeheartedly for making my life possible --
{\it grazie molto e non ti dimentico mai}.
I also thank my brother Kevin and his family for
providing constant support and encouragement.
 
I gratefully acknowledge the University of Melbourne for its financial
support.
 
\vspace{3in}
 
{\hspace{90mm} September 25, 1997}
 
{\hspace{90mm} Melbourne}
 
\newpage
 
\section*{\centerline{\huge\bf Abstract}}
\addcontentsline{toc}{section}{\bf Abstract}
 
\vspace{25mm}
 
\noindent
A majority of quasar surveys have been based on criteria which assume
strong blue continua or a UV-excess. Any amount of dust along the
line-of-sight is expected to drastically extinguish the optical/UV
flux leading to a selection bias.
Radio surveys however should suffer no bias against
extinction by dust.
Recently, a large complete sample of radio-selected quasars has become
available (the `Parkes sample').
A majority of these sources exhibit optical--to--near-infrared continua
that are exceedingly `red', very unlike those of quasars selected optically.
The purpose of this thesis, broadly speaking, is to explore the
problem of incompleteness in
optical quasar surveys due to obscuration by dust, and to
interpret the relatively `red' continua
observed in the Parkes quasar sample.
 
The first part of this thesis explores the observational consequences
of an intervening cosmological dust component.
A preliminary study explores the effects of different foreground dust
distributions (on galaxy-cluster scales
to the visible extent of normal galaxies) on obscuration of
background sources.
Numerical simulations of dusty-galaxies
randomly distributed along the
line-of-sight with simplified assumptions are then performed, and
implications on optical counts of quasars and absorption-line statistics are
explored.
This foundation is extended by considering
the effects of more complicated models of
foreground obscuration where the dust content evolves with redshift.
The Parkes sample is used to constrain
evolutionary and physical properties of dust in intervening systems.
The contribution of line-of-sight galactic dust
to the reddening observed in this sample is also constrained.
 
The second part examines the continuum properties of Parkes quasars
in the framework of a number
of absorption and emission mechanisms to assess the
importance of extinction by dust.
Three classes of theories are explored:
`intrinsically red' AGN emission models,
dust extinction models, and
host-galaxy light models.
Simple models are developed and tested against the available data.
Several new correlations between spectral properties are predicted
and identified observationally.
For the dust model, we explore the effects of dust on soft
X-rays and compare our predictions with ROSAT data.
Possible physical dust properties are constrained.
I then consider the possibility that a
`red' stellar component from the host galaxies
contributes to the observed reddening.
This contribution is quantified using a novel spectral fitting
technique.
Finally, an observational study of
near-infrared polarisation is presented
to distinguish between two models for the reddening:
the intrinsic `synchrotron emission model', and the dust model.
Combined with spectral and photometric data,
these observations are used to constrain
various emission and dust absorption models.
 
\vspace*{\fill}
\newpage
 
 
\section*{\huge\bf{Preface}}
\addcontentsline{toc}{section}{\bf Preface}
 
\vspace{25mm}
 
\noindent
Some of {\it my}
work which has been published, or is due to be published,
and that which is {\it not} my own, but is extensively used and of substantial
importance in this thesis
is outlined as follows:
 
\vspace{5mm}
 
\noindent
\underline{{\it Chapter 1:}} this chapter involves an introduction and
review and thus is not original.
 
\noindent
\underline{{\it Chapter 2:}} to be published as Masci \& Francis (1997):
\\Masci,F.J. \& Francis,P.J.,
{\it ``Diffuse Dust and Obscuration
of the Background Universe.''}
Publications of the Astronomical Society of Australia, (1997, submitted).
\\The Cluster-QSO two-point angular correlation data used in
section~\ref{spatclust}
is from the
literature.

\noindent
\underline{{\it Chapter 3:}} published as Masci \& Webster (1995):
\\Masci,F.J. \& Webster,R.L.,
{\it ``Dust Obscuration in the Universe.''} Publications of the
Astronomical Society of Australia, {\bf 12}, p.146 (1995).
\\The initial stages of the
derivation for the probability distribution
function in dust optical depth follows from Wright (1986).
Except where specifically stated in this and all chapters that follow,
all photometric and spectroscopic
observations (and their reduction)
for the Parkes quasar sample
are the result of the
following collaborations: Drinkwater \etal (1997) and Francis \etal (1997).
 
\noindent
\underline{{\it Chapter 4:}} to be published as Masci \& Webster (1997):
\\Masci,F.J. \& Webster,R.L.,
{\it ``Constraining
the Dust Properties of Galaxies Using Background Quasars''} (1997,
in preparation).
\\The parameterisation assumed for our evolutionary dust model is
partially based on previous
indirect studies of the global evolution in heavy metals.
 
\noindent
\underline{{\it Chapter 5:}} various aspects of the material presented in
section~\ref{synchmodel} was published in
Masci \& Webster (1996):
\\Masci,F.J. \& Webster,R.L.,
{\it ``Red Blazars'': Evidence Against a Synchrotron Origin.} In
Proceedings of IAU Conference 163: `Accretion Phenomena and Related
Outflows.' ASP Conference Series, {\bf 121}, p.764 (1996).
\\The mathematical formalism used to model the synchrotron process
(section~\ref{synchshap}),
and explore diagnostics for X-ray absorption (section~\ref{Theory})
is from the
literature.
The observational tests presented in
sections~\ref{ewweak},~\ref{emewdust} and~\ref{Balmpks}
were devised in
collaboration with Dr. P. Francis, and will appear in Francis \etal (1997).
Soft X-ray data for Parkes quasars was provided by Siebert \etal (1997),
and my analysis will appear in Masci \etal (1998; in preparation).
 
\noindent
\underline{{\it Chapter 6:}} this chapter consists of original research and
will be published as Masci \& Webster (1998):
\\Masci,F.J. \& Webster,R.L.,
{\it ``Host Galaxy Contribution to the Colours
of `Red' Quasars.''} Monthly Notices of the Royal Astronomical Society,
(1998; submitted).
 
\noindent
\underline{{\it Chapter 7:}} Dr. P. Francis assisted
generously in the polarimetry
observations.
A loose description of the reduction procedure was provided by Dr. J. Bailey
of which no published material exists.
Modelling of the polarimetry data is based on extensions of
existing models from
a number of independent studies.
 
\vspace*{\fill}
 
\tableofcontents
 
\listoffigures
\addcontentsline{toc}{section}{\bf List of Figures}
 
\listoftables
\addcontentsline{toc}{section}{\bf List of Tables}

\chapter{Introduction}
\pagenumbering{arabic}

\vspace{1mm}
\leftskip=4cm

{\it ``Undoubtedly one of the greatest difficulties, if not the
greatest of all, in the way of obtaining an understanding of the real
distribution of the stars in space, lies in our uncertainty about the
amount of loss suffered by the light on its way to the observer.''} 

\vspace{1mm}

\hfill {\bf --- Jacobus Cornelius Kapteyn, 1909} 

\leftskip=0cm

\section{Background}

Quasars are the most luminous and exotic objects known in the
universe and are the only objects observable in large numbers
at sizeable redshifts.
The most distant quasars known at present have redshifts of
$\sim5$, corresponding to epochs at which the universe was less than
10\% of its present age. 
The standard model (Lynden-Bell, 1969) of a possible supermassive 
($\sim10^{8}M_{\odot}$) black hole
fueled
by an accretion disk, provides a laboratory for
studying the physical processes of matter under extreme conditions.

Quasars are of great importance in astronomy. The fact that they
can be seen at early epochs in the universe provides constraints
on models for galaxy formation and evolution of
the large scale structure. 
Recently, Katz \etal (1994) have shown that the inferred quasar
number density at high redshifts can be easily accommodated in
hierarchical galaxy formation theories such as CDM 
in which galaxies form first followed by larger scale structures.
As shown by Loeb (1993), 
the existence of quasars at high redshift implies rapid
collapse and very efficient cooling of high mass, dark matter haloes
to form black holes on relatively short timescales of $\simlt 1$Gyr.

High redshift quasars provide background light sources to
probe the nature and matter content of the 
intervening medium. 
This has been achieved through two means.
First, 
the existence of galaxies, proto-galaxies and
gas clouds absorb the light from background quasars giving rise 
to absorption-line features in quasar spectra. 
Absorption-line studies have provided important constraints 
on a wealth of physical parameters, including evolution
of the neutral hydrogen content in the universe.
Second, gravitational lensing of background quasars
where light is deflected by intervening matter, can be used 
to probe mass concentrations in galaxies, haloes
and on galaxy cluster scales.
Quasars therefore remain an important tool in cosmology for
examining the distribution and evolutionary history 
of matter in the universe.

Despite recent attempts for improving selection techniques for finding quasa-
rs, a majority of quasar surveys have been based on criteria 
which assume strong blue continua or a UV-excess (eg. Schmidt \& Green, 1983).
Such surveys however are believed to be seriously incomplete 
as indicated by the increasing numbers of quasars being discovered 
in X-ray and
radio surveys (see Crampton, 1991 for a review).
Optical studies are biased against quasars with 
weak optical continuum emission. 
Strong evidence for a
population of optically-weak quasars has recently been
provided
by optical follow-up observations of complete radio-selected samples
(eg. Kollgaard \etal 1995 and references therein;
Webster \etal 1995).
A significant proportion of these sources also have optical continuum slopes 
much redder than those quasars selected by standard optical techniques.
We hypothesise that 
the most probable explanation is reddening by dust (Webster \etal 1995).

Any given amount of dust along the line-of-sight is expected to 
drastically extinguish the observed UV-optical flux leading to a 
selection bias. 
Such dust
may be located in the local quasar environment, somewhere along
the line-of-sight; for example in intervening dusty galaxies, or
a combination of both.
If a majority of quasars are obscured by line-of-sight
dust, say in intervening galaxies, then studies which use
bright optical quasars to detect absorption lines and
gravitational lens systems will have underestimated their true
abundance. Therefore, an understanding of the effects of dust
on quasars is essential.
Whatever the mechanism for the optical reddening, if the bulk of quasars
remain undetected, then all statistical studies of evolution
in the early universe may be seriously incomplete.

In this thesis, I will explore two issues: 
the problem of incompleteness in
optical quasar surveys due to obscuration by dust,
and the interpretation of the relatively red optical 
continua observed in a complete sample of radio quasars (the `Parkes sample'). 
The amount of dust involved, both along the
line-of-sight and in the quasar environs is investigated.
A number of predictions of the dust reddening hypothesis
for Parkes quasars are presented and compared with observations.
A more specific outline of the thesis follows.

\subsection{Plan of this Thesis}

This thesis is divided into two parts. Part I (Chapters 2, 3 and 4) is
concerned with the observational consequences of an intervening 
cosmological dust component and uses the Parkes sample to constrain 
the amount of dust involved. 
Chapter 2 explores the dependence of the spatial extent of
foreground dust distributions on obscuration of background sources. 
We show that `large-scale' diffusely distributed dust 
(eg. on galaxy cluster scales)
is more effective at obscuring background sources than dust
confined to the visible extent of normal galaxies.
In Chapter 3, we simulate 
obscuration by dust located in galaxies randomly distributed along the
line-of-sight.
We use a range of parameters that may describe the dust properties of galaxies
to explore implications on optical counts of quasars.
In Chapter 4, we explore the effects of more 
complicated models of dusty line-of-sight galaxies 
where the dust content evolves. 
Implications on 
quasar and absorption-line statistics are investigated.
We use the Parkes sample to constrain
evolutionary and physical properties of dust in intervening systems. 
The contribution of line-of-sight galactic dust
to the reddening observed in this sample is quantified. 

Part II (Chapters 5, 6 and 7) examines 
the continuum properties of Parkes quasars in the 
framework of a number of absorption and emission 
mechanisms to assess the importance of extinction by dust. 
Chapter 5 explores two classes of theories to explain the
reddening observed: `intrinsically red' AGN emission models,
and dust extinction models.
Simple models are developed and tested against the available data. 
For the dust model, we explore the effects of dust on soft 
X-rays and compare our predictions with soft X-ray data for Parkes sources. 
Possible physical dust properties are also discussed.
In Chapter 6, we consider the possibility that a 
`red' stellar component from the host galaxies
of Parkes quasars contributes to the observed reddening. 
We quantify this contribution using a new, unbiased spectral fitting method.
In Chapter 7, we present a near-infrared polarisation study of Parkes quasars 
to distinguish between two models for the reddening: 
the synchrotron emission model and the dust model. 
Combined with spectral and photometric data, 
the observations are used to constrain 
various emission and dust absorption models. 

All key results are summarised in Chapter 8.
Some avenues for further research suggested from my 
work on Parkes quasars is also discussed.

\section{The Study of Cosmic Dust}

The first indications that photometric observations of astronomical
objects were affected by dust date back to the 1930s when    
stellar observations by Trumpler (1930) and
Stebbins \etal (1934) showed that distant parts of the galaxy were
obscured by interstellar dust. 
Since then, observations have established that dust is an important
constituent of galaxies.
Its presence has significantly contributed to our understanding
of star formation rates and
on the evolution of galaxy populations with cosmic time (Franceschini
\etal 1994; De Zotti \etal 1995).
Recent observations of local galaxies that are suspected of having
intense regions of star formation are also observed to be very dusty
(see Rieke \& Lebofsky, 1979 for a review). Dust is crucial in star
forming regions in that it strongly absorbs the UV radiation
responsible for molecular dissociations and provides the site for
the formation of $\rm{H_{2}}$ - the most abundant molecule in
star forming clouds. Dust also controls the temperature
of the interstellar medium (ISM). The initial cooling of
dense molecular clouds through dust-gas interactions
enhances
gravitational collapse allowing star formation to occur. 

The main questions about dust refer generally to its composition,
grain size and temperature. Our knowledge primarily comes from
observations of the ISM of our Galaxy and simulations in the laboratory.
The grains forming cosmic dust consist of silicates, graphite,
dirty ices and heavy metals and range in size from
$0.005$-$1\mu$m (Draine, 1981).
Grains typical of the ISM of our galaxy form from the condensation
of heavy elements initially synthesised in supernovae and
grow by accretion in dense molecular clouds, depending crucially on
the environment in the ISM. 
Grains are expected to be efficiently destroyed by
supernova shocks in the ISM of galaxies (McKee \etal 1987).
Recently however, studies have shown that
molecular star forming clouds
can supply the total ISM dust in less than $10^{8}$ years,
at a rate faster than it is destroyed by supernovae (Wang, 1991a and
references therein).

Dust absorbs the UV-optical radiation in star-forming regions
and re-radiates absorbed energy in the far-infrared part of the
spectrum ($10\mu$m-1mm). Observations by the Infrared
Astronomy Satellite (IRAS) have shown that radiation by dust  
from galaxies is comparable to what we see in the visual band 
(Soifer, Houck \& Neugebauer, 1986).
Since dust in mostly made of heavy elements, its evolution
is directly
connected to chemical evolution in the ISM of galaxies.
Consequently, observations at infrared wavelengths have
contributed significantly towards our understanding of
the evolutionary properties of galaxies and their contribution
to the integrated background radiation (Wang 1991a; 1991b).

\section{The Interaction of Radiation with Dust and 
Observational Diagnostics}

The two major processes in which dust grains can alter the transmission
of electromagnetic radiation are absorption and scattering.
These effects,
collectively referred to as extinction, are the best
studied property of cosmic dust since they can be accurately determined
over a range of wavelengths.
A determination of the extinction properties are essential for 
correcting astronomical 
measurements for the effects of
dust.
Other very important diagnostics of dust 
are its emission properties, both in spectral
features and continuum emission in the infrared, 
and its polarisation properties.
These properties are briefly discussed below. 
An excellent and thorough analysis of the physics can be found in
Hoyle \& Wickramasinghe (1991).

\subsection{Extinction}
\label{extd}

Extinction of radiation at optical wavelengths 
by dust is largely due to
scattering, with the effects of absorption
becoming important at shorter wavelengths. 
Light from a source shining 
through dust along the line-of-sight has its flux 
reduced according to
\begin{equation}
F_{\lambda}\,=\,F_{\lambda 0}e^{-\tau_{\lambda}},
\label{exttau}
\end{equation}
where $F_{\lambda 0}$ is the flux that would be received at Earth
in the absence of dust, $F_{\lambda}$ is the actual
flux observed and $\tau_{\lambda}$ is defined as the
optical depth at the wavelength observed.
This is derived by solving the radiative transfer equation for
a homogeneous slab of dust.
The extinction in magnitudes is related to the optical depth
by $A_{\lambda}=2.5\log(F_{\lambda }/F_{\lambda 0})\equiv1.086\tau_{\lambda}$.
The extinction law is defined by specified values of $\tau_{\lambda}$
at a range of wavelengths and can generally be expressed as
\begin{equation}
\tau_{\lambda}\,=\,C\,f(\lambda),
\label{taulam}
\end{equation}
where $C$ relates to the amount of dust 
and the function $f(\lambda)$ depends on the optical properties of 
grains.

\begin{figure}
\vspace{-2.5in}
\plotonesmall{1}{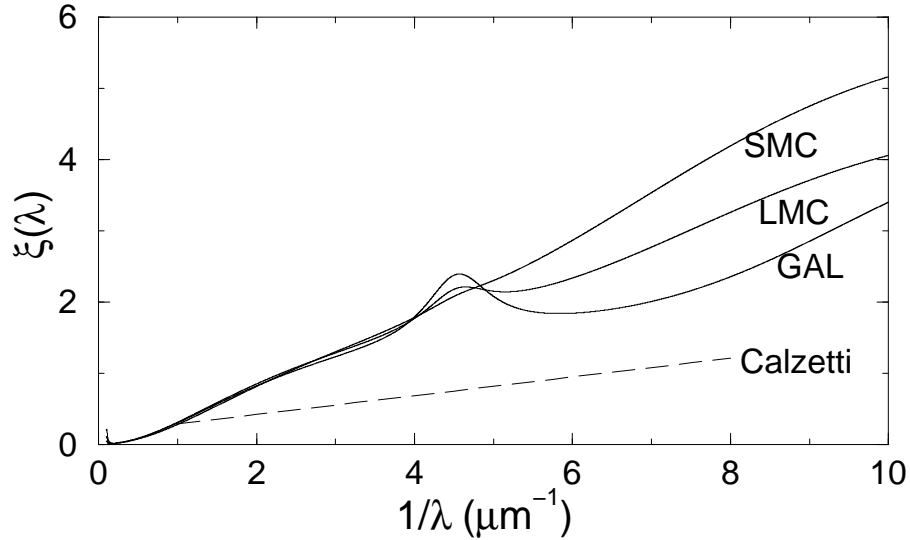}
\vspace{-2.5in}
\caption[Extinction curves of the Milky Way, LMC, and SMC]{Empirical fits to the
extinction curves of the Milky Way, LMC, and SMC from Pei (1992).
The extinction $\xi$ is measured relative to that observed in the 
$B$-band ($\equiv\tau_{\lambda}/\tau_{B}$).
Also shown (dashed) is the extinction law derived by
Calzetti \etal (1994) for external galaxies.}
\label{exteg}
\end{figure}

Extinction laws in our galaxy are primarily determined from observations 
of stars by comparing the spectral energy
distribution of a reddened star with that of 
an unreddened star of similar spectral-type.
The galactic interstellar extinction laws of
Savage \& Mathis (1979) and Seaton (1979) are commonly used 
to model the effects of dust and correct
astronomical observations. 
Pei (1992) has obtained excellent empirical 
fits for extinction curves derived from previous studies for 
the Milky Way, LMC and SMC.
These are shown in Fig.~\ref{exteg}, 
where the extinction is measured relative to
that observed in the $B$-band, ie.  
$\xi(\lambda)=\tau_{\lambda}/\tau_{B}$. 
The striking feature is the almost universal behaviour at wavelengths 
$\lambda>2500$\AA$\,$.
The ``mean'' interstellar extinction curve derived in our galaxy 
varies by a factor of $\sim10$ in extinction 
over the range $1000{\rm\AA}<\lambda<2\mu$m and
exhibits both 
continuous extinction and a strong feature known as the ``2175\AA$\,$ bump''
over which the extinction varies appreciably. This feature
decreases in strength going from the Milky Way to the LMC and SMC.
This bump is believed to be due to graphite, however its origin
is not as yet well understood. 

It is remarkable that a
recent study by Calzetti \etal (1994) finds no evidence for a 2175\AA$\,$ 
bump feature in the extinction curves of nearby galaxies (see dashed 
curve in Fig.~\ref{exteg}).
These authors also find the extinction laws for a number 
of galaxies to be considerably flatter as a function of
wavelength in the UV, with extinction measures smaller by factors 
of $\simgt3$ at $\lambda\simlt 1590$\AA$\,$ than those in the Milky Way. 
This behaviour is commonly referred to as ``gray extinction''.
They attribute the ``grayness'' as being due to their method of derivation.
Their method relies on examining correlations
between the Balmer line ratio ($H\alpha/H\beta$; see section~\ref{Balmpks}) 
and optical
continuum reddening. Each of these quantities is expected to
arise from different physical regions in the galaxies: the hydrogen
lines from hot ionizing stars associated with dusty regions, and the continuum
from older stellar populations which may have had time to drift away from
the dustier regions. Thus, the continuum measurements may be weighted 
mostly by regions of lowest extinction.
The absence of the 2175\AA$\,$ bump feature has been interpreted
as being either due to
effects of scattering of light within extended dust-regions
into the line-of-sight (therefore
reducing the effective extinction in the UV), 
or to a chemical composition of grains different from that
in the Milky Way. 

There have been many efforts devoted to the understanding
of the composition and optical properties of dust grains
giving rise to the various extinction laws, in particular
that of the Milky Way (Mathis, Rumpl \& Nordsiek, 1977; Drain \& Lee, 1984). 
These authors have studied different chemical mixtures to 
reproduce the Milky Way extinction law and as yet no definite agreement
exists.
Draine \& Lee (1984) find that for a $1:1.12$ graphite and silicate
mixture, the extinction (or the function $f(\lambda)$ in Eqn.~\ref{taulam})
at optical wavelengths varies as a function of wavelength
as $f(\lambda)=\lambda^{-n}$, where $n\sim1$-2.
This is derived using the optical properties of grains
and the theory of light scattering which involves
a solution of Maxwell's equations.
The index $n$ depends on the
characteristic grain size ``$a$'', and its dependence has been studied via 
the extinction efficiencies $Q_{ext}(a,\lambda)$ of spherical 
dielectric grains (eg. Greenberg, 1968).
In the optical--to--near-infrared, the extinction law in our galaxy approaches
the characteristic behaviour $\lambda^{-1}$. 
This is expected when
the sizes of individual grains become comparable with the 
wavelength of light ($\lambda\sim a$), where the extinction efficiency is
given by $Q_{ext}\propto\lambda^{-1}$.
For comparison, when the grain sizes are very small compared with
the wavelength ($a\ll\lambda$), the extinction efficiency varies as 
$Q_{ext}\propto\lambda^{-4}$,
known as the Rayleigh scattering regime. When $a\gg\lambda$, $Q_{ext}$
levels off asymptotically to about twice the geometric cross-section 
of the grain. 
Hence, larger grain sizes
produce a flatter dependence with wavelength. 

Approaching higher energies such as X-rays, and depending on the grain size,
grains become somewhat transparent.
Using graphite and silicate
mixtures, Laor \& Draine (1993) show that the extinction
decreases by at least two orders of magnitude from 0.1-10keV 
(see Fig.~\ref{Xext}).
At energies $\simgt1$keV, the dust extinction law becomes the same
as if all grains were composed of neutral atoms in the gas phase
rather than possibly being ionized. In this case, atomic X-ray absorption will
dominate the optical depth.
At soft energies ($\simlt 0.5$keV), the X-ray absorption cross-section
is dominated mostly by H and He which, given a
large neutral fraction, can attenuate the X-ray flux by more than
80\%, even for relatively low values of the 
hydrogen column density (ie. $N_{H}\simgt10^{18}{\rm cm}^{-2}$). 
At $\simgt1$keV however, absorption
becomes dominated by metals (primarily the K-edges of oxygen and carbon
and to a lesser extent Ne, Mg, Si, S and Fe - Morrison \& McCammon, 1983).
See section~\ref{Theory} for more details.
Metals in dust grains can be shielded from the effects of X-ray absorption, 
however,
modifications of
only a few percent
to their absorption cross-section are expected, even at soft X-ray energies 
(Fireman, 1974).

\subsection{Other Diagnostics}
\label{otherdiag}

In addition to general extinction over a continuous wavelength range,
the presence of spectral absorption features can be used
to confirm the presence of dust, and also
provide information regarding the compositions and nature of the grains. 
Apart from the 2175\AA$\,$ bump in our galaxy, strong $9.7\mu$m
and $18\mu$m absorption/emission features are almost always seen in our
ISM and the ISM of a number of external galaxies when 
the dust is optically thin to its own emission, ie. $\tau_{10\mu{\rm m}}\ll 1$.
These features have been attributed to silicate grains with 
sizes $a\simlt 3\mu$m (Laor \& Draine, 1993). These authors also show that 
their absence in the
near-infrared spectra of active galaxies (including luminous quasars) 
provides important constraints on the chemical and physical properties of dust
in such sources (see section~\ref{qsodust} for more details).

Infrared (IR) continuum emission from dust grains can arise through
two means. First, by absorption and heating by UV photons and
subsequent re-radiation at infrared wavelengths and second, by
steady state emission or fluorescence. The latter process occurs
for far-infrared wavelengths, $\lambda>100\mu$m.
An important diagnostic of this emission is in the
determination of dust temperatures and/or masses. 
These can be calculated by
assuming that dust grains are in thermal equilibrium with their 
surroundings so that the emission can be approximated
as a blackbody. 
The blackbody radiation law can be written 
\begin{equation}
B(\nu,T_{d})\,=\,\frac{2h\nu^{3}}{c^{2}}\frac{1}{\exp(h\nu/kT_{d}) - 1}\,\,\,
{\rm erg\,cm^{-2}sec^{-1}Hz^{-1}ster^{-1}}.
\label{BBlaw}
\end{equation}
Given measurements of the infrared flux
at two or more frequencies, the dust temperature $T_{d}$ can be determined by
fitting equation 1.3.
In the determination of dust masses however, we define an
additional quantity called the dust emissivity, $\kappa_{\nu}$, 
with units ${\rm cm^{2}}$ per
gram of dust. From galactic observations at sub-mm wavelengths, 
$\kappa_{\nu}\propto\nu^{2}$ (Chini \& Kr\"{u}gel, 1994).
As an example, let us consider
an IR-source at some redshift $z$, whose observed IR flux is due to
thermal dust emission. The dust mass, assuming the dust is optically thin to
its own emission with effective temperature $T_{d}$ is
given by
\begin{equation}
M_{dust} = \frac{S_{\nu_{o}}D_{L}^{2}}{(1+z)\kappa_{\nu_{r}}
B(\nu_{r},T_{d})},
\label{Mdust1}
\end{equation}
where $\nu_{o}$ and $\nu_{r}$ are the observed and rest 
frame frequencies respectively, $\kappa_{\nu_{r}}$ is the dust 
emissivity defined above, $D_{L}$ the luminosity distance
and $S_{\nu_{o}}$ the observed flux. The factor $(1+z)$ is due
to the decrease in frequency bandpass in the observer's frame
with increasing redshift.
Depending on the temperature $T_{d}$ and emission frequency $\nu$,
there are two limiting behaviours for the
blackbody radiation law (Eqn.~\ref{BBlaw}) which are commonly used to
simplify calculations.
One limit
is the well known Rayleigh-Jean's 
law and is applicable in the low frequency limit:
\begin{equation}
B(\nu,T_{d})\,\approx\,2\left(\frac{\nu}{c}\right)^{2}kT_{d}
\,{\rm\hspace{1cm} for\hspace{2mm}}\,\frac{h\nu}{kT_{d}}\ll 1.
\label{lownu}
\end{equation}
In the high frequency limit, we have Wien's law: 
\begin{equation}
B(\nu,T_{d})\,\approx\,\frac{2h\nu^{3}}{c^{2}}\exp\left(-\frac{h\nu}{kT_{d}}
\right)\,
{\rm\hspace{1cm} for\hspace{2mm}}\,\frac{h\nu}{kT_{d}}\gg 1.
\label{highnu}
\end{equation}
 
As a general rule, the blackbody law allows us to
approximate the effective dust temperature through the relation:
\begin{equation}
T_{d}\,\simeq\,2898\left(\frac{\lambda_{max}}{\mu{\rm m}}\right)^{-1}{\rm K},
\label{tdlmax}
\end{equation}
where $\lambda_{max}$ is the
wavelength at which the emission peaks (ie. where $B(\nu,T_{d})$ is a 
maximum).
If dust is to survive, it is estimated that grains
cannot exceed a temperature of $\sim1500$K after which
evaporation or sublimation will occur.
Thus, we expect dust to emit only at wavelengths 
$\lambda\simgt2\mu$m.
Evidence for hot dust approaching the limiting 
temperature $T_{d}\sim1500$K is
believed to exist in the environments of
quasars and other active galaxies, giving
rise to the characteristic near-IR bump
observed in the range $1-3\mu$m
(Sanders \etal 1989).

Polarisation of radiation produced by scattering or
transmission through aligned dust particles
provides a useful diagnostic for determining the
size distribution of dust grains, the conditions under which
grains can be aligned (eg. magnetic field geometry) and most importantly,
the geometry of scattering regions (Whittet, 1992 and references therein).
In the Milky Way, where
the most accurate determinations of dust-polarisation have
been made, the polarisation is of order a few percent. 
The polarisation properties of dust are less well
determined than those of extinction, however, optical polarisation coincident 
with extinction or IR emission may confirm 
that the latter processes are due to dust.

To summarise, the difficulties in drawing a consistent picture
of the interaction of dust grains with radiation
lie in their chemical and physical properties, such as composition
and grain size distribution. These properties, which are primarily determined
from extinction measures, are 
very sensitive to the geometrical
distribution of dust in galaxies and hence subject to
considerable uncertainty.

\section{The Cosmological Distribution of Dust}

Infrared and optical studies of local galaxies have provided a good
description of their dust content, 
however, the study of dust at high redshift has yet to produce
conclusive results.
Despite many efforts for detecting high redshift ($z>1$)
galaxies in the optical,
it is suspected that large numbers are obscured by dust
associated with intense star formation activity
(Djorgovski \etal 1993; Franceschini, 1994).
Our knowledge regarding high redshift dust (at $z>2$) has 
primarily come from two sources. 
First, by direct imaging of intrinsically luminous, high redshift 
radio galaxies at far-IR to mm wavelengths,
and second, from studies of QSO absorption line systems.
The observational status is given below.

\subsection{Dust in Local Galaxies}
\label{localga}

Various morphological and spectral studies of nearby galaxies have
demonstrated that the bulk of their IR luminosity originates from
extended emission by dust.
The local far-infrared luminosity density is estimated to be about 
one-third of the optical luminosity density (eg. Saunders \etal 1990).
This emission has lead to estimates of galactic dust masses in the range
$10^{7}$-$10^{9}M_{\odot}$, ie. $\simlt0.5\%$ of the total mass in a
typical galaxy (Rieke \& Lebofsky, 1979).
Furthermore, Zaritsky (1994a) 
has provided strong evidence for
diffuse dust haloes extending to distances $\sim60h_{75}^{-1}$ kpc 
from the
centers of galaxies.
This was provided by studies of the colours of distant galaxies seen through
the haloes of nearby spirals, where background galaxies at smaller
projected separations were statistically redder than those in outer regions. 
If such a diffuse component proves to be common, then the local dust content
is expected to be greater by almost an order of magnitude 
than previously estimated
using IR emission alone.

In cases where IR emission is not observed, measures of the
dust optical depth using surface 
photometry can provide a useful diagnostic for determining the
dust content of galaxies. 
This method is based on statistical studies of the
dependence of optical surface brightness on the viewing
angle of a spiral disk.
For a transparent disk, the surface brightness
depends sensitively on inclination, while for an
opaque ``dusty'' disk it does not (Holmberg, 1958). This allows limits
to be placed on the dust optical depth. 
One caveat however is that this method is limited to
galaxies of relatively high surface brightness.
Recent studies predict central $B$-band optical depths for spiral
galaxies in the range $0.5\simlt\tau_{B}\simlt2$
(Giovanelli \etal 1994 and references therein). 
In E/SO galaxies, extinction is not as well studied, though not
believed to be as large.
Goudfrooij \etal (1994a; 1994b; 1995) suggest there is a diffuse
component with $0.1\simlt\tau_{B}\simlt0.9$.
These optical depth properties still remain an open question.

\subsection{High Redshift Dust and Evolution}
\label{hzdev}

Luminous radio galaxies can be studied out to large cosmological distances.
Recently, a number of radio galaxies at redshifts $z\simgt3$ have been
observed to emit a strong far-IR to mm continuum.
If the emission is assumed to be thermal emission from dust, then
dust masses comparable to those of local galaxies
are implied (Chini \& Kr\"{u}gel, 1994 and references therein).
The detection of dusty objects at $z\sim4$ is 
puzzling. One might expect that the majority of 
objects at high redshifts
(corresponding to epochs only $5-10\%$ of the age of the universe) still
contain a large reservoir of primordial gas.
The presence of dust implies an early episode of star formation and
provides a step towards understanding the interstellar medium, and
hence formation and evolution of galaxies at high redshift.

Strong evidence for the existence of dust at high redshift is
provided by studies of QSO absorption line systems.
The highest neutral hydrogen column density systems known
as the damped Ly-$\alpha$ absorbers are believed to be
the progenitors of present day galactic disks.
Observations of trace metals such as Ni, Cr and Zn in these
systems have provided an opportunity to study chemical and dust
evolution in galaxies at high redshift 
(Meyer \& Roth, 1990; Pettini \etal 1994; 1997). 

The relative strengths of various metal ions in the
spectra of background quasars have been used to infer the amount
of metal depletion in the gas phase and hence the amount of dust 
assuming depletion onto grains.
Recent studies of metal absorption line systems towards redshifts
$z\sim2$ have indicated metallicities which are $\sim10\%$ solar
and dust-to-gas ratios less than $8\%$ of the galactic value
(Meyer, Welty and York, 1989; Pettini \etal 1994).
This constraint on the dust-to-gas ratio is also in
agreement with studies by Fall \etal (1989) who compared the
spectra of quasars that have damped Ly-$\alpha$ with those
that do not. They found that quasars with damped Ly-$\alpha$ absorption
in their spectra had
spectral slopes somewhat redder than those without absorption. 

The lower metal abundances in QSO absorption systems 
relative to those in the galaxy have
been interpreted as evidence for less chemical enrichment at high $z$.
It is important to note however that such measurements may
be subject to considerable observational bias.
Existing observations of QSOs with damped Ly-$\alpha$ absorption lines
are drawn mostly from optically-selected, magnitude-limited samples
and hence, it is possible than an unrecognised population
of highly reddened QSOs may escape detection due
to heavy obscuration by dust in damped Ly-$\alpha$ systems. 
Thus, present QSO absorption line studies may only preferentially
pick out chemically unevolved gas.
Fall \& Pei (1993) have attempted to correct for this bias, estimating that
the true metal and dust content may be on average 2-3 times
greater than that deduced observationally.

There seems to be considerable difficulty in
applying the results of QSO absorption line studies and
high redshift radio galaxy surveys to the study
of primeval galaxies and evolution of their ISM. 
It may be some time before large enough, complete
statistical samples 
can be compiled so to obtain a consistent 
picture of chemical and dust evolution in the universe. 
At present, we must resort to dynamical modelling
of stellar processes in galaxies to gain some insight into this
problem 
(eg. Wang 1991a and references therein).

\section{Dust and Quasars}

In my study of the effects of dust on the properties
and identification of quasars in the optical, I shall
investigate two possibilities for its location. 
First, I will consider
extrinsic or line-of-sight dust associated
with foreground galaxies and second, intrinsic dust or 
dust physically associated with the quasars themselves.
A brief review of the status on each of these studies
is given below.

\subsection{Obscuration of Quasars by Intervening Dust}

Studies investigating the effects of intervening, 
cosmologically distributed dust on background quasars were 
initially motivated by the apparent lack of high 
redshift quasars in the optical (McKee \& Petrosian, 1974; 
Cheney \& Rowan-Robinson, 1981). 
Based on observations of individual
quasar spectra, colours
and counts as a function of redshift, it was concluded that
uniformly distributed intergalactic dust as distinct
from galactic dust was not likely to be
a large source of extinction. From thereon, modelling
of this problem assumed all dust to
be situated in clumps or hypothesised dusty galaxies 
along the line-of-sight.

Various theoretical studies by Ostriker \& Heisler (1984); 
Heisler \& Ostriker (1988), Fall \& Pei (1989, 1993) and
Wright (1986, 1990) conclude that at least $70\%$
of bright quasars at redshifts $z\simgt3$ may be obscured
by intervening galactic dust in the optical.
A more refined treatment was undertaken by Wright (1990) who
stressed the importance of galaxy ``hardness'' 
on the properties of a quasar sample subject 
to both flux and colour selection effects. 
The term ``hardness'' depends on the assumed spatial distribution
for dust optical depth through an individual galaxy.
For optical depth profiles which are smooth or ``fuzzy'' around the
edges (eg. an exponential or King model),
the ``hardness''
is defined by the magnitude of
the {\it central} optical depth.
For central $B$-band optical depths $\tau_{B}\gg1$, galaxies were defined
as being ``hard-edged'' up to a typical scale radius, while
for $\tau_{B}\simlt1$, galaxies
were defined as ``soft-edged''.
For ``sharp-edged'' optical depth profiles (eg. an optical depth
which is uniform throughout a finite projected area and zero outside - as
considered by Ostriker \& Heisler, 1984),
the hardness is defined by the magnitude of this optical depth.
Wright (1990) found that galaxies which are
``soft'' around the edges will cause many background quasars
to `appear' reddened without actually removing them 
from a flux-limited sample.

It is important to note that these studies are
extremely model dependent. The most crucial parameters are 
the number of absorbing galaxies involved and their dust properties
(eg. individual optical depths and dust spatial extent).
These parameters are presently known to within a factor of three.
Further studies to determine these parameters are therefore necessary 
before plausible conclusions 
regarding the effects of
foreground dust on optical observations of 
high redshift quasars can be made.

If published estimates of the effects of foreground dust
on optical studies of quasars
are correct,
then implications for present cosmological studies are immense.
Studies which use bright optical quasars
to count absorption line systems and gravitational
lenses may have severely underestimated their true abundance.
Studies involving damped Ly-$\alpha$ absorption systems
to determine a wealth of physical
parameters such as chemical composition, dust and gas content, and
star formation rates at high redshift,
are also likely to be affected (Fall \& Pei, 1993; Pei \& Fall, 1995). 
If substantial numbers of quasars remain undetected,
then a significant fraction of the 
observed $\gamma$- and X-ray background
may arise from dust 
obscured quasars (Heisler \& Ostriker, 1988). More importantly, 
these quasars would 
significantly contribute to the UV photon density at high redshifts,
thereby altering the nature and physical state of the intergalactic medium at
early epochs.

\subsection{Dust in the Quasar Environment}
\label{qsodust}

There are two major lines of study investigating the 
existence and properties of dust 
in the neighbourhood of quasars.
The first class of studies is based on the observed IR emission.
It is suggested that the near-IR ``bump'' at $1-3\mu$m
and spectral turnover at $\lambda>100\mu$m in a majority of
optically-selected quasars
can be attributed to emission by dust
(Neugebauer \etal 1979; McAlary \& Rieke, 1988; Sanders \etal 1989).
In fact, about a third of the total luminosity
from quasars is emitted in the $1-100\mu$m range (Sanders \etal 1989).
The most attractive explanation for this emission is
thermal radiation from dust heated by a central UV-optical continuum source. 
This is supported by a significant
lack of variability and polarisation in the IR compared to the 
emission properties at $\lambda < 1\mu$m (Sitko \& Zhu, 1991).
The IR continuum emission can be fitted with
a variety of models: optically thin dust in a spherical
distribution (Barvainis, 1987), 
optically thick dust in a torus with equatorial optical 
depths $A_{V}\sim100$ on scales less than
a few hundred parsecs
(Pier \& Krolik, 1992), or dust in a highly warped disk on
much larger scales (Sanders \etal 1989).
These studies are consistent with 
a significant number of quasars
having their optical emission substantially reduced,
preventing their detection in the optical.

The second class of studies involve searching for dust from
differential extinction
of emission lines in the optical-UV spectra
of quasars (Draine \& Bahcall, 1981; MacAlpine, 1985; Netzer \& Laor, 1993).
MacAlpine (1985) and  Netzer \& Laor (1993)
conclude that much of the absorbing dust
is likely to be embedded within the narrow
line emitting gas (the NLR) on scales no larger than about a kiloparsec, ie.
on scales a little larger than that of a possible dusty torus.
Dust in the NLR is expected to modify the physical conditions
of the emitting gas and consequently, the continuum spectral slope
(see MacAlpine, 1985 for a review).

Together with studies of the IR continuum, these studies
have provided strong constraints on the possible location for
dust in quasars and other lower luminosity active galactic 
nuclei (AGN).
For quasars, it is predicted that dust grains are 
heated to high evaporation
temperatures. Barvainis (1987) has estimated an ``evaporation
radius'' greater than $\sim1$pc, ie. scales
much larger than the hypothesised broad line emitting region (BLR). 
Furthermore, the characteristic turnover at $\lambda\sim100\mu$m
has been shown to imply that a significant fraction of 
the thermal IR radiation may be emitted
from scales $\simgt100$pc (Edelson, Malkan \& Rieke, 1987).
This leaves two possible options for the location of the
dust: a torus and/or NLR clouds, or the quasar host galaxy.

There is remarkable similarity between
quasars and other nearby, less luminous AGN 
such as Seyferts.
The growing importance of dust in
Seyfert galaxies 
may suggest that at least some quasars 
should also exhibit similar properties.
Recently, optically thick dust
with azimuthal symmetry such as a torus, has been invoked to
explain the difference between broad and narrow-line AGN
in the framework of unified schemes (Antonucci, 1993).
The distribution of this material in a torus can
obscure the central UV-optical continuum source 
and BLR from direct view, while
emission from the larger scale NLR is unaffected.
This is supported by spectropolarimetric observations
showing broad emission lines in
polarized light in narrow line (type-II) Seyfert galaxies. 
Further evidence for a torus-shaped dust region is
provided by the observation of ionization cones in several
Seyfert galaxies, demonstrating that the gas in the host galaxy
sees an anisotropic ionizing source. 
Evidence for dust tori in high
redshift quasars is scarce. However, spectropolarimetric
observations revealing the presence
of a hidden BLR, combined with a reddened UV-optical continuum
has been found in two quasars
(Hines \& Wills, 1992 and Wills \etal 1992).

\subsection{Optically Dust-Obscured Quasars}
\label{dustobscu}

One drawback in the above studies investigating ``intrinsic'' dust, 
is that almost all are based on optically-selected quasar
samples. 
Optical studies are expected to be
heavily biased against significant extinction by dust. 
These inevitably show that dust is
present and an important constituent of quasars, however, it is possible that 
a large population remains
undetected optically due to obscuration by dust.
Evidence for such a population may be provided by studies at 
wavelengths where {\it no} bias against dust obscuration is expected.
There are three possible wavelength regimes of interest:
The infrared, X-rays and the radio. I shall consider
each in turn below.
 
Near or far-infrared surveys may provide a
useful means of detecting dust obscured quasars.
The ideal strategy is to look for
intrinsically luminous point like sources in the IR.
Follow-up IR-spectroscopy is then required for
further identification of standard characteristics such as 
broad emission lines.
The ultraluminous infrared sources detected by 
the Infrared Astronomy Satellite (IRAS)
are likely candidates for dust-obscured quasars. These sources have IR 
luminosities comparable to those of quasars 
($L_{IR}\simgt10^{11}L_{\odot}$)
and there is debate about whether such objects 
host buried quasars (Hines \etal 1995 and references therein).
Evidence for such a population
has been presented by Hill \etal (1987) and Sanders \etal (1988).
Moreover, a large abundance of faint, possibly dust-obscured quasars
is suggested by the IRAS selected sample of
Low \etal (1989). In this study, FeII emission was shown to
be prominent in quasars with strong IR luminosity 
relative to their optical emission, suggesting a medium significantly
enriched in metals and dust.  
Through an investigation of their
radio properties, Lonsdale \etal (1995)
claim
that it is physically plausible for dust enshrouded quasars to power
the IR emission observed from IRAS sources.

X-ray selected samples may provide stronger evidence
for an optically obscured quasar population.
Soft X-rays ($\simlt0.5$keV) are subject to considerable absorption by hydrogen
and heavy metals (see section~\ref{Theory}), and hence soft X-ray surveys
are expected to be heavily biased.  
Hard X-rays ($\simgt5$keV) however
penetrate the dust, and absorption may become
significant only when the gas column density is sufficiently high, 
ie. $\simgt10^{23}{\rm cm}^{-2}$ (Awaki \etal 1990).
Stocke \etal (1991) found a factor of almost two
orders of magnitude dispersion in the X-ray-to-optical ratios
of an X-ray selected sample as compared to those selected optically.
This scatter may be caused by variable dust extinction in the optical
and in fact supports the claim by
McDowell \etal (1989) for a possible a correlation 
between X-ray-to-optical flux ratio
and $B-K$ colour.

There have been numerous X-ray spectroscopic 
studies claiming soft X-ray absorption in excess of 
that expected from the galaxy in the spectra of radio-selected quasars 
(Madejski \etal 1991; Wilkes \etal 1992; Elvis \etal 1994).
Absorptions corresponding to $\simgt10^{22}$ atoms ${\rm cm}^{-2}$
were deduced in a majority of cases.
These studies however failed 
to detect any associated extinction by dust in the optical.
A correlation is expected at some level, however as will be 
discussed in section~\ref{Theory}, these separate processes may 
critically depend on the physical conditions, locations and geometry
of the absorbing gas and dust. 

Optical identification of radio-selected sources
probably provides the best technique for
detecting dust-obscured quasars, being subject to fewer
possible selection effects than surveys based at X-ray or IR wavelengths.
Criteria for selecting quasars in the radio is primarily based on their
characteristic ``flat'' spectral energy distribution at
GHz frequencies; $\alpha\simlt0.5$ where $S_{\nu}\propto\nu^{-\alpha}$. 
Most of the 
bright (ie. $\simgt0.1$Jy) radio surveys based on this criterion
have $\simgt70\%$ of sources identified with quasars which have 
optical properties extremely similar to those selected optically. 
More than 10\% also comprise sources
which are highly polarised and strongly variable at radio to
optical wavelengths. These sources belong 
to the ``Blazar'' class of AGN (Angel \& Stockman, 1980). 
There have been very few radio-surveys specifically aimed towards finding 
quasars. Most surveys did not include criteria based on radio
spectral-slope. Both steep and flat-spectrum sources were selected, where
the former comprised the majority of sources and were usually identified 
with galaxies.
For instance, an optical study of a complete radio sample with 
$S_{2.7{\rm GHz}}>0.1$Jy
by Dunlop \etal (1989)
found a quasar fraction $\simlt30\%$. 
Their spectroscopic identifications however are also significantly
incomplete. 

\begin{figure}
\vspace{-2.5in}
\plotonesmall{0.9}{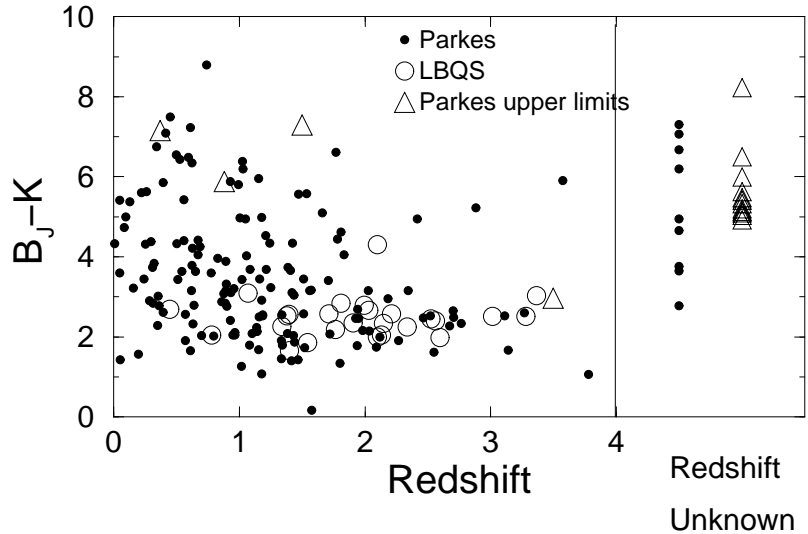}
\vspace{-2.3in}
\caption[Observed $B-K$ colour as a function of redshift]
{Distribution in $B_{J}-K$ colour as a function of redshift
for quasars from a complete sample of radio-loud, flat spectrum
sources (filled circles; Drinkwater \etal 1996),
and a subsample of optically selected quasars from the
Large Bright Quasar Survey (LBQS-open circles; Hewett \etal 1995).
Sources with unknown redshifts are shown at right.} 
\label{colvsz}
\end{figure}

Recently, the
largest and most complete radio-selected quasar sample 
has been compiled by Drinkwater \etal (1997), initially
selected from the Parkes Catalogue.
All sources have flat radio spectra ($\alpha<0.5$)  
and 2.7GHz fluxes $S>0.5$Jy at the epoch of the Parkes survey.
Based on a high identification rate in the optical and near-infrared,
this study may provide crucial evidence for
a substantial population of dust-reddened quasars. 
A broad and flat distribution in optical--to--near-IR colour with
$2\leq B-K\leq10$ as compared to $B-K\sim2-3$ for an
optically selected sample was found (see Fig.~\ref{colvsz}).
It is important to note however that only $\sim10\%$ 
of sources in optical-quasar surveys are radio-loud. 
It is uncertain then whether the distribution of reddening in 
this sample is applicable to samples of radio-quiet QSOs. 
If so, then this sample predicts that existing
optical quasar-samples may be only $\sim20\%$ 
complete (Webster \etal 1995).

\section{Radio-Selected, Optically Reddened Quasars}

Radio-selected quasars have optical characteristics very similar
to those selected optically. Recent studies however
have found one exception:
as seen from Fig.~\ref{colvsz}, the large spread in
optical--near-IR colour suggests that not all quasars can be characterised
by the lower envelope cutoff of $B-K\sim2.5$.
This section gives a brief review of the 
previous work aimed at
disentangling the nature of this ``red'' population.

In previous studies, flat-spectrum radio sources
where no optical counterpart was detected were often 
classified as ``Empty Fields''. Deep optical
and near-infrared 
observations in fact showed that many were very red in optical--near-IR colour.
They were characterised by spectral indices, $\alpha\simgt2$, where
$S_{\nu}\propto\nu^{-\alpha}$ (Spinrad \& Smith, 1975; Rieke, Lebofsky \&
Kinman, 1979;
Bregman \etal 1981; Rieke \etal 1982).
For comparison, optically-selected quasars (see Fig.~\ref{colvsz}) are
characterised by optical--near-IR spectral indices $\sim0.2-0.3$.

Possible causes for the relatively red optical--to--near-IR continua 
have been discussed, 
but no consensus exists. 
Some authors have claimed that since a majority of these 
flat-spectrum radio sources were
also associated with ``blazar-like'' activity, the redness
may be a characteristic of the synchrotron emission mechanism. 
(Cruz-Gonzalez \& Huchra 1984; Bregman \etal 1981;
Rieke \etal 1982).
Other explanations such as optical continuum reddening by dust
somewhere along the line-of-sight have been suggested, but
evidence has been scarce. 
Some of the observational tests
(eg. Ledden \& O'Dell, 1983) 
included searching for 
metal absorption-line systems, X-ray
absorption, or effects of gravitational lensing if the
absorbing material lies in an unrelated intervening system.
In a study by Ledden \& O'Dell (1983) and more recently by 
Kollgaard \etal (1995), 
evidence for soft X-ray absorption 
was presented for a few of the reddest sources. 
The numbers however were 
too low to draw any firm conclusion. 

The difficulty in obtaining a self-consistent picture of
radio-loud optically reddened quasars 
lies in the lack of
sufficient knowledge of the sources themselves.
There appear to be no additional characteristics which
correlate with optical reddening that may provide a hint
about the physics.

More than 40\% of the 323 sources in the complete, flat spectrum
radio sample of Drinkwater \etal (1997) have  
optical--near-IR 
colours $B-K\,\simgt\,5$. This corresponds to optical--near-IR
spectral indices $\alpha\,\simgt\,1.8$, very much redder than
those of optically selected quasars where $\alpha\sim0.2-0.3$. 
At present, it remains unclear whether these sources 
represent one extreme of the
flat spectrum radio population or a fundamentally new class
of AGN. Their faint identifications and red colours suggest 
that significant numbers of
quasars may be severely 
under-represented in current optical surveys.
The last chapters of this thesis shall explore possible
physical mechanisms to explain the anomalous properties 
of sources in this sample.

\newpage
\part{Cosmologically Distributed Dust}
\newpage
\chapter{Diffuse versus Compact Dust Distributions} 

\vspace{1mm}
\leftskip=6cm
 
{\it Who can number the clouds by wisdom\\
Or who can pour out the bottles of heaven,\\
When the dust runneth into a mass,\\
And the clods cleave fast together?\\} 
 
\vspace{1mm}
 
\hfill {\bf --- Job, 38:37}
 
\leftskip=0cm

\section{Introduction}

There are number of studies claiming that 
dust in foreground galaxies has a substantial effect
on the colours and counts of optically selected quasars
(Ostriker \& Heisler, 1984; Heisler \& Ostriker,
1988; Fall \& Pei, 1992 and Wright, 1990).
It is estimated that
more
than 50\% of bright quasars at a redshift of $z\sim3$ may be obscured by dust
in intervening galaxies and hence missing from optical samples. 
These studies
assumed that dust was confined
only within the visible extent of normal massive galaxies.
However, distant populations such as faint field galaxies and quasars
may also be observed through foreground diffuse dust distributions.
Such distributions 
may be associated with galaxy clusters and extended galactic haloes.

A truly diffuse, intergalactic dust distribution is ruled out based
on the counts of quasars and reddening as a function of redshift
(eg. Rudnicki, 1986; Ostriker \& Heisler, 1984).
Such observations indicate that if a significant amount of dust
exists, it must be patchy and diffuse with relatively low
optical depth so that quasars will appear reddened without 
being removed from flux limited samples.

Galaxy clusters provide a likely location for `large-scale' diffusely
distributed dust.
Indirect evidence is provided 
by several studies which reported large deficits of distant quasars
or clusters of galaxies behind nearby clusters (Boyle \etal 1988; 
Romani \& Maoz, 1992
and references therein).
These studies propose that extinction by intracluster dust is the
major cause.
Additional evidence for diffuse dust distributions is
provided by observations of massive local galaxies
where in a few cases, dust haloes extending to scales $\simgt50$kpc
have been confirmed 
(Zaritsky, 1994 and Peletier \etal 1995; see the review in 
section~\ref{localga}).

It is possible that a uniformly distributed dust component
exists in the intergalactic medium (IGM). 
Galactic winds associated with prodigious star formation 
activity at early epochs may have provided a likely source of  
metal enrichment and hence dust 
for the IGM (eg. Nath \& Trentham, 1997).
Observations of metal lines in Ly$\alpha$ absorption systems of
low column density ($N_{HI}\simlt10^{15}{\rm cm^{-2}}$)
indeed suggest that the IGM was enriched to about
$Z\sim0.01 Z_{\odot}$ by redshift $z\sim3$ (Womble \etal 1996;
Songaila \& Cowie, 1996). 
A source of diffuse dust may also have been provided 
by an early generation of pre-galactic stars 
(ie. population III stars) associated with
the formation of galactic haloes (McDowell, 1986).
Later in this thesis, it will be shown that possible reddening from 
uniformly distributed IGM dust 
is limited by observations of radio-selected quasars. 
Since radio-selected quasars should have 
no bias against reddening by dust (see
section~\ref{dustobscu}),
such a 
component must be of sufficiently low optical depth to avoid producing 
a large fraction of `reddened' sources at high redshift (see Fig.~\ref{colvsz}).

In this chapter, we show that a given quantity
of dust has a much greater effect on the 
background universe when diffusely distributed.  
We shall investigate the effects of diffuse dust
from, firstly, the existence of a diffuse component 
in galaxy clusters and secondly,
from a hypothesised uniformly distributed component in the IGM.

This chapter is organised as follows: 
in the next section, we explore the 
dependence of background source counts 
observed through a given mass of dust
on its spatial extent. 
In section~\ref{spatclust}, we investigate 
the spatial distribution of dust optical depth 
through galaxy clusters and its effect on the counts 
and colours of background sources. 
Section~\ref{didu} explores the 
consequences if all dust in the local universe were assumed
uniformly distributed in the IGM.
Further implications are discussed in section~\ref{disdc} and all results are
summarised in section~\ref{conctwo}.
All calculations use a Friedmann cosmology with
$q_{0}=0.5$ and Hubble parameter $h_{50}=1$ where 
$H_{0}=50h_{50}\, \rm km\,s^{-1}\,Mpc^{-1}$.

\section{Compact versus Diffuse Dust Distributions} 
\label{scale}

In this section, we explore 
the dependence of 
obscuration of background sources on the spatial
distribution of a given mass of dust. 
For simplicity, we assume the dust to be associated with
a cylindrical face-on disk with 
uniform spatial dust mass density. 
We quantify the amount of obscuration
by investigating the number of background sources 
behind our absorber that are 
missed from an optical flux-limited sample.

The fraction of sources missing to some luminosity $L$
relative to the case where there is  
no dust extinction is simply $1-N_{obs}(>L)/N_{true}(>L)$, where
$N_{obs}(>L)$
represents the observed number of sources in the presence of dust.
For a uniform dust optical depth $\tau$, $N_{obs}(>L)
\equiv N_{true}(>e^{\tau}L)$.
For simplicity, we assume that background sources are described by a 
cumulative luminosity function that follows a power-law: 
$\Phi\propto L^{-\beta}$, where $\beta$ is the slope.
With this assumption, the fraction of background sources missing
over a given area when viewed through our dusty absorber with 
uniform optical depth $\tau$ is given by
\begin{equation}
\label{fmisu}
f_{miss}(\tau)\,=1\,-\,e^{-\beta\tau}.
\end{equation}
If the `true' number of background sources per unit solid angle is 
$n_{true}$, then the total number of background sources 
lost from a flux limited sample within 
the projected radius $R$ of our absorber can be written:
\begin{equation}
\label{Nlost}
N_{lost}(\tau)\,=\,n_{true}\frac{\pi R^{2}}{D^{2}}(1\,-\,e^{-\beta\tau}),
\end{equation}
where $D$ is the distance of the absorber from us.

To investigate the dependence
of background source counts on the spatial dust distribution, 
we need to first 
determine the dependence of $\tau$ in Eqn.~\ref{Nlost} 
on the spatial extent $R$
for a {\it fixed} mass of dust $M_{d}$.
This can be determined from the individual properties of grains as follows.
The extinction optical depth at a wavelength $\lambda$ through a
slab of dust composed of grains with uniform radius $a$ is defined as
\begin{equation}
\label{tl}
\tau_{\lambda}\,=\,Q_{ext}(\lambda,a)\pi a^{2}n_{d}l_{d},
\end{equation}
where $Q_{ext}$ is the extinction efficiency which depends on the grain
size and dielectric properties, $n_{d}$ is the number density of grains
and $l_{d}$ is the length of the dust column along the line-of-sight.
Assuming our cylindrical absorber (whose axis lies along the line-of-sight) 
has a uniform dust mass density: 
$\rho_{dust}=M_{d}/\pi R^{2}l_{d}$, where $R$ is
its cross-sectional radius, 
we can write,
$n_{d}=\rho_{dust}/\frac{4}{3}\pi a^{3}\rho_{d}$,
where $\rho_{d}$ is the mass density of an individual grain.
We use the extinction efficiency $Q_{ext}$ in the $V$-band as
parameterised by Goudfrooij \etal (1994) for a graphite and silicate
mixture (of equal abundances) 
with mean grain size $a\sim0.1\mu$m characteristic of the
galactic ISM.
The value used is $Q_{ext}(V,0.1\mu m)=1.4$. 
We use a galactic extinction curve to convert
to a $B$-band extinction measure, where typically
$\tau_{B}\sim1.3\tau_{V}$ (eg. Pei, 1992).
Combining these quantities,
we find that the $B$-band optical depth, 
$\tau_{B}$,
through our 
model absorber can be written in terms of its dust mass and 
cross-sectional radius as follows:
\begin{equation}
\label{to}
\tau_{B}\,\simeq\,1\left(\frac{M_{d}}{10^{8}M_{\odot}}\right)
\left(\frac{R}{20{\rm kpc}}\right)^{-2}
\left(\frac{a}{0.1\mu{\rm m}}\right)^{-1}
\left(\frac{\rho_{d}}{2{\rm gm}\,{\rm cm}^{-3}}\right)^{-1},
\end{equation}
where we have scaled to a dust mass and radius typical of local 
massive spirals and ellipticals (eg. Zaritsky, 1994).
This measure is consistent with mean optical depths
derived by other means (eg. Giovanelli \etal 1994 and references therein). 

From Eqn.~\ref{to}, we see that the dust optical depth through
our model absorber for a {\it fixed} dust mass varies in terms of 
its cross-sectional radius $R$ as 
$\tau\propto 1/R^{2}$. 
For the nominal dust parameters in Eqn.~\ref{to}, 
the number of sources missed behind our model absorber (Eqn.~\ref{Nlost}) 
can be written
\begin{equation}
\label{NlostR}
N_{lost}(<R)\,=\,N_{true}(<20{\rm kpc})\,R_{20}^{2}(1-e^{-\beta R_{20}^{-2}}),
\end{equation}
where $R_{20}=R/20{\rm kpc}$ and 
\begin{equation}
N_{true}(<20{\rm kpc})\,\equiv\,n_{true}\frac{\pi(20{\rm kpc})^{2}}{D^{2}}
\label{Ntrues}
\end{equation}
is the `true' number of background sources falling
within the projected scale radius $R=20$kpc.

From the functional forms of Eqns.~\ref{Nlost} and~\ref{NlostR}, there
are two limiting cases:
\begin{enumerate}
\item
For optical depths $\tau\gg1/\beta$, the factor 
$1-e^{-\beta\tau}$ in Eqn.~\ref{Nlost} is of order unity. 
This corresponds to values of $R$ such that $0<R_{20}\ll\beta^{1/2}$
for the nominal parameters in Eqn.~\ref{to}.
For values of $R$ in this range, we have $N_{lost}\propto R^{2}$ and the 
obscuration of background sources will depend most strongly on $R$.

\item
For $\tau\ll 1/\beta$ or equivalently $R_{20}\gg\beta^{1/2}$,
$N_{lost}$ will approach a constant limiting value,
independent of the dust extent $R$. From Eqn.~\ref{NlostR}, this
limiting value can be shown to be 
$N_{lost}(<R)\,=\,N_{true}(<20{\rm kpc})\beta$.
\end{enumerate}

\begin{figure}
\vspace{-3.6in}
\plotonesmall{1}{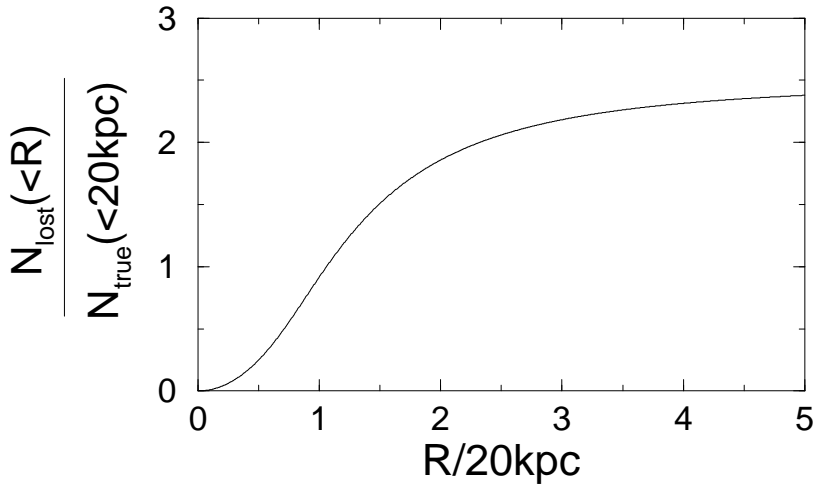}
\vspace{-2.4in}
\caption[Dependence of background source deficit on radial dust extent]
{The number of sources missing from an optical flux limited sample
behind a face-on dusty disk
with {\it fixed}
mass $M_{dust}=10^{8}M_{\odot}$ as a function of its radial extent $R$.
This number is normalised against the `true' number of background sources
(in the absence of the absorber) that fall within the solid angle
$\pi(20{\rm kpc})^{2}/D^{2}$ subtended by our nominal
radius of 20kpc at distance $D$. See Eqn.~\ref{NlostR}.}
\label{NlostRgal}
\end{figure}

As a simple illustration, we show in Fig.~\ref{NlostRgal} the
dependence of the number of background sources 
missing behind our model dust absorber on
$R$, for a fixed dust mass of
$10^{8}M_{\odot}$ as defined by Eqn.~\ref{NlostR}.
We assumed a cumulative luminosity function slope of $\beta=2.5$,
typical of that for luminous galaxies and quasars. 
From the above discussion, we see that when 
$R_{20}\simgt\beta^{1/2}$, ie. $R\simgt30{\rm kpc}$
(or when $\tau\simlt 1/\beta=0.4$), 
the obscuration
will start to 
approach its maximum value and remain approximately constant as 
$R\rightarrow\infty$. 

We conclude
that when dust becomes diffuse and extended on a scale such that
the mean optical depth $\tau$ through the distribution
satisfies $\tau<1/\beta$, where $\beta$ is the cumulative luminosity
function slope of background sources, obscuration 
will start to be important and is maximised for $\tau\ll1/\beta$.
The characteristic spatial scale at which this occurs  
will depend on the dust mass through Eqn.~\ref{to}.
For the typical grain values in Eqn.~\ref{to},
this characteristic radius
is given by
\begin{equation}
R\,\simeq\,31\left(\frac{\beta}{2.5}\right)^{1/2} 
\left(\frac{M_{d}}{10^{8}M_{\odot}}
\right)^{1/2}{\rm kpc}. 
\label{radiuscale}
\end{equation}
The simple model in Fig.~\ref{NlostRgal} shows that the
obscuration of background sources due 
to a normal foreground galaxy will be most effective if dust
is distributed over a region a few times its optical radius.
This prediction may be difficult to confirm observationally
due to possible contamination from light in the galactic absorber.
In the following
sections, we explore
two examples of possible diffuse 
dust distributions on relatively large scales that can be explored 
observationally.

\section{Diffuse Dust in Galaxy Clusters}
\label{spatclust}

A number of studies
have attributed the existence of large deficits of 
background sources behind nearby galaxy clusters as 
due to extinction by dust. 
Bogart \& Wagner (1973) found that
distant rich Abell clusters were anticorrelated on the sky with nearby ones.
They argued for a mean extinction of $A_{V}\approx0.4$mag
extending to $\sim2.5$ times the optical radii of the nearby clusters.
Boyle, Fong \& Shanks (1988)
however claimed a 
$\sim30\%$ deficit of background quasars within $4'$ of clusters
consisting of tens of galaxies.
These authors
attribute this to an extinction $A_{V}\approx0.15$mag, and
deduce a dust mass of $\sim10^{10}M_{\odot}$ within 0.5Mpc 
of the clusters. 
Romani \& Maoz (1992) found that optically-selected quasars
from the V\'{e}ron-Cetty \& V\'{e}ron (1989) catalogue avoid
rich foreground Abell clusters.
They also found deficits of $\sim30\%$
out to radii
$\sim5'$ from the clusters, and postulate a
mean extinction, $A_{V}\approx0.4$mag.

The numbers of background sources behind clusters is also expected to
be modified by 
gravitational lensing (GL) by the cluster potential.
Depending on the intrinsic luminosity function of the background population,
and the limiting magnitude to which the sources are detected, GL 
can cause either an enhancement or a deficit in the number of 
background sources. 
The GL effect has been used to explain various 
reports of overdensities of both optically and radio-selected
quasars behind foreground clusters
(Bartelmann \& Schneider, 1993; Bartelmann \etal 1994;
Rodrigues-Williams \& Hogan, 1994;
Seitz \& Schneider, 1995).  
The reported overdensities for {\it optically-selected} QSOs 
are contrary to the studies above where 
anticorrelations with foreground 
clusters are found.
These overdensities however are claimed on angular scales
$\sim10'-30'$ from the cluster centers, considerably
larger than scales on which most of the underdensities have been claimed,
which are of order a few arcminutes. 
One interpretation is that
dust obscuration bias may be greater towards cluster centers due 
to the presence of greater quantities of dust.
On the other hand, the reported anticorrelations on small
angular scales can perhaps be explained by the optically 
crowded fields, where QSO identification may be difficult.
At present, the effects of clusters on background source counts still
remains controversial. 

More direct evidence for the existence of intracluster dust
was provided by Hu \etal (1985) and 
Hu (1992) who compared the Ly-$\alpha$ flux from
emission line systems in ``cooling flow'' clusters with
Balmer line fluxes at optical wavelengths.  
Extinctions of
$A_{V}\sim0.2-1$mag towards the cluster centers
were found, in good agreement with estimates
from the quasar deficits above.
Intracluster dust has been predicted to give rise to
detectable diffuse IR emission (eg. Dwek \etal 1990).
The most extensive search was conducted by Wise \etal (1993), who
claimed to have detected excess diffuse 60-100$\mu$m emission at the 2$\sigma$
level from a number of rich Abell clusters.
They derived dust temperatures in the range 24-34 K and
dust masses $\sim10^{10}M_{\odot}$ within radii of $\sim1$Mpc.
Recently,
Allen (1995) detected strong X-ray absorption and optical 
reddening in ellipticals situated at the centers of rich cooling flow
clusters, providing strong evidence for dust. 
These studies indicate that intracluster dust is
certainly present, however, 
the magnitude of its effect on producing background source
deficits remains a controversial issue. 

In this section, we
give some predictions that
may be used to further constrain cluster dust properties, or
help determine the dominant mechanism (ie. GL or extinction) by which clusters
affect background observations. 

\subsection{Spatial Distribution of Cluster Dust?}

X-ray spectral measurements show the presence of hot, metal enriched
gas in rich galaxy clusters with $\sim0.5-1$ solar metallicity.
This gas is believed to be both
of galactic and primordial
origin (ie. pre-existing IGM gas), with the bulk of metals being 
ejected from cluster galaxies (see Sarazin, 1986 for a review).
Ejection from galaxies may occur abruptly
through collisions between the cluster galaxies, `sudden' ram-pressure
ablation, or through continuous ram-pressure stripping by
intracluster gas (eg. Takeda \etal 1984). 
The lack of significant amounts of dust (relative to what should have
been produced by stellar evolution) and interstellar gas in cluster
ellipticals provides evidence for a mass loss process.
On the other hand, in ellipticals that avoid dense cluster
environments, significant quantities 
of neutral hydrogen, molecular gas and dust 
have been detected (eg. Lees \etal 1991).

If the dust-to-gas ratio of intracluster gas 
in rich clusters were similar to that of the Milky Way, then a radial
gas column density of typically $\simgt10^{22}{\rm cm}^{-2}$
with metallicity $Z=0.5Z_{\odot}$ would produce an extinction 
$A_{V}\simgt4$mag.
This however is much greater than the value observed. 
The likely reason for the deficiency of dust in the intracluster medium is its
destruction by thermal sputtering in the hot gas, a process which operates on
timescales $t_{sputt}\simlt10^{8}n_{-3}a_{0.1}$ yr, where
$n_{-3}=n_{H}/10^{-3}{\rm cm}^{-3}$ is the gas density and
$a_{0.1}=a/0.1\mu{\rm m}$ the grain radius (Draine \& Salpeter, 1979). 
Dust injection timescales from galaxies
is typically of order a Hubble time 
(eg. Takeda \etal 1984) and 
hence, grains are effectively destroyed, with only the most recently
injected still surviving and providing possibly some measurable 
extinction.

The spatial distribution in dust mass density remains a 
major uncertainty. A number of authors have shown that
under a steady state of continuous injection from
cluster galaxies, destruction by thermal sputtering
at a constant rate, and assuming instantaneous mixing with the hot gas,
the resulting mass density in dust will be of order 
\begin{equation}
\label{rhodust}
\rho_{dust}\,\sim\,
10^{-31}\left(\frac{a}{0.1\mu{\rm m}}\right)
\left(\frac{Z_{d}}{0.01}\right)h_{50}\,\,{\rm gm}\,{\rm cm}^{-3},
\end{equation}
(eg. Dwek \etal 1990)
where $Z_{d}$ is the injected dust-to-gas mass ratio, assumed to be
equal to the mean value of the galactic ISM, $Z_{d}\simeq0.01$ (Pei, 1992). 
According to this simple model,
the dust mass density is independent of gas density and position in the
cluster.
If we relax the assumption of instantaneous mixing of dust with
the hot gas however, so that the spatial distribution of gas is
different from that of the injected dust, the radial distribution
of dust can significantly differ from uniformity throughout a cluster.
Such a non-uniform spatial dust distribution
may be found in clusters exhibiting cooling flows. 
If, as suggested by Fabian \etal (1991), most of the cooled gas remains
cold and becomes molecular in cluster cores, then
a relatively large amount of dust may also form, resulting in a
dust distribution which peaks within the central regions. 

We explore the radial dependence of extinction
optical depth through a cluster,
and the expected deficit in background sources
by assuming that dust is diffusely distributed and follows a
spatial density distribution:
\begin{equation}
\label{rhodR}
\rho_{dust}(R)\,=\,\rho(0)\left[1 - \left(\frac{R}{R_c}\right)^2\right]^{-n},
\end{equation}
where $R_c$ is a characteristic radius which we fix and 
$n$ is our free parameter. 
Eqn.~\ref{rhodR} with $n=3/2$ is the usual King profile which
with $R_c\simeq0.25$Mpc, represents a good approximation to 
the galaxy distribution in clusters.
Thus for simplicity, we keep $R_c$ fixed at $R_c=0.25$Mpc and vary $n$.
To bracket the range of possibilities in the distribution of 
intracluster dust, we shall consider the range $0\,<\,n\,<\,3/2$. 
$n=0$ corresponds to the simple case where 
$\rho_{dust}(R)=\rho(0)=$constant, which may describe a situation
where injection of dust is balanced by its destruction by hot gas 
as discussed above.
The value $n=3/2$ assumes that dust
follows the galaxy distribution. This profile may arise 
if grain destruction by a similar distribution of 
hot gas were entirely absent. 

\subsection{Spatial Distribution of Dust Optical Depth 
and Background Source Deficits}

To model the spatial distribution of optical depth,
we assume that intracluster dust is distributed within a 
sphere of radius $R_{max}$. 
The central dust mass density $\rho(0)$ in Eqn.~\ref{rhodR}
is fixed by assuming
values for the total dust mass $M_{dust}$ and $R_{max}$ such that 
\begin{equation}
\label{Mdust2}
M_{dust}=\int^{R_{max}}_0{4\pi R^{2}\rho(R)\,dR}.
\end{equation}
We assume a cluster dust radius of $R_{max}=1$Mpc, which represents
a radius containing $\simgt90\%$ of the virial mass 
of a typical dense cluster characterised by 
galactic velocity dispersion $\sigma_{V}\sim800{\rm km s}^{-1}$ 
(Sarazin, 1986 and references therein).   
We assume 
that the total dust mass within $R_{max}=1$Mpc 
is $M_{dust}=10^{10}M_{\odot}$. This value is
consistent with that derived from extinction measures by 
Hu, Cowie \& Wang (1985), IR emission detections by Wise \etal (1993) and
theoretical estimates of 
the mean intracluster dust density as given by Eqn.~\ref{rhodust}. 

Using Eqn.~\ref{tl}, the $B$-band optical depth through our
spherical intracluster dust distribution at some projected distance
$r$ from its center can be written: 
\begin{equation}
\label{tint}
\tau_{B}(r)\,=\,\frac{3Q_{ext}}{4a\rho_{d}}\int^{2(R_{max}^{2}-r^{2})^{1/2}}
_{0}{\rho_{dust}\left([r^{2}+R'^{2}]^{1/2}\right)\,dR'}, 
\end{equation}
where $\rho_{dust}(R)$ is our assumed radial density distribution 
(Eqn.~\ref{rhodR}) and $\rho_{d}$ the mass density of an individual dust grain. 
For a uniform dust density ($\rho_{dust}(R)=\rho_{dust}(0)=$constant),
and our assumed values of $R_{max}$ and $M_{dust}$ given above,
the radial dependence in dust optical depth can be written:
\begin{equation}  
\label{toclr}
\tau_{B}(r)\,=\,\tau_{B}(0)\left[1-
\left(\frac{r}{R_{max}}\right)^{2}\right]^{1/2}, 
\end{equation}
where $\tau_{B}(0)$ is the optical depth through the center of our cluster,
which with 
grain properties characteristic of the galactic ISM, will scale as
\begin{equation}
\label{tocl}
\tau_{B}(0)\,\simeq\,0.06\left(\frac{M_{dust}}{10^{10}M_{\odot}}\right)
\left(\frac{R_{max}}{{\rm Mpc}}\right)^{-2}
\left(\frac{a}{0.1\mu{\rm m}}\right)^{-1}
\left(\frac{\rho_{d}}{2{\rm gm}\,{\rm cm}^{-3}}\right)^{-1}.
\end{equation}
This value is about a factor three times lower than estimates of the
mean extinction derived from the deficit of QSOs
behind foreground clusters (eg. Boyle \etal 1988), and that implied by the
Balmer decrements of Hu \etal (1985).
For a fixed dust mass of $10^{10}M_{\odot}$ however, we can achieve
larger values for the central optical depth by steepening the 
radial dust-density distribution profile, determined by the slope $n$
in Eqn.~\ref{rhodR}. 

\begin{figure}
\vspace{-3.5in}
\plotonesmall{1}{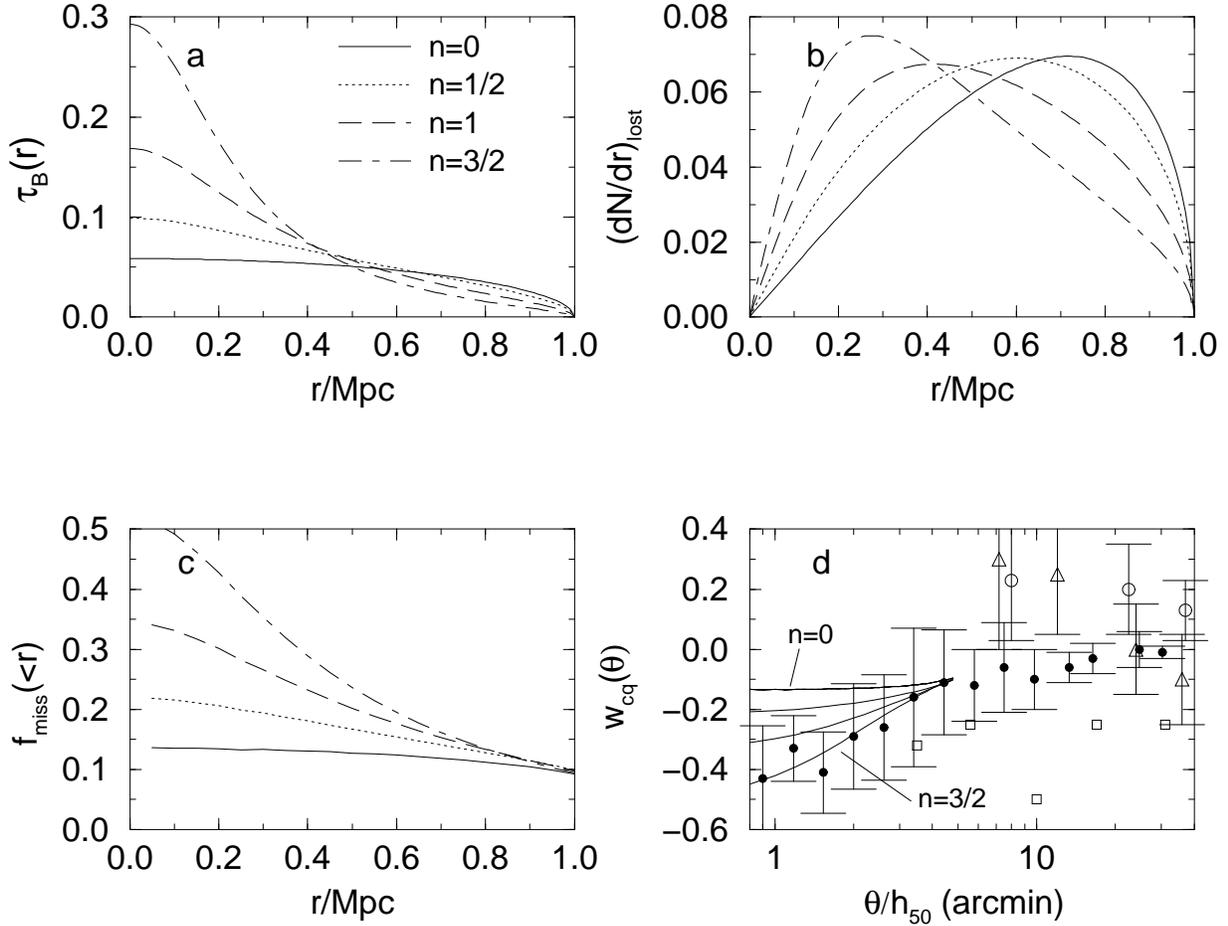}
\vspace{-1in}
\caption[Extinction by a `rich'
galactic cluster]
{{\bf a.} Optical depth as a function of projected cluster radius for
various dust density distributions as parameterised by Eqn.~\ref{rhodR}.
{\bf b.} Differential number of background sources lost from a flux-limited 
sample (arbitrary scale). 
{\bf c.} Total fraction of background sources missing to some $r$ 
and {\bf d.} Cluster-QSO two-point angular correlation function,
filled circles: Boyle \etal (1988), squares: Romain \& Maoz (1992),
open circles: Rodrigues-Williams \& Hogan (1994), triangles: 
Rodrigues-Williams \& Hawkins (1995).} 
\label{clusters}
\end{figure}

Fig.~\ref{clusters}a shows the optical depth
as a function of projected cluster radius
for the cases $n=0$, 0.5, 1, and 1.5.
The case $n=0.5$ approximately corresponds to
the model of Dwek \etal (1990), which included effects of mild
sputtering by hot gas in order to fit for the observed IR emission
from the Coma cluster. 
As shown, the case $n=0$ ($\rho_{dust}(R)=$constant) predicts that 
the dust optical depth should be almost 
independent of projected radius $r$.
Within all projected radii, the optical depths predicted by 
our diffuse dust model lie in the range $0<\tau_{B}<0.3$.
Turning back to the discussion of section~\ref{scale} where we show that 
background obscuration by diffuse dust reaches its maximum 
for $\tau_{B}<1/\beta$, these optical depths satisfy this
condition for $\beta\simlt2.5$, typical of luminous 
background galaxies and quasars. 

We now explore the effects of these models 
on background source counts as a function of projected cluster radius.
We first give an estimate of the projected radius 
at which the numbers of background sources lost from a flux-limited
sample is expected to be a maximum.
This is determined by investigating the dependence in the differential 
number of
sources missing, $dN_{lost}$, within an interval ($r,r+dr$) as a 
function of projected radius $r$. 
From Eqn.~\ref{Nlost}, this 
differential number will scale as 
\begin{equation}
\label{dNlost}
dN_{lost}(r)\,\propto\,r\,dr\left[1 - e^{-\beta\tau(r)}\right],
\end{equation} 
where $\tau(r)$ is given by Eqn.~\ref{tint}.
Fig.~\ref{clusters}b plots $dN_{lost}/dr$ as a function of $r$ for 
our various models, where we have assumed $\beta=2.5$.
Thus from observations, 
an identification of the projected radius at which the 
background source deficit peaks can be used to  
constrain the spatial distribution of intracluster dust. 

The cumulative fraction of background 
sources missing within a projected cluster
radius is given by 
\begin{equation}
\label{Nlostltr}
f_{miss}(<r)\,=\,\frac{N_{lost}(<r)}{N_{true}(<r)}\,=\,
\frac{1}{\pi r^{2}}\int^{r}_{0}{2\pi r'\left[1-e^{-\beta\tau(r')}\right]\,dr'}. 
\end{equation}
This fraction is shown in Fig.~\ref{clusters}c. 
As expected, the $n=3/2$ model which contains the largest amount of dust
within the inner few hundred kiloparsecs predicts the strongest
trend with $r$, while the opposite is predicted if the dust density 
were completely uniform. 
These predictions can be compared with a number of existing studies of the
observed two-point angular correlation function between clusters and 
optically-selected QSOs. 
This function is usually defined as 
\begin{equation}
\label{wtheta} 
w_{cq}(\theta)\,=\,\frac{\langle N_{obs}(<\theta)\rangle} 
{\langle N_{ran}(<\theta)\rangle}\,-\,1, 
\end{equation}
where $\langle N_{obs}(<\theta)\rangle$ is the average number
of observed cluster-QSO pairs within an angular radius $\theta$ and
$\langle N_{ran}(<\theta)\rangle$ is that expected in a random
distribution.
For our purposes, $\langle N_{ran}(<\theta)\rangle$ can be
replaced by the ``true'' number of cluster-QSO pairs expected in the
absence of dust,
and hence, we can re-write Eqn.~\ref{wtheta} as 
\begin{equation}
\label{wthetalost}
w_{cq}(\theta)\,\equiv\,-\frac{\langle N_{lost}(<\theta)\rangle}
{\langle N_{true}(<\theta)\rangle}\,=\,-f_{miss}(<\theta). 
\end{equation}

We compare our models with a number of studies of $w_{cq}(\theta)$
for optically selected QSOs in Fig.~\ref{clusters}d.
These studies considerably differ from each other in the selection
of the QSO and cluster samples, and as seen, both 
anticorrelations and correlations on different angular scales 
are found.
The former have been interpreted in terms of extinction by intracluster
dust, while the latter with the GL phenomenon.
In most cases, the reported overdensities are too large to be 
consistent with GL models given our current knowledge
of cluster masses and QSO distributions.

It is interesting to note that the studies which have reached the
smallest angular scales ($\simlt5'$) are also those in which anticorrelations
between QSOs and foreground clusters have been reported. 
This can be understood in terms of a larger dust concentration
and hence extinction towards cluster centers.
These studies however may not be free of selection effects,
such as in the detection of QSOs from the visual inspection of
objective prism plates. 
From a cross-correlation analysis of galactic stars with their cluster
sample however, 
Boyle \etal (1988) found that such selection effects are minimal.

The maximum dust radial extent assumed in our models, 
$R_{max}=1$Mpc, corresponds to angular scales $\sim5'$ at the mean 
redshift of the clusters ($\langle z_{c}\rangle\sim0.15$) used in
these studies. 
Thus, as shown in Fig.~\ref{clusters}d, 
our model predictions only extend to $\sim5'$.
As shown in this figure, the $n=3/2$ model which corresponds to the
case where the dust density is assumed to follow the galaxy distribution,
provides the best fit to the Boyle \etal (1988) data. 
We must note that this is the only existing study performed to 
angular scales $\sim1'$ with which we can compare our models. 
Further studies to such scales are necessary to 
confirm the Boyle \etal result, and/or 
provide a handle on any selection effects.  

\subsection{Summary}

To summarise, we have shown that for a plausible value of the
dust mass in a typical rich galaxy cluster, 
obscuration of background sources will be most
effective if dust is diffusely distributed on scales $\sim1$Mpc.
This conclusion is based on our predicted optical depths
($\tau_{B}<0.3$) satisfying our condition for 
`maximum' obscuration: $\tau_{B}<1/\beta$ 
(see section~\ref{scale}), where typically
$\beta\simlt2.5$ for luminous background galaxies and QSOs. 

We have explored the spatial distribution in dust optical depth
and background source deficits expected through a 
typical rich cluster by assuming different 
radial dust density profiles. 
These predictions can be used to constrain cluster dust properties.
A dust density distribution with $n=3/2$ (Eqn.~\ref{rhodR}) appears
to best satisfy the `small scale' cluster-QSO angular correlation study
of Boyle \etal (1988). 
 
\section{Diffuse Intergalactic Dust?}
\label{didu}

There have been a number of studies claiming that
the bulk of metals in the local universe had already formed by  
$z\sim1$ (eg. Lilly \& Cowie, 1987; White \& Frenk, 1991; Fall \& Pei, 1995). 
Similarly, models of dust evolution in the galaxy show that the
bulk of its dust content was formed in the first few billion years
(Wang, 1991).
These studies suggest that 
the global star formation rate peaked at epochs $z\simgt2$ 
when the bulk of 
galaxies were believed to have formed.
Supernova-driven winds at early epochs
may thus have provided 
an effective mechanism by which
chemically enriched material and dust were dispersed into the IGM.
As modelled by Babul \& Rees (1992), such a mechanism
is postulated to be crucial in the evolution of the `faint blue'
galaxy population observed to magnitudes $B\sim28$.
Nath \& Trentham (1997) also show that this mechanism could explain 
the recent detection of
metallicities $Z\sim0.01Z_{\odot}$
in low density Ly$\alpha$ absorption systems at $z\sim3$. 
Another source of diffuse IGM dust may have been provided from 
an epoch of population III star formation associated with the
formation of galactic haloes
(eg. McDowell, 1986).

What are the effects expected on background sources if all dust formed
to the present day was
completely uniform and diffuse throughout the IGM?
In this section, we show that 
such a component will have a low optical depth and have an
insignificant effect 
on the colours of background sources, but will be high enough
to significantly bias their number counts 
in the optical.

\subsection{Comoving Dust Mass Density}

To explore the effects of a diffuse intergalactic dust component, we need to 
assume a value for the mean mass density in dust in the local universe.
This density must not exceed the total mass density in heavy metals at the
present epoch. 
An upper bound for the local mass density in metals (hence dust)
can be derived from the assumption that the {\it mean} metallicity of the
local universe is typically: $Z\sim\Omega_{metals}/\Omega_{gas}\sim0.01$ 
(ie. the ratio of elements heavier than helium to total gas mass), 
as found from galactic
chemical evolution models (eg. Tinsley, 1976) and abundance observations
(Grevesse \& Anders, 1988). 
Combining this with the upper bound in the baryon density predicted from
big-bang nucleosynthesis (Olive \etal 1990) where 
$\Omega_{gas}\simlt\Omega_{baryon}<0.06h_{50}^{-2}$, it is apparent that
\begin{equation}
\label{upperlim}
\Omega_{metals}(z=0)\,<\,6\times10^{-4}h_{50}^{-2}.
\end{equation}

Let us now compute the total mass density in dust used in
previous studies that modelled the effects of dust in 
individual galaxies on background quasars.
Both Heisler \& Ostriker (1988) and Fall \& Pei (1993) modelled
these effects by assuming that dust in each galaxy
was distributed as an exponential disk with scale radius 
$r_{0}\simeq30$kpc and central face-on optical depth, $\tau_{B}=0.5$. 
The comoving mean mass density in dust (relative to the critical density) 
in these studies,
given a comoving galaxy number density
$n_{0}=0.002h_{50}^{3}{\rm Mpc}^{-3}$, can be shown to be
\begin{equation}
\label{HOdust}
\Omega_{dust0}\,\simeq\,7.3\times10^{-6}h_{50}
\left(\frac{n_{0}}{0.002{\rm Mpc}^{-3}}\right) 
\left(\frac{r_{0}}{30{\rm kpc}}\right)^{2}\left(\frac{\tau_{B}}{0.5}\right) 
\end{equation}
(see section~\ref{ev}).
This is consistent with the constraint in Eqn.~\ref{upperlim}.
Thus, as a working measure, we assume the comoving mass
density defined by Eqn.~\ref{HOdust} in the calculation that follows. 

\subsection{Obscuration by Diffuse Intergalactic Dust}
 
If the dust mass density given by Eqn.~\ref{HOdust} is assumed 
uniformly distributed and constant 
on comoving scales to some redshift,
the $B$-band optical depth through a dust sheet of width d$l$ at redshift $z$
in an observer's frame can be written (see Eqn.~\ref{tl})
\begin{equation}
\label{dtau}
d\tau_{B}\,\simeq\,7.2\times10^{-6}h_{50} 
\left(\frac{\Omega_{dust0}}{7.3\times10^{-6}}\right)
\left(\frac{a}{0.1\mu{\rm m}}\right)^{-1}
\left(\frac{\rho_{d}}{2{\rm gm}\,{\rm cm}^{-3}}\right)^{-1} 
\left(\frac{{\rm d}l}{{\rm Mpc}}\right)(1+z),
\end{equation}
where we have assumed dust properties characteristic of the galactic ISM.
The factor $(1+z)$ is due to our assumption of a $1/\lambda$
dependence for the dust extinction law. This arises
from the fact that light received in the $B$-band 
corresponds to light of wavelength $\lambda_{B}/(1+z)$ at redshift $z$,
which consequently suffers greater extinction.
With $dl/dz=6000h_{50}^{-1}(1+z)^{-5/2}$Mpc (for a $q_{0}=0.5$ and 
$\Lambda=0$ cosmology), the total mean optical depth
to some redshift in an observer's $B$-band will scale as
\begin{equation}
\label{tautotone} 
\tau_{B}(z)\,\simeq\,0.1\left(\frac{\Omega_{dust0}}{7.3\times10^{-6}}\right)
\left(\frac{a}{0.1\mu{\rm m}}\right)^{-1}
\left(\frac{\rho_{d}}{2{\rm gm}\,{\rm cm}^{-3}}\right)^{-1}
\left[1-(1+z)^{-1/2}\right].
\end{equation}
This represents the total optical depth if all dust in the 
intervening galaxy model of Heisler \& Ostriker (1988) were
assumed uniformly distributed throughout the universe.

Assuming dust is uniformly distributed to $z=2$, the {\it observed} 
$B$-band optical
depth from Eqn.~\ref{tautotone} will be of order
\begin{equation}
\label{tauz2}
\tau_{B}({\rm U})(z=2)\,\simeq\,0.04.
\end{equation}
Using a galactic extinction law (Pei, 1992), 
this corresponds to an extinction in $B-R$ colour of 
$E_{B-R}\sim0.02$mag.
Thus, if background faint field galaxies and QSOs are observed through
a uniform intergalactic dust distribution, their observed colours are
not expected to be significantly affected. 
We now show however that the numbers of sources missing at such redshifts 
could be significantly greater than that claimed by previous studies which 
assume all dust to be associated with
massive galaxies alone.

If dust to some distance $D$ covers an area of sky $A$
and hence, has covering factor $C_{d}=A/4\pi D^{2}$,
the number of
background sources lost from a flux-limited sample can be estimated from
Eqn.~\ref{Nlost}. 
In general, the number of background sources at some redshift
lost from
an area of sky with dust covering factor $C_{d}$ will scale as: 
\begin{equation}
\label{Nlostgen}
N_{lost}(\tau,z)\,\propto\,C_{d}f_{miss}(\tau,z), 
\end{equation}
where $f_{miss}\equiv 1-e^{-\beta\tau_{B}(z)}$
is the fraction of sources missing per unit area. 
For a completely uniform dust distribution, 
$C_{d}({\rm U})=1$, and to 
redshift $z=2$, $f_{miss}\simeq10\%$ for $\beta=2.5$.  

If dust were confined to individual galaxies along the
line-of-sight however, their covering factor, 
assuming they follow a  
Poisson distribution is typically
$C_{d}\simeq\bar{N}_{z}\exp{(-\bar{N}_{z})}$,
where $\bar{N}_{z}$ is the mean number of absorber intersections 
to redshift $z$:
\begin{equation}
\label{Nz1}
\bar{N}_{z}\,\simeq\,0.01h_{50}^{2}
\left(\frac{n_{0}}{0.002{\rm Mpc}^{-3}}\right)
\left(\frac{r_{0}}{30{\rm kpc}}\right)^{2}\left[(1+z)^{1.5} - 1\right]
\end{equation}
(see Appendix C).
We have scaled to the nominal parameters assumed in the
intervening galaxy model of Heisler \& Ostriker (1988) (hereafter HO). 
In this model, we find a covering
factor of only $C_{d}({\rm HO})\simeq0.04$ to $z=2$.
We can estimate the mean effective optical depth {\it observed} in the
$B$-band through an individual absorber to $z\simeq2$ in the HO model
by using the formalism of section~\ref{scale}. 
For a {\it fixed} mass of dust, Eqn.~\ref{to} implies that the
product of the area (or covering factor) and optical depth of 
a dust distribution:
$\tau\times C_{d}$, is a constant, depending 
on grain properties and dust mass alone.
Using this relation,
the {\it observed} effective absorber optical depth to $z\simeq2$ 
in the HO model, $\tau_{B}({\rm HO})$, 
can be estimated by scaling from our values 
of $\tau_{B}({\rm U})$ and $C_{d}({\rm U})$ 
above for uniformly distributed dust:
\begin{equation}
\tau_{B}({\rm HO})\,\simeq\,\tau_{B}({\rm U})C_{d}({\rm U})/C_{d}({\rm HO})
\,=\,\frac{0.04\times1}{0.04}\,=1.
\label{HOopt}
\end{equation}
Using this value, the fraction of background sources
missed by obscuration from an individual absorber is `effectively' 
$f_{miss}=1-\exp(-2.5\times1)\simeq91\%$. 
Combining these results, we find using Eqn.~\ref{Nlostgen} that
the number of sources missing at $z\simeq2$ due to a uniform
foreground dust distribution to be greater by a factor of 
$\frac{1\times0.1}{0.04\times0.91}\,\sim\,3$
than that predicted by Heisler \& Ostriker (1988). 

We must note that this estimate makes no                      
allowance for possible evolution in dust content. 
Effects of foreground diffuse dust on source counts           
at $z>2$ may be significantly reduced 
if appreciable evolution has occured.
Effects of models where the dust content evolves will be considered in
Chapter 4.

\subsection{Summary}

We conclude that the existence of a significant amount of
diffusely distributed dust (eg. with mass density 
on comoving scales of order that 
observed in local galaxies) can enhance the
number of background sources missing in optical samples. 
Due to its relatively large covering factor, diffuse dust predicts 
a reduction in optical counts at $z>2$ about three 
times greater than
that claimed by previous studies.

The colours of background sources
are not expected to be significantly affected.
This implies that the use of background populations 
to measure a diffuse IGM dust component will be extremely difficult. 

\section{Discussion}
\label{disdc}

In this section, we discuss some further uncertainties 
and implications regarding the existence of diffuse dust in the universe.  

First, the effects of intracluster dust on background sources
critically depends 
on the amount of dust present, and its spatial distribution. 
Regardless of the mechanism by which grains are
injected into the intracluster medium from galaxies,
it is possible that a 
significant fraction are destroyed in the injection
process.
Significant amounts of hot gas are also believed to exist in the
ISM of cluster ellipticals (eg. Forman \etal 1985). 
This gas
is expected to destroy grains on timescales
$\simlt10^{8}$yr (Draine \& Salpeter, 1979), 
much shorter than injection timescales.
Such destruction mechanisms can thus prevent the
formation of significant quantities
of dust.

It is possible that
the spatial distribution of intracluster dust is not `diffuse'
and uniformly distributed, but inhomogeneous.
For example, Fabian \etal (1991) propose that if most of the cooled
gas resulting from cluster cooling flows remains cold and becomes
molecular, then this may provide suitable conditions for 
large amounts of dust to form.
A clumpy dust distribution that follows cooling flow filaments
may result, reducing the effective dust covering factor
and hence background source deficit. 
These issues need to be addressed
before attributing such deficits
to extinction by dust.

The existence of a smooth IGM dust component also remains
a major uncertainty.
Due to their deep gravitational potential, Margolis \& Schramm (1977) 
showed that it is unlikely for supernovae-driven winds to 
expel significant quantities 
of dust from a massive galaxy to large scales.
For low mass galaxies however (eg. dwarfs), Babul \& Rees
(1992) show that this mechanism can be effective. 
Such systems are believed to comprise a majority of the
`faint-blue population' which show an excess $\sim20-30$ times that 
predicted from non-evolving galaxy models for $B>24$ (eg. Tyson, 1988). 
Simulations 
based on star-formation rates that assume yields in metallicity from
local observations, predict that the
amount of metals (and hence dust, assuming a fixed fraction of
metals condense into grains at a constant rate) produced from such a population 
will be smaller than local estimates by an order of magnitude
(eg. White \& Frenk, 1991). 
If the only source of IGM dust was from these `low-mass' galaxies, 
then the total optical depth to $z\simgt2$ would
be insignificant, and effects
on the background universe would be minimal. 

\section{Conclusions}
\label{conctwo}

In this chapter, we have shown that dust is more effective
at obscuring background sources
when diffuse or extended.
We find that obscuration of
background sources by a given dust distribution with
optical depth $\tau_{B}$ will be most effective 
when $\tau_{B}<1/\beta$, where $\beta$ is the 
cumulative luminosity function slope of the sources. 

We have explored the effects of diffuse dust from, firstly, galaxy clusters 
and secondly, from a hypothesised uniform IGM component.
By assuming different radial dust density profiles
in a typical rich cluster,
we have predicted the optical depth and background source deficit
as a function of projected cluster radius. 
These predictions can be compared with future observations 
to constrain the properties of intracluster dust.
Our predicted optical depth measures ($\tau_{B}\simlt0.3$) 
satisfy the above criterion ($\tau_{B}<1/\beta$) for background luminous
QSOs and
galaxies.  
Existing studies claiming anticorrelations in the
distribution of QSOs with foreground clusters down to scales $\sim1'$
are consistent with a dust density profile that follows
the galaxy distribution. 

As a further illustration,
we have explored the effects of a diffuse IGM dust component
with cosmic mass density equal to that observed 
in local galaxies. 
Assuming this density is constant on
comoving scales to $z=2$, 
we find
a deficit in background sources about three 
times greater than that predicted 
assuming dust in normal galaxies alone.

The `diffuseness' of the dust is the key parameter which we claim
determines the effectiveness of obscuration of the background universe.  
Although such dust distributions may be difficult to detect,
we must not neglect their possible presence. 
Further studies of spatial dust distributions, 
preferably via 
the counts and colours of background sources will be essential
in confirming our predictions.
\chapter{Modelling the Obscuration of Quasars by Foreground Dusty Galaxies} 

\vspace{1mm}
\leftskip=4cm
 
{\it Images of broken light which dance before me\\
Like a million eyes,\\
\ldots They tumble blindly as they make their way across the universe.\\} 
 
\vspace{1mm}
 
\hfill {\bf --- Across the Universe,}

\hfill {\bf John Lennon \& Paul McCartney, 1970}
 
\leftskip=0cm

\section{Introduction}

As early as 1930, Trumpler showed that optical observations of
distant stellar clusters were affected by dust, which he suggested was in a
thin absorbing layer in the galactic plane.
Similarly, sources at large redshift might be severely affected by 
dust in intervening galaxies, thus biasing our knowledge of the distant 
universe.  An historical review of cosmological studies of obscuration by dust
is given by Rudnicki (1986).

We are motivated by recent
observations of a large radio-selected sample of quasars 
(Drinkwater \etal 1997),
which is optimal for the study of the effects of dust. 
A large fraction of the sources 
have optical--to--near-infrared colours and optical continuum slopes
much redder than
is characteristically assumed in optical searches for quasars. Webster
\etal (1995) have argued
that the reddening is due to dust, although the location of the
dust is uncertain.  There are two obvious possibilities: either 
the dust is in the local quasar environment or else it is 
extrinsic and lies along the 
line-of-sight.
In this and the following chapter, we investigate the effects of an intervening dust
component which is located in galaxies 
along the line-of-sight.

The effect of intergalactic dust on observations of objects at
cosmological distances has been discussed by 
Ostriker \& Heisler (1984); Heisler \& Ostriker (1988);
Fall \& Pei (1989; 1992)
and Wright (1986; 1990). These authors 
show that the line-of-sight
to a high-redshift quasar has a high probability of intercepting a
galaxy disk, particularly if the dusty disk is larger than the optical 
radius of the galaxy.  The principle issue in these calculations is that
realistic dust distributions, which are ``soft'' around the edges,
will cause many quasars to `appear' reddened without actually removing 
them from a flux limited sample. There has been little evidence
for a population of reddened quasars in the past.  This has been considered
a strong constraint on models postulating that
dust might obscure a large fraction of the
high-redshift universe.  The new observations of Drinkwater \etal (1997)
remove this constraint, and will provide a distribution of reddening 
as a function
of redshift against which intervening models can be measured.

Ostriker \& Heisler (1984) 
and Heisler \& Ostriker (1988), have suggested that intervening dusty
galaxies would individually produce enough extinction to remove
background quasars from a flux-limited sample. 
In order to avoid the problem of generating too many 
(unobserved) reddened quasars,
these authors suggest that the dusty regions are `hard-edged':
either a quasar is reddened out of the sample or it is not reddened at
all.
They find that existing quasar
observations (at optical, X-ray and radio wavelengths) are consistent with
significant obscuration setting in by $z\sim 3$. They estimate that more
than $80\%$ of bright quasars at this redshift may be obscured by dust
in intervening galaxies and hence missing in optically-selected samples.

Similar conclusions were reached by 
Fall \& Pei (1989; 1992) who modelled
the obscuration of quasars by dust in damped Ly-$\alpha$ absorption
systems. These intervening systems
have column densities of neutral hydrogen in excess of
$10^{20}\,\rm cm^{-2}$ and are believed to be the progenitors of
present day galactic
disks. Fall \& Pei (1992) estimate 
that $10\%$-$70\%$ of bright quasars at $z=3$ are
obscured by dust in damped Ly-$\alpha$ systems.
They find that the `true' comoving density of bright
quasars can exceed the observed comoving density by factors of up to 4
at $z=3$ and by more than an order of magnitude at $z=4$.

Wright (1986) has described a numerical method to compute the total
optical depth in intervening dusty galaxies. In a later paper,
Wright (1990) examined the relationship
between the reddening observed in
the colours of optically-selected quasars and total $B$-band optical depth 
along the line-of-sight. He finds that
individual galaxies with ``soft'' edges and
large central optical depths (compared to those previously assumed) 
are required to satisfy the range of colours observed.
This model also predicts a reduction in the numbers of
quasars in optical flux-limited samples similar to that previously claimed. 

In almost all of the studies above, 
only a relatively small region of parameter space defining the
dust properties of galaxies has been explored. 
Most of the parameters assumed
describe galaxies which are dustier than
those observed locally (eg. Giovanelli \etal 1994; Byun 1993).
In this chapter, we repeat 
these calculations using a range of parameters
that may describe the dust 
properties of galaxies.
We modify a statistical method
introduced by Wright (1986) to calculate the 
distribution in net optical depth in dust
as a function of redshift.
Implications of the results of our simulations on optical QSO number counts, 
and on the recently detected reddening in the
radio-quasar sample of Drinkwater \etal (1997) are explored. 

This chapter is organised as follows: 
in section~\ref{oddi}, we describe the method used to compute the
optical depth distribution. 
Section~\ref{mpres} presents the 
galaxy parameters for four different models and the 
results of our calculations. 
Section~\ref{simplepl} investigates the effects of intervening
galactic dust on the counts of quasars in the optical. 
We also quantify the bias introduced from using `optically-selected'
quasars to infer the amount of dust reddening in the line-of-sight.
A comparison of our predictions with
observations in the radio sample of Drinkwater \etal (1997) is given in
section~\ref{crpq}. 
All results are summarised in section~\ref{concthree}.
Throughout, all calculations assume a Friedmann cosmology with
$q_{0}=0.5$ and Hubble parameter $h_{100}=1$ where 
$H_{0}=100h_{100}\,\rm km\,s^{-1}\,Mpc^{-1}$.

\section{Optical Depth Distribution}
\label{oddi}

We explore the effect of dust in intervening galaxies 
on quasar observations by computing the probability 
distribution of optical depth $\tau$ as a function of redshift along 
any random line-of-sight.
The optical depth for photons emitted at a redshift $z$ is the
sum of optical depths from all galaxies intercepted to redshift $z$. 
We define $\tau$ to be the total optical depth that is
encountered by the emitted photons which are 
observed at $z=0$ in the $B$-passband ($\lambda\simeq4400$\AA).

The method used 
is based on Wright (1986). We initially follow Wright
by modelling the universe as a series of concentric thin shells.  
The observed extinction $\tau$ between redshifts 0 and $z$ can be 
represented as a sum of extinctions over intermediate redshifts,
\begin{equation}
\tau (0,z)=\tau (0,z_{1})+\tau (z_{1},z_{2})+\cdots +\tau (z_{n},z).
\label{tsum}
\end{equation}
For a uniform dust distribution, the $\tau$'s in each redshift bin are 
all definite numbers, but if the absorption is due to dusty galaxies then
each $\tau$ is treated as an independent random variable. The probability
density function $p(\tau\,|\,0,z)$ for optical depth $\tau$ over some redshift
range $0\rightarrow z$ 
can thus be written as a repeated convolution of probabilities
$p(\tau\, |\,z_{i},z_{i+1})$ over the intermediate redshift bins:
\begin{equation}
p(\tau\, |\,0,z)=p(\tau\, |\,0,z_{1}) \otimes\cdots \otimes p(\tau
\,|\,z_{n},z),
\label{tconv}
\end{equation}
where $\otimes$ represents convolution.
A Fourier transform method
can then
be used to calculate $p(\tau\, |\,0,z)$ in terms of
model dependent galaxy parameters to be discussed later. We wish to obtain an
explicit expression for $p(\tau\, |\,0,z)$ that may be computed for any
set of model parameters. 
Details of the derivation are as follows.

If we denote $p_{i}(\tau)\equiv p(\tau\, |\,z_{i},z_{i+1})$ as the
probability density that a photon passing through the $i^{th}$ shell
encounters an optical depth $\tau$, then the Fourier transform,
$\tilde{p}_{i}(s)$ of $p_{i}(\tau)$ is defined
\begin{equation}
\tilde{p}_{i}(s)=\int {e^{2\,i\,\pi \,s\,\tau
}}\,p_{i}(\tau) \,d\tau.
\label{pis} 
\end{equation}
By convolving the $n$ thin shells as in Eqn.~\ref{tconv} and taking the
Fourier transform of this convolution, we have the following 
\begin{equation}
\tilde{p}(s\,|\,0,z)=\tilde{p}(s\,|\,0,z_{1}) \otimes\cdots \otimes
\tilde{p}(s\,|\,z_{n},z),
\label{psconv}
\end{equation}
\begin{equation}
\Rightarrow\,\,  \tilde{p}(s\,|\,0,z)=\int {e^{2\,i\,\pi \,s\,\tau}}\,p(\tau
\,|\,0,z) \,d\tau.
\label{psz}
\end{equation}
Since the shells are considered to be thin ($\tau\sim 0$), the
probability that the optical depth $\tau$ within any one shell is
nonzero is small. Hence for the $i^{th}$ shell we have
$$
p_{i}(\tau)\approx 1\quad\hbox{for}\; \tau\sim 0
$$
\begin{equation}
\Rightarrow\,\,  \tilde{p}_{i}(s) \sim \int {e^{2\,i\,\pi \,s\,\tau}}\,d\tau
 = 1 \quad\hbox{from Eqn.~\ref{pis}}.
\label{pistwo}
\end{equation}
Thus we may write using Eqn.~\ref{psconv},
\begin{eqnarray}
\nonumber
\ln\tilde{p}(s\,|\,0,z) & = &\sum_{i=1}^{n}\ln\tilde{p}_{i}(s)\approx
\sum_{i=1}^{n} \bigl[\tilde{p}_{i}(s) - 1\bigr]\\ 
\nonumber
\\
 & = &\sum_{i=1}^{n} \biggl[\int {e^{2\,i\,\pi \,s\,\tau}}\,p_{i}(\tau)
\,d\tau 
-
\int {p_{i}(\tau) \,d\tau}\biggr]
\nonumber
\\
 & = &\sum_{i=1}^{n} \int {(e^{2\,i\,\pi \,s\,\tau}-1)\,p_{i}(\tau)
\,d\tau}.
\label{pssum}
\end{eqnarray}
Taking the limit where the shell size reduces to zero ($\Delta z\rightarrow 0$
and hence $n\rightarrow \infty$) we have from Eqn.~\ref{pssum},
\begin{equation}
\ln\tilde{p}(s\,|\,0,z) = \int_{0}^{z}{dz'\,\int{(e^{2\,i\,\pi \,s\,\tau}-1)\,p(\tau,z') \,d\tau}}.
\label{psint}
\end{equation}
The function $p(\tau,z')$ in the integrand of Eqn.~\ref{psint} gives the 
probability density distribution for $\tau$ for some interval 
$z'\rightarrow z'+dz'$. Thus, the quantity $p(\tau,z')\,d\tau\,dz'$ gives the
probability that the optical depth $\tau$ lies within the range
$\tau,\,\tau + d\tau$ for the interval $dz'$.

Our method for calculating
the final result $p(\tau\, |\,0,z)$ now differs from Wright
(1986).
We wish to express $p(\tau,z')$ in Eqn.~\ref{psint} in terms of observables,
such as the mean number and optical depth of individual
galaxies within any interval $\Delta z$. 
We can do this by noting that the mean optical depth for the
interval $dz$ in which $\tau$ lies within $\tau,\,\tau+d\tau$
can be written 
\begin{equation}
\bar{\tau}_{dz}=\tau\,p(\tau,z)\,d\tau\,dz.
\label{bartd}
\end{equation}
We can also write this mean optical depth in terms of the mean galaxy
number, $d\bar{n}$, within $dz$ and their individual optical depths.
Assuming each galaxy within $dz$ has a uniform dust distribution with
optical depth $\tau_{0}$, so that $p_{gal}(\tau)=\delta (\tau - \tau_{0})$,
we have
\begin{equation}
\bar{\tau}_{dz}=d\bar{n}\,\tau_{0}\,\delta (\tau - \tau_{0})\,d\tau.
\label{bartn}
\end{equation}
The mean number of galaxies, $d\bar{n}$, along a line of sight 
at some redshift within the range
$z\rightarrow z+dz$, is defined by (Weinberg, 1972)
\begin{equation}
d\bar{n} = \sigma\,n_{0}\,\frac{c}{H_{0}} (1+z)(1+2q_{0}z)^{-1/2}dz,
\label{nz}
\end{equation}
where $\sigma$ is the cross-sectional area of a typical face-on galaxy,
and $n_{0}$ the local comoving number density of galaxies which is assumed to
be constant.  In other words we will assume a non-evolving galaxy population.
Combining Eqns.~\ref{bartd},~\ref{bartn} and ~\ref{nz} we have, 
for any interval $dz$,
\begin{equation}
p(\tau,z)\,d\tau\,dz\,\approx\,\sigma\,n_{0}\,\frac{c}{H_{0}} (1+z)
(1+2q_{0}z)^{-1/2}\tau_{0}\,\delta (\tau - \tau_{0})\,\frac{d\tau}{\tau} dz.
\label{ptz}
\end{equation}
Substituting Eqn.~\ref{ptz} into Eqn.~\ref{psint} and evaluating the integral
with respect to $\tau$, we arrive at
\begin{equation}
\ln\tilde{p}(s\,|\,0,z)=\sigma\,n_{0}\,\frac{c}{H_{0}}\int_{0}^{z}{dz'\,(1+z')
(1+2q_{0}z')^{-1/2}(e^{2\,i\,\pi \,s\,\tau_{0}(z')}-1)}.
\label{lnpsz}
\end{equation}

Equation~\ref{lnpsz} holds for the case of uniform galactic disks
where the optical depth of an individual galaxy, $\tau_{0}$, is constant
throughout the disk. Note that $\tau_{0}$ in Eqn.~\ref{lnpsz} is a function
of redshift $z$. 
This is a consequence of the increase in absorber rest-frame frequency 
with increasing redshift.
The observed optical depth in a fixed bandpass will therefore have a 
redshift dependence that exactly corresponds to the frequency dependence
of dust extinction in the absorber frame. 

For galaxies with a non-uniform dust distribution, the optical depth is
a function of the impact parameter $r$, so that $\tau_{0}(z)$ in 
Eqn.~\ref{lnpsz} 
is replaced by $\tau(r,z)$, where $r$ is the distance from the galaxy's center.
Following Wright (1986) and as confirmed observationally by
Zaritsky (1994), we will model the absorbers as 
exponential galactic disks,
where the optical depth through a face-on disk decreases exponentially with
distance $r$ from the center;
\begin{equation}
\tau(r,z)\, =\, \tau_{0}(z)\,e^{-r/r_{0}},
\label{trz}
\end{equation}
where $r_{0}$ is a characteristic galactic radius and $\tau_{0}(z)$, now,
is the value of $\tau$ through the centre of the galaxy $(r=0)$.
This profile is
supported by similar observations for the light distribution in nearby
spirals (Freeman, 1970), and the existence of                   
near-exponential radial metallicity gradients
(Vila-Costas \& Edmunds, 1992; Oey \& Kennicutt, 1993).
Given the exponential profile (Eqn.~\ref{trz}), the cross-section $\sigma$ in
Eqn.~\ref{lnpsz} 
is replaced by an integral over $r$.
Making these replacements, Eqn.~\ref{lnpsz} becomes
$$
\ln\tilde{p}(s\,|\,0,z)\, =\, n_{0}\,\frac{c}{H_{0}}\int_{0}^{z}{dz'\,(1+z')
(1+2q_{0}z')^{-1/2}}\,
$$
\begin{equation}
\times \int_{0}^{\infty}{(\exp\,[2\,i\,\pi \,s\,\tau_{0}(z')\,e^
{-r/r_{0}}]-1)\,2\pi rdr}.
\label{lnpstwo} 
\end{equation}

Furthermore, we will allow the disks of galaxies to be tilted with
respect to the line of sight by some random inclination angle
$(\frac{\pi}{2}-\theta)$ ($\theta$ being the angle between the plane of
the disk and the plane of the sky). To introduce tilts
we will need to consider a random inclination factor $\mu$, where
\begin{equation}
\mu\,=\,\cos\,\theta\,\rm{\hspace{5mm}}(0\,\leq\,\mu\,\leq\,1).
\label{mu}
\end{equation}
As a consequence, optical depths $\tau(r,z)$, will be increased by this
factor and cross-sectional areas, $\sigma$, decreased by the same amount.
Thus with the substitutions:
$$
\tau(r,z)\rightarrow\frac{\tau(r,z)}{\mu}\,=\,\frac{\tau_{0}(z)}{\mu}
\,e^{-r/r_{0}}
$$
$$
\rm{and\hspace{5mm}}\,\sigma\rightarrow\mu\,\sigma
$$
and averaging over $\mu$ where $\mu$ is randomly distributed between 0
and 1, Eqn.~\ref{lnpstwo} becomes
\newpage
$$
\ln\tilde{p}(s\,|\,0,z)\,=\, n_{0}\,\frac{c}{H_{0}}\int_{0}^{z}{dz'\,(1+z')
(1+2q_{0}z')^{-1/2}}\,
$$
\begin{equation}
\times\int_{0}^{\infty}{\int_{0}^{1}{(\exp\,[2\,i\,\pi
\,s\,\frac{\tau_{0}(z')}{\mu}\,e^{-r/r_{0}}]-1)\,\mu d\mu}\,2\pi
rdr}.
\label{pszthree}
\end{equation}
With the change of variables $y=e^{-r/r_{0}}$ and $t=1/\mu$ and
rearranging terms, we reach the final expression
$$
\ln\tilde{p}(s\,|\,0,z)=\,-2\,\tau_{g}\int_{0}^{z}{dz'\,(1+z')(1+2q_{0}z')^{-1/2}\,}
$$
\begin{equation}
\times\int_{0}^{1}{dy\,\frac{\ln y}{y}\left[E_{3}(-2\,i\,\pi\,s\,\tau_{0}(z')
\,y)-\frac{1}{2}\right]},
\label{pszfour}
\end{equation}
where the constant $\tau_{g}$ defined as 
\begin{equation}
\tau_{g}\,=\,n_{0}\,\pi r_{0}^{2}\,\frac{c}{H_{0}}
\label{tg1}
\end{equation}
is a model dependent parameter discussed in the next section and
\begin{equation}
E_{3}(-2\,i\,\pi\,s\,\tau_{0}(z')\,y)\equiv
\int_{1}^{\infty}{\frac{e^{(2\,i\,\pi\,s\,\tau_{0}(z')\,y)\,t}}{t^3}\,dt} 
\label{E3}
\end{equation}
is a standard mathematical function termed the ``Exponential Integral''.

Equation~\ref{pszfour} can be calculated numerically and the inverse Fourier
transform of $\tilde{p}(s\,|\,0,z)$ gives the required probability
density distribution function for the optical depth, $p(\tau\,|\,0,z)$,
as a function of $z$, i.e.
\begin{equation}
p(\tau\,|\,0,z)\,=\,\int_{-\infty}^{\infty}{e^{-2\,i\,\pi \,s\,\tau}\,\tilde{p}(s\,|\,0,z)\,ds}.
\label{ptf}
\end{equation}

To summarise, we have followed Wright (1986) 
(Eqns.~\ref{tsum} - \ref{pssum}) by
dividing the universe into a series of concentric shells, each with a
distribution of galaxies. Using a Fourier transform method and
convolving the effect due to each shell, we have obtained the
probability density function for the total optical depth $\tau$ due to
intervening galaxies in the range $0\rightarrow z$. 
Once the probability distribution function $p(\tau\,|\,0,z)$ is
calculated from Eqns.~\ref{pszfour} - \ref{ptf} 
(for some set of galaxy parameters),
we can then determine the probability that the total optical depth is
some given value $\tau$ along any line-of-sight to redshift $z$.

\section{Model Parameters and Results}
\label{mpres}

Our model depends on
three parameters which describe the 
characteristics of the intervening galaxies. 
$n_0$ and $r_0$ are included in $\tau_{g}$
(Eqn.~\ref{tg1}), which gives the average
number of intersections in a Hubble length of a light ray within $r_{0}$
of a galaxy's centre.
The third parameter, $\tau_{B}$, is the
dust opacity at the center of an individual absorber. In this work we  
assume the optical depth follows a simple linear dependence on frequency, thus
\begin{equation}
\tau_{0}(z)\,=\,\tau_{B}\,(1+z),
\label{tzb}
\end{equation}
where $\tau_{B}$ is the $B$-band optical depth at the
center of a local galaxy ($z=0$) and 
$\tau_{0}(\nu_{0})\propto \nu_{e}=\nu_{0}(1+z)$.  More complex
models where the extinction curve includes a $2200$\AA~ 
bump, as in the Milky Way
are considered by Heisler \& Ostriker (1988).  However recent work
(Calzetti \etal 1994) finds no evidence of such a feature in the 
extinction curves of nearby galaxies.

Table~\ref{tpi} gives the 
values of three parameters for each of four models
to be considered.
Model 1 uses the values of Heisler \& Ostriker (1988). Their value of 
$\tau_{g}=0.2$ was chosen to yield a sky covering fraction of
galaxies which is consistent with the percentage of quasars detected with
damped Ly-$\alpha$ absorption systems along the line-of-sight. 
For the nominal value $n_{0}=0.02h_{100}^{3}{\rm Mpc}^{-3}$, 
they find (from Eqn.~\ref{tg1}) $r_{0}\simeq33$ kpc, which 
is large compared to values typical of present day spirals 
($r_{0}\sim 4-5$ kpc, Freeman, 1970).
Heisler \& Ostriker
have set $\tau_{B}=0.5$ from a study of the range of absorption in
galaxies by Phillips (1986).

\begin{table}
\begin{center}
\begin{tabular}{|l|c|c|c|c|}
\hline
Model & $n_{0}(h_{100}^{3}{\rm Mpc^{-3}})$ & 
$r_{0}$(kpc) & $\tau_{g}$ & $\tau_{B}$
\\ \hline
\hline
1. Heisler \& Ostriker (1988)& 0.02& 33& 0.2& 0.5\\
2. Wright (1990)& 0.005& 33& 0.05& 2\\
3. Our model& 0.02& 10& 0.0188& 2\\
4. Our model& 0.01& 5& 0.0023& 2\\
\hline
\end{tabular}
\caption{Adopted model parameters for calculation of the distribution functions
$p(\tau|z)$.}
\label{tpi}
\end{center}
\end{table}

Model 2 uses the values of Wright (1990) who studied correlations
between reddening and obscuration of background quasars. This model
contains the same total
amount of dust as Model 1 but the dust is concentrated
in a smaller number of more opaque clouds. In other words, by making
$\tau_{B}$ larger, one approaches the opaque, hard-edged disk
limit.

The values in Models 3 and 4 are believed to be more representative 
of local galaxies.
In both of these models $\tau_{B}=2$, 
consistent with lower limits for spiral galaxies derived by Disney
and Phillipps (1989), Byun (1993) and Giovanelli \etal (1994). Model 3
has a characteristic radius of $r_{0}=10$ kpc, while Model 4
represents a minimal model with $r_{0}=5$ kpc and
$n_{0}=0.01h_{100}^{3}{\rm Mpc^{-3}}$.

\subsection{Results and Analysis}
\label{resennoev}

\begin{figure}
\vspace{-3in}
\plotonesmall{1}{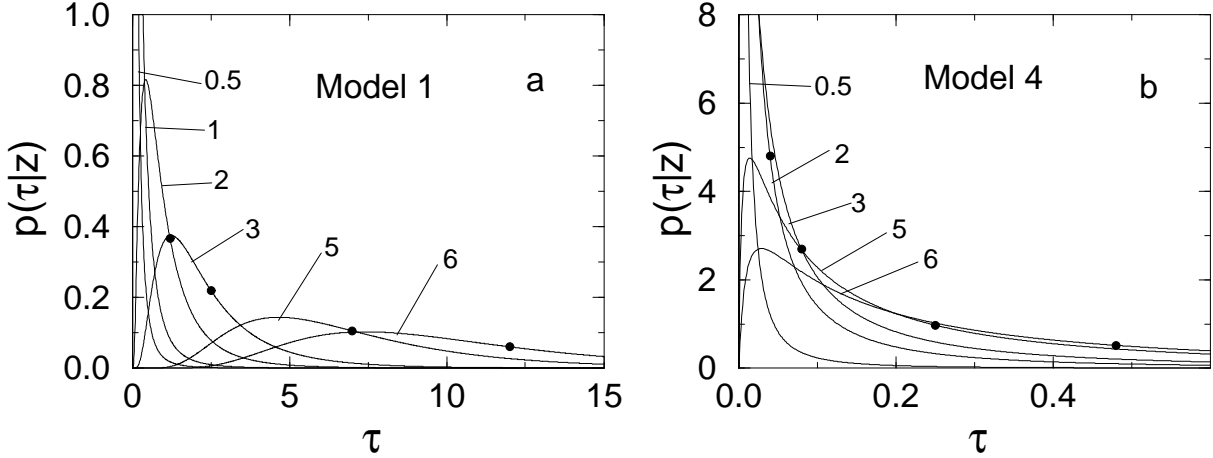}
\vspace{-2.5in}
\caption[Optical depth probability distribution functions]{
{\bf a.} Optical depth probability distribution functions 
$p(\tau\,|\,z)$ for Model 1 (see Table~\ref{tpi}) where each curve
corresponds to the redshift indicated. 
The horizontal axis is the total optical
depth $\tau$ observed in the $B$ band. The median point for each curve (for
$z\geq 2$) is indicated by a dot.
{\bf b.} Same as (a) but for Model 4. 
}
\label{pdfs1}
\end{figure}

For each set of galaxy parameters ($n_0$,$r_0$,$\tau_{B}$), 
distribution functions $p(\tau\,|\,z)$ for the
total optical depth have been computed from Eqns.~\ref{pszfour} - \ref{ptf} 
in redshift intervals of 0.5 up to $z=6$.
Figures~\ref{pdfs1}a and b 
show the optical depth probability density distributions 
at different redshifts for Models 1 and 4 respectively.
As discussed, these refer to two extreme scenarios. 
The curves in Fig.~\ref{pdfs1}a closely resemble those computed by Heisler
and Ostriker (1988) (see their figure 1), except that the median
points differ. This is due to Heisler and Ostriker adopting an
extinction curve for the dust which involves the $2200$\AA~ feature. The
probability that the total optical depth to 
some redshift $z$ lies within the interval
$0\rightarrow\tau_{max}$ is given by the area under the normalised
curve:
\begin{equation}
P(0\,\leq\,\tau\,\leq\,\tau_{max}\,|\,z)\,=\,\int_{0}^{\tau_{max}}
{p(\tau\,|\,z)\,d\tau}.
\label{ptr}
\end{equation}
For each curve in Fig.~\ref{pdfs1}, a dot is drawn to indicate the median
point where the curve integrated over $\tau$ equals $\frac{1}{2}$.
Physically, this means that for a specific dust model and redshift $z$, 
we should expect
that in a radio-quasar survey 
(expected to be unbiased against reddening by dust),
at least 50\%
of sources with redshifts $\simgt z$ 
should suffer extinctions corresponding to
$\tau\,\geq\,\tau_{median}$.

\begin{figure}
\vspace{-3in}
\plotonesmall{1}{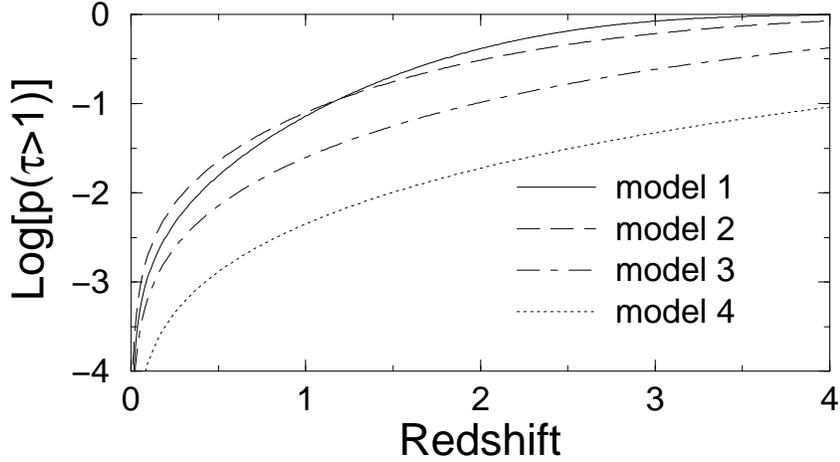}
\vspace{-2.5in}
\caption[Probability that $\tau_{dust}>1$ as a function of redshift]{Probability that the total optical depth $\tau$ towards some 
redshift $z$ is greater than
1 as a function of $z$. The vertical axis represents $\log_{10}p(\tau > 1)$. 
Model 1 (solid curve), Model 2 (dashed), Model 3 (dot-dashed) and
Model 4 (dotted),
(see Table~\ref{tpi}).
}
\label{ptgt1}
\end{figure}

Figure~\ref{ptgt1} 
shows the probability that the total optical depth towards some
redshift $z$ is greater than 1, computed from the distribution functions
for Models 1, 2, 3 and 4 in Table~\ref{tpi}. 
The probability $P(\tau\,>\,1)$ up to a redshift $z$
is the fraction of the sky to that redshift which has 
at least $\tau\,=\,1$. 
Model 1 predicts that all sources with $z>4$ should have at least
$\tau\,=\,1$, whereas the percentage drops to $80\%$ and $50\%$ for
Models 2 and 3 respectively. The obscuration in Model 4 becomes significant
for redshift $z>6$, where more than $30\%$ of 
sources are predicted to encounter an optical 
depth of at least $\tau\,=\,1$. 

As shown in Fig.~\ref{ptgt1}, 
Model 1 corresponding to ``soft''-edged galactic disks predicts that
obscuration should be less severe by as much as
a factor of 2 at redshifts $z<1$ than that of the
``harder''-edged disk Model 2. 
For redshifts $z>1$, the situation reverses: optical depths 
predicted by Model 1 exceed those predicted by Model 2. 
This can be explained by the values of the parameters defining these models 
in Table~\ref{tpi}.
The ``harder'' edged galaxies with higher central optical depths in 
Model 2 are more effective in causing significant obscuration at lower 
redshifts ($z<2$). Light emitted from a source is more likely to suffer
reddening once it encounters a harder-edged disk than for a softer-edged 
disk (Model 1), even though Model 1 contains a greater number of galaxies per
unit redshift interval.
At redshifts $z>2$, obscuration becomes more dependent on the number of
galaxies intercepting a light ray (the parameter $\tau_{g}$ in 
Table~\ref{tpi}). 
Model 1 with four times as many galaxies along the line of sight,
therefore dominates the optical depth to high redshifts. 

\begin{figure}
\vspace{-3in}
\plotonesmall{1}{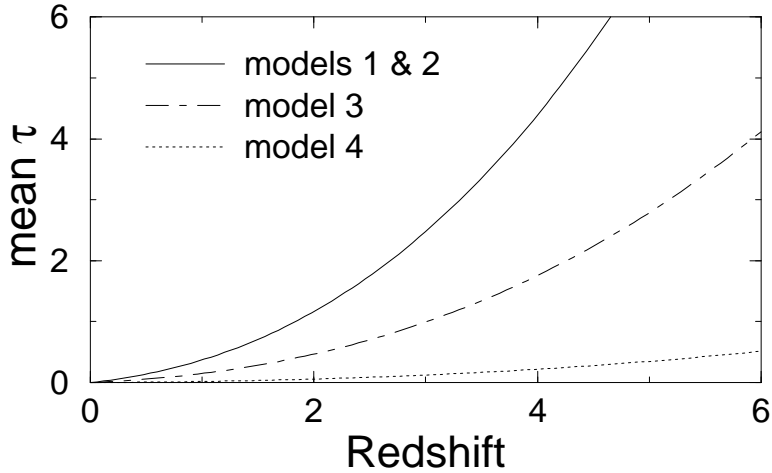}
\vspace{-2.5in}
\caption[Mean optical depth as a function of redshift]{
Mean optical depth as a function of redshift along any random
line of sight for Models 1 and 2 (solid curve-both these models contain
the same amount of dust), Model 3 (dot-dashed) and Model 4 (dotted).
}
\label{meantz}
\end{figure}
 
The mean optical depth to a given $z$ provides a handle on making the
comparison between the dust content specified by each set of model
parameters in Table~\ref{tpi}. 
Heisler \& Ostriker (1988) have shown that the
mean optical depth to a redshift $z$ scales as the following product:
\begin{equation}
\bar{\tau}(z)\,=\,0.8\,\tau_{g}\,\tau_{B}\left[(1+z)^{5/2} -
1\right].
\label{tmean}
\end{equation}
Equation~\ref{tmean} is plotted in Fig.~\ref{meantz} 
for the models listed in 
Table~\ref{tpi}. 

\section{Effect on Quasar Number Counts}
\label{simplepl}

We will give a simple illustration of
the effects of dust in intervening foreground galaxies
on quasar counts in the optical. We assume that 
the `true' number of 
quasars (ie. in a `dust-free' universe) at a given redshift follows a
power-law: 
\begin{equation}
N_{true}(>L)\,=\,N_{true}(>L_{l})\,\left(L\over L_{l}\right)^{-\beta},
\label{pl1} 
\end{equation}
where $N_{true}(>L_{l})$ is the total number of quasars observed at fixed $z$
with fluxes
greater than some limiting flux $L_{l}$, and $\beta$ is the
cumulative luminosity function slope.

In a dusty universe, quasars observed through a uniform extinction  
$\tau$ have their true fluxes  reduced by a factor
$e^{-\tau}$. 
The total number of sources that will be reddened below some flux limit
$L_{l}$ (and hence removed from a sample) can therefore be written: 
$N_{dust}(<L_{l})=N_{true}(>L_{l})-N_{true}(>e^{\tau}L_{l})$. 
Using Eqn.~\ref{pl1}, we can rewrite this as: 
$N_{true}(>L_l)[1-e^{-\beta \tau}]$.
Since there is a probability $p(\tau\,|\,z)$ of encountering an
optical depth $\tau$, 
the total fraction of sources lost from a sample 
(relative to a dust-free universe) at some redshift is:
\begin{equation}
f_z = \frac{N_{dust}(<L_l)}{N_{true}(>L_l)}\,=
\, 1-\int_0^{\infty} p(\tau\,|\,z) e^{-\beta \tau} d\, \tau. 
\label{fz}
\end{equation}
This fraction scales in terms of the model galaxy parameters as follows:
$$
f_z\,\simeq\,1-\exp\left[-\tau_{g}\tau_{B}^{2/5}\beta^{2/5}((1+z)^{11/6}-1)\right]
$$
\begin{equation}
{\rm\hspace{-15mm}}\sim\,\tau_{g}\tau_{B}^{2/5}\beta^{2/5}
\left[(1+z)^{11/6}-1\right],
\label{fzapp}
\end{equation}
where the last step follows when the argument of the exponential is small, which
for typical values of $\beta=2.5$ and $\tau_{B}=1$,
is sufficiently accurate for $\tau_{g}\simlt0.02$ and redshifts $z<2$. 
We see from Eqn.~\ref{fzapp} that 
the obscuration of background sources 
by line-of-sight galactic dust has a stronger dependence
on the covering factor parameter $\tau_{g}$ (Eqn.~\ref{tg1}),
than on the central absorber optical depth $\tau_{B}$ (see section~\ref{resan} 
for more details). 
The fraction of sources missing from some flux limit 
will increase with increasing $\beta$; in other words, 
a steeper
luminosity function implies that a greater fraction of sources will
be lost from a flux-limited sample.

\begin{figure}
\vspace{-3in}
\plotonesmall{1}{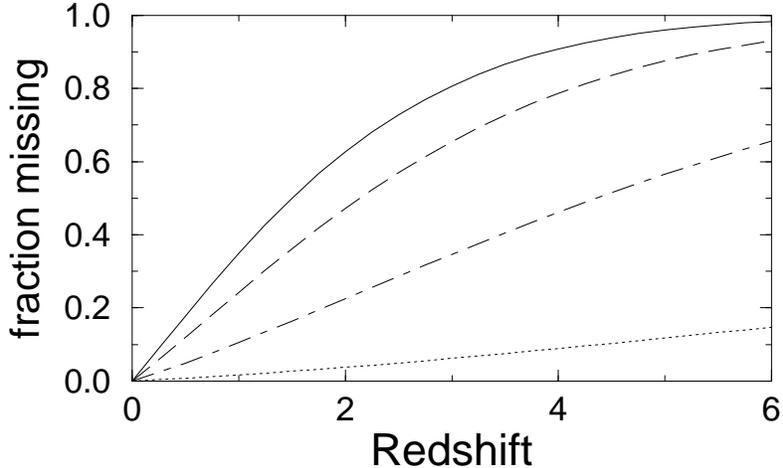}
\vspace{-2.5in}
\caption[QSO fraction missing from a flux-limited sample vs. $z$] 
{Lower bound on the fraction of QSOs missing from a flux-limited sample
at redshift $z$ as a function of $z$ due to our
intervening dust models (see Eqn.~\ref{fz}). 
Each curve is defined as in Fig.~\ref{ptgt1}.
}
\label{fmissz}
\end{figure}

From surveys of quasars selected in the $B$-band, 
the slope $\beta$ at the bright end
of the cumulative luminosity function (typically for $m_{B}\simlt19$) 
lies in the range $2.5-3.0$ (Boyle \etal 1990).
We have chosen the lower bound $\beta\,=\,2.5$. 
Figure~\ref{fmissz} plots 
$f_{z}$, which with $\beta\,=\,2.5$, is a lower limit
on the fraction of `bright' quasars missing 
from a $B$-band flux limited sample. 
These sources may 
be detected as optical `empty fields' in 
radio surveys.
The prediction for each of our four models is shown.
Model 1 predicts that almost no quasars should be observed at $z>4$
within some flux limit, while for model 4, 
we expect at least 5\% and $15\%$ of quasars at redshifts $z>2$ and $z>6$ 
respectively to have been missed.  

\subsection{Dust Reddening and Bias in Optically-Selected Samples} 
\label{redbiasopt}

The mean optical depths as a function of redshift illustrated 
in Fig.~\ref{meantz}
represent `true' optical depths associated with
line-of-sight dusty galaxies.
Since optically-selected quasars are expected to be biased against
significant amounts of dust in their line-of-sight,
the predictions in Fig.~\ref{meantz} 
are expected to differ considerably from reddening estimates 
deduced from optically-selected samples. 
Due to the relatively large dispersion about the mean optical  
depth to any redshift as shown in the probability
distribution functions (PDFs) of Fig.~\ref{pdfs1}, 
only background sources which
have suffered sufficiently low optical depths
will be reddened such as to remain within the flux 
limit of a sample.
Thus, this observational bias
will lead to 
an effective `observed' mean optical depth that is considerably smaller
than the `true' optical depths modelled above. 
In this section, we present a simple model
to quantify this bias. 

In general, the mean optical depth to any redshift is defined as 
\begin{equation}
\langle\tau\rangle\,=\,\int^{\infty}_{0}\tau p(\tau|z)\,d\tau,
\label{tgen}
\end{equation}
where $p(\tau|z)$ is the probability distribution function 
describing the dust model.
To give an estimate of ``observed'' optical depth as inferred from observations
of the reddening in an optically-selected quasar 
population, 
we must take into account the effects of the ``finite'' flux limit imposed. 
To do this, we require information regarding 
the quasar luminosity distribution. 
To give a simple illustration, we shall assume a power-law
luminosity function like above (Eqn.~\ref{pl1}). 
We require an effective ``observed''
PDF for the optical depth to the optically 
detected sources. For a power-law luminosity function, 
this observed PDF is given by $p_{o}(\tau|z)=e^{-\beta\tau}p(\tau|z)$, where 
$e^{-\beta\tau}$ is the probability that a source which is
reddened by amount $\tau$, will remain within the
flux of a sample (see above) and $p(\tau|z)$ is the ``true'' 
optical depth probability distribution modelled in section~\ref{oddi}. 
From Eqn.~\ref{tgen} and requiring that $p_{o}(\tau|z)$ be normalised, 
the ``observed'' optical depth
to optically detected sources at some redshift can be written:
\begin{equation}
\langle\tau\rangle_{obs}\,=\,\frac{\int_{0}^{\infty}
\tau e^{-\beta\tau}p(\tau|z)\,d\tau}
{\int_{0}^{\infty}e^{-\beta\tau}p(\tau|z)\,d\tau}.
\label{tobs}
\end{equation}
This can be shown to scale in terms of the dust model parameters as 
follows: 
\begin{equation}
\langle\tau\rangle_{obs}\,\sim\,\tau_{g}\beta^{-2/5}(\tau_{B}^{2/5} + 1)^{-1}
\tau_{B}^{2/5}\left[(1+z)^{11/6} - 1\right].
\label{tobsa}
\end{equation}

\begin{figure}
\vspace{-3in}
\plotonesmall{1}{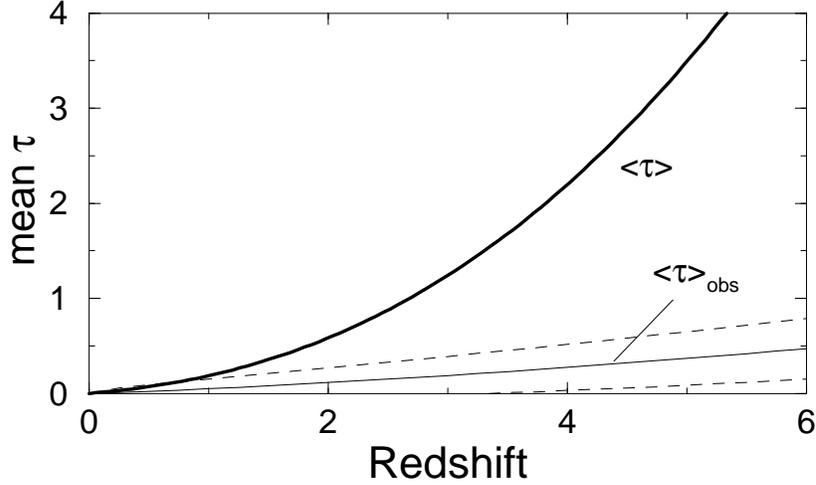}
\vspace{-2.5in}
\caption[``True'' vs. ``Observed'' mean optical depth
as a function of $z$]{
``True'' mean $B$-band optical depth (thick line) and ``observed'' optical
depth (thin line)
as a function of redshift. The observed trend is that 
inferred from an optically-selected QSO sample assuming a power-law 
cumulative LF with slope $\beta=2.5$.
The dust parameters assumed are:
$n_{0}=0.02h_{100}^{3}{\rm Mpc^{-3}}$, $r_{0}=15$kpc and $\tau_{B}=1$. 
The region enclosed by the two dashed lines represents the $1\sigma$
spread about the mean observed reddening curve.
The $1\sigma$ value in the true mean optical depth (not shown) 
is about equal to the mean itself at all redshifts.
}
\label{ot}
\end{figure}

Figure~\ref{ot} shows the ``true'' and ``observed''
model optical depths as a function
of redshift. A quasar (cumulative) luminosity function slope of $\beta=2.5$
and dust parameters $n_{0}=0.02h_{100}^{3}{\rm Mpc^{-3}}$, 
$r_{0}=15$kpc and $\tau_{B}=1$
are assumed.
The $1\sigma$ spread is shown for $\langle\tau\rangle_{obs}$ only,
while for the ``true'' optical depth $\langle\tau\rangle$, 
the $1\sigma$ value (where 
$\langle\tau\rangle_{1\sigma}=\langle\tau\rangle\pm1\sigma$)
is almost equal to the mean
itself at all redshifts.
We see that the mean observed reddening to optically detected
quasars can be less by up to a factor of 3 than the true mean at redshifts $z>1$
and a factor of 6 at redshifts $z>2$.
Comparing Eqns.~\ref{tmean} and ~\ref{tobsa}, the
ratio $\langle\tau\rangle_{true}/\langle\tau\rangle_{obs}$ 
will be independent of the covering factor parameter $\tau_{g}$,
and will only slightly depend on the central galaxy optical depth $\tau_{B}$.
We conclude that the use of optically 
selected quasars as a means to measure the dust content along the line-of-sight
can lead to results that are grossly in error.

\section{Comparison with `Reddened' Parkes Quasars} 
\label{crpq}

Let us consider whether the reddening observed in the radio-selected quasar 
sample of Drinkwater \etal (1997) (hereafter `Parkes quasars') 
can all be attributed to dust in galaxies that may happen to 
lie along the line-of-sight. 
From Fig.~\ref{colvsz} in section~\ref{dustobscu}, 
two aspects of the observations are relevant: 
(i) overall, more than $50\%$ of the sample is reddened with
colours $B_{J}-K_{n}\,>\,3.5$, and
(ii) the distribution in reddening (with median $B_{J}-K_{n}\simeq3.5$) is
relatively uniform with redshift to $z\sim3$.

Assuming the generic $1/\lambda$ extinction law, and that the intrinsic
(unabsorbed) colour of Parkes quasars is
equivalent to the mean value $B_{J}-K_{n}\simeq2.3$ 
found for optically-selected quasars
(which show relatively small scatter; see Fig.~\ref{colvsz}), 
the observed $B$-band
optical depth can be written:
\begin{equation}
\tau\,\simeq\,1.14(B_{J}-K_{n} - 2.3).
\label{colopt}
\end{equation}
Observed colours of $B_{J}-K_{n}\,>\,3.5$ therefore correspond to
optical depths $\tau\simgt1.4$ in the $B$-band. 

As shown in Fig.~\ref{ptgt1}, model 4 (our `minimal' model) 
predicts that the number of reddened quasars 
exhibiting $\tau\simgt1$ is small. 
For example at $z\simlt2$, 
there is almost no predicted reddening.  From the observations, as many
as $50\%$ of such quasars show significant reddening. Thus, 
dust model 4
cannot account for all the observed reddening.
The most stringent test for model 1 (our `maximal' model) 
is at low redshift ($z\simlt1$), 
where a reddening of $\tau<1$ is predicted
for all but
a few percent of quasars (see Fig.~\ref{ptgt1}).  
More than $50\%$ of Parkes quasars observed at $z\simlt1$
however appear strongly reddened.
The predicted number of reddened quasars to $z\sim3$ is 
closer to the observed value, but the dependence of reddening with 
redshift is inconsistent with that observed. 

We conclude that if the observed reddening 
is to be totally explained by 
dust in galaxies in the line-of-sight, then 
more complicated models are required. 
An example is the inclusion of 
evolution in galactic dust which has been neglected in this chapter. 
A more detailed model is presented and constrained 
using the data in the next chapter. 

\section{Conclusions}
\label{concthree}

In this chapter, we have modelled the effects of obscuration due to dust
in galaxies randomly distributed along the line-of-sight as a function
of redshift.
We have modelled galaxies as
randomly inclined disks with exponential dust profiles. 
As an initial investigation,
no allowance has been made for possible evolution in galactic dust content
with redshift.

Our main results are:
\\\indent 1. Flux-limited, optically selected quasar surveys are strongly
biased against reddening by line-of-sight dust. 
Optically selected quasars are expected to exhibit 
only a few percent of the total reddening
to any redshift, and do not provide a handle on the
total dust content of the universe. 
Our `minimal' dust model predicts that at least 10\% of
quasars at redshifts $z>3$
may have been under-represented in optical surveys.
\\\indent 2. The distribution of observed reddening in quasars
in the radio-selected sample of Drinkwater \etal (1996)
cannot be fully explained by a non-evolving 
distribution of dusty galaxies along the line-of-sight.
The observed dependence of reddening on redshift is inconsistent with our
predictions.
\\\indent 3. The models depend on a range of 
parameters which will require detailed 
fitting of the observations.  
The observed reddening distribution 
as a function of redshift for Parkes quasars may provide 
a strong constraint on possible models.
In the next chapter, a detailed comparison of more
complicated models
with the data is presented.
 
The predictions of this chapter can be best confirmed using radio-selected
quasar samples which should have no bias against obscuration by dust.
If models of dusty line-of-sight galaxies are correct, then a large
fraction of radio-selected quasars at high redshift must appear
heavily reddened. 
If they are to be completely ruled out, then
current radio surveys will have to identify virtually all highly reddened
sources as low redshift galaxies or quasars that are intrinsically 
reddened. 
\chapter{Time Evolution of Galactic Dust and 
the Obscuration of Quasars} 

\vspace{1mm}
\leftskip=3cm
 
{\it All space, all time,\\
The stars, the terrible perturbations of the suns,\\
Swelling, collapsing, ending, serving their longer, shorter use,\\
Ever the mutable,\\
Ever materials, changing, crumbling, recohering \ldots \\} 
 
\vspace{1mm}
 
\hfill {\bf --- Song Of Myself} 

\hfill {\bf Walter Whitman, 1855}
 
\leftskip=0cm

\section{Introduction}

As shown in Chapter 3, the distribution of observed reddening in quasars 
in the radio-selected sample of Drinkwater \etal (1997) (hereafter the
`Parkes sample')
cannot be fully explained by a non-evolving 
distribution of dusty galaxies along the line-of-sight.
The observations (see Fig.~\ref{colvsz})
indicate a mean reddening which scales as a function of
redshift
as $\bar{\tau}\propto(1+z)^{0.1}$.
This dependence is
considerably `flatter' than the mean optical depth predicted from
models where the dust content is assumed {\it not} to evolve where
$\bar{\tau}\propto(1+z)^{2.5}$ (see section~\ref{resennoev}).
By introducing evolution in
dust content, it may be possible to reproduce the
`flat' reddening behaviour as a function of redshift in the Parkes sample.
This chapter explores the effects 
of more complicated models where galactic dust content evolves. 

Cosmic evolution in dust content to redshifts $z\sim3$ 
is indirectly suggested by numerous claims of reduced 
chemical enrichment at $z\simgt2$ relative to local estimates.
This is provided by observations of trace metals
and their relative abundances in QSO
absorption-line systems to $z\sim 3$ (Meyer \& Roth, 1990;
Savaglio, D'Odorico \& M\"{o}ller, 1994; Pettini $et\,al$. 1994; Wolfe
$et\,al$. 1994;
Pettini \etal 1997; Songaila, 1997),
which are thought to arise
from intervening clouds or the haloes and 
disks of galaxies. These studies indicate
mean metallicities $\simeq 10\%$ and $\simlt1\%$ 
solar at $z\sim2$ and $z\sim3$ respectively, and dust-to-gas ratios
$\simlt8\%$ of the galactic value at $z\sim2$.
These estimates are consistent with simple global evolution models of 
the star formation rate and evolution of gas
in the universe 
(Pei \& Fall, 1995).
If the observed metallicities in QSO absorption systems are common, 
then their interpretation
as galactic disks implies that substantial evolution
has taken place since $z\sim 3$.
If dust follows a similar trend (as expected if dust follows
heavy metal production from stellar processes), then one may expect
the effects of
obscuration to high redshift to be reduced relative to
non-evolving predictions.

In this chapter, we investigate the reddening effects of
dust located in galaxies along the line-of-sight 
using models where the dust content evolves.
Implications of obscuration on 
QSO and absorption line statistics are explored.
We use optical and near-infrared
observations in the Parkes sample to constrain
evolutionary and physical properties of dust in intervening systems. 
Using existing observations of
galactic dust properties, we also quantify the 
contribution of dust in external systems to the reddening observed
in this sample. 

This chapter is organised as follows: The next section briefly describes the 
generalised model. Section~\ref{mprev} 
presents the model predictions. Implications of 
dust obscuration on observations of background quasars and QSO absorption 
line systems are explored in section~\ref{impev}. 
In section~\ref{conmodp}, we fit our model to 
the Parkes quasar data. Comparisons of our constrained parameters with 
existing observations are made in section~\ref{compexo}.
Possible uncertainties are discussed in section~\ref{evmodd} 
and all results are summarised in section~\ref{concfour}. 
All calculations assume a Friedmann cosmology with $q_{0}=0.5$,
and Hubble parameter $h_{50}=1$ 
where $H_{0}=50h_{50}\, \rm km\,s^{-1}\,Mpc^{-1}$
(unless otherwise stated).
 
\section{The Evolutionary Dust Model}
\label{briefdes}

We calculate the distribution in total dust optical depth from model
galaxies along the line-of-sight as a function of redshift by
following the method presented in Chapter 3. 
This method does not include the effects of evolution. We shall generalise
this model by considering the possibility of evolution in the dust
properties of galaxies.

We assume the following properties for individual
absorbing galaxies. As assumed in Chapter 3, we model 
galaxies as randomly tilted exponential disks,
where the
optical depth through a face-on disk decreases exponentially
with distance $r$ from the center;
\begin{equation}
\tau(r,z)\,=\,\tau_{0}(z)\,e^{-r/r_{0}}.
\label{expr}
\end{equation}
$r_{0}$ is a characteristic radius and $\tau_{0}(z)$,
the value of $\tau$ through the center of the galaxy $(r=0)$.
The redshift dependence of $\tau_{0}$ is due to the increase in 
absorber rest frame frequency with redshift. 
 
Since the 
observed reddening in Parkes sources (with which we shall
compare our model) is measured in terms
of a $B-K$ colour excess, we require an extinction law that
covers the optical to near-infrared wavelength range.
In this study, we assume an extinction law which represents an average 
of a number of laws computed for our galaxy and several external galaxies
by Rieke \& Lebofsky (1985), Knapen \etal (1991) and
Jansen \etal (1994). This law applies
for passbands $U$ to $K$ where 
the optical depth scales as frequency as
$\tau\propto\nu^{1.5}$. Thus, the optical depth in an observer's frame 
through an absorber at redshift $z$ ($\tau_{0}(z)$ in Eqn.~\ref{expr}) 
will follow the dependence:
\begin{equation}
\tau_{0}(z)\,=\,\tau_{B}\,(1+z)^{1.5},
\label{tz}
\end{equation}
where $\tau_{B}$ is 
the $B$-band optical depth through the
center of a local galaxy ($z=0$).
 
\subsection{Evolution} 
\label{ev}
 
Equation~\ref{tz} must be modified if the dust content in each
galaxy is assumed to evolve with cosmic time. 
The optical depth seen through the
center of a single absorber at some redshift, $\tau_{0}(z)$, 
depends
on the quantity of
dust formed from past stellar processes.
For simplicity, we assume all galaxies form simultaneously, maintain a
constant space density,
and increase in dust content at a rate that is 
uniform throughout.
 
We parameterise evolution in dust content by following
simulations of the formation of heavy metals in the
cold dark matter scenario of galaxy formation by  
Blain \& Longair (1993a,1993b).
These authors assume that galaxies form
by the coalescence of gaseous protoclouds through
hierarchical clustering as prescribed by Press \& Schechter (1974).
A fixed fraction of the mass involved in each merger event is converted
into stars, leading to the formation
of heavy metals and dust.
It was assumed that 
the energy liberated
through stellar radiation was absorbed
by dust and re-radiated into the far-infrared. 
They found that such radiation can contribute substantially
to the far-infrared background intensity from which they constrain 
a model
for the formation of heavy metals
as a function of cosmic time.
Their models show that the comoving density of heavy metals 
created by some redshift $z$, given that star formation
commenced at some epoch $z_{SF}$ follows the form 
\begin{equation}
\Omega_{m}(z)\,\propto\,\ln\left({1+z_{SF}\over1+z}\right),
{\rm\hspace{6mm} where}\,z<z_{SF}.
\label{omegaZ}
\end{equation}
 
We assume that a fixed fraction of heavy metals condense
into dust grains so that
the comoving density in dust, $\Omega_{d}(z)$,
follows a similar dependence as Eqn.~\ref{omegaZ}.
The density in dust relative to the present closure density
in $n_{0}$ exponential
disks
per unit comoving volume is given by 
\begin{equation}
\Omega_{d}\,=\,\frac{n_{0}M_{d}}{\rho_{c}},
\label{omegaD}
\end{equation}
where $\rho_{c}=3H_{0}^{2}/8\pi G$ and 
$M_{d}$ is the dust mass in a single exponential disk. 
This mass can be estimated using Eq.7-24 from Spitzer (1978)
where the total density in dust, $\rho_{d}$,
is related to the extinction $A_{V}$ along a path length $L$ in kpc
by 
\begin{equation}
\langle\rho_{d}\rangle\,=\,1.3\times10^{-27}\rho_{g}
\left(\frac{\epsilon_{o} + 2}{\epsilon_{o} - 1}\right)\,(A_{V}/L).
\label{rhod}
\end{equation}
$\rho_{g}$ and $\epsilon_{o}$ are the density and
dielectric constant of a typical dust grain respectively 
and the numerical factor has
dimensions of
${\rm gm\,cm}^{-2}$ - see Spitzer (1978).
Using the exponential profile (Eqn.~\ref{expr}) where
$\tau(r)\propto A_{V}(r)$
and integrating along cylinders, 
the
dust mass in a single exponential 
disk can be found in terms of the model parameters
$\tau_{B}$ and $r_{0}$.
We find that the comoving density in dust at some redshift scales as
\begin{equation}
\Omega_{d}(z)\,\propto\,\tau_{B}(z)\,n_{0}\,r_{0}^{2},
\label{omD2}
\end{equation}
where $\tau_{B}(z)$ is the central $B$-band
optical depth and $r_{0}$ the dust scale radius 
of each disk.
Thus, the central optical depth, $\tau_{B}(z)$, in any
model absorber at some redshift
is directly proportional to the mass density in dust or 
heavy metals
as specified by Eqn.~\ref{omegaZ};
\begin{equation}
\tau_{B}(z)\,\propto\,\ln\left({1+z_{SF}\over1+z}\right).
\label{tbz}
\end{equation}
 
\begin{figure}
\vspace{-3in}
\plotonesmall{1}{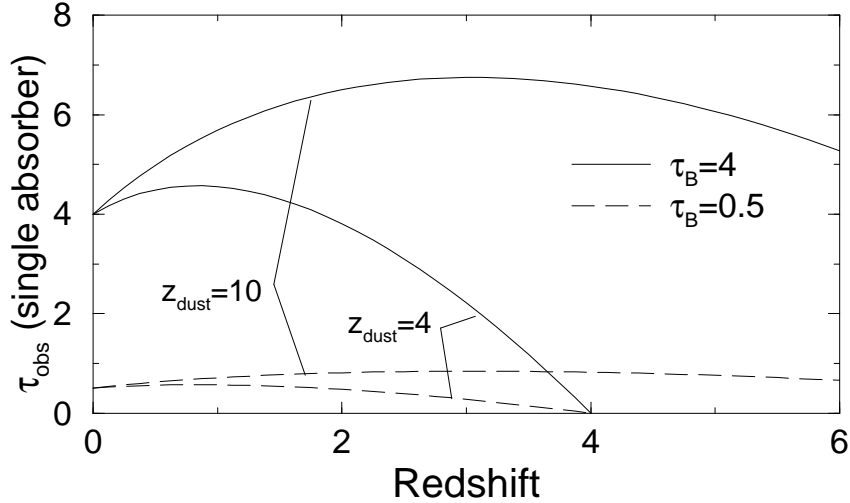}
\vspace{-2.6in}
\caption[``Observed'' optical depth through a single model absorber]{
Optical depth in an observer's $B$-bandpass as a function 
of redshift through a single model absorber defined by Eqn.~\ref{tboz}.
$\tau_{B}$ is the rest frame central $B$-band optical depth and 
$z_{dust}$ the dust formation epoch.
}
\label{singleabs}
\end{figure}

The redshift dependence in optical depth observed in the fixed
$B$-bandpass due to a {\it single} absorber now involves two factors.
First, the extinction properties of the dust as defined by Eqn.~\ref{tz} 
and second, its evolution specified by Eqn.~\ref{tbz}.
We replace $z_{SF}$ (the star formation epoch) by a hypothesised ``dust
formation epoch'' $z_{dust}$.
By convolving Eqns.~\ref{tz} and~\ref{tbz}, and requiring
that locally; $\tau_{0}(z=0)\,=\,\tau_{B}$,
the ``observed'' optical depth through a {\it single} absorber at some redshift
$z<z_{dust}$ now takes the form: 
\begin{equation}
\tau_{0}(z)\,=\,\tau_{B}(1 + z)^{1.5}\left[1 - 
{\ln(1+z)\over\ln(1+z_{dust})}\right].
\label{tboz}
\end{equation}
Fig.~\ref{singleabs} illustrates 
the combined effects of evolution and increase in 
observed frame $B$-band extinction with redshift defined by Eqn.~\ref{tboz}.
The extinction
initially increases with $z$ due to a decrease in corresponding 
rest
frame wavelength. Depending on the value for $z_{dust}$, it then
decreases due to evolution in dust content. This latter effect dominates
towards 
$z_{dust}$.

The characteristic galactic dust radius $r_{0}$ defined in Eqn.~\ref{expr} 
is also given a redshift dependence in the sense that galaxies
had smaller dust-haloes at earlier epochs.
The following evolutionary form is adopted:
\begin{equation}
r_{0}(z)\,=\,r_{0}\,(1+z)^{\delta},{\rm\hspace{7mm}}\,\delta < 0,
\label{rozev}
\end{equation}
where $\delta$ gives the rate of evolution and $r_{0}$ is 
now a `local' scale
radius. 
Evolution in radial dust extent is suggested by
dynamical models of
star formation in an initially formed protogalaxy
(Edmunds, 1990 and references therein).
These studies show that the star formation rate and
hence metallicity in disk galaxies has a radial dependence
that decreases outwards at all
times. 
It is thus quite plausible that galaxies have an evolving effective
`dust radius' which follows chemical enrichment from stellar
processes.
 
Our parameterisation for evolution in galactic dust (Eqns~\ref{tbz} 
and~\ref{rozev}) 
is qualitatively similar to the
`accretion models' for chemical evolution of Wang (1991),
where the effects of grain destruction by supernovae and
grain formation in molecular clouds is taken into account.
This model is also consistent with empirical age-metallicity
relationships inferred from stellar spectral observations
in the Galaxy (Wheeler, Snedin \& Truran 1989), and models
of chemical evolution on a cosmic scale implied 
by absorption-line observations of quasars 
(Lanzetta \etal 1995; Pei \& Fall, 1995). 
 
\section{Model Parameters and Results}
\label{mprev}

\subsection{Model Parameters}
\label{param} 

Our model depends on
four independent parameters which describe the 
characteristics and evolutionary properties of intervening galaxies. 
The parameters defined `locally' are: the comoving number density 
of galaxies; $n_{0}$, the characteristic dust radius; $r_{0}$ and
dust opacity $\tau_{B}$ at the center of an individual absorber. 
The evolution in $\tau_{B}$ and $r_{0}$ is defined by Eqns~\ref{tbz} 
and~\ref{rozev} respectively. Parameters defining their evolution are
$\delta$ for $r_{0}$, and the 
`dust formation epoch' $z_{dust}$ for $\tau_{B}$.  
Both $n_0$ and $r_0$ have been conveniently combined into the 
parameter $\tau_{g}$ where
\begin{equation}
\tau_{g}\,=\,n_{0}\,\pi r_{0}^{2}\,{c \over H_{0}},
\label{tg2}
\end{equation}
with ${c \over H_{0}}$ being the Hubble length.
This parameter is proportional to the number 
of galaxies and mean optical depth introduced 
along the line-of-sight (see section~\ref{resan}).
It also represents a
`local' covering factor in dusty galactic disks - the fraction of sky
at the observer covered in absorbers.

In all calculations, we assume a fixed value for $n_{0}$.
From Eqn.~\ref{tg2}, any evolution in the comoving
number density $n_{0}$ is included in the evolution parameter $\delta$
for $r_{0}$ (Eqn.~\ref{rozev}).
Thus in general, $\delta$ represents 
an effective evolution parameter for both $r_{0}$ and $n_{0}$. 
Our model is therefore specified
by four parameters: 
$\tau_{g}$, $\tau_{B}$, $\delta$ and $z_{dust}$.

\subsection{Results and Analysis}
\label{resan}

\begin{figure}
\vspace{-2.3in}
\plotonesmall{1}{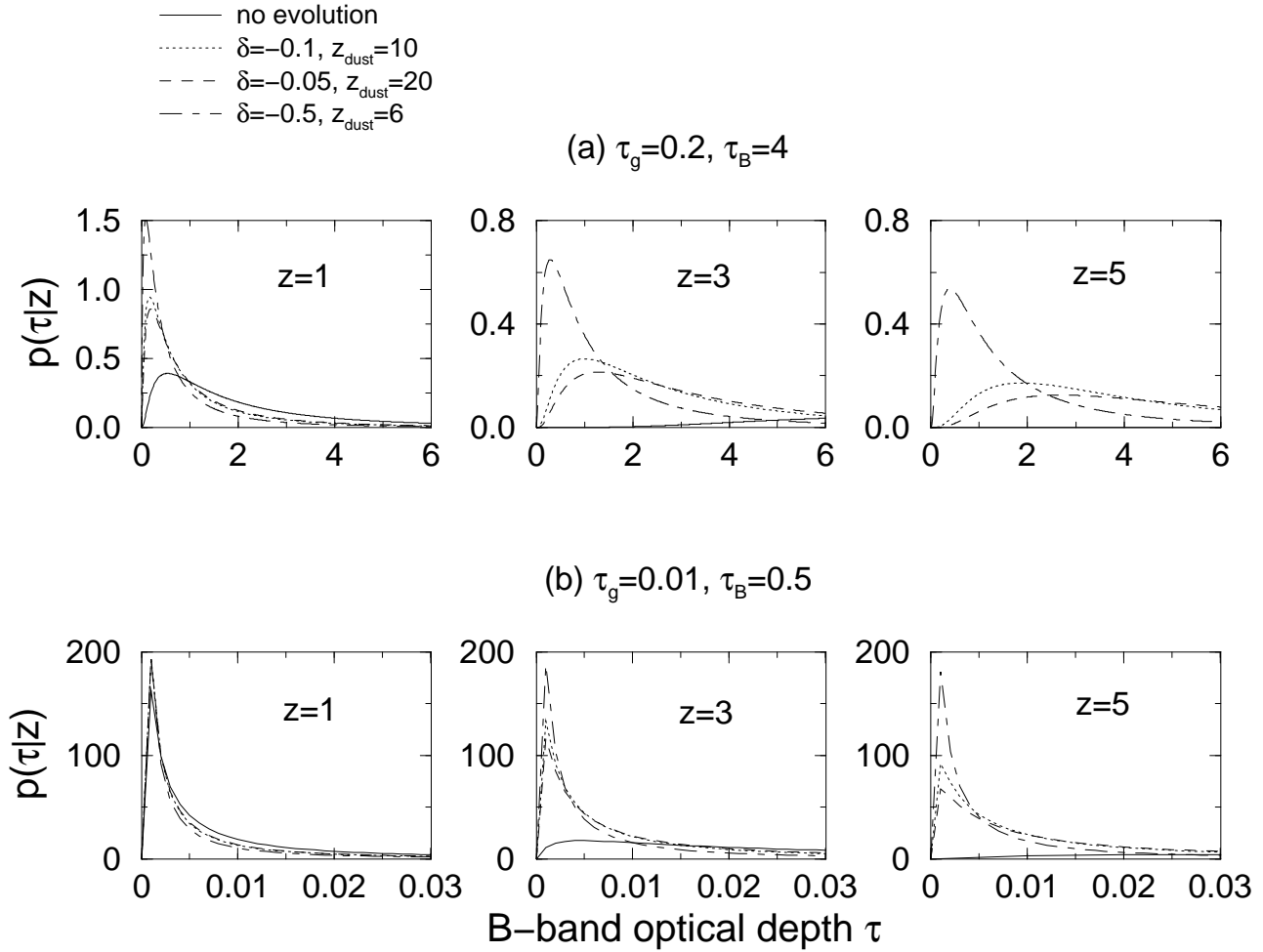}
\vspace{-1.5in}
\caption[Optical depth probability distribution functions in evolutionary model]{Optical depth probability distribution functions $p(\tau|z)$
to redshifts $z=$1,3 and 5, where $\tau$ is the total optical depth
observed in the $B$-band. 
Two different sets of galaxy parameters ($\tau_{g}$, $\tau_{B}$) are considered:
(a) (0.2,4) and (b) (0.01,0.5) (see section~\ref{param}). 
For each of these, we show four evolutionary models specified by
($\delta$, $z_{dust}$). `No evolution' corresponds to $\delta=0$ and
$z_{dust}=\infty$ and the `Strongest evolution' to 
$\delta=-0.5$ and $z_{dust}=6$.}
\label{pdfs2}
\end{figure}

Using the formalism of Chapter 3 (ie. Eqns~\ref{pszfour} - \ref{ptf}), 
Fig.~\ref{pdfs2} shows
probability density functions $p(\tau\, |\,z)$ for the
total optical depth up to redshifts $z=$1, 3 and 5 for two sets of
galaxy parameters ($\tau_{g}$, $\tau_{B}$), and four sets of 
evolutionary parameters ($\delta$, $z_{dust}$) for each. 
The values ($\tau_{g}$, $\tau_{B}$) were chosen
to bracket the range available from existing observations.
In studies investigating the effects of intervening galaxies on
background quasars, Heisler \& Ostriker (1988) chose
$\tau_{g}=0.2$ and $\tau_{B}=0.5$, with 
$\tau_{g}$ (Eqn.~\ref{tg2}) corresponding to a scale radius $r_{0}=33$kpc,
for a nominal 
comoving density $n_{0}=0.002h_{50}^{3}{\rm Mpc^{-3}}$. 
On examination of the 
literature, Fall \& Pei (1992) 
however considered the ranges: $0.005\leq\tau_{g}\leq0.05$ and 
$0.5\leq\tau_{B}\leq5$. This range for $\tau_{B}$ is
consistent with values found by
Rix \& Rieke (1993),
Giovanelli \etal (1994), Valentijn (1994) and Xu \& Buat (1995). 
 
The values ($\delta$, $z_{dust}$) were 
chosen to cover a range of evolution strengths for $r_{0}$ and
$\tau_{B}$ respectively. 
To cover a plausible range of dust formation epochs,
we consider $5\leq z_{dust}\leq20$,
consistent
with a range of galaxy `formation' epochs predicted by existing 
theories of structure formation (eg. Peebles, 1989).
The upper bound $z_{dust}=20$ corresponds to 
the star formation epoch considered in the galaxy formation models 
of Blain \& Longair (1993b).
The values considered for $\delta$ ($-0.5<\delta<-0.05$) 
may be compared with those implied by 
observations of the space density of
metal
absorption systems from QSO spectra as a
function of redshift (Sargent, Boksenberg \& Steidel, 1988; Thomas \&
Webster 1990).
A decrease in absorber numbers is often implied, suggesting fewer 
(or less) chemically enriched regions at high redshift.
Present estimates on the evolution
of absorber numbers however are poorly constrained.
Thomas \& Webster (1990) have combined several datasets increasing
absorption redshift ranges to give
strong constraints on evolution models.
For C{\small IV} absorption, 
their values for the evolution parameter $\gamma$ (where
$dN/dz\propto(1+z)^{\gamma}$-see section~\ref{evp} for more details)
correspond in our model, at the $1\sigma$ level to
$\delta=-0.85\pm0.5$ and $-1.7\pm1$
for rest frame equivalent widths
$W_{0}\simgt0.3$\AA$\,$ and $W_{0}\simgt0.6$\AA$\,$ 
respectively.

The area under any normalised 
curve in Fig.~\ref{pdfs2} gives the fraction
of lines-of-sight to that redshift which have optical
depths within some interval $0\rightarrow\tau_{max}$. 
Towards high
redshifts, we find that 
obscuration depends most sensitively on the parameter $\tau_{g}$, in
other words, on the covering factor of absorbers
(Eqn.~\ref{tg2}). 
Figure~\ref{pdfs2} shows that as the amount
of dust at high redshift increases, ie., as $\delta$ and $z_{dust}$ 
increase,
the curves show little horizontal shift
towards larger optical depths from $z=1$ to $z=5$.
A significant shift becomes 
noticeable however for the weaker evolution cases, and is largest 
for `no evolution' (solid lines). 
This behaviour is further investigated below. 

In order to give a clearer comparison between the 
amount of obscuration and strength of evolution implied by our model
parameters ($\tau_{g},\,\tau_{B},\,\delta,\,z_{dust}$), 
we have calculated the mean and variance in
total optical depth as a function of redshift.
Formal derivations of these quantities are
given in 
Appendix C. Here we are mainly concerned with their general dependence
on the model parameters. 
 
A useful quantity first worth noting is the number of galaxies
intercepted along the line-of-sight. In a $q_{0}=0.5$ ($\Lambda=0$)
universe, the average number of intersections
within a scale length $r_{0}$ of a galaxy's center by a light ray
to some redshift is given by 
\begin{equation}
\bar{N}(z)\,=\,\left({2\over3+4\delta}\right)\tau_{g}
\left[(1+z)^{1.5\,+\,2\delta}
- 1\right].
\label{nzev}
\end{equation}
Where $\delta$ and $\tau_{g}$ are defined in Eqns~\ref{rozev}
and~\ref{tg2} 
respectively. 
 
The mean optical depth due to intervening dusty galaxies to some
redshift $z$ is found to scale as
\begin{equation}
\bar{\tau}(z)\,\sim\,\tau_{g}\,\tau_{B}\,{(1+z)^{2\delta+3}\over
2\delta+3}\,\left[1 - {\ln(1+z)\over(2\delta+3)\ln(1+z_{dust})}\right],
\label{tauscale}
\end{equation}
where $\tau_{B}$ is the optical depth at the center of a local galaxy.
An exact expression for $\bar{\tau}$ follows for the no
evolution case:  
\begin{equation}
\bar{\tau}(z)\,=\,0.67\,\tau_{g}\,\tau_{B}\left[(1+z)^{3} -
1\right].
\label{taunoev}
\end{equation}

A more convenient measure of the obscuration is the scatter or variance 
in optical
depth about the mean to some redshift. 
The mean optical depth has a simple linear dependence on the parameters
$\tau_{g}$ and $\tau_{B}$ and thus gives no indication of the degree
to which each of these parameters contributes to the scatter. 
As seen from the probability distributions
in Fig.~\ref{pdfs2}, 
there is a relatively large scatter about the mean optical depth 
to any
redshift. 
The variance is found to scale as
\begin{equation}
\sigma^{2}_{\tau}(z)\,\sim\,\tau_{g}\,{\tau_{B}}^{2}\,
{(1+z)^{2\delta+4.5}\over2\delta+4.5}\,{\left[1 -
{\ln(1+z)\over(2\delta+4.5)\ln(1+z_{dust})}\right]}^{2}.
\label{varev}
\end{equation}
In the no evolution case, the variance is given by the exact expression:
\begin{equation}
\sigma^{2}_{\tau}(z)\,=\,
0.44\,\tau_{g}\,{\tau_{B}}^{2}\,\left[(1+z)^{4.5} - 1\right].
\label{varnoev}
\end{equation}
From Eqn.~\ref{varev} or ~\ref{varnoev}, 
it is seen that the strongest dependence 
of the variance is on the central absorber optical depth $\tau_{B}$.
Larger values of $\tau_{B}$ (corresponding to `harder-edged' disks), 
are
likely to introduce considerable scatter amongst random individual
lines of sight, even to relatively low redshift.  

\begin{figure}
\vspace{-2.3in}
\plotonesmall{0.9}{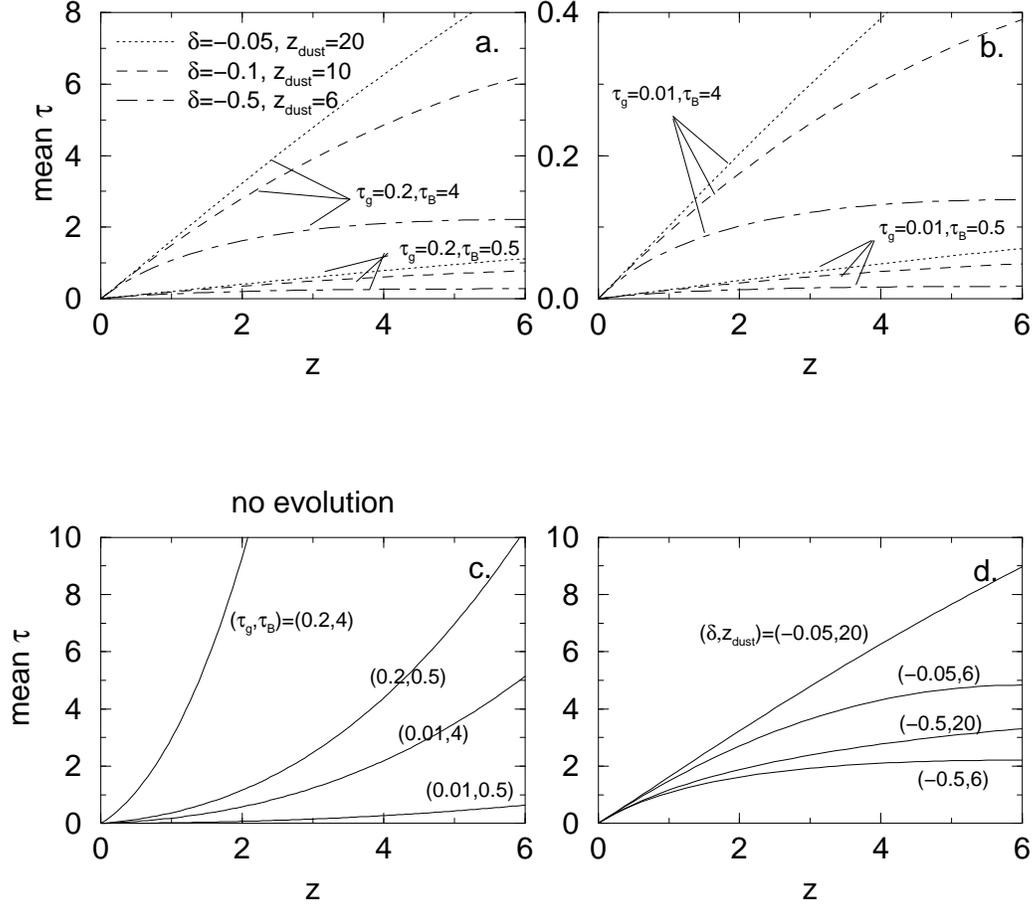}
\vspace{-1.1in}
\caption[Mean optical depth as a function of $z$ in evolutionary model]{
Behaviour in mean reddening,
$\langle\tau\rangle$, as a function of
redshift for a range of model parameters ($\tau_{g}$, $\tau_{B}$)
and ($\delta$, $z_{dust}$).
(a) For ($\tau_{g},\tau_{B}$)=(0.2,4) and (0.2,0.5), (b) Same as (a) but for
$\tau_{g}=0.01$,
(c) Redshift dependence of mean reddening in {\it no evolution} model
for a range of parameters ($\tau_{g}$, $\tau_{B}$).
(d) Scaling of the mean reddening with respect to the evolutionary
parameters ($\delta$, $z_{dust}$). ($\tau_{g}$, $\tau_{B}$)
are fixed at (0.2,4).}
\label{meantaus}
\end{figure}

In Fig.~\ref{meantaus}, we show how the mean optical depth varies as a
function of redshift for a range of evolutionary parameters.
`Strong evolution' is characterised by $\delta=-0.5$, $z_{dust}=6$
(dot-dashed curves),
as compared to the `no', `weak' and
`moderate' evolution cases indicated.
The mean optical depth
flattens out considerably towards high redshift in the strong evolution case, 
and 
gradually steepens
as $\delta$ and $z_{dust}$ are increased.
Note that no such flattening is expected in mean reddening for the
no evolution case (Fig.~\ref{meantaus}c).
The mean optical depth to redshifts $z\simgt1$ in evolution models can be 
reduced by factors of at least four, even for low to moderately low
evolution strengths. 

\begin{figure}
\vspace{-2.8in}
\plotonesmall{1}{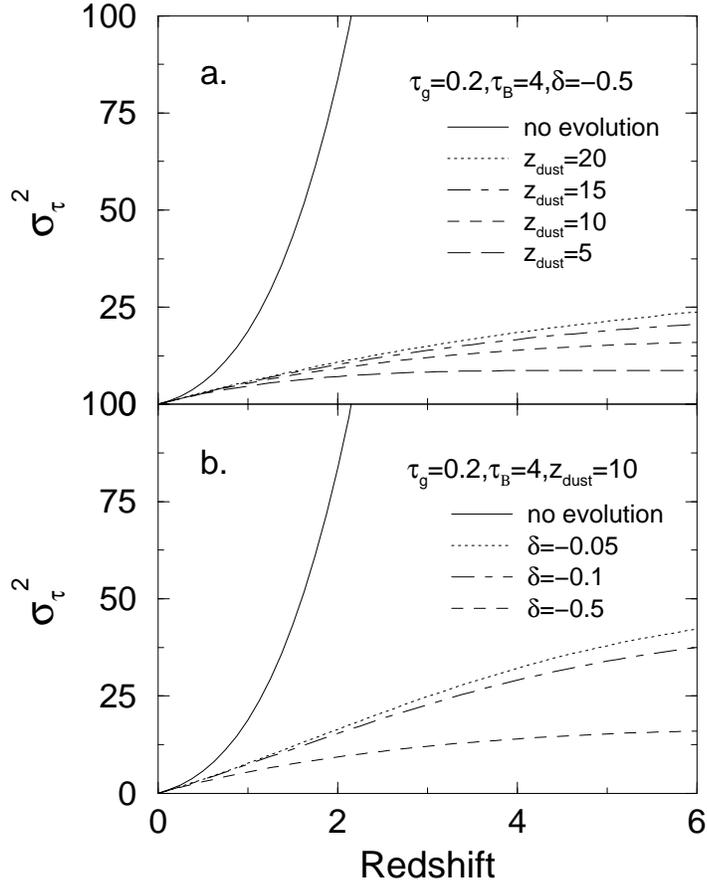}
\vspace{-1.5in}
\caption[Variance in optical depth as a function of redshift]{
Variance ($\sigma_{\tau}^{2}$) in optical depth
as a function of redshift showing scaling with respect to
the evolution parameters ($\delta$, $z_{dust}$). ($\tau_{g},\tau_{B}$)
are fixed at (0.2,4). (a) $\delta$ fixed at -0.05 and
$z_{dust}$ is varied.
(b) $z_{dust}$ fixed at 10 and $\delta$ is varied.
}
\label{vars}
\end{figure}

Figure~\ref{meantaus}d 
shows the scaling of mean optical depth with respect to
the evolutionary parameters. 
It is seen that obscuration depends most sensitively on the parameter
$\delta$, which controls the rate of evolution in galactic dust scale radius 
$r_{0}$.
A similar trend is followed in Fig.~\ref{vars}, which shows the dependence of
variance in optical depth on evolution as a function of redshift,
for fixed ($\tau_{g}$, $\tau_{B}$). Considerable
scatter is expected if the dust radius of a typical galaxy evolves
slowly with cosmic time as shown for the `weakest' 
evolution case $\delta=-0.05$ in Fig.~\ref{vars}. 

Our main conclusion is that the inclusion of evolution in dust content,
by amounts
consistent with other indirect studies can dramatically reduce
the redshift dependence of total reddening along the line-of-sight
to $z\simgt1$, contrary 
to previous non-evolving predictions. 

\section{Implications}
\label{impev}

This section explores some consequences of obscuration by 
intervening galactic dust as predicted by our generalised
model above. We investigate two issues: First, implications 
on QSO absorption line statistics due to obscuration of 
background QSOs by dust in absorption systems
and second, on evolution of the comoving quasar number density with redshift. 

\subsection{Quasar Absorption Line Systems} 
\label{qabsd}

Due to their strength, the heavy element absorption systems 
in QSO spectra studied in depth are the
Mg{\small II} and C{\small IV} 
doublet lines (Tytler \etal 1987; Sargent, Boksenberg
\& Steidel, 1988).
A major step in understanding the origin of these systems is to
tie the properties of the absorbers to the properties of galaxies.
It has been found that $\sim80\%$ of known Mg{\small II} absorbers are 
associated with galaxies, confirming the hypothesis that
metal line systems do in fact arise from intervening galaxies
(Bergeron \& Boisse, 1991 and references therein).

One of the major concerns in QSO absorption line surveys 
is the existence of a selection bias due to obscuration by dust.
As discussed by Pettini \etal (1994), this bias is based 
on the fact that metal line absorption systems are usually detected 
in quasars which need to be relatively bright
in the optical (ie.$B\simlt19$) to enable accurate
spectroscopic confirmation.
Those absorbers that are significantly dusty and
hence presumably metal rich may obscure some of the
quasars and hence escape detection.
This will lead to an underestimate of inferred
absorber numbers and hence
mean metallicity towards high $z$.
 
In this section, we compute the distribution 
in equivalent width (EW) of metal absorption line systems predicted 
by our evolving dust-galaxy model.
We also estimate the
fraction of metal line systems that may have been missed
from existing surveys due to obscuration by dust.

We assume absorbers to be associated with {\it spherical} haloes 
where the dust optical depth through an absorber 
follows the exponential profile Eqn.~\ref{expr}. 
Allowing for non-uniformity, the number of absorbers along the line-of-sight
is assumed to follow a Poisson distribution. In other words, 
the probability that an absorber is intersected within a distance $r$
of its center along a line-of-sight to redshift $z$ is
given by:
\begin{equation}
P(<r,z)\,=\,\bar{N}(r,z)\exp\left[-\bar{N}(r,z)\right], 
\label{prob}
\end{equation}
where $\bar{N}(r,z)$ represents the mean number of intersections
within $r$.
In a complete sample, unbiased with respect to obscuration by dust,
we shall define $\bar{N}=\bar{N}_{true}(r,z)$ where
\begin{equation}
\bar{N}_{true}(r,z)\,=\,n_{0}\pi r^{2}\left({c\over
H_{0}}\right)\int_{0}^{z}{(1+z')
\over (1+2q_{0}z')^{1/2}}\,dz'
\label{Nvszgen}
\end{equation}
(see Appendix C). 
As before, we assume a direct proportionality between heavy metal 
abundance and the amount of dust extinction.
To convert from optical depth through a single absorber to an absorption line
EW measure, we make use of the ``local'' calibration: 
\begin{equation}
W_{0}({\rm C{\small IV}})\,\simeq\,(5-15)E_{B-V}{\rm\AA\,mag^{-1}}, 
\label{EWvsEBV}
\end{equation}
found along a number of galactic and LMC sight lines
(Ostriker, Vogeley \& York, 1990 and references therein). 
We shall follow these authors who used Eqn.~\ref{EWvsEBV}
to infer the amount of dust reddening associated with QSO absorption
line systems. 
We assume the mean value 
$W/E_{B-V}\simeq10{\rm\AA\,mag^{-1}}$ in our calculations.

The fraction of background QSOs at a redshift $z_{qso}$ exhibiting
absorption lines 
with rest frame EWs greater than some value $W_{0}$,
is simply the fraction of QSOs whose line-of-sight passes
within some radius $r$ of an exponential dust profile such that
$W>W_{0}$ for radii $<r$.
Converting our rest frame central $B$-band optical depth $\tau_{B}$
into $E_{B-V}$ assuming a galactic extinction law, Eqn.~\ref{tbz}
can be re-written:
\begin{equation}
\label{Ez}
E_{B-V}(z)\,=\,(0.21)\left[1 - 
{\ln(1+z)\over\ln(1+z_{dust})}\right]. 
\end{equation}
Our exponential dust profile at any $z$ which includes evolution in the 
scale radius $r_{0}$ can be written
\begin{equation}
\label{Er}
E_{B-V}(r,z)\,=\,E_{B-V}(z)\exp\left[-\frac{r}{r_{0}(1+z)^{\delta}}\right]. 
\end{equation}
$z_{dust}$ and $\delta$ are our evolutionary model parameters as 
defined in section~\ref{param}.
Combining Eqns.~\ref{EWvsEBV},~\ref{Ez} and~\ref{Er}, we find
that an absorber at redshift $z$ will give rise to an absorption line with 
EW greater than some value 
$W_{0}$ if it is intercepted within a radius: 
\begin{equation} 
\label{rW}
r(W_{0},z)\,=\,-r_{0}(1+z)^{\delta}\ln\left(\frac{W_{0}}{2.1\left[
1-{\ln(1+z)\over\ln(1+z_{dust})}\right]}\right). 
\end{equation}
The ``true'' (or unbiased) mean number
of absorber intersections
to a redshift $z_{qso}$ (Eqn.~\ref{Nvszgen}),
which yield EWs $>W_{0}$ can now be written
\begin{equation}
\label{Nz2}
\bar{N}_{true}(W_{0},z_{qso})\,=\,n_{0}\pi\left({c\over
H_{0}}\right)\int_{0}^{z_{qso}}{r(W_{0},z')^{2}(1+z')^{1+2\delta
}\over (1+2q_{0}z')^{1/2}}\,dz', 
\end{equation}
where $r(W_{0},z')$ is defined by Eqn.~\ref{rW}. 

To determine the fraction of QSOs with absorption line EWs$>W_{0}$ 
in an optical flux-limited QSO sample, we must account for the bias
due to obscuration by dust. In other words, the mean number of
absorption systems intercepted to `bright' optically selected QSOs is expected
to be less than $\bar{N}_{true}$ defined by Eqn.~\ref{Nz2}. 
Let us define the ``observed'' mean number of intersections 
as
\begin{equation}
\label{Nf}
\bar{N}_{obs}\,=\,f\bar{N}_{true}, 
\end{equation}
where $f$ is the `dust bias' factor.
To some redshift and for a fixed absorber optical depth $\tau_{B}$,  
the ``true'' (unbiased) 
mean optical depth, $\bar{\tau}_{true}$,
is directly proportional to the mean number of absorbers 
intercepted $\bar{N}_{true}$ (cf. Eqns.~\ref{nzev} and~\ref{taunoev}), 
since both depend linearly on the covering factor parameter $\tau_{g}$. 
For a fixed QSO distribution, this implies that $\bar{\tau}_{obs}\propto\bar{N}_{obs}$
will also be true, where $\bar{\tau}_{obs}$ is the
``observed'' mean optical depth to optically detected sources
(see section~\ref{redbiasopt}).
Comparing the functional forms of $\bar{N}_{true}$, 
$\bar{\tau}_{true}$ and $\bar{\tau}_{obs}$
(eg. Eqns.~\ref{nzev},~\ref{tmean} and ~\ref{tobsa}),
one can easily (and crudely) guess an approximate functional form for 
$\bar{N}_{obs}$. Given this,
it can be shown that for redshifts $z\simlt3$ and optical 
depths $\tau_{B}\simlt1$,
the bias factor $f$ will scale as:  
\begin{equation}
\label{bias}
f\,=\,\frac{\bar{N}_{obs}}{\bar{N}_{true}}\,\simeq\,
\frac{\bar{\tau}_{obs}}{\bar{\tau}_{true}}.
\end{equation}
For a given set of absorber dust parameters and QSO luminosity function 
slope, $f$ can be easily computed from the observed and
true mean optical depth following the formalism in section~\ref{redbiasopt}. 

\begin{figure}
\vspace{-3.8in}
\plotonesmall{1}{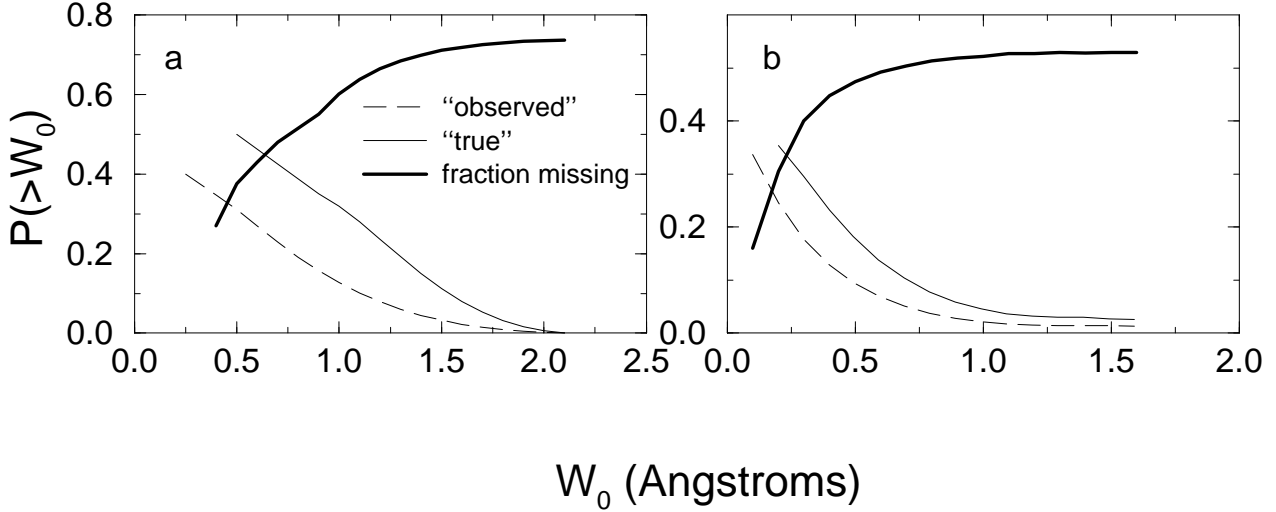}
\vspace{-2.1in}
\caption[Predicted equivalent width distribution and incompleteness in 
metal absorption line studies]{
Cumulative distribution of rest frame equivalent width for metal line
absorption systems (thin lines) where dashed and solid lines
refer to the `observed' and `true' prediction (ie. with and without the
effects of dust obscuration bias in optical flux-limited QSO samples)
respectively. Thick lines represent the fraction of absorbers
with EWs $>W_{0}$ that are missing from a flux-limited sample as a
function of $W_{0}$.
The range of $W_{0}$ considered is typical of that observed for 
C{\small IV} and Mg{\small II}
absorption.
A mean QSO redshift of $\bar{z}=3$ is assumed. Absorbers are modelled with
dust parameters: $(r_{0},n_0,\tau_{B})=(30{\rm kpc},
0.002h_{50}^{3}{\rm Mpc^{-3}},1)$.
(a) The predictions assume no evolution in these parameters.
(b) Same as (a), except that `moderate' 
evolution in dust properties is assumed
where $(\delta,z_{dust})=(-1,5)$.
}
\label{metalEWdist}
\end{figure}
 
To summarise, the ``true'' (unbiased) fraction of QSOs with absorption
line EWs $>W_{0}$ is computed
by combining Eqns.~\ref{prob},~\ref{rW} and~\ref{Nz2}, 
while the ``observed'' fraction (corrected for
dust obscuration) involves use of the bias factor $f$ (Eqns.~\ref{Nf} 
and~\ref{bias}), where $\bar{N}$ is replaced by $\bar{N}_{obs}$ in
Eqn.~\ref{prob}.

Figure~\ref{metalEWdist} shows the cumulative EW distribution
of a typical metal absorption line (eg. either C{\small IV} or Mg{\small II})
expected from a population
of quasars at $z_{qso}\sim3$ in our model.
The dust parameters assumed for model absorbers are an exponential scale
radius $r_{0}=30$kpc and central optical depth $\tau_{B}=1$ 
$(E_{B-V}\sim0.2)$. A comoving number density of 
$0.002h_{50}^{3}{\rm Mpc^{-3}}$
is assumed, similar to that inferred for local galaxies. 
Fig.~\ref{metalEWdist}a refers to {\it no evolution} in these
parameters with $z$, while Fig.~\ref{metalEWdist}b assumes a 
`moderate' evolution
in both $r_{0}$ and $\tau_{B}$.
Each plot shows the true distribution expected in a complete sample
unbiased with respect to obscuration by dust and that expected in an
optical flux-limited QSO sample.
The latter assumes a QSO cumulative luminosity function 
slope $\beta=2.5$. 
 
The heavy solid lines represent the fraction
of absorption lines with EWs$>W_{0}$ missing from a
flux-limited QSO sample relative to the `true' number expected
in the absence of dust obscuration.
As expected, the incompleteness in the number of absorption lines
as a function of EW increases towards the highest EWs.
This can be explained by lines-of-sight passing 
close to an absorber's center  
where relatively more dust is intercepted in its 
assumed exponential dust profile. 
Thus, if the dust distribution in QSO absorbers follows an
exponential-type law, then current absorption line surveys are more likely
to probe their outer regions (at $r\simgt r_{0}\sim10-30$kpc).
The inclusion of moderate amounts of evolution in
absorber properties (Fig.~\ref{metalEWdist}b),
only slightly reduces 
the incompleteness in absorber numbers.

Since the dust properties assumed for model absorbers in 
Fig.~\ref{metalEWdist} are at the upper limits of existing observations,
we conclude
that current metal absorption line surveys
may have missed up to 40\% of lines with EWs$\simgt0.5$\AA.
For comparison, Fall \& Pei (1995) find that up to to 60\% 
of damped Ly$\alpha$ systems at redshifts $z\simlt2$ 
may have been missed due to
obscuration by associated dust. These latter systems
however may correspond to a totally
different class of absorber. If they are to be interpreted
as arising from the gaseous disks of galaxies,
then they may also probe
the dustiest regions,
where obscuration is more efficient. 

\subsection{Quasar Evolution and Statistics}

There are numerous observations suggesting that the
space density
of bright quasars declines beyond $z\approx3$
(Sandage, 1972; Schmidt, Schneider \& Gunn, 1988).
This has been strongly confirmed from various 
luminosity function (LF) estimates to $z\sim4.5$ (Hartwick \& Schade, 1990;
Pei, 1995 and references therein), where 
the space density is seen to decline by at least an order of magnitude
from $z=3$ to $z=4$.
Heisler \& Ostriker (1988) speculate that the decline may be due to
obscuration by intervening dust, which reduces the number of quasars
observed by ever-increasing amounts towards high $z$. 
The results of Fall \& Pei (1993) however show that the observed
turnover at $z\sim2.5$ and decline thereafter may still exist once
the effects of intervening dust (mainly associated with damped Ly$\alpha$
systems) are corrected for. 
Since no evolution in dust content was assumed in either of these studies,
we shall further explore the effects of intervening dust on 
inferred quasar evolution using our evolutionary galactic dust model.

Since we are mainly interested in ``bright'' quasars ($M_{B}\simlt-26$)
at high redshifts, a single power-law for the observed LF should suffice:
\begin{equation}
\label{oLF}
\phi_{o}(L,z)\,=\,\phi_{\ast o}(z)L^{-\beta-1},
\end{equation} 
where $\beta\simeq2.5$.
This power law model immensely simplifies the relation 
between observed and ``true'' LFs (corrected for obscuration by dust).
In the presence of dust obscuration, the observed LF can be written in
terms of the true LF, $\phi_{t}$ as follows:
\begin{equation}
\label{oLFtLF}
\phi_{o}(L,z)\,=\,\int^{\infty}_{0}d\tau\,\phi_{t}(e^{\tau}L,z)
e^{\tau}p(\tau\,|\,z)
\end{equation}
(see section~\ref{simplepl}), where
$p(\tau\,|\,z)$ is our optical depth
probability distribution for evolving intervening dust (see Chapter 3). 
The extra factor of $e^{\tau}$ in Eqn.~\ref{oLFtLF}
accounts for a decrease in luminosity 
interval $dL$ in the presence of dust. 
Eqns.~\ref{oLF} and~\ref{oLFtLF} imply that the true LF can be written 
\begin{equation}
\label{tLF}
\phi_{t}(L,z)\,=\,\phi_{\ast t}(z)L^{-\beta-1},
\end{equation}
and the ratio of observed to true LF normalisation as
\begin{equation} 
\label{norm}
\frac{\phi_{\ast o}(z)}{\phi_{\ast t}(z)}\,=\,\int^{\infty}_{0}d\tau\,
e^{-\beta\tau}p(\tau\,|\,z). 
\end{equation}

The observed comoving density of quasars brighter than some
absolute magnitude limit $M_{lim}$ as a function of redshift is
computed by integrating the LF:
\begin{equation}
\label{No}
N_{o}(z\,|M_{B}<M_{lim})\,=\,\int^{\infty}_{L_{lim=L(M_{lim})}}dL\,\phi_{o}(L,z).
\end{equation} 
Thus, the true comoving number density $N_{t}$, can be easily
calculated by replacing $\phi_{o}$ in Eqn.~\ref{No} by
$\phi_{t}\equiv(\phi_{\ast t}/\phi_{\ast o})\phi_{o}$ leading to the
simple result:
\begin{equation}
\label{Nt}
N_{t}(z\,|M_{B}<M_{lim})\,\simeq\,\left(\frac{\phi_{\ast o}(z)}{\phi_{\ast t}(z)}
\right)N_{o}(z\,|M_{B}<M_{lim}),
\end{equation} 
where the normalisation ratio is defined by Eqn.~\ref{norm}. 
 
\begin{figure}
\vspace{-3in}
\plotonesmall{1}{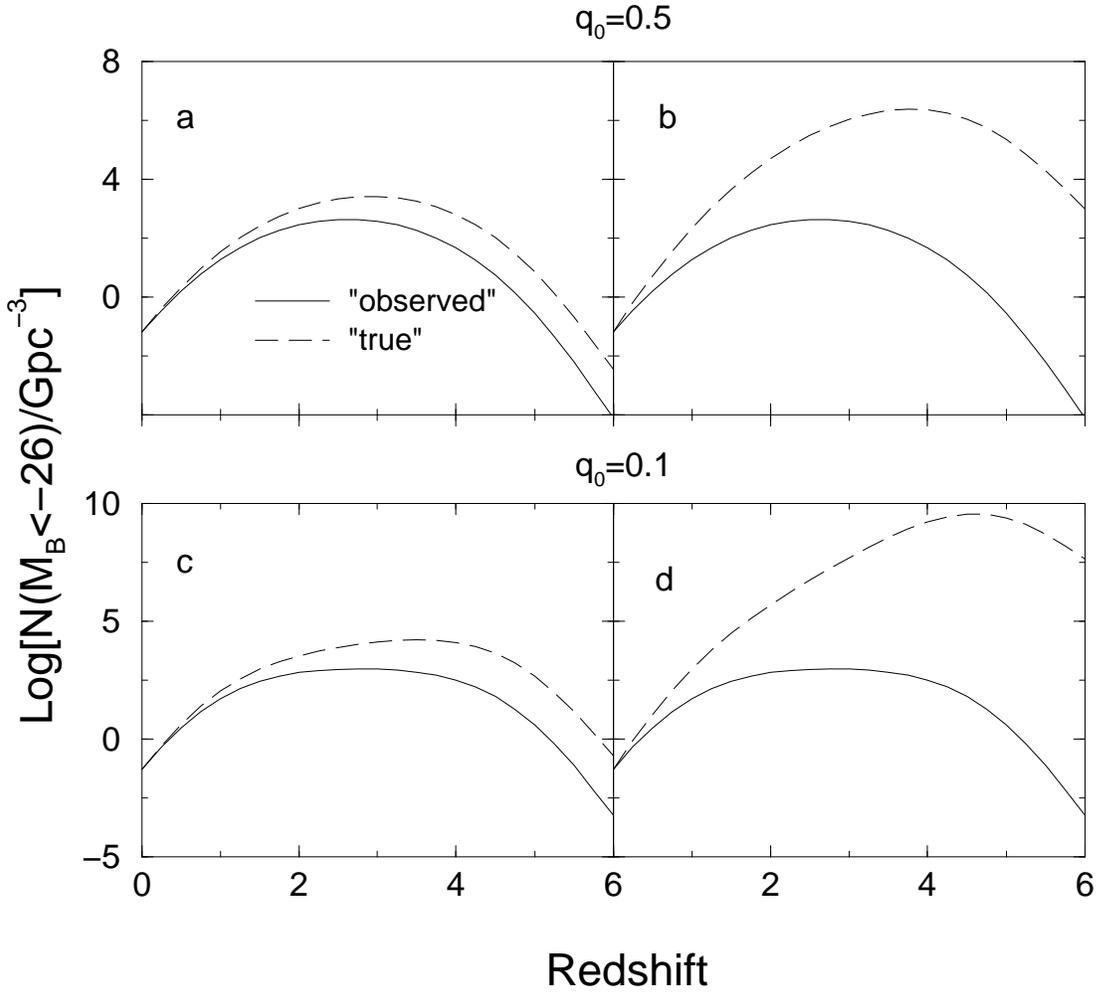}
\vspace{-0.8in}
\caption[Comoving number density of quasars with $M_{B}<-26$ as a function of redshift]{ 
Comoving number density of quasars with $M_{B}<-26$ 
as a function of redshift. Observed trends (solid curves)
are taken from the empirical fits of 
Pei (1995) while dashed curves corrects these trends 
for obscuration by dust. These are predicted assuming 
our
evolving intervening
galactic dust model with $\tau_{B}=1$ and $r_{0}=10$kpc (Figs a
and c) and $\tau_{B}=3$ and $r_{0}=30$kpc (Figs b and d). 
In all cases, we have assumed relatively ``weak'' evolution in 
dust content with $z$, defined by the parameters: 
$z_{dust}=20$ and $\delta=-0.05$. 
}
\label{comov}
\end{figure}

Figure~\ref{comov} shows both the observed and true comoving 
density of bright quasars (with $M_{B}<-26$) as a function of redshift. 
The true comoving density in all cases was determined by assuming relatively
`weak' evolution in the dust properties of intervening galaxies. 
Two sets of galactic dust parameters for each $q_{0}$ defined by
$(\tau_{B},r_{0})=(1,10{\rm kpc})$ (Figs a and c) and 
$(\tau_{B},r_{0})=(3,30{\rm kpc})$
(Figs b and d) are assumed. We shall refer to these as our 
``minimal'' and ``maximal'' dust models respectively which brackets the
range of parameters observed for local galaxies. 

Comparing the `true' QSO redshift distribution with that observed,
two features are apparent.
First, the true number density vs. $z$ relation has qualitatively
the same behaviour as that observed.
No flattening or increase in true quasar numbers with $z$ is apparent. 
Second, there appears to be a shift in the redshift, $z_{peak}$,
where the quasar density peaks. This shift is
greatest for our maximal dust model where $z_{peak}$ is increased
by a factor of almost 1.5 relative to that observed. 
This implies that the bulk of quasars may have formed at earlier
epochs than previously inferred from direct observation.
 
Our predictions for QSO evolution, corrected for obscuration
by `evolving' intervening dust differs
enormously from that predicted by Heisler \& Ostriker (1988).
The major difference is that these authors neglected evolution in dust
content with $z$. As shown in Fig.~\ref{meantaus}, 
non-evolving models lead to a rapid increase
in dust optical depth with $z$ and hence this will explain 
their claim of a continuous increase in the true QSO space density
at $z>3$.
As shown in Fig.~\ref{comov}, the inclusion of even a low-to-moderately low 
amount of evolution in dust content dramatically reduces the excess number
of quasars at $z>3$ than predicted by Heisler \& Ostriker (1988). 

We find that there is no significant difference in the characteristic
timescale, $t_{QSO}$ for QSO formation at $z>z_{peak}$, where
\begin{equation}
\label{time}
t_{QSO}\,\simeq\,\left(\frac{N}{\stackrel{.}N}\right)_{z>z_{peak}}\,\sim\,
1.5\,{\rm Gyr}, 
\end{equation} 
is found for both the observed and dust corrected results in Fig.~\ref{comov}.
We conclude that the decline in space density of bright QSOs
at redshifts $z>3.5$ is most likely to be real and an artifact
of an intrinsic rapid turn-on of the QSO population with time. 
This is consistent with estimates of evolution inferred from
radio-quasar surveys where no bias from 
dust obscuration is expected 
(eg. Dunlop \& Peacock, 1990). 

An increased space
density of quasars 
at redshifts $z>3$ predicted by correcting for dust obscuration
has implications for theories of structure formation in the Universe.
Our minimal dust model 
(Figs.~\ref{comov}a and c) predicts that the true space density
can be greater by almost two orders of magnitude than that observed, while our
maximal dust model (Figs.~\ref{comov}b and d) predicts this
factor to be greater than 5 orders of magnitude. 
These predictions can be reconciled with the quasar number densities
predicted from hierarchical galaxy formation simulations involving 
cold-dark matter
(eg. Katz \etal 1994). 
It is found that there are $>10^{3}$ times
potential quasar sites at $z>4$ (associated with high density peaks) 
than required from current observations.
Such numbers can be 
easily accommodated by our predictions if a 
significant quantity of line-of-sight dust is present. 

To summarise, we have shown that with the inclusion of moderate
amounts of evolution in dust content with $z$, the bias due
to dust obscuration will not be enough to flatten the true 
redshift distribution of bright quasars beyond $z=3$. 
A significant excess however (over that observed) in quasar numbers is 
still predicted.
 
\section{Constraining Model Parameters}
\label{conmodp}

Using the model above, we now constrain our galactic and evolutionary
dust parameters using an observed
reddening distribution. We shall use optical and near-infrared 
observations of a complete sample of `flat spectrum' radio sources  
(Drinkwater \etal 1997; hereafter, the `Parkes sample') 
which is optimal for the effects of dust to high
redshift.
 
\subsection{Observed Reddening Distribution}
\label{obsreddi}
 
The Parkes sample represents a complete subsample 
of the Parkes Catalogue,
with sources chosen to have flat radio spectra ($\alpha<0.5$ where
$S_{\nu}\propto\nu^{-\alpha}$ in the range 2.7-5 GHz) and 2.7 GHz
fluxes $S>0.5$ Jy.
Optical 
identifications show that $\approx70\%$ of the 323 
sources exhibit quasar-like
characteristics, typical of bright, flat spectrum radio sources. 
About 8\% of sources remain optically unidentified to $B_{J}\sim22.5$, and
are classified as `empty fields'.
 
Figure~\ref{colvsz} in section~\ref{dustobscu} shows the 
distribution in $B_{J} - K_{n}$ colour as a function of redshift. 
For comparison, a sample of 28 optically-selected QSOs drawn from the Large
Bright QSO Survey (hereafter the LBQS, Hewett \etal 1995) is also shown. The
Parkes 
sample contains a substantial population of exceedingly
reddened quasars with colours lying in the range
$1<B_{J}-K_{n}<10$. The optically-selected QSOs are strongly clustered
around the value $B_{J}-K_{n}\approx2.5$. 
Approximately 60\% of Parkes sources have colours redder
than the lower envelope defined by the LBQS sample and almost all 
of these have $B_{J}$ magnitudes fainter than the LBQS survey limit, 
$B_{J}\sim19$. We argue (see Webster \etal 1995) 
that the reddening
is due to dust,
and that optically-selected quasars 
are strongly biased against dust-reddened
sources.
We use this observed reddening distribution as a function
of redshift to constrain the parameters of our model.

We must note that the observed $B_{J}-K_{n}$ colours in Fig.~\ref{colvsz}
may be subject to
random errors of $\sim1$ mag and possible systematic errors of $\sim0.5$ mag
(Drinkwater \etal 1997). These primarily arise from uncertainties in 
$B_{J}$ magnitudes, which are based on uncalibrated COSMOS and APM
scans of UK Schmidt plates.
These errors are not expected to 
depend on redshift, and hence
are unlikely to significantly affect our results. 

\subsection{Fitting the Data}

The galactic {\it central} optical depth $\tau_{B}$ we
refer to in our model only applies to that
observed in the $B$-bandpass.
In our case, the observed  
reddening is measured in terms 
of $B_{J}-K_{n}$ colour, with the bandpasses $B_{J}$ and $K_{n}$ 
centered on wavelengths
4750\AA$\,$ and 2.1$\mu m$ respectively.
The corresponding measure of extinction will thus be in terms of a colour
excess $E_{B_{J} - K_{n}}$ defined as
\begin{equation}
E_{B_{J} - K_{n}}\,=\,A(B_{J})\,-\,A(K_{n})\,=
\,1.086\,(\tau_{B} - \tau_{K}),
\label{extBK}
\end{equation}
where $A(B_{J})$ and $A(K_{n})$ 
represent extinctions in magnitudes and $\tau_{B}$,
$\tau_{K}$ are equivalent optical depth measures. 
Using our assumed extinction law $\tau_{\nu}\propto\nu^{1.5}$ 
(see section~\ref{briefdes}),
we find that
$E_{B_{J} - K_{n}}\approx0.97\tau_{B}$.
We replace the galactic central optical depth $\tau_{B}$ in our models above
by this quantity, so that all measures of obscuration will now be
referred in terms of the color excess $E_{B_{J}-K_{n}}$.
 
In order to calculate the amount of reddening suffered by these sources as
a function of redshift,
we need a
measure of the `intrinsic' or unreddened $B_{J}-K_{n}$ colour distribution. 
Since the `bluest' Parkes quasars are no bluer than those in the LBQS
(see Fig.~\ref{colvsz}), we assume the 
unreddened colour in all Parkes quasars at all redshifts 
to be the mean colour observed in the LBQS distribution: $B_{J}-K_{n}\sim2.5$.
This is justified from our modelling in Chapter 3 where `bright' 
optically-selected quasars were shown to be strongly biased against
significant reddening by dust.
We therefore assume that the scatter in the colours of Parkes sources
relative to those in the LBQS is all due to extinction by dust in galaxies 
that may happen to lie along the line-of-sight of otherwise normal
blue quasars.
 
Adopting the colours of LBQS sources as
the `unreddened' colours of Parkes quasars 
is a conservative assumption. The small dispersion
in LBQS colours is most likely due to a 
selection bias and hence may not fully represent the true unreddened colour
distribution of Parkes quasars. 
In other words, since LBQS quasars were all selected with well 
defined selection criteria in the optical (eg. Morris \etal 1991), this
may have resulted in their very uniform optical-to-near-IR
properties.
The spread in $B_{J}-K_{n}$ colour for
LBQS quasars is consistent with systematic errors from 
the $B_J$ and $K_{n}$ photometry (P. Francis, private
communication), and not due to variations in intrinsic source
properties.
The `true' unreddened colour distribution for
radio-quasars however may include some scatter due to variation in intrinsic
properties.
For example, dust associated with the 
Parkes quasars themselves
would 
introduce considerable scatter.
Thus, due to these possibilities, the parameters 
$\tau_{g}$ and $\tau_{B}$ (see section~\ref{param}) describing 
the normalisation in dust content we constrain 
are to be interpreted as upper
limits.
 
We now use the Parkes sample to determine the range of
dust properties of local galaxies as specified
by the parameters $\tau_{B}$ and $\tau_{g}$--respectively the central
optical depth and 
galaxy covering factor,
and their evolution specified by $z_{dust}$ and $\delta$ respectively.
From the definition for $\tau_{g}$ (Eqn.~\ref{tg2}), the
comoving number density $n_{0}$ is held fixed, so that
our results for $\tau_{g}$ will only be used to constrain 
$r_{0}$-the local radial dust extent.
 
To fit these parameters, we compare the
observed reddening distribution in Fig.~\ref{colvsz} with our
model distributions to any redshift
(see Fig.~\ref{pdfs2}) using the likelihood ratio test.
The advantage of this test is that it does not require binning
of the source redshifts or extinctions $E_{B_{J}-K_{n}}$. 
The method we use is as follows.
We define $P_{i} = P(E_{i}|z_{i})$ as the
probability density that the $i^{th}$ quasar 
observed with redshift $z_{i}$
is reddened by its observed amount, $E_{i}$. This
is calculated from our theoretical probability distribution
for a given set of model parameters which we wish to constrain. 
The likelihood function or probability that a model represented by some set of 
parameters $\tau_{B},\tau_{g},z_{dust},\delta$ would result
in reproducing the observed reddening state of all quasars in
question is defined by
\begin{equation}
L(\tau_{B},\tau_{g},z_{dust},\delta)\,=\,\prod^{N_{Q}}_{i=1}P_{i},
\label{likli}
\end{equation}
where $N_{Q}$ is the total number of quasars in the sample.
The best fit parameters maximise $L(\tau_{B},\tau_{g},z_{dust},\delta)$ 
or more conveniently $\ln L$. The $1\sigma$ and $2\sigma$ 
confidence limits on {\it one} single parameter are given by the extrema of the
likelihood contours at 61\% and 14\% of $L_{max}$ respectively
(Press \etal 1994).

\subsection{Results}
\label{pres}
 
Results of the likelihood analysis and $1\sigma$ and $2\sigma$ confidence 
limits on each parameter
are presented in Table~\ref{tab1} for two values of $q_{0}$.
From Eqn.~\ref{tg2}, the local dust scale radius $r_{0}$ is given by
\begin{equation}
r_{0}\,\simeq\,162\,\tau_{g}^{1/2}
\left(\frac{n_{0}}{0.002h_{50}^{3}{\rm Mpc^{-3}}}\right)^{-1/2}h_{50}^{-1}
\,\,{\rm kpc},
\end{equation}
where we have scaled to the average local galactic space density
$n_{0}=0.002h_{50}^{3}{\rm Mpc^{-3}}$, consistent with 
that derived from luminosity function estimates by 
Efstathiou \etal (1988) and Loveday \etal (1992).
The values for $r_{0}$ in Table~\ref{tab1} 
were derived assuming this value for $n_{0}$. 
We find that the model fit is most sensitive to the 
parameter $\tau_{B}$, and least sensitive to $z_{dust}$, 
as shown by the relatively
small and large uncertainties in 
these parameters respectively in Table~\ref{tab1}.
 
\begin{table}
\vspace{-1in}
\begin{center}
\begin{tabular}{|c|c|c|c|}
\hline
\multicolumn{1}{|c|}{} & \multicolumn{3}{|c|}{$q_{0}=0.5$}\\
\hline
parameter & Best & $1\sigma$ & $2\sigma$\\
\hline
\hline
$\tau_{B}$ & 1.5 & 0.8-2.2 & 0.5-2.7\\
$r_{0}(h_{50}^{-1} {\rm kpc})$ & 166 & 112-203 & 60-254\\
$z_{dust}$ & 1.2 & 7.6 & 14.5\\
$\delta$ & $-2.2$ & $-3.1$-$-0.25$ & $-7.3$-$-0.14$\\
\hline
\multicolumn{1}{|c}{} & \multicolumn{3}{c|}{$q_{0}=0$}\\
\hline
\hline
$\tau_{B}$ & 1.6 & 1.1-2.4 & 0.7-2.9\\
$r_{0}(h_{50}^{-1} {\rm kpc})$ & 171 & 108-223 & 61-267\\
$z_{dust}$ & 0.9 & 6.9 & 13.2\\
$\delta$ & $-1.6$ & $-2.1$-$-0.16$ & $-6.2$-$-0.06$\\
\hline
\end{tabular}
\caption[Best fit parameters with $1\sigma$ and $2\sigma$ ranges]{Best fit parameters with $1\sigma$ and $2\sigma$ 
ranges. For $z_{dust}$, the $1\sigma$ and $2\sigma$ values 
correspond to upper limits only.} 
\label{tab1}
\end{center}
\end{table}
 
\begin{figure}
\vspace{-2.2in}
\plotonesmall{1}{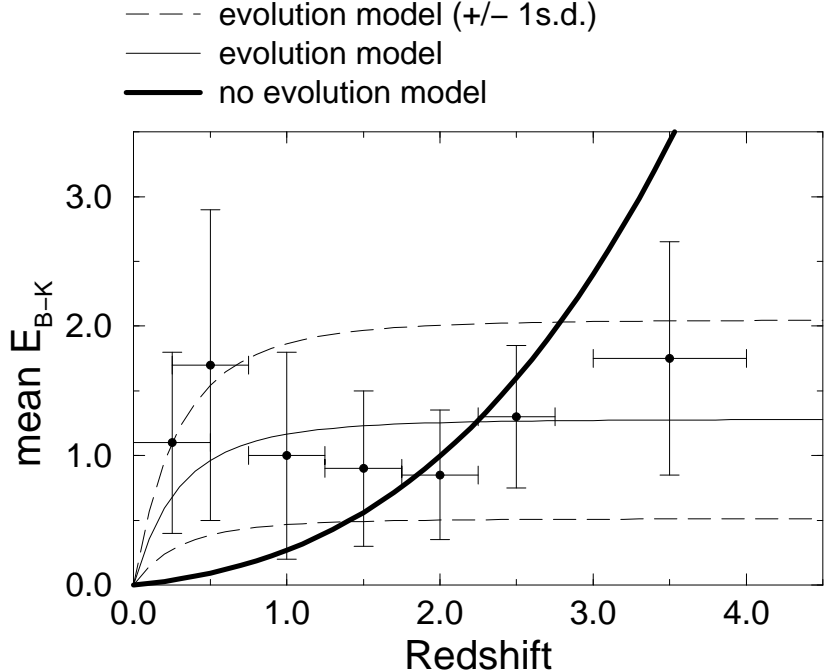}
\vspace{-2.6in}
\caption[Evolutionary model fit to mean reddening of Parkes quasars]
{Mean reddening of Parkes quasars as a function of redshift
(filled circles).  
Vertical bars represent $1\sigma$ values of the 
observed reddening distribution (Fig.~\ref{colvsz}) 
at the redshifts indicated. 
Horizontal bars are redshift bin windths.
Thin-solid and dashed curves represent
the mean reddening and $1\sigma$ range
respectively for our best fit {\it evolution} model with
$q_{0}=0.5$ (see Table~\ref{tab1}). 
Thick curve represents the best fit  
{\it no-evolution} model
(see section~\ref{pres}).}
\label{modfit}
\end{figure}

We have computed the mean
reddening
$\bar{E}_{B_{J}-K_{n}}$ and corresponding $1\sigma$ range in
the `observed' reddening distribution at a range of redshifts.
These are shown as points and vertical bars respectively in Fig.~\ref{modfit}.
Also shown is the best fit {\it evolution} model (thin solid lines) using
the parameters for 
$q_{0}=0.5$ from Table~\ref{tab1}.
The best fit {\it no-evolution} model ($\delta=0,\,z_{dust}=\infty$)
is shown by the thick solid line. This model requires 
$\tau_{g}\simeq0.2$ and $\tau_{B}=0.5$.
The evolutionary model provides a better fit to these observations.
In terms of the $\chi^{2}$ statistic for the
fits in Fig.~\ref{modfit}, the (4 parameter) evolution and 
(2 parameter) no-evolution model
correspond to $\chi^{2}_{min}=2.1$ and 11.5
respectively.
The significance of these values is $\sim92\%$ and $10\%$ respectively.
In other words, if the reddening is truly described by the 
evolution model, then $\sim8\%$ of the time a set of observations
will give a $\chi^{2}$ greater than 2.1.

The redshift dependence of the observed reddening is 
considerably weak, scaling as $\sim(1+z)^{0.15}$ at $z\simgt1$ as 
compared to $\sim(1+z)^{2.5-3}$ for the non-evolving prediction.
We conclude that if
the dust content of 
intervening galaxies is unevolving,
then it is not possible to fully explain
the redshift dependence of reddening
in Parkes quasars using the 
properties of local galaxies alone.
Implications and comparisons of the results in 
Table~\ref{tab1} with existing observations are discussed in the
the next section.
 
\section{Comparison With Existing Observations}
\label{compexo}

In this section, we first present theoretical 
constraints on the amount of dust implied by studies of the
local mass density in metals, observed limits on dust-to-gas
ratios and the far-IR background radiation. These results
are then compared with our parameter fits in Table~\ref{tab1}. 
In the final section, we compare our constrained {\it evolution} parameters
in Table~\ref{tab1} with studies of chemical evolution of 
QSO absorption-line systems. 

\subsection{Independent Constraints On $\tau_{B}$ and $r_{0}$} 

One of the major uncertainties in quantifying the effects of 
line-of-sight galactic 
dust on background quasars is the dust properties of the 
absorbing galaxies involved. 
In our model, the `local' parameters defining these properties are
the central optical depth $\tau_{B}$ and scale radius $r_{0}$.
We will constrain these parameters
using independent studies of the local mass density in metals, 
observed galactic dust-to-gas ratios, and
the total amount of dust required to explain 
the extragalactic far-infrared background. 
 
From the formalism in section~\ref{ev} and assuming typical grain
properties: $\rho_{g}=2{\rm gm}\,{\rm cm}^{-3}$ 
and $\epsilon_{0}=4$ (Laor \& Draine, 1993),
the local mean mass density in dust relative to the critical density
$\rho_{c}\equiv 3H_{0}^{2}/8\pi G$ is
given by
\begin{equation}
\label{omegalocal}
\Omega_{dust}(z=0)\,\simeq\,(7.3\times 10^{-6})h_{50}^{-2}
\left(\frac{n_{0}}{0.002h_{50}^{3}{\rm Mpc}^{-3}}\right)
\left(\frac{r_{0}}{30{\rm kpc}}\right)^{2}
\left(\frac{\tau_{B}}{0.5}\right),
\end{equation} 
where we have scaled to the nominal values 
$n_{0}=0.002h_{50}^{3}{\rm Mpc}^{-3}$,
$r_{0}=30{\rm kpc}$ and $\tau_{B}=0.5$.

An upper bound for 
the local mean mass density in heavy metals was derived
from predictions of the far-infrared background radiation intensity 
in hierarchical galaxy formation simulations by Blain \& Longair (1993ab). 
These authors derived the upper bound 
\begin{equation}
\label{BL}
\Omega_{metals}(z=0)\,\simlt\,2.5\times 10^{-5}h_{50}^{-2}. 
\end{equation} 
Since only a few percent of metals are believed to be locked up in
dust grains in the local environment, the local mass density in dust
must inevitably satisfy this constraint. 

A lower limit on $\Omega_{dust}$ can be derived using the relation 
$\Omega_{dust}=kl\Omega_{HI}$ from Pei (1992).
$k$ is the dimensionless dust-to-gas ratio and the parameter $l$
is a quantity depending on the physical and optical properties
of dust grains. 
Adopting the value $k=0.08$ which is a minimum value typical of the SMC, 
the value $l=0.013$ which is a minimum value
for a mixture of graphite and silicates 
(Pei, 1992) and $\Omega_{HI}(z=0)\simeq2\times10^{-4}h_{50}^{-2}$ 
from 21cm observations (Rao \& Briggs, 1993), we have the lower bound 
\begin{equation}
\label{LB}
\Omega_{dust}(z=0)\,\simgt\,2.08\times10^{-7}h_{50}^{-2}. 
\end{equation}
Combining Eqns.~\ref{omegalocal},~\ref{BL} and~\ref{LB} we have:
\begin{equation}
\label{constraint1}
0.03\,\simlt\,\left(\frac{n_{0}}{0.002h_{50}^{3}{\rm Mpc}^{-3}}\right)
\left(\frac{r_{0}}{30{\rm kpc}}\right)^{2}
\left(\frac{\tau_{B}}{0.5}\right)\,\simlt\,3.4.
\end{equation}

The dust properties of our model galaxies can also be constrained from
measurements of the far-IR background radiation as measured from the
DIRBE and FIRAS experiments on the cosmic background explorer (COBE).
If all dust to some redshift $z_{max}$ is heated to a temperature
$T_{d}$, it would make the following contribution to the mean
intensity of background radiation at an observed frequency $\nu_{o}$
(eg. Longair, 1993):
\begin{equation}
\label{intensity}
I_{\nu_{o}}\,=\,\frac{3cH_{0}}{8\pi G}\int^{z_{max}}_{0}dz\,
\frac{\Omega_{dust}(z)\kappa_{\nu_{e}}B_{\nu_{e}}(T_{d})}{(1+z)^{2}
(1+2q_{0}z)^{1/2}}. 
\end{equation} 
$\kappa_{\nu_{e}}$ and $B_{\nu_{e}}$ are the dust opacity and 
Planck function at the emitted frequency $\nu_{e}=\nu_{o}(1+z)$ 
respectively.
Typically, $\kappa_{\nu_{e}}\simeq(4.5\times10^{-24})\nu_{e}^{2}
{\rm cm}^{-2}{\rm gm}^{-1}$ at sub-mm wavelengths (Draine, 1985) and
$B_{\nu}$ is defined by Eqn.~\ref{BBlaw}.   
$\Omega_{dust}(z)$ is the comoving density of dust as a function 
of redshift and hence
includes
evolutionary effects. $\Omega_{dust}(z)$ is 
derived from Eqn.~\ref{omegalocal} 
with the replacements: $r_{0}\rightarrow r_{0}(z)$ (Eqn.~\ref{rozev}) and
$\tau_{B}\rightarrow\tau_{B}(z)$ (Eqn.~\ref{tbz}). 
We assume a dust temperature of $T_{d}=20$K, which is a lower
bound for grains in the galaxy emitting at 
sub-mm wavelengths (Greenberg, 1971). 
Our evolutionary dust parameters defining $\Omega_{dust}(z)$ have
been fixed at $z_{dust}=6$ ($\equiv z_{max}$ in Eqn.~\ref{intensity}) 
and $\delta=-0.5$, corresponding
to `moderate' evolution in dust properties. 

Current limits for the intensity of background radiation from the
DIRBE and FIRAS experiments are
\begin{equation}
\label{cobe}
I_{\nu_{o}}\,\simlt\,4\times10^{-17}\,\,{\rm ergs}\,{\rm s}^{-1} 
{\rm cm}^{-2} {\rm Hz}^{-1} {\rm sr}^{-1}
\end{equation}
at wavelengths from $50\mu$m to 1cm (Mather \etal 1993). 
Combining Eqns.~\ref{intensity} and~\ref{cobe} leads to
\begin{equation}
\label{cobelim}
h_{50}^{-1}\left(\frac{n_{0}}{0.002h_{50}^{3}{\rm Mpc}^{-3}}\right)
\left(\frac{r_{0}}{30{\rm kpc}}\right)^{2}
\left(\frac{\tau_{B}}{0.5}\right)\,\simlt\,42.
\end{equation}
This upper bound is 
somewhat less conservative
than that given by Eqn.~\ref{constraint1}.
An agreement however can be achieved if
a higher dust temperature ($T_{d}$) and/or weaker evolution in dust
content with $z$ is assumed in our derivation leading to Eqn.~\ref{cobelim}. 

With the nominal value $n_{0}=0.002h_{50}^{3}{\rm Mpc}^{-3}$, the
constraints defined by Eqns.~\ref{constraint1} and~\ref{cobelim}
are consistent with the range of parameters $0.5\simlt\tau_{B}\simlt4$
and $5\simlt(r_{0}/{\rm kpc})\simlt30$ derived from
previous studies (Giovanelli \etal 1994; Peletier \etal 1995;
Heisler \& Ostriker, 1988 and Zaritsky, 1994). 
 
There appears to be remarkable agreement between our
constrained value of the central $B$-band optical depth, $\tau_{B}$,
using the Parkes data (Table~\ref{tab1}) 
with that deduced from photometric observations of nearby spirals.
As discussed in section~\ref{resan}, $\tau_{B}$ primarily determines
the scatter in reddening at any redshift. 
Our best fit values for $r_{0}$ however
argues against the hypothesis that all reddening
is due to dusty galaxies in the line-of-sight.
Recent studies of the distribution of dust in local spirals
find an upper limit to the exponential
scale radius of $r_{0}\sim30$kpc (Zaritsky, 1994; Peletier \etal 1995).
Since $r_{0}$ is directly related to the dust covering factor
or normalisation parameter $\tau_{g}$ (Eqn.~\ref{tg2}),
if $r_{0}=30$kpc then 
the number density $n_{0}$ will have to be increased by a factor of $\sim30$
to satisfy the reddening observed in the Parkes sample.
Nonetheless, the absorber covering factor ($\propto n_{0}r_{0}^{2}$)
required to explain the fraction
of reddened sources is 
inconsistent with both of the independent constraints on local 
dust content 
(Eqns.~\ref{constraint1} and~\ref{cobelim}) derived above.

We conclude that while the `scatter' in reddening of Parkes sources
is consistent with current estimates of
individual galactic optical depths,
the fraction of reddened sources cannot be explained using the dust
covering factor provided by
present-day galaxies. 
The observed fraction of reddened sources is underestimated by at least an
order of magnitude if galaxy dust properties at the upper limits of present
observations are assumed. 

\subsection{Evolutionary Parameters}
\label{evp}
 
Our results for $\delta$ in Table~\ref{tab1}, 
which describes evolution in 
the cross-section for dust absorption (Eqn.~\ref{rozev}), 
can be compared to that implied by studies of
the evolution of metal absorption-line systems in quasar 
spectra.
These systems are thought to arise in
gas associated with galaxies and their haloes and it is not
unreasonable to assume that 
such systems also contain dust.
In other words, such a proportionality assumes that
dust grains are made of heavy metals and that grain growth 
proceeds in rough proportion as heavy elements
are produced from nucleosynthesis.
 
In general, evolution in the number of metal absorption
line systems per unit $z$, that takes into account
effects of cosmological expansion, is parameterised as follows;
\begin{equation}
\frac{dN}{dz}\,=\,\frac{c}{H_{0}}n_{z}\pi r_{0}(z)^{2}
(1+z)(1+2q_{0}z)^{-1/2}.
\label{linev}
\end{equation}
Evolution, such as a reduction in absorber numbers with redshift,
can be interpreted as either a decrease in the comoving
number density $n_{z}$, or effective cross-section $\pi r_{0}(z)^{2}$.
With our assumption of 
a constant comoving density $n(z)=n_{0}$, and
an evolving dust scale radius $r_{0}$ as
defined by Eqn.~\ref{rozev}, we have
$dN/dz\propto(1+z)^{\gamma}$, where $\gamma=1+2\delta$ for 
$q_{0}=0$ and $\gamma=0.5+2\delta$ for $q_{0}=0.5$.
Hence for {\it no evolution}, $\gamma=1$ and 0.5 for $q_{0}=0$ and
$q_{0}=0.5$ respectively. 

No agreement yet exists in the value of the
evolution strength $\gamma$
corresponding to
different ions detected, most notably the 
prominent transitions  
C{\small IV}$\lambda\lambda$1548, 1551 and 
Mg{\small II}$\lambda\lambda$2795, 2802. 
Due to a limited wavelength range at which Mg{\small II} systems
can be detected, little is known on their evolution properties
for $z>1.5$.
For C{\small IV} systems however, 
evolution has been confirmed and 
in particular, the highest
rest frame equivalent width systems ($W_{0}\simgt0.6$\AA)
appear to show the most rapid decrease in number
towards high redshifts.
(Sargent, Steidel \& Boksenberg 1988; Thomas \& Webster 1990).
It is likely that these systems are those
associated with dust rather than the lower equivalent width systems 
($W_{0}\simlt0.3$\AA)
in which the trend appears consistent 
both with and without evolution at the $1\sigma$ level.
 
Our best fit values for $\delta$ in Table~\ref{tab1} 
correspond to $\gamma=-3.9$
and $-2$ for $q_{0}=0.5$ and $q_{0}=0$ respectively.
Thomas \& Webster (1990) have combined several datasets
and found that 
for C{\small IV} absorption systems with rest frame equivalent widths
$W_{0}\simgt0.6$\AA, $\gamma=-2.8\pm2$ at the $1\sigma$ level.
This value is consistent
with our estimates above, supporting the notion that
the highest equivalent width systems may be a useful
tracer of
the dust content in the universe, rather than the
more numerous less chemically enriched systems at redshifts $z>1$.
 
We now compare our constrained evolutionary model 
with recent determinations of the heavy element abundance
in damped Ly-$\alpha$ absorption 
systems and the Ly-$\alpha$ forest to $z\sim3$.
The damped Ly-$\alpha$ systems are interpreted as the 
progenitors of galactic disks (Wolfe \etal 1986), and recent 
studies by Pettini \etal (1994; 1997)
deduce metal abundances and dust-to-gas ratios at $z\sim1.8-2.2$
that are
$\sim 10\%$ of the local value.
The Lyman forest systems however are more numerous, and 
usually correspond to gas columns
$>10^{7}$ times lower than those of damped Ly-$\alpha$ absorbers. 
High resolution metal-line observations by Songaila (1997)
deduce metallicities $\simlt1.5\%$ solar at $z\sim2.5-3.8$. 
See below for more details.
 
The metallicity at any redshift $Z(z)$, is generally
defined as the mass fraction of heavy metals relative
to the total gas mass: $Z(z)=\Omega_{m}(z)/\Omega_{g}(z)$.
At all redshifts, we will assume a constant dust-to-metals ratio, 
$\Omega_{d}(z)/\Omega_{m}(z)$, where a fixed fraction of
heavy elements is assumed to be condensed into dust grains.
Therefore the metallicity $Z(z)$, relative to the local solar 
value, $Z_{\odot}$, can be written: 
\begin{equation}
\frac{Z(z)}{Z_{\odot}}\,=\,\frac{\Omega_{d}(z)}{\Omega_{d}(0)}
\frac{\Omega_{g}(0)}{\Omega_{g}(z)}.
\label{mdg}
\end{equation}
From the formalism in section~\ref{ev}, 
the mass density in dust
relative to the local density, $\Omega_{d}(z)/\Omega_{d}(0)$,
can be determined and is found to be 
independent of the galaxy properties $r_{0}$ and $\tau_{B}$,
depending only on
our evolution parameters, $\delta$ and $z_{dust}$.
This is given by
\begin{equation}
\frac{\Omega_{d}(z)}{\Omega_{d}(0)}\,=\,\left[1 - 
\frac{\ln(1+z)}{\ln(1+z_{dust})}\right](1+z)^{2\delta}.
\label{omegev}
\end{equation}
The gas ratio, $\Omega_{g}(0)/\Omega_{g}(z)$, is adopted
from studies of the evolution in gas content of damped Ly-$\alpha$ systems.
These systems are believed to account for at least $80\%$
of the gas content in the form of 
neutral hydrogen at redshifts $z\simgt2$ (Lanzetta \etal 1991).
We adopt the empirical fit of Lanzetta \etal (1995), who find that the 
observed evolution in $\Omega_{g}(z)$ is
well represented by $\Omega_{g}(z)=\Omega_{g}(0)\exp(\alpha z)$,
where $\alpha=0.6\pm0.15$ and $0.83\pm0.15$ for
$q_{0}=0$ and $q_{0}=0.5$ respectively.
 
\begin{figure}
\vspace{-2.7in}
\plotonesmall{1}{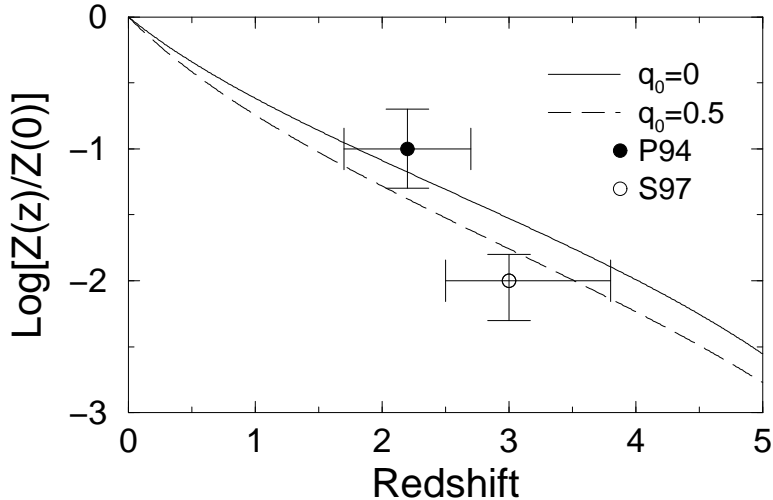}
\vspace{-3in}
\caption[Observed and predicted relative metallicity vs. redshift]{
$1\sigma$ upper limits in the relative metallicity as a function of redshift
(assuming $1\sigma$ upper limits in the evolution parameters $z_{dust}$
and $\delta$ from Table~\ref{tab1}) predicted from our model fit to the
Parkes data (see section~\ref{pres}).
The filled and open data points
with $1\sigma$ error bars represent mean observed values
from Pettini \etal (1994) and Songaila (1997) respectively.
}
\label{metfit}
\end{figure}

A direct proportionality between dust and heavy metal abundance
implies that since $\Omega_{d}=0$ for redshifts $z\geq z_{dust}$
(the dust formation epoch), the metallicity 
$Z(z\geq z_{dust})$ is also zero.
Due to our relatively low `best fit' estimates for $z_{dust}$
in Table~\ref{tab1} compared to $z\sim2-3$ where previous
metallicity measurements have been made,
we have used our $1\sigma$ upper limits
for $z_{dust}$ and $\delta$.
Using these parameters, the curves in Fig.~\ref{metfit} show
the relative metallicity given by Eqn.~\ref{mdg} 
as a function of redshift. 
For comparison, the mean metallicities
$Z\approx 0.1Z_{\odot}$ and $Z\approx 0.01Z_{\odot}$ 
observed in damped Ly-$\alpha$ systems at 
$z\approx2.2$ and the Lyman forest at $z\simgt2.5$ respectively
are also shown. 
The value $Z_{2.2}\approx 0.1Z_{\odot}$ 
was determined from an average of Zn{\small II} measurements 
in a sample of 17 damped Ly-$\alpha$ systems with $1.8\simlt z\simlt 3$
by Pettini \etal (1994).
The value $Z_{3}\approx 0.01Z_{\odot}$ was deduced 
from high resolution Keck observations of C{\small IV} and Si{\small IV}
absorption towards 7 quasars with $2.5\simlt z_{em}\simlt 3.8$ by
Songaila (1997) (whose result was converted using $q_{0}=0.5$, $h_{50}=1$). 

As shown in Fig.~\ref{metfit}, our predicted $1\sigma$ upper limits
for the metallicity agrees well with the observed estimates of
$Z(z\sim2.2)\approx 0.1Z_{\odot}$ and 
$Z(z\sim3)\approx 0.01Z_{\odot}$.
This consistency implies that possible
selection biases in observations from which 
estimates of the metallicity are based are likely to be minimal.
The major bias as noted by Pettini \etal (1994)
is that the absorption systems are selected from quasar surveys which
need to be relatively bright in the optical (ie. $B\simlt19$).
Those systems that are significantly dusty and 
hence presumably metal rich may obscure some of the
quasars and hence escape detection. 
This may lead to a possible underestimate in the inferred
mean metallicity.
In the radio-selected sample from which our model is constrained
however, no bias due to dust obscuration is expected.
We must note however that the reddening data for the 
sample of Parkes quasars 
used in this study may 
still suffer from possible selection effects, and not provide a useful 
measure of evolution. 
These are discussed in the next section.
 
\section{Discussion}
\label{evmodd}
 
This section discusses possible uncertainties
that may account for 
the lack of a redshift dependence in $B-K$ colour to $z\sim4$
other than that expected from reddening by 
{\it evolving} intervening dust. 

Effects of incompleteness in redshift determinations and/or
photometry (primarily $K_{n}$) in the Parkes sample
is unlikely to account this observation. 
The available redshifts represent $\sim65\%$ of all sources
independent of optical magnitude to $B_{J}\sim23$.
As shown in Francis \etal (1997), the distribution of 
$B-K$ colours
for sources with and without measured redshifts is extremely similar and
thus, there is no bias with respect to $B-K$ colour.
While the sample is complete in $B_{J}$, the completeness in $K_{n}$ is
$\approx60\%$. 
This is unlikely to introduce significant bias against `red', 
`high redshift' sources as more than 90\% with $z\simgt2$ 
are measured in $K_{n}$.
The few
optically unidentified sources with $B_{J}\simgt23$ and unknown redshift 
however
(see right panel in Fig.~\ref{colvsz}) 
may steepen
the observed colour-redshift relation if they reside at $z\simgt3$ 
and their colours are exceedingly red.
Aside from these sources, 
we conclude that the distribution in colour as a function of redshift in 
Fig.~\ref{colvsz} closely represents an 
unbiased subsample with respect to $B-K$ and redshift.

The `flat' redshift dependence of mean reddening for Parkes sources
may be an artifact from the use of 
$B-K$ colour itself. This measure may not
be a useful indicator of dust reddening and
may be biased due to intrinsic source properties.
Evolutionary effects, 
such as a stronger increase in 
$B$-band luminosity relative to that in the near-infrared
towards high redshift may be present, masking any increase in reddening
expected if an 
appreciable amount of
line-of-sight dust were present. 
Since a large
fraction of sources
display quasar-like characteristics,
stronger evolution in $B$-band luminosity than
in $K$ may arise if the former depends more sensitively on evolution 
of the central AGN. 
Despite numerous claims for an evolving mean $B$-band luminosity
in studies
of {\it optically-selected} quasars
(Boyle \etal 1988), evidence for a similar behaviour in the near-infrared
is uncertain.
Evolution in the optical--near-infrared properties of radio-selected quasars
is yet to be explored. 
 
Is it possible that there is insufficient
sampling of the number of lines-of-sight towards high redshift 
from which a reasonable conclusion can be drawn? 
In other words, since only $\sim 15$\%
of Parkes sources
have $z\simgt2$,  
they may not be representative of
`true', dusty lines-of-sight towards high redshift. 
The decrease in the number of sources with $z\simgt2$ is undoubtably 
due to the relatively bright flux limit
$S_{2.7{\rm GHz}}\geq0.5$Jy for this
sample. At high redshift, only
radio-luminous sources are selected,
and hence,
a relatively lower space density is expected due
to the steepness of the radio luminosity function
(Dunlop \& Peacock, 1990).
Optical identification of sources 
in surveys of sufficiently low radio flux limit
may improve the statistics at high redshift and provide a stronger test.
 
The results of our constrained model critically depend on the dust properties
assumed for individual galaxies.
For instance, is it reasonable to give galaxies an exponential
dust distribution? Such a distribution is expected to give
a dust covering factor to some redshift considerably larger
than if a clumpy distribution
is assumed (Wright, 1986).
A clumpy dust distribution (for spirals
in particular) is expected, as dust is
known to primarily form in dense, molecular star-forming
clouds (Wang, 1991 and references therein).
As noted by Wright (1986), ``cloudy disks'' 
with dust in optically-thick clumps
can reduce the
effective cross section for dust absorption by a factor of five and hence, are
less efficient at obscuring background quasars.
Such an assumption will thus severely weaken the 
intervening dust hypothesis 
for the anomalous optical--near-IR colours of Parkes quasars.

\section{Summary and Conclusions}
\label{concfour}
 
In this chapter, we have modelled the optical depth in dusty
galaxies along the line-of-sight as a function of redshift
assuming evolution in dust content.
Our model depends on four parameters which specify the 
dust properties of local galaxies and their evolution:
the exponential dust scale radius $r_{0}$, central $B$-band optical
depth $\tau_{B}$, ``evolution strength'' $\delta$ where
$r_{0}(z)=r_{0}(1+z)^{\delta}$, and $z_{dust}$ - a hypothesised
dust formation epoch.
 
Our main results are:
\\\indent 1. For evolutionary parameters 
consistent with existing studies of the evolution of metallicity
deduced from
QSO absorption line systems, 
a significant ``flattening'' in the mean and variance
in $B$-band optical depth
to redshifts $z>1$ is expected.
The mean optical depth to $z\simgt1$
is smaller by at least a factor of 3 compared to
non-evolving model predictions. Obscuration by dust is not
as severe as shown in previous studies if effects of evolution are accounted 
for.
\\\indent 2. By allowing for even moderately low amounts of evolution,
line-of-sight dust is not expected 
to significantly affect existing optical studies of QSO evolution.
Correcting for dust obscuration, evolving dust models
predict the `true' (intrinsic) space density of bright quasars to
decrease beyond $z\sim2.5$, as observed, contrary to 
previous non-evolving models where a continuous monotonic
increase was predicted.
\\\indent 3. We have explored the implications on 
QSO absorption line statistics using
this model, mostly in relation to the 
prominent Mg{\small II} and C{\small IV}
systems. 
We find that up to 40\% of metal lines with equivalent widths 
$\simgt0.5$\AA$\,$ (the typical median) may have
been missed in absorption line surveys by dust in systems
obscuring background QSOs. 
\\\indent 4. We have compared our models with optical and near-IR
observations of a complete radio-selected quasar sample (the `Parkes sample').
The mean reddening in $B-K$ colour as a function of redshift is extremely flat
and cannot be explained by models
where the dust content of galaxies is non-evolving.
This redshift dependence however is easily reproduced
with evolutionary parameters consistent with other studies.
We require a {\it sky covering factor} in dusty absorbers to some redshift
that scales as $\sim(1+z)^{0.15}$, as compared to $\sim(1+z)^{1.5}$ 
for non-evolving models in a $\Omega=1$ universe.
\\\indent 5. The fraction of Parkes sources reddened at any redshift
cannot be explained using the dust properties
of present-day galaxies. 
A covering factor in dusty absorbers more than two orders of magnitude
greater than that predicted assuming average properties of local
galaxies is required.
With galaxy properties at the upper limits of present 
observations ($n_{0}\simeq0.002h_{50}^{3}{\rm Mpc}^{-3}$ and 
$r_{0}\simeq30$kpc),
the fraction of reddened sources is underestimated by at least an
order of magnitude.
\\\indent 6. The {\it total amount} of `galactic-type'
dust required to explain the reddening in Parkes sources is inconsistent
with theoretical constraints derived from the local mass density in metals,
observed limits on dust-to-gas ratios, and the far-IR background radiation. 
We conclude that the observed reddening cannot be {\it totally} due to
dust located in extrinsic systems that may happen to lie in the
line-of-sight of otherwise `normal' blue quasars. 

\newpage
\part{The Reddening Mechanism in Parkes Quasars}
\newpage

\chapter{The Reddening Mechanism in Parkes Quasars} 

\vspace{1mm}
\leftskip=4cm
 
{\it ``Where the telescope ends, the microscope begins.\\ 
Which of the two has the grander view?''\\}
 
\vspace{1mm}
 
\hfill {\bf --- from Les Miserables,}

\hfill {\bf Victor Hugo, 1862}
 
\leftskip=0cm

\section{Introduction}

Recently, a large complete sample of radio-selected
quasars has been compiled by Drinkwater \etal (1997), initially
selected from the Parkes catalogue.
All sources have flat radio spectra ($\alpha_{2.7-5{\rm GHz}}<0.5$; where
$f_{\nu}\propto\nu^{-\alpha}$) and 2.7GHz fluxes $S>0.5$ Jy.
Based on spectroscopic observations and 
a high identification rate in the optical and near-infrared,
this study has revealed a large number of quasars with 
very red
optical--to--near-infrared colours and optical continuum slopes.
As shown in Fig.~\ref{histos}, a broad distribution in colours
and optical slopes with 
$2\leq B_{J}-K_{n}\leq 10$ and $-1\leq\alpha_{opt}\leq4$ is observed.
For comparison (the bottom histograms), optically-selected quasars
show a relatively small scatter
about the mean values $B_{J}-K_{n}\simeq2.5$ and $\alpha_{opt}\simeq0.3$.
From here on, the sources in the Drinkwater \etal sample will be referred
to as ``Parkes'' sources and {\it all} spectral slopes $\alpha$ are defined
as $f_{\nu}\propto\nu^{-\alpha}$.
A subscript will be used to refer to the waveband over which $\alpha$ is
defined.

The principal question that remains is the following:
What is the physical mechanism for the large scatter observed in  
optical--to--near-IR continuum slopes for Parkes sources?
The theories can be divided into three major classes: intrinsically red
AGN emission models, dust extinction
models, and host galaxy emission models. 
In this chapter we shall primarily explore the first two classes, and
consider host galaxy light models in Chapter 6.

\subsection{Intrinsically Red Emission Models}

Intrinsically red emission models include those where the
reddening in continuum slope is an intrinsic property of 
the AGN emission. A model which has attained considerable popularity
in explaining the anomalous properties of Parkes quasars is the synchrotron
model (eg. Serjeant \& Rawlings, 1996). 
Early studies have interpreted the relatively steep
optical--to--near-IR continua of a number of flat spectrum radio quasars
as due to an intrinsically ``red'' synchrotron component characteristic
of such sources (eg. Bregman \etal 1981; Rieke, Lebofsky \& Wi\'{s}niewski, 
1982 and Cruz-Gonzalez \& Huchra, 1984). 
In the framework of unification models for radio-loud AGN,
flat spectrum radio sources are believed to be due to an
enhancement of self-absorbed synchrotron emission from compact regions
in a jet by relativistic beaming along the line-of-sight. 
If such a beamed component extends into the near-IR and optical, and
is intrinsically red, then it could explain the wide scatter in
$B_{J}-K_{n}$ colours observed. 

Other intrinsic red emission models involve mechanisms where the
optical/UV continuum is intrinsically weak, such as 
the weak ``blue-bump'' quasars 
identified by McDowell \etal (1989), or where a  
steep pure thermal emission spectrum is added, say from dust, dominating in the
near-IR.

\subsection{Dust Extinction}

Recently, Webster \etal (1995) have suggested that the observed spread
in continuum slopes could be due to extinction by varying amounts 
of dust
along the line-of-sight and not an intrinsic emission property of the quasars. 
Such dust could be located in regions close to the quasar itself, such
as in the host galaxy, in the narrow line region or intercloud medium
(eg. Wills \etal 1993). 
It could also be located in unrelated extrinsic systems
such as intervening galaxies that happen to lie along the line-of-sight.
As shown in section~\ref{compexo} however,
if intervening galaxies with
dust properties and space densities 
at the upper limits of present observations are assumed,
then $\simlt10\%$ of the observed colour spread
can be accounted for.
This does not completely rule out an extrinsic origin for the extinction
as an undetected population of dusty absorbers could still be present. 
It has been known for some time that line-of-sight absorption
by neutral hydrogen and helium of the Lyman continuum
due to both
the Lyman forest and Lyman-limit systems can significantly reduce the
optical-UV flux observed (eg. M{\o}ller \& Jacobsen, 1990).
Such a mechanism however requires high source
redshifts ($z>4$) to produce any observable effect.

Based on the early discoveries of red radio-quasars and the classical 
optical ``empty-field'' sources in previous radio samples (see Chapter 1), 
only one previous study has presented possible evidence for the dust
extinction hypothesis. 
Ledden \& O'Dell (1983) found that these 
``optically quiet'' and red, compact radio
sources are on average 
significantly weaker X-ray emitters than the optically luminous,
bluer sources.
These authors concluded that absorption by gas and dust
is a likely explanation.
Since then, there have also
been various claims of excessive soft X-ray absorption
associated with these 
sources (eg. Kollgaard \etal 1995). The results however 
are extremely model dependent and based on small statistical samples.

\begin{figure}
\vspace{-2.8in}
\plotonesmall{0.8}{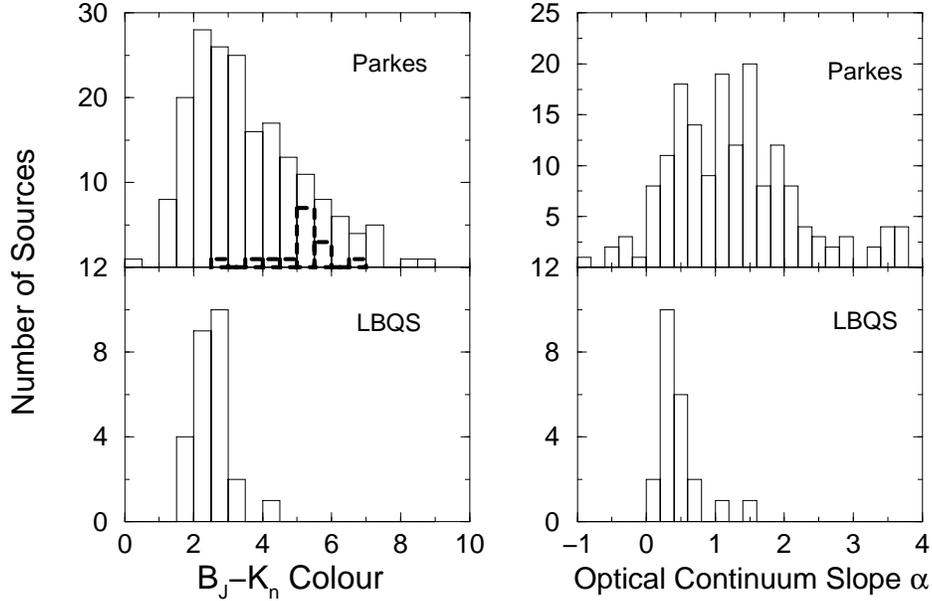}
\vspace{-1in}
\caption[Distribution of $B_{J}-K_{n}$ and optical continuum slope]{Distribution of $B_{J}-K_{n}$ colours (left) and optical
continuum slopes (right), where $f_{\nu}\propto\nu^{-\alpha}$.
Optical slopes are from continuum fits within 
$3200{\rm\AA}\simlt\lambda_{obs}\simlt8000$\AA$\,$ (see Francis \etal 1997). 
Top panels show results for all {\it spatially unresolved} Parkes sources in
the optical and near-IR. For comparison,
the lower panels show distributions for a sub-sample of bright
optically selected quasars (from the LBQS sample; Francis, 1996).
The dashed histogram is for sources with 
lower limits in $B_{J}-K_{n}$.}
\label{histos}
\end{figure}

\vspace{5mm}
\noindent
It remains unclear whether the red Parkes quasars represent one extreme
of the radio-loud AGN population due to intrinsic or 
extrinsic effects, or a fundamentally new class of AGN.
An understanding of the origin of the spread in continuum slopes could
provide a clue.
Most importantly, a determination of the reddening mechanism could allow
one to assess the long standing problem of incompleteness in 
quasar number counts introduced by optical selection effects. 
Webster \etal (1995) claim that the observed spread in $B-K$ for Parkes quasars 
implies that at least 50\% of quasars may have been
missed from
optical surveys.
This conclusion is only valid however, if the reddening mechanism
operating in Parkes quasars is also applicable to radio-quiet quasars which
dominate the optical samples.

In this chapter, we try to understand the spread in continuum 
slopes and colours in the framework of the two classes of theories 
discussed above: intrinsically red emission models and dust models. 
We will develop simple models and look for correlations in observed
properties that may distinguish between these two models. 
For the dust model, we shall explore the effects of dust on soft
X-rays and compare our predictions with soft X-ray data for Parkes sources.
Implications for possible physical dust properties 
are also discussed. 

This chapter is organised as follows:
In the next section, we review the completeness and 
properties of sources in the Drinkwater \etal (1997) sample. 
In section~\ref{synchmodel}, we explore 
predictions and implications of the
synchrotron reddening model. 
Other possible `intrinsically red' emission models are explored
in section~\ref{otherints}.
Preliminary observational tests of the dust model are 
investigated in section~\ref{pretd}. 
In section~\ref{sxabsp}, we explore the effects of dust (and associated gas)
on soft X-rays and compare our predictions
with available ROSAT data for Parkes quasars.
In section~\ref{impsurv}, we explore the implications of soft X-ray absorption
on quasar X-ray surveys. 
Implications, possible physical dust properties
and further predictions are discussed in section~\ref{disredm}.
All results are summarised in section~\ref{concfive}.

\section{The Sample}

The Parkes Half-Jansky Flat Spectrum Sample contains 323 sources
and is described in detail by Drinkwater \etal (1997). 
The completeness in terms of optical spectra and optical and
near-infrared photometry is described by Francis \etal (1997). 
The sample was initially selected according to the following:
\\\indent i. Radio fluxes at 2.7 GHz $f_{2.7}>0.5$ Jansky
\\\indent ii. Radio spectral 
indices between 2.7 and 5 GHz of $\alpha<0.5$ where 
$f_{\nu}\propto\nu^{-\alpha}$ 
\\\indent iii. Galactic latitude $|b|>20^{\circ}$ 
\\\indent iv. B1950 declinations: $-45^{\circ}<\delta<+10^{\circ}$.

Optical and near-IR photometric observations show that a 
majority of the sources are unresolved as is common for radio samples
selected on the basis on flat radio spectra at high frequencies.
33 ($\simeq10\%$) of the 323 sources however appear extended in 
$B_{J}$ or $K_{n}$.
Optical spectra of these sources show features 
characteristic of those seen in normal nearby galaxies.
Since we are primarily interested in the properties of 
quasars, we ignore these extended sources in this chapter. 

Optical spectra are available for a sub-sample of 187 non-extended
Parkes sources from
which optical continuum slopes
can be reliably measured. Approximately 90\% of these sources
have broad emission lines characteristic of
normal QSOs and the remainder have featureless
continua typical of BL-Lacs
(see Francis \etal 1997). 
The sources span the redshift range $0<z<3.9$ (see Fig.~\ref{colvsz}) and
have optical magnitudes $12<B_{J}<25$. 

Optical spectra are available for $\sim65\%$ of Parkes sources,
independent of optical magnitude to $B_{J}\sim22.5$ - the
limit of UK Schmidt and Palomar sky survey plates.
24 sources remain unidentified
with $B_{J}>22.5$ and do not have optical spectra. 
As shown in Francis \etal (1997), 
the distribution of $B_{J}-K_{n}$ colours 
for the spectroscopically identified quasars 
is similar to that of the whole sample of
unresolved Parkes sources shown in Fig.~\ref{histos}. 
The spectroscopically confirmed quasars therefore represent
an unbiased subsample with respect to $B_{J}-K_{n}$ colour. 

\section{The Synchrotron Model}
\label{synchmodel}

The synchrotron reddening mechanism is the first and most popular 
of the class of intrinsic `red' emission models we shall explore.
Although poorly defined, this model makes a number
of simple predictions of the relationship between reddening 
and other observables such as strength of radio core dominance
and emission line equivalent width.
In this section, we test these predictions with available data.
We also discuss limits on the steepness of a synchrotron radiation 
spectrum imposed by the theory of synchrotron emission.

\subsection{Radio Core Dominance}
\label{rcd}

Flat spectrum radio sources selected at high frequencies are normally
associated with compact radio cores, which often imply brightness
temperatures in excess of the inverse Compton limit, $T_{B}\sim10^{12}$K,
allowed for incoherent synchrotron emission (eg. Readhead, 1994).
This, together with observations at other wavelengths
suggest the emission is boosted by
relativistic bulk motion along the line-of-sight.
A parameter used to quantify the amount of
relativistic beaming of the radio emission is via the
the ratio of `core' (presumably beamed) flux to `extended' (unbeamed) flux, or 
`dominance' of the radio-core. 
If the red $B-K$ colours of Parkes quasars are due to a 
beamed synchrotron component dominating in the near-infrared, which is an
extension of the radio, then one would expect they correlate
with the dominance of radio-core emission.

\begin{figure}
\vspace{-3.5in}
\plotonesmall{1}{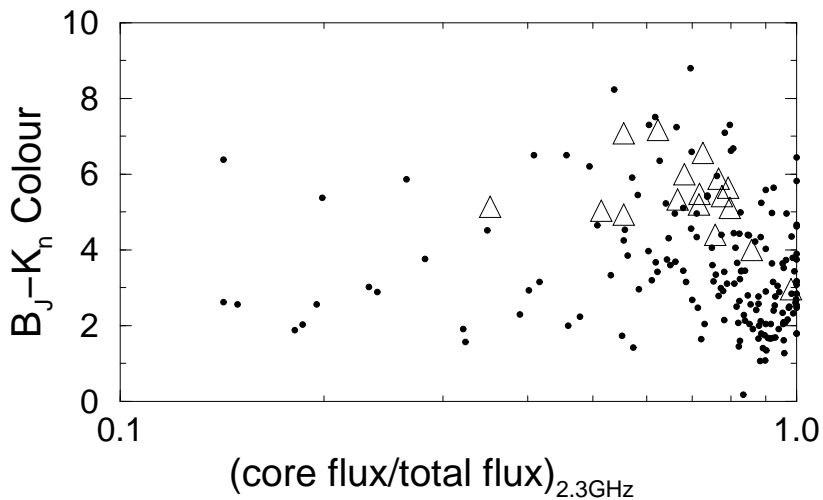}
\vspace{-2.4in}
\caption[$B_{J}-K_{n}$ colour as a function of radio-core dominance]{$B_{J}-K_{n}$ colour as a function the ratio of (core)
flux on $0.1''$ scales to total flux at 2.3GHz (a measure of the `visibility'). 
Only Parkes sources spatially unresolved in $B_{J}$ or $K_{n}$ are shown.
Triangles represent lower limits on $B_{J}-K_{n}$ colour.}
\label{vis}
\end{figure}

We quantify the dominance of the radio-core at 
some frequency as the ratio of `core' flux to `total' (core plus
extended) flux. This quantity is usually referred to as
the visibility ($V$) - a measure of the ratio of correlated fluxes
on different 
scales using 
an interferometer.
Core flux measurements on angular scales of $\sim0.1''$ for all
Parkes sources were obtained in a study by Duncan \etal (1993).
These authors measured visibilities at 2.3GHz using the 
two-element Parkes-Tidbinbilla (284km baseline) interferometer.
The visibilities are plotted against $B_{J}-K_{n}$ colour in Fig.~\ref{vis}.
Uncertainties in $V$ and $B_{J}-K_{n}$ are typically no more than 10\% and
1 magnitude respectively. 
Sources with $V>1$ (due to observational uncertainties) have been set to 
$V=1$ in Fig.~\ref{vis}. 
Even though the core components are unlikely to be fully resolved on
$\sim0.1''$ scales, a large number of sources have $V<0.8$, implying
their total flux is resolved within measured 
uncertainties. 

The colours do not correlate with the degree of prominence of the
radio core component as expected if the near-IR emission were beamed
along with that in the radio.
In particular, it can be seen 
that on average, the most core-dominated (unresolved) 
quasars comprise
a larger population of blue sources.
The result in Fig.~\ref{vis} is thus inconsistent with 
first order predictions of the synchrotron model.
This however must not be taken as conclusive evidence,
as the
intrinsic properties of the beamed radio and near-IR components 
can to some degree be independent of one another, and 
contain considerable scatter.
This will reduce any expected correlation. 

\subsection{Emission Line Equivalent Widths}
\label{ewsynch}

The synchrotron model for reddening predicts an anticorrelation between
the equivalent widths (EW) of emission lines and $B-K$ colour. 
This anticorrelation can be qualitatively understood as follows:
a synchrotron (possibly beamed) 
component must make a larger contribution to the near-IR
flux than in the optical if a `red' $B-K$ colour is observed.
Thus for a specific emission line, as the $B-K$ colour
increases, the (synchrotron) continuum level increases and the
EW decreases. 
This effect will be strongest for lines of longest rest frame 
wavelength relative to an observer's $B_{J}$ bandpass.
This prediction strictly 
assumes that the beamed component
is independent of the photoionizing mechanism
so that emission line
{\it fluxes} (hence EWs) will not vary with $B-K$ colour
to produce significant scatter in any expected anticorrelation.

\begin{figure}
\vspace{-3.5in}
\plotonesmall{1}{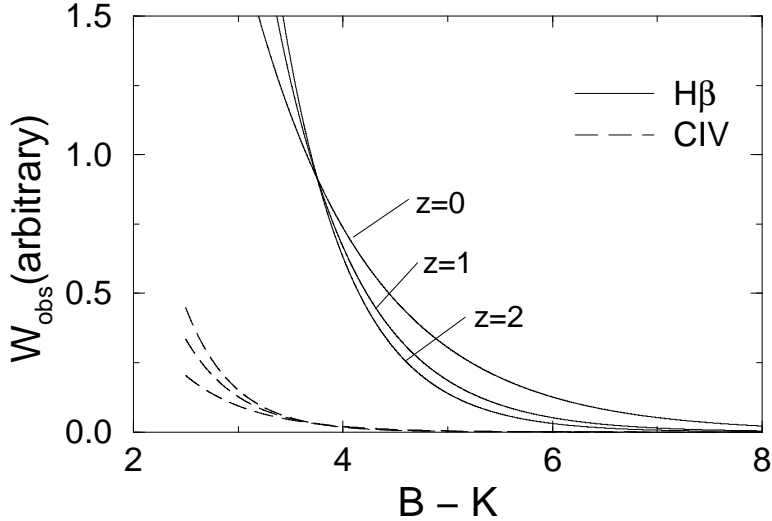}
\vspace{-2.55in}
\caption[Prediction for $B-K$ vs. equivalent width in
the synchrotron model]{Predicted relations between observed 
emission line equivalent width and $B-K$ colour in the synchrotron 
model for sources at redshifts: 0,1, and 2. 
Predictions are the lines: 
H$\beta$ ($\lambda=4861$\AA; solid lines) and
C{\small IV} ($\lambda=1549$\AA; dashed lines).}
\label{EWpredsynch}
\end{figure}

Fig.~\ref{EWpredsynch} shows the predictions of a simple model
for EW vs $B-K$ colour.
This model assumes an observed power-law
$B$ to $K$ continuum whose slope varies due to a variation 
in $K$ flux alone. 
The following dependence of observed frame EW on $B-K$ slope
$\alpha_{BK}$ is predicted: 
\begin{equation}
W_{obs}(\lambda_{l})\,\propto\,(1+z)
\left[\frac{(1+z)\lambda_{l}}{\lambda_{B}}\right]^{2-\alpha_{BK}},
\label{synchEW}
\end{equation}
where $\lambda_{B}\simeq4400$\AA~ and 
the continuum slope extrapolated from $K$ to $B$
can be written:
\begin{equation}
\alpha_{BK}\approx 0.572(B-K)-1.148.
\label{slopcol}
\end{equation}
The redshift dependence is due to two factors, first, the
characteristic 
\\$1+z$ increase in observed-frame EW and second, due to
the redshifting of the actual emission line. 
This latter effect implies stronger
anticorrelations with increasing source redshift.

\begin{figure}
\vspace{-3.4in}
\plotonesmall{1}{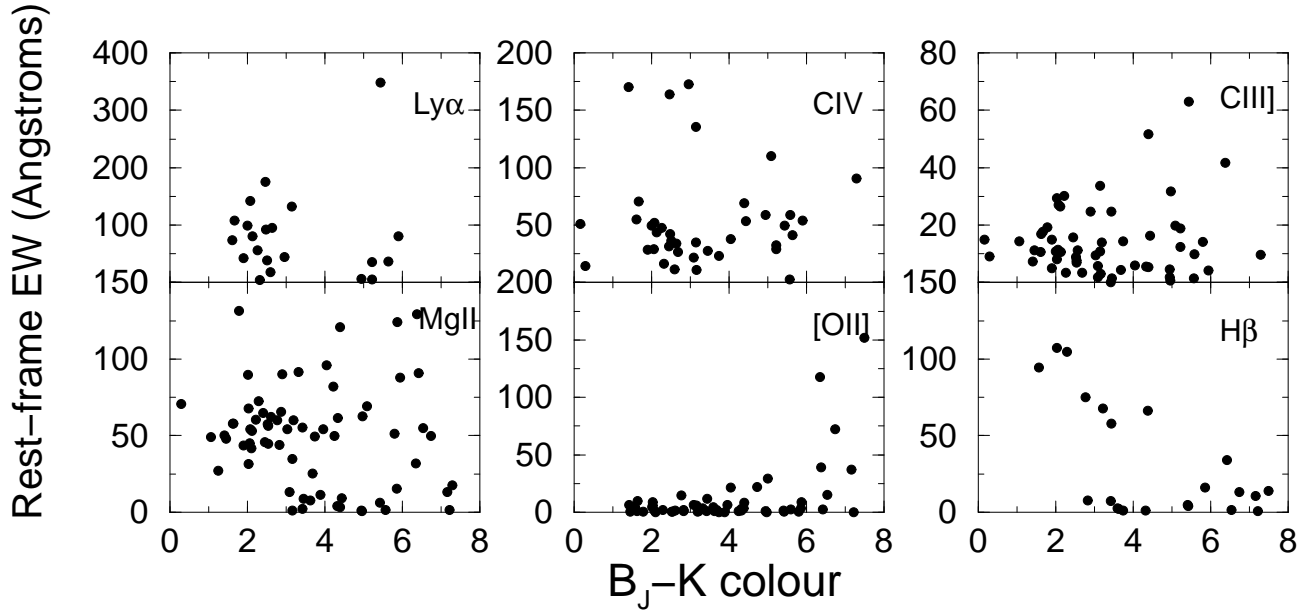}
\vspace{-2in}
\caption[Observed emission line equivalent widths as a function of $B-K$]{Emission line equivalent widths as a function of  
$B_{J}-K$ colour for Parkes quasars. The broad (permitted) lines are 
Ly$\alpha$ (1216\AA), 
H$\beta$ (4861\AA),
CIV (1549\AA),
CIII] (1909\AA) and 
MgII (2798\AA). The only narrow forbidden line is [OII] (3727\AA). 
Sources with spatially extended structure (ie. galaxies) have been omitted.
See Francis \etal (1997) for more details.}
\label{EWdata}
\end{figure}

Figure ~\ref{EWdata} shows the 
rest frame EWs of various emission lines as a function of  
$B_{J}-K$ colour for all Parkes sources unresolved in $K_{n}$ or $B_{J}$. 
The method used to estimate these EWs can be found in Francis \etal (1997).
For most lines, 
no significant anticorrelation between EW and $B_{J}-K$ colour is observed.
Random errors in measured EWs are typically
of order a few percent and hence, are unlikely 
to contribute significant scatter in Fig.~\ref{EWdata}.
As discussed in section~\ref{obsreddi} however, 
random and systematic uncertainties in $B_{J}-K$ colours may be 
up to one magnitude.
Nonetheless, we find that similar trends are observed if we 
replace $B_{J}-K$ colour
by optical continuum slope (which correlate with 
each other; see Fig.~\ref{optslopevscol}).

The range in measured EWs (particularly those for MgII
and CIII]) extend to relatively low values.
These are believed to be those sources
where the
beamed component   
contributes strongly in both the near-IR and optical, 
forming an extension of the classical BL-Lac subclass.
The redness in this population may be 
purely described by the beamed synchrotron model.
Comparing 
figures ~\ref{EWdata} and ~\ref{EWpredsynch} however, we conclude 
that the data for the higher EW Parkes quasars is
broadly incompatible with predictions of the synchrotron model. 

\subsection{Synchrotron Spectral Shapes}
\label{synchshap}

The theory of synchrotron emission makes a
specific prediction for the range of near-infrared to optical
spectral slopes that can be acheived.
To explore these and compare them with the continuum slopes of
Parkes quasars,
we first review the 
relevant formulae and physical parameters which determine 
the observed spectral energy
distribution (see eg. Pacholczyk, 1970).

\subsubsection{Background}

First, the instantaneous power per unit frequency emitted by a {\it single} 
relativistic particle of energy $E$, charge $e$ and rest mass $m$
moving in a uniform magnetic field $B$ with an arbitrary pitch angle
$\alpha$ is
\begin{equation}
P_{s}(\nu,E,\alpha,B)\,=\,\frac{\sqrt{3}e^{3}B}{mc^{2}}
\sin{\alpha}\,F(\nu/\nu_{c}),
\label{singlep}
\end{equation}
where
\begin{equation}
F(x)\,=\,x\int^{\infty}_{x}K_{5/3}(\eta)\,d\eta,
\label{Fx}
\end{equation}
with $K_{5/3}(\eta)$ representing the modified Bessel function of order $5/3$.
$\nu_{c}$ is the critical frequency defined by
\begin{equation}
\nu_{c}\,=\,\frac{3eE^{2}B}{4\pi m^{3}c^{5}}\sin{\alpha},
\label{nuc}
\end{equation}
where for electrons:
\begin{equation}
\nu_{c}\,\sim\,1.6\times10^{13}(B/{\rm G})
\sin{\alpha}(E/{\rm GeV})^{2}\,{\rm Hz}.
\label{nuce}
\end{equation}
The single particle radiation spectrum (\ref{singlep}) has its
maximum at a frequency
\begin{equation}
\nu_{max}\,\simeq\,0.29\nu_{c}.
\label{numax}
\end{equation}
The limiting cases at high ($\nu\gg\nu_{c}$) and low ($\nu\ll\nu_{c}$) 
frequencies for this radiation spectrum as defined by \ref{Fx}
are respectively as follows:
\begin{equation}
F(x)\,\propto\,x^{1/2}\exp{(-x)}{\hspace{3mm}\rm{for}\hspace{3mm}} x\gg 1
\label{limithnu}
\end{equation}
$$
\propto\,x^{1/3}{\hspace{9mm}\rm{for}\hspace{3mm}} x\ll 1.
$$

In general, the relativistic particles are likely to have a
a distribution of energies. If we denote the number density
distribution as $N(E)dE$, between energies $E_{min}$ and $E_{max}$,
the total synchrotron spectrum will then be given by a convolution of this
distribution with Eqn.~\ref{singlep}:
\begin{equation} 
P_{tot}(\nu,\alpha,B)\,=\,\int^{E_{max}}_{E_{min}}P_{s}(\nu,E,\alpha,B)N(E)dE.
\label{distp}
\end{equation}
For simplicity, 
calculations often assume that every particle has the same pitch angle $\alpha$,
moving in a uniform magnetic field $B$. If however a random
distribution of field strengths and pitch angles were present, then
one must also integrate over these quantities. 

\subsubsection{Limits on Spectral Slopes}

For an assembly of particles described by a power-law energy distribution 
$N(E)\propto E^{-s}$, it can be shown that $P_{tot}(\nu)$ will also 
behave as a power law ($\propto\nu^{-\alpha_{s}}$) 
but with spectral index $\alpha_{s}\,=\,\frac{s-1}{2}$. 
Power-law energy distributions are a natural consequence
of the currently favoured particle acceleration mechanisms
involving diffusion across shock fronts (eg. Blandford \& Eichler, 1987).
Different physical processes are believed to determine the characteristic
value of $\alpha_{s}$ observed, however there is a well
determined lower limit on
$\alpha_{s}$.
For ``optically thin'' synchrotron radiation discussed above, a
{\it minimum} value of $\alpha_{s}=-1/3$ is expected since this is the
spectral index associated with a single particle at low frequencies
(see Eqn.~\ref{limithnu}). 
A more conservative lower limit of $s\geq 2$ and hence $\alpha_{s}\geq 0.5$
however is expected if the distribution $N(E)\propto E^{-s}$
is to be a solution of the continuity equation 
in the steady state case (O'Dell, Scott \& Stein, 1987). 
This limit is also imposed by the theory of diffusive shock acceleration
(eg. Drury, 1983).

Our interest is primarily in the upper limit to the steepness of a 
synchrotron radiation spectrum in the optical to near-infrared.
At present this remains somewhat arbitrary since it sensitively 
depends on the physics assumed for the particle acceleration mechanism. 
It has been shown that there is no difficulty in accounting
for spectra much steeper than the lower limit for $\alpha_{s}$
discussed above through steepening from synchrotron losses during the
acceleration process (Heavens \& Meisenheimer, 1987; Drury, 1983; Fritz, 1989). 
Such studies were motivated by the discovery of a number of 
extragalactic radio sources exhibiting an abrupt cutoff in their spectra
at near-IR wavelengths (mostly around $\sim10^{14}$Hz)
(Rieke \etal 1979; Neugebauer \etal 1979; Bregman \etal 1981;
Rieke, Lebofsky \& Wi\'{s}niewski, 1982).
Similar high frequency cutoffs were also detected in the
radio hotspots of a number of ``classical'' double-lobed radio galaxies 
(Meisenheimer \etal 1989).
These sources displayed high polarisations and uniform power-law             
dependences from radio to near-IR frequencies.
From such uniformity, it was concluded 
that all particles contributing 
to the emission in this range were produced by the same acceleration mechanism.
The cutoffs, which are almost as sharp as exponential in frequency,
were then interpreted as a consequence of relativistic particles undergoing 
synchrotron losses during their acceleration process,
which affects the highest energy particles severely. 

There have been many attempts at quantifying the steepest
turnover possible in a synchrotron radiation spectrum via
simulations of particle acceleration (eg. Fritz, 1989). 
The physics however remains poorly known, 
leading to vastly conflicting results.

There does however exist a strict limit to how steep a cutoff
can be achieved in the synchrotron spectrum of any relativistic particle
distribution. 
The convolution in Eqn.~\ref{distp} will always lead to a smoothing of
sharp features in the particle energy spectrum. 
This ensures that 
any observed cutoff in the synchrotron radiation spectrum will not be
steeper than the {\it least} steepest cutoff in either of the two
functions $P_{s}$ and $N(E)$ in Eqn.~\ref{distp}. 
We can immediately conclude that for observed radiation spectra
having a more abrupt frequency cutoff than the single particle spectrum
(Eqn.~\ref{singlep}), no physically plausible energy distribution
of particles can correspond to it alone. 
The steepest possible spectral 
cutoff (and hence slope) will therefore be that characteristic of a
single particle, or ``monoenergetic'' particle distribution: 
$N(E)\propto\delta(E-E_{o})$.
As shown by Eqn.~\ref{limithnu}, this will result in a spectrum which
becomes exponential towards high frequencies. 

\subsubsection{Monoenergetic Model Fits}
\label{monofits}

\begin{figure}
\vspace{-3in}
\plotonesmall{1}{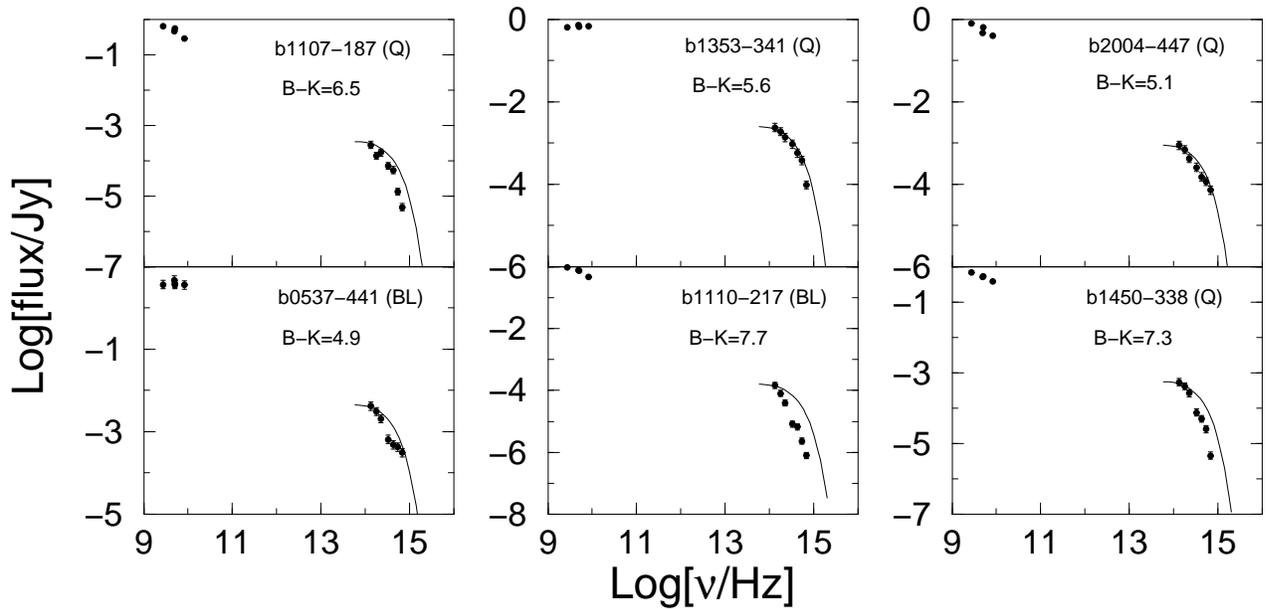}
\vspace{-1.7in}
\caption[Comparisons with the 
monoenergetic synchrotron model]{Multifrequency data (dots) for some of the reddest
Parkes sources. Low frequency data correspond to radio measurements
at 2.7-8.4GHz, and high frequency data are contemporaneous
photometric measurements
from $B_{J}$ to $K_{n}$. Source identifications are shown in brackets.
Solid lines represent the monoenergetic model prediction (essentially
the synchrotron spectrum of a single particle).}
\label{multi}
\end{figure}

Since the single particle spectrum 
(hereafter SPS; Eqn.\ref{singlep}) involves 
the steepest spectral turnover observable, we shall compare this with
the spectra of a number of the ``reddest'' Parkes sources.
The reddest sources are likely to exhibit spectra which cutoff abruptly
from the near-IR to optical.
We have chosen four quasars and two BL-Lacs with $B_{J}-K_{n}\simgt5$, and 
with contemporaneous photometric measurements in
$B_{J},V,R,I,J,H$ and $K_{n}$ obtained by P. Francis in April 1997 
(see Table~\ref{monmag}). 
These measurements, together with 
other available data, mostly in the radio at 2.7-8.4GHz are shown 
in Fig.~\ref{multi}.

As defined by Eqns.~\ref{numax} and~\ref{nuc}, 
the SPS has its
maximum at a frequency $\nu\simeq0.29\nu_{c}$. This frequency can thus also
be regarded as a turnover frequency, $\nu_{T}$ after which the spectrum
becomes exponential for $\nu\gg\nu_{T}$. 
Since the only effective parameter which determines
the SPS is $\nu_{c}$, this can be fixed by the
observed turnover frequency: $\nu_{T}\simeq0.29\nu_{c}$. 

In order to fit the spectra, we therefore need to determine $\nu_{T}$.
Due to the lack of sufficient far to near-IR data for our sources,
it is difficult to unequivocally determine the value of the
turnover frequency in each case. 
We can however make use of the fact that the values of $\nu_{T}$
found from previous observations of high frequency cutoffs 
in synchrotron dominated radio sources (see above) 
always seem to lie in the range $3\times10^{13}-10^{14}$Hz ($10-3\mu$m).
As a working measure, we therefore assume in each source the fixed value
$\nu_{T}=6\times10^{13}$Hz. 
For this value of $\nu_{T}$, we have computed the high frequency
monoenergetic radiation spectrum expected, with the
normalisation being the only free variable in each source. This prediction 
is represented by solid lines in Fig.~\ref{multi}. 

\begin{table}
\vspace{-0.3in}
\leftskip=-0.3in
\begin{tabular}{|l|c|c|c|c|c|c|c|c|c|}
\hline
Source & ID & Redshift & $B_{J}$ & $V$ & $R$ & $I$ & $J$ & $H$ & $K_{n}$\\ 
\hline
\hline
PKS0537$-$441 & BL & 0.893 & 17.935 & 17.342 & 16.924 & 16.348 & 14.76 & 13.85 &13.02 \\
PKS1107$-$187 & Q  & 0.497 & 22.444 & 21.098 & 19.295 & 18.736 & 17.46 & 17.21 &15.95 \\
PKS1110$-$217 & BL &  ?    & 24.410 & 23.010 & 21.570 & 21.100 & 19.06 & 17.82 &16.72 \\
PKS1353$-$341 & Q  & 0.223 & 19.199 & 17.456 & 16.738 & 15.940 & 15.20 & 14.38 &13.61 \\
PKS1450$-$338 & Q  & 0.368 & 22.525 & 20.396 & 19.388 & 18.688 & 16.99 & 16.03 &15.23 \\ 
PKS2004$-$447 & Q  & 0.250 & 19.586 & 18.737 & 18.176 & 17.349 & 16.48 & 15.46 &14.50 \\
\hline
\end{tabular}
\caption[Contemporaneous photometry for red quasars and BL-Lacs]{Contemporaneous photometry for `red' Quasars (Q) and BL-Lacs (BL).} 
\label{monmag}
\end{table}

We see that three of the
bluest spectra observed (with $B_{J}-K_{n}\simlt5.5$) 
agree very well with this prediction,
while in the remaining cases, 
the observed spectrum appears a little steeper. 
In particular, 
given the full possible range: $3\times10^{13}\simlt\nu_{T}\simlt 10^{14}$Hz, 
the monoenergetic spectrum predicts colours in the range
$3.5\simlt B_{J}-K_{n}\simlt6$. These are bluer than those observed in 
the reddest Parkes quasars with $B_{J}-K_{n}>6$ in Fig.~\ref{multi}. 
As shown in the
colour distribution of Fig.~\ref{histos}, these sources comprise
$\sim10\%$ of all unresolved Parkes sources with measured $B_{J}-K_{n}$
colours.

We must emphasise that the slope of the
SPS at high frequencies
sensitively depends on the turnover frequency $\nu_{T}$.
If instead $\nu_{T}\ll 10^{13}$Hz were found, then the almost exponential
behaviour of the SPS at high frequencies will predict considerably steeper
slopes than those observed.
The observed slopes would then satisfy the monoenergetic upper limit 
prediction, and
it would then be possible to find a `realistic' particle distribution
to describe the observations.

Given the nominal range for $\nu_{T}$ from previous
studies, we conclude that the
optical--to--near-IR continuum slopes of the reddest 
Parkes quasars are more or less
consistent and in some cases moderately
redder than the monoenergetic prediction.
These spectral slopes are as extreme as possible as
allowed by simple synchrotron theory.
Even particle energy distributions that are monoenergetic in nature are
strictly speaking not possible under any 
stochastic acceleration process.
Only power-law energy distributions are a natural consequence of 
conventional acceleration mechanisms (eg. Blandford \& Eichler, 1987).
Their statistical nature will never lead to distributions
exhibiting sharp features or ones that are close to monoenergetic. 
Due to the sensitive dependence of the SPS on $\nu_{T}$ however,
multifrequency observations at $\nu<10^{14}$Hz are crucially required
to confirm 
the result in Fig.~\ref{multi}.
At this stage, we conclude (speculatively) that the spectral slopes 
are inconsistent with a `pure' synchrotron mechanism alone.

\subsection{Summary}

Above, we have explored some simple first order predictions
of the synchrotron reddening model and compared them with available data.
We find that $B_{J}-K_{n}$ colours do not correlate with either
the strength of radio core emission or the EWs of most broad lines. 
These results are contrary to that expected if the colours were purely
due to a beamed synchrotron component dominating the near-IR flux.
Guided by previous observations of high frequency spectral 
cutoffs in synchrotron dominated radio sources,
we find that the reddest Parkes quasars (with $B_{J}-K_{n}\simgt5$)
exhibit slopes which have to be
as extreme as possible as allowed by synchrotron theory.
Unrealistic particle energy distributions are required in most cases.
These findings {\it do not} provide conclusive evidence
against the synchrotron model, although they suggest that
all of the observed reddening cannot 
be purely attributed to this mechanism alone. 

\section{Other `Intrinsically Red' Emission Models}
\label{otherints}

In this section, we explore the consequences of two other
reddening models based on intrinsically red emission: First, a 
model where the redness is due to intrinsically weak optical/UV emission,
and second, a model involving thermal emission from dust
dominating in the near-infrared. 

\subsection{`Weak' Optical/UV Emission?} 
\label{ewweak}

At present, there is considerable
evidence suggesting that the optical-UV continua
and source of photoionizing flux in a majority
of quasars arises from a common mechanism.
Thermal processes, such as optically thin 
free-free emission from hot gas or optically thick emission 
from an accretion disk are currently most favoured (Barvainis, 1993 and
references therein).
Regardless of the nature of the photoionizing source,
we can test a simple hypothesis using the EWs of emission lines 
and
colours observed.
This hypothesis states that the redness in continuum slopes
may be due to an intrinsically variable photoionizing
optical/UV continuum source. 
A decrease in the photoionizing flux
beyond the Lyman-limit ($\lambda<912$\AA) is expected to 
depress the optical ($B_{J}$)
continuum significantly if both arise from a common mechanism.
McDowell \etal (1989) identified such a class of quasars with weak
`blue bump' spectra relative to both the infrared and soft X-rays.
These sources also tended to have large $B-K$ colours.

If the redness in Parkes quasars were due to such a mechanism, then  
a decrease in the photoionizing flux 
from the central AGN also implies a decrease in the flux to be
reprocessed by the surrounding gas. To first order, 
we then expect
a decrease
in the flux emitted by such gas and hence the EWs of emission lines.
As a consequence, a strong anticorrelation between the EW of a 
particular emission line
and $B_{J}-K$ colour (or optical continuum slope) is expected.

\begin{figure}
\vspace{-3in}
\plotonesmall{1}{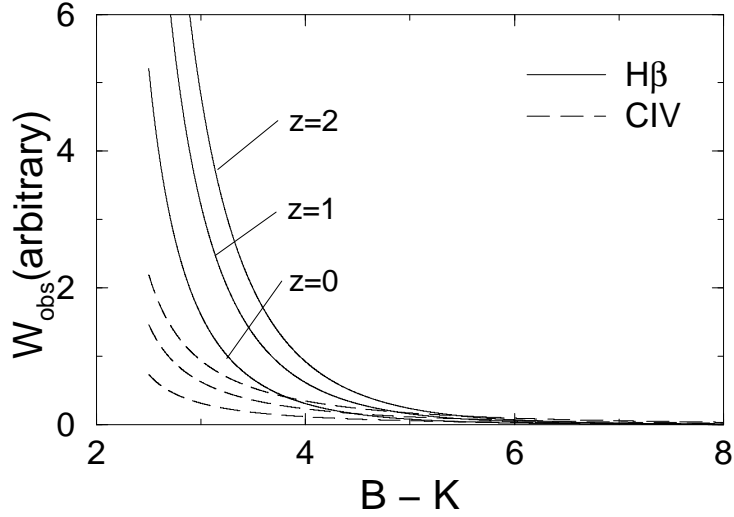}
\vspace{-2.8in}
\caption[Prediction for $B-K$ vs. equivalent width in
the `weak blue-bump' model]{Predicted relations between observed
emission line equivalent width and $B-K$ colour in the `weak blue bump' 
model for sources at redshifts: 0,1, and 2.
Predictions are for the lines:
H$\beta$ ($\lambda=4861$\AA; solid lines) and
C{\small IV} ($\lambda=1549$\AA; dashed lines).}
\label{EWpredphot}
\end{figure}

To explore the strength of this correlation, we have used a simple
model which assumes:
a power-law optical-UV and hence photoionizing
continuum, and, that
all ionizing photons available to the gas are involved in an
ionization process so that $L_{line}\propto$ number of photons
with $\lambda<912$\AA$\,$ (see Osterbrock, 1989). 
Figure ~\ref{EWpredphot} shows the predictions
of such a model for the H$\beta$ and C{\small IV} lines. 
With the additional assumption that a power-law extrapolated between
$K$ and $B$ can be extended into the far-UV,
the observed EW of an emission line
with rest wavelength $\lambda_{l}$ from a source at redshift $z$
will scale as
\begin{equation}
W_{obs}(\lambda_{l})\,\propto\,(1+z)\lambda_{l}^{2}
\frac{1}{\alpha_{BK}}\left(\frac{\lambda_{l}}{912{\rm \AA}
}\right)^{-\alpha_{BK}},
\label{EWform}
\end{equation}
where $\alpha_{BK}$ is the $K$ to $B$ (and hence photoionizing) continuum
slope (see Eqn.~\ref{slopcol}).

This model may be an oversimplification for two reasons:
first, a power-law continuum from $K$ to $B$ may not extend
into the far-UV and second, the intensity in an individual emission line
may not be strictly proportional to the ionizing continuum flux.
A proportionality between emission line and ionizing flux
is more likely to be observed for lines
dominated by recombination processes alone, such as those resulting from 
Balmer and Lyman series transitions in H and He.
Evidence for this was provided by Shuder (1981), where strong
correlations between line fluxes and optical-UV continuum luminosity
in a large AGN sample was found. 
Photoionization model predictions however show that emission lines
resulting from heavy elements (eg. C,N,O,Mg)
depend on physical conditions such as temperature and density in a 
very complicated way (eg. Netzer, 1990).
Similar correlations involving such lines are
not immediately expected. 
Nevertheless, if emission lines are primarily due to reprocessing 
of UV photoionizing radiation by surrounding gas, we must expect
on average a partial correlation between the two. 
Such correlations however may be avoided if the physical
properties of emission line regions (such as 
densities or covering factors) scale with photoionizing flux
such as to overcome variations in photoionizing flux. 
As discussed by Francis \etal (1997), such compensating effects
require physically unmotivated fine tuning of fundamental AGN properties.

An examination of the observed trend in EW versus $B_{J}-K_{n}$ colour in
Fig.~\ref{EWdata} shows that for most lines, 
no significant anticorrelation is present.
This is incompatible with the first order 
predictions of the `weak blue bump' model
shown in Fig.~\ref{EWpredphot}. 
We conclude that such a mechanism is unlikely to account for the
diversity in continuum slopes observed.

\subsection{Emission from Hot Dust} 

It can be shown that the redness in near-IR to optical continuum slope
{\it cannot} 
arise from a pure {\it thermal} radiation mechanism
associated
with emission by dust.
Assuming the near-IR were due to thermal reprocessing by dust, 
we shall investigate whether the range of continuum slopes observed is 
consistent with temperatures attainable by grains.
Near-IR emission by heated dust is believed to peak at
rest wavelengths $\lambda\simgt1.5\mu$m for a dust
sublimation temperature $T_{d}\sim1750$K, 
characteristic of graphite and silicate grains (Laor \& Draine, 1993). 
As a simple estimate, we can predict the spectral slope $\alpha$
(where $f_{\nu}\propto\nu^{-\alpha}$) in the high frequency (Wien) 
limit expected from a thermal dust radiation spectrum.
Given that in the Wien limit, the observed flux at frequency $\nu_{o}$ from a 
source at redshift $z$ scales as 
$f_{\nu_{o}}\propto\nu_{o}^{3}\exp\left[-(1+z)h\nu_{o}/kT_{d}\right]$,
the spectral index $\alpha$ ($\equiv-d\ln{f}/d\ln{\nu}$) 
at some wavelength $\lambda_{o}$
can be written
\begin{equation}
\alpha\,\simeq\,-3\,+\,8
(1+z)(T_{d}/1750{\rm K})^{-1}(\lambda_{o}/\mu{\rm m})^{-1}.
\label{alpha1}
\end{equation}

To achieve spectral indices at wavelengths $\lambda_{o}<1\mu$m of
$\alpha_{\nu}<3$ (or equivalently $B-K<8$), we require dust
temperatures $T_{d}\simgt2330(1+z)$K. 
For $z=1$, typical for Parkes sources, this implies $T_{d}\simgt4500K$. 
Such values are considerably greater
than the sublimation temperature, $T_{d}\sim1750$ for dust grains 
composed of graphite
and silicates. 
The upper limit $T_{d}<T_{sub}\sim1750$K in Eqn.~\ref{alpha1} will result in
slopes at $\lambda_{o}<1\mu$m of $\alpha_{\nu}\simgt13$ or colours
$B_{J}-K_{n}>20$ (for $z=1$) if thermal dust emission was dominant. 
This limit is obviously much greater than the range of 
colours observed (see Fig.~\ref{histos}). 
We conclude that `pure' thermal emission from dust is unlikely to
account for the red colours observed. 
If this mechanism were dominating in the near-IR, then 
the observed continuum slopes require
grain temperatures in excess of the maximum 
attainable by grains after which they sublime. 
Measurements of the polarisation level at near-IR wavelengths
may provide a clue as to the nature of the emission 
mechanism
(see Chapter 7).

\section{Preliminary Tests of the Dust Model}
\label{pretd}

While the reddening in weak-lined BL-Lac type sources may 
be purely intrinsic in origin (eg. beamed synchrotron), 
the data for Parkes quasars is broadly inconsistent with this mechanism. 
Intrinsically red (possibly beamed) emission is unlikely to dominate
into the optical/UV of these sources without significantly affecting 
their emission line
EWs. What is required is a mechanism that reddens by making the
high frequency emission fainter relative to the near-IR. 
In the following sections, we shall explore the possibility that extinction 
by dust
contributes to the reddening observed.

\subsection{Emission Line Equivalent Widths}
\label{emewdust}

As discussed above, the 
observed
trends between EW and $B_{J}-K_{n}$ colour 
shown in Fig.~\ref{EWdata} 
are generally inconsistent with the 
strong anticorrelations predicted by both the
`synchrotron' and `weak bump' emission models. 

For most {\it broad} lines, no significant 
anticorrelations are observed. 
This result favours a mechanism where broad emission line
and continuum fluxes are reddened equally, and is consistent with reddening
of both components by dust. 
Since dust is not expected to survive in the BLR 
(eg. Laor \& Draine, 1993), 
this can be explained by the presence of a dust screen
in front of the BLR and central continuum source somewhere along the
line-of-sight. 

As shown in Fig.~\ref{EWdata}, 
the EW of the {\it narrow} line [OII] (the 
strongest narrow line that can be measured), is seen to
actually {\it correlate} with $B_{J}-K_{n}$ colour.
This may be due to patchy dust obscuration in the NLR.
Depending on geometry, there is the possibility that the NLR gas will see
a stronger central ionizing source than we do if the
latter is obscured from direct view
by some cloud of dust.
More spectroscopic data however is required to strengthen this
explanation.

We conclude that the EW versus reddening data of Fig.~\ref{EWdata}
is consistent with that expected if extinction by dust 
were equally affecting
broad line and continuum fluxes. 

\subsection{Optical Continuum Slope versus $B-K$ Colour}
\label{optvscol}
 
\begin{figure}
\vspace{-3in}
\plotonesmall{1}{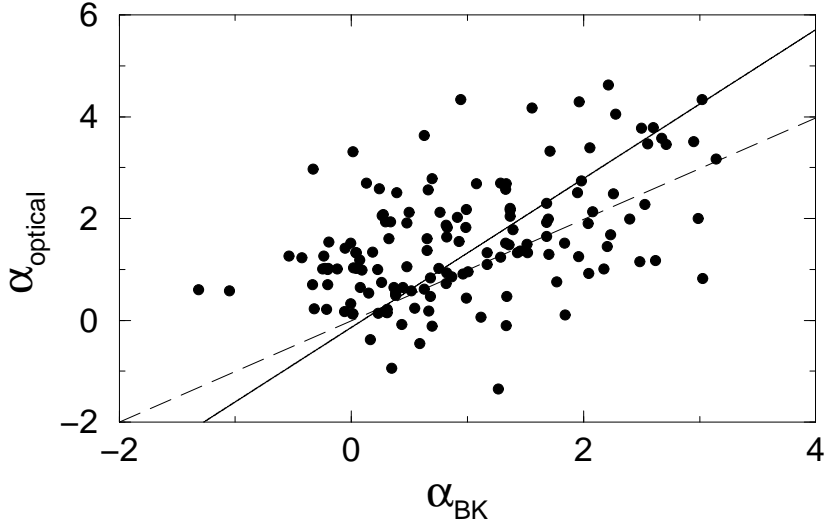}
\vspace{-2.5in}
\caption[Optical spectral slope as a function of near-IR--to--optical continuum slope]{
Optical continuum slope ($\alpha_{optical}\simeq\alpha_{0.3-0.8\mu m}$) 
as a function of 
near-IR--to--optical continuum slope 
($\alpha_{BK}\equiv\alpha_{0.44-2.2\mu m}$) where 
$f_{\nu}\propto\nu^{-\alpha}$ for Parkes quasars. 
The solid line represents 
the prediction of a simple dust model which assumes
an intrinsic QSO slope of $\alpha_{int}=0.3$ and a $1/\lambda$ dust extinction
law. The dashed line is the line of equality, $\alpha_{optical}=\alpha_{BK}$. 
}
\label{optslopevscol}
\end{figure}

In the preceding sections, we have mainly been concerned with interpreting
the relatively large spread in $B-K$ colours of Parkes quasars.
Compared to 
optically selected quasars, a
large spread is also observed in optical
($\sim3200$-8000\AA$\,$ observed frame) 
continuum slope (see Fig.~\ref{histos}).
This implies that the reddening extends throughout the rest frame optical-UV
and must also be explained.
As shown in Fig.~\ref{optslopevscol}, 
a weak correlation between 
optical continuum slope
and $B-K$ slope ($\alpha_{BK}$) is observed.
Results of a Spearman's rank correlation 
test show that this is significant at the 95\% confidence level. 
The scatter in optical continuum slope in Fig~\ref{optslopevscol} appears
to be greater by almost a factor of two than that in 
power-law slope interpolated
between $B_{J}$ and $K_{n}$.
This is consistent with that predicted from reddening by dust. 
Due to its near-inverse wavelength dependent nature, extinction
by dust introduces a characteristic spectral curvature that increases
progressively towards shorter wavelengths.

To explore this in more detail, we have predicted
the relation between optical and $B-K$ slope  
expected using a simple dust model.
The dust optical depth at any ``observed'' wavelength $\lambda_{obs}$
is assumed to
follow a $1/\lambda$ type behaviour: 
\begin{equation}
\tau_{obs}\,=\,\tau_{B}(1+z)
\left(\frac{\lambda_{obs}}{\lambda_{B}}\right)^{-1},
\label{dustzlaw}
\end{equation}
where the redshift dependence accounts
for the characteristic decrease in
rest frame wavelength
with redshift,
and $\tau_{B}$ is the rest frame optical depth at $\lambda_{B}=4400$\AA.
We have also assumed
an intrinsic ``unreddened'' optical continuum slope of 
$\alpha_{int}=0.3$ - a typical
mean value observed in optically selected quasars (Francis \etal 1996). 
The predicted correlation is shown as the solid line 
in Fig.~\ref{optslopevscol} 
and is independent of both the normalising factor 
$\tau_{B}$, and redshift in Eqn.~\ref{dustzlaw}.
This prediction agrees with $>2\sigma$ significance 
with the general trend observed.

Equation~\ref{dustzlaw} makes a simple prediction for the redshift
dependence of reddening observed.
If the dust is assumed to be physically associated with the
sources themselves, then as implied by Eqn.~\ref{dustzlaw},
an increase in extinction in an observer's frame
and hence continuum slope
with source redshift is expected.
No such increase however is observed (eg. see Fig.~\ref{colvsz}).
The lack of an increase in observed reddening with
redshift may be due to a compensation 
by intrinsic colour evolution in the source emission,
or evolution in the dust content itself. 

The observed correlation between optical and $B-K$ continuum
reddening also argues against the synchrotron model 
whereby the redness is due to
intrinsically red (possibly beamed) 
emission dominating in the near-IR.
If the reddening in $B-K$ were purely due to
such a component superimposed on a normal `blue'
QSO spectrum in the optical, then a correlation with $\alpha_{opt}$ is not
immediately expected. 
As discussed in section~\ref{ewsynch}, a red beamed component
is also unlikely to contribute            
significantly in the optical
without drastically reducing emission line EWs.            
The fact that a large fraction of 
optical continuum slopes are slightly redder than those
interpolated from $B_{J}$--to--$K_{n}$ (ie. those above the dashed line in 
Fig.~\ref{optslopevscol}),
is inconsistent with this model.
The reddening in $B-K$ colour
for Parkes quasars must be connected in some way with that observed
at shorter wavelengths - ie. the optical-UV.
Dust offers the simplest and best explanation.

\subsection{Balmer Decrements}
\label{Balmpks}

Emission line ratios are the most widely used indicators of 
reddening by dust (eg. Draine \& Bahcall, 1981). 
For Parkes quasars, most of the emission lines are too
weak to provide any reliable information.
In a few of the higher quality spectra however, Balmer decrements, defined
as the flux ratio: ${\rm H}\alpha(\lambda=6561{\rm \AA})/
{\rm H}\beta(\lambda=4861{\rm \AA})$ have been measured. 
Since very few spectra extend far enough to the red to show the
H$\alpha$ emission line for redshifts $z\simgt0.5$, all
spectra with the H$\alpha$ line were visually inspected and only
quasars with H$\alpha$ velocity widths $>2000{\rm km}\,{\rm s}^{-1}$
(at FWHM) were used. 

\begin{figure}
\vspace{-3in}
\plotonesmall{1}{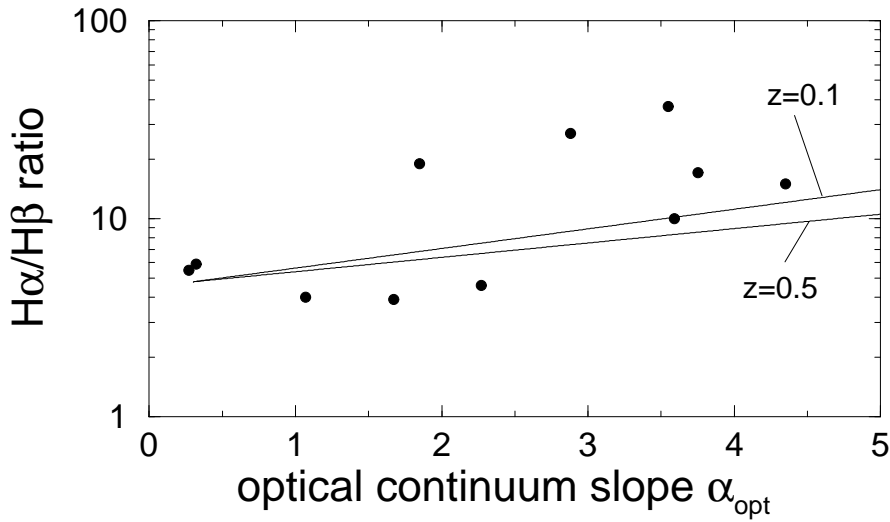}
\vspace{-2.65in}
\caption[Observed Balmer decrements as a function of optical slope]{Balmer decrements $({\rm H}\alpha/{\rm H}\beta)$ as a 
function of optical continuum power-law slope $\alpha_{opt}$.  
The lines represent the prediction of a simple dust model for
two dust redshifts. This assumes 
an unabsorbed QSO slope of $\alpha_{opt}=0.3$ and a $1/\lambda$ extinction
law.}
\label{Balmer}
\end{figure}

The Balmer decrements are shown as a function of optical 
continuum slope $\alpha_{opt}$ ($f\propto\nu^{-\alpha_{opt}}$) in 
Fig.~\ref{Balmer}.
The reddest quasars are seen to have extremely large Balmer decrements,
while the four bluest quasars have decrements of $4.8\pm0.9$
consistent with the average value 
$\langle{\rm H}\alpha/{\rm H}\beta\rangle\sim4.5$ 
{\it observed} 
for optically-selected quasars (Netzer, 1990 and references therein).
The observed correlation (significant at the 95\% level),
is consistent with a simple dust model that assumes an intrinsic
(unabsorbed) optical slope of $\alpha_{opt}=0.3$ and a $1/\lambda$
extinction law.
We must note that the statistics in 
Fig.~\ref{Balmer} are too low from which 
to make the dust model conclusive 
on the basis of emission line ratios alone.

\subsection{Summary}

In this section, we have presented three pieces of observational
evidence which argue in favour of the dust reddening mechanism. 
First, the fact that continuum slopes do not correlate
with the EWs of most broad lines is consistent with a 
scenario where dust reddens both continuum and broad line emitting regions
equally.
Second, optical continuum slopes correlate with $B_{J}-K_{n}$ colours
in such a way that is consistent with the predictions of a simple dust model.
Third, a correlation is observed between the few available measures
of Balmer decrements and optical continuum slope.
These last two findings provide the strongest possible evidence for the dust
model.

\section{Soft X-ray Absorption}
\label{sxabsp}

As reviewed in section~\ref{dustobscu}, 
there have been numerous studies reporting
the presence  of
soft ($\simlt2$keV) X-ray absorption in excess of that expected from the galaxy 
towards radio quasars.
As discussed by Elvis \etal (1994) however, 
such studies have found very little evidence for 
associated optical reddening by dust.
Strong evidence for associated MgII and soft X-ray absorption
in a number of radio loud quasars has been confirmed 
(eg. Mathur \etal 1994), though evidence for associated optical extinction
in these sources is weak.
A possible explanation for this contradiction is that 
since X-ray absorption estimates are derived from
spectra which require relatively large X-ray counts, 
such studies may be biased against those sources 
with low counts due to strong X-ray (and hence optical) absorption. 

If strong optical extinction by dust is known apriori however, 
then 
associated absorption of soft X-rays is expected to be present at some level. 
Evidence for excess soft X-ray absorption in a number of optically
reddened radio quasars has been presented by Kollgaard \etal (1995).
These authors however claim that their results 
are strongly model dependent, and
also consistent
with explanations other than that due to absorption by 
associated gas and dust.
Their statistics are also too low from which to draw any reasonable
conclusion.
Strong evidence for an association
was recently reported by Puchnarewicz \etal (1996)
for a large sample of Seyfert 1s and quasars.
These authors found a correlation between optical reddening and
optical-to-soft X-ray continuum slope that was consistent with
absorption by dusty ``cold'' gas with column densities 
$>10^{21}{\rm cm}^{-2}$ and approximately Galactic dust abundance.  

Unfortunately, very little spectroscopic X-ray data
for ``red'' Parkes quasars exists from which
absorption gas columns can be constrained.
We do however have access to soft X-ray ROSAT broadband data
for a majority of sources in the Drinkwater \etal sample. 
Based on predictions involving
optical-to-soft X-ray continuum slope, 
we will use such data to further test the dust hypothesis.
In this section, we first 
present some predictions of the effects of dust (and associated gas) 
on soft X-rays at energies 0.1 - 10keV.
We then use the broadband soft X-ray data
to investigate whether X-ray
absorption is present at a level consistent with that implied by
the observed optical reddening.

\subsection{Theoretical Investigation}
\label{Theory}

The degree of X-ray absorption by metal enriched gas in the range 
0.1-10keV primarily depends on two independent physical parameters:
first, on the column density of absorbing gas in the line-of-sight
and second, on the effective absorption cross-section of the
heavy metal species present.
Cross-sections for heavy metals with $Z>2$ are $\sim5-10$ times greater
than that for neutral hydrogen and helium. Thus, the HI absorbing column
required for a given amount of absorption can be less if the metal 
abundance is high.
In the case of the galactic ISM where the metal abundance is typically
$\simlt1$\% relative to hydrogen by mass (Tinsley, 1976),
hydrogen and helium are responsible for almost all of the absorption
at energies $\simlt1$keV.
At higher energies however, where absorption is dominated by metals, the 
absorption is significantly reduced (see below). 

As pointed out by Fireman (1974), condensation of heavy metals into dust
grains will shield atoms from X-ray absorption and hence reduce
their effective cross-section.
For a typical galactic ISM grain radius $\sim0.15\mu$m,
the decrease in soft X-ray attenuation is predicted to be 
moderate to moderately low, 
being reduced at most by $\sim10\%$ at 0.3keV and $\sim4\%$ at 1keV.
At higher energies, dust grains essentially become transparent 
(Morrison \& McCammon, 1983).

We now predict the amount of X-ray absorption expected for a given
optical dust extinction measure. 
Before we do so however, we need a relation defining the total
gas column density
(on which X-ray absorption estimates are based) 
in terms of the optical extinction. 
For simplicity, we assume the relationship between mean total gas column
and optical reddening as derived empirically from Ly-$\alpha$
absorption measurements in the galaxy (Bohlin \etal 1978):
\begin{equation}
N({\rm HI+H_{2}})_{tot}\,\simeq\,5.8\times10^{21}E(B-V)
\,{\rm cm}^{-2}{\rm mag}^{-1}, 
\label{ext_col}
\end{equation}
where 
$E(B-V)$ is the extinction in $B-V$ colour. 
This relation is also
consistent with empirical estimates of the dust-to-gas ratio in the
SMC and LMC by Bouchet \etal (1985) and Fitzpatrick (1985). 
It must be emphasised that Eqn.~\ref{ext_col} is only based on measures
of the total {\it hydrogen} column density, while as discussed
above, X-ray absorption also depends on the abundance of heavy
metals both in the gas phase and locked up in dust.
The relative abundance of these components however is $\simlt0.1N({\rm HI})$,
and hence will make neglible contribution to the numerical factor in
Eqn.~\ref{ext_col}. 

Let us assume that absorption 
occurs from dusty gas at redshift $z_{d}$
along the line-of-sight to a source. Then assuming a $1/\lambda$
dust extinction law, we have in the absorber rest frame:
$E(B-V)\simeq0.25(1+z_{d})^{-1}A_{V}({\rm obs})$, 
where $A_{V}({\rm obs})$ is the extinction
measured in an observer's $V$ passband. 
We can thus re-write Eqn.~\ref{ext_col} as follows:
\begin{equation}
N_{H}(A_{V},z_{d})\,\simeq\,(1.5\times10^{21})(1+z_{d})^{-1}A_{V}({\rm obs})\,
{\rm cm}^{-2}{\rm mag}^{-1}.   
\label{newNh}
\end{equation}
We shall adopt the X-ray absorption cross section 
as a function of energy as derived numerically by Morrison \& McCammon (1983)
for a gas with galactic ISM metal abundances.
Denoting the cross-section at some ``observed'' 
energy $E_{o}$ by $\sigma_{(1+z_{d})E_{o}}$ (which has units:
${\rm cm^{2}}$ per H atom), the fraction of X-ray flux attenuated
in an observer's frame can be written:
\begin{equation}
f(E_{o})\,=\,1\,-\,\exp{\left[-\sigma_{(1+z_{d})E_{o}} 
\,N_{H}(A_{V},z_{d})\right]},
\label{atten}
\end{equation}
where $N_{H}$ is defined by Eqn.~\ref{newNh}. 

\begin{figure}
\vspace{-2.5in}
\plotonesmall{1}{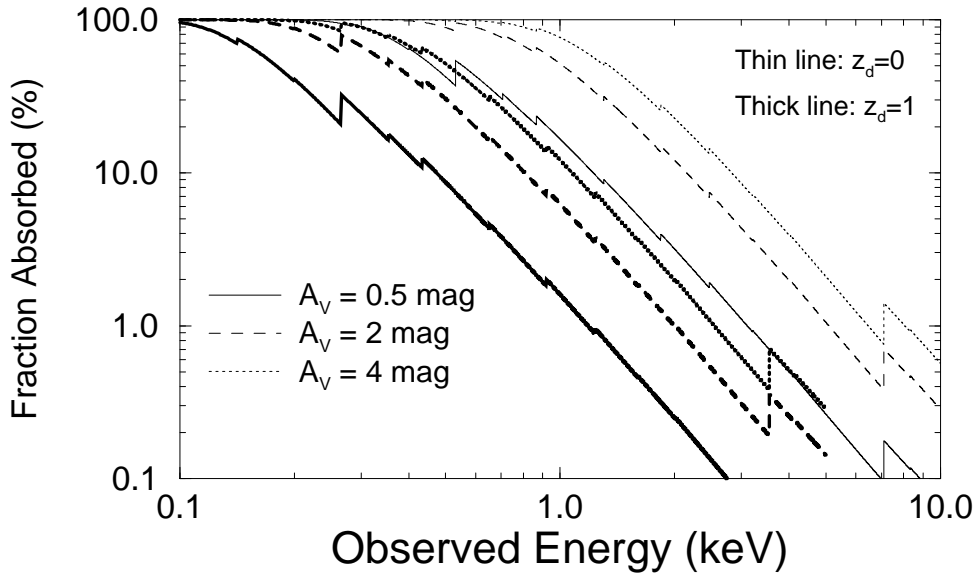}
\vspace{-2.7in}
\caption[X-ray attenuation as a function of observed energy]{X-ray attenuation as a function of observed energy
for a gas-dust absorber at redshifts $z_{d}=0$ and $z_{d}=1$ (thin and
thick lines respectively) and three dust extinction measures $A_{V}$.
Galactic ISM gas columns and metal abundances are assumed. 
}
\label{Xext}
\end{figure}

This fraction 
is shown in Fig~\ref{Xext} for a gas-dust absorber at two redshifts $z_{d}$
and three values of the observed extinction 
$A_{V}({\rm obs})$.
As can be seen, the attenuation is strongest at energies $<1$keV which is
predominately due to hydrogen. If most of this hydrogen were ionized
or absent however, then given a galactic heavy metal abundance $\sim0.1N(H)$, 
the attenuation at 1keV will be reduced by $\sim80\%$. 
Another apparent feature is that the observed X-ray attenuation at some energy 
is expected to decrease with increasing absorber redshift $z_{d}$. 
This is due to the characteristic decrease in effective 
absorption cross-section,
$\sigma(E)$, with energy in the absorber rest frame.  
The implications of this result on soft X-ray QSO surveys will be discussed
in section~\ref{impsurv}.

We must emphasize that the extinction-column density relation
defined by Eqn.~\ref{newNh} on which Fig.~\ref{Xext} is based,
only holds for the specific conditions characteristic of the
galactic ISM. 
The amount of soft X-ray absorption associated with optical extinction by
dust crucially depends on a number of physical properties of the
absorbing gas-dust mixture. 
It may be possible in some cases to have strong X-ray absorption and
neglible optical extinction as observed in number 
of high redshift radio quasars by Elvis \etal (1994).
Such a situation may be explained by any of the following:
A metal and hence dust
abundance significantly lower than that observed in the local ISM,
temperatures that are too high for dust to survive, 
only large grains exist which cause little reddening.
The opposite however, where one observes only strong optical extinction
is not immediately expected since dust is always known to form from
accretion of heavy metals in relatively cold molecular environments
where X-ray absorption will be inevitable (eg. Wang, 1991a). 
As an initial working hypothesis, our predictions in the next section 
shall assume an extinction-to-gas column conversion
relation and metal abundances characteristic 
of the galactic ISM.

\subsection{Analysis of ROSAT Data}
\label{rosatan}

If the relatively large spread in optical continuum slopes of Parkes
quasars is due to reddening by dust, then a number of
simple predictions involving
optical--to--soft X-ray and near-infrared--to--soft X-ray
continuum slope can be made.
In this section, we first discuss the model predictions 
and assumptions and then compare
them with available ROSAT broadband data.

\subsubsection{Predictions \& Assumptions}

If a source at redshift $z$ is observed through a given amount of dust
and gas situated at redshift $z_{d}\leq z$, then some observed 
optical--to--X-ray
power-law slope, say $\alpha_{BX}$ (where $f_{\nu}\propto\nu^{-\alpha}$), 
can be written in terms of the 
corresponding intrinsic (unabsorbed) value and extinction optical depths
as follows:
\begin{equation}
\alpha\left[\nu_{B},\nu_{X}\right]_{obs}\,=\,
\alpha\left[\nu_{B}^{'},\nu_{X}^{'}\right]_{int} + 
\frac{\tau[(1+z_{d})\nu_{X}]-\tau[(1+z_{d})\nu_{B}]}{\ln{(\nu_{X}/\nu_{B})}},
\label{alphaOX}
\end{equation}
where in the source rest frame, $\nu_{B}^{'}=(1+z)\nu_{B}$ and
$\nu_{X}^{'}=(1+z)\nu_{X}$. 
The optical depths $\tau$ are evaluated at frequencies 
in the absorber rest frame. 
As implied by Eqn.~\ref{atten}, the extinction optical depth at some
X-ray frequency in the absorber frame is defined as  
$\tau_{X}=\sigma\left[(1+z_{d})\nu_{X}\right]
N(H)_{tot}$,
where $\sigma$ is the effective absorption cross-section per H atom, and
$N(H)_{tot}$ the total atomic column density in the line-of-sight.
Using the definition for $N(H)_{tot}$ from Eqn.~\ref{newNh} and
assuming a $1/\lambda$ type extinction law,
we can write the absorber rest frame optical depths in 
Eqn.~\ref{alphaOX} as follows: 
\begin{equation}
\tau_{X}\,\simeq\,\sigma\left[(1+z_{d})\nu_{X}\right]
(1.5\times10^{21})(1+z_{d})^{-1}E_{obs}(B-K),
\label{optdepths}
\end{equation}
$$
\tau_{B}\,\simeq\,1.14E_{obs}(B-K),  
$$
where $E_{obs}(B-K)$ is the colour excess measured in an observer's frame.
This is defined as
\begin{equation}
E_{obs}(B-K)\,=\,(B-K)_{obs} - (B-K)_{int}, 
\label{ebminusk}
\end{equation}
$$
{\hspace{1.8cm}}\approx\,\alpha_{obs}({\rm opt})\,-\,\alpha_{int}({\rm opt})
$$
where $(B-K)_{int}$ is the intrinsic (unabsorbed) colour.
The last step follows from our assumption of a $1/\lambda$ extinction law
and an optical slope measured between $\sim3000-8000$\AA$\,$ in the
{\it observed} frame (with $f_{\nu}\propto\nu^{-\alpha}$).
We shall also consider
the observed 
near-IR(K)-to-X-ray continuum slopes, 
$\alpha_{KX}$, of Parkes quasars in our analysis.
These can be defined by a relation 
analogous to Eqn.~\ref{alphaOX} with
$\tau_{B}$ replaced by
$\tau_{K}$ in the absorber rest frame where
\begin{equation}
\tau_{K}\,\simeq\,0.23E_{obs}(B-K). 
\label{Koptdepth}
\end{equation}
We can see from Eqns.~\ref{alphaOX} - \ref{ebminusk} that
for given values of $(B-K)_{int}$ and $\alpha_{BX}$(int),  
the dust model will predict a specific correlation between
observed colour $(B-K)_{obs}$ and optical-to-X-ray (or near IR-to-X-ray)
continuum slope.

As an initial working measure, we assume for simplicity
the intrinsic (unabsorbed) slopes: 
$(B-K)_{{\rm int}}$ (or $\alpha_{int}({\rm opt})$), $\alpha_{BX}$(int)
and $\alpha_{KX}({\rm int})$ to be fixed at the mean values found for
optically selected quasars. 
As argued in section~\ref{redbiasopt}, 
this choice is based on our claim that optically-selected
quasars are expected to be strongly biased against 
significant absorption by dust. 
The values assumed are discussed as follows:

First, from the relatively small scatter observed in $B-K$ colour for
quasars in the LBQS sample (see Fig.~\ref{histos}), we assume the mean value 
$(B-K)_{{\rm int}}\simeq2.3$.
Similarly, for unabsorbed optical continuum slope, we assume the mean
value $\alpha_{int}({\rm opt})=0.3$.
A number of studies have also found a relatively 
small scatter in the optical--to--soft X-ray (1keV) flux ratio of
optically-selected quasars 
(eg. Kriss \& Canizares, 1985; Wilkes \etal 1994; La Franca
\etal, 1995). The distribution in these flux ratios indicates a mean value 
$\alpha_{BX}({\rm int})\sim1.5$ and does not significantly differ
between radio loud and radio quiet quasars 
(Wilkes \& Elvis, 1987; Green \etal 1995).    
This value is also consistent with that implied
by a composite SED
for radio loud (optically-selected) quasars as 
derived by Elvis \etal (1994).  
From this composite, we find a $K$-to-1keV continuum slope 
of $\alpha_{KX}({\rm int})\sim1$. 
This composite also indicates that for source redshifts $z\approx0-3$,
the intrinsic slopes $\alpha_{BX}$ and $\alpha_{KX}$
will be approximately constant in an observer's frame.
This is
consistent with the redshift independence in optical--to--soft X-ray spectral
index found for both radio loud and radio quiet quasars in the LBQS sample
(Green \etal 1995).
Thus, our dust model predictions (based on 
Eqns.~\ref{alphaOX} - \ref{Koptdepth}) will assume 
the following intrinsic slopes:
\begin{equation}
(B-K)_{{\rm int}}\,=\,2.3,
\label{intslopes}
\end{equation}
$$
\alpha_{int}({\rm opt})\,=\,0.3
$$
$$
\alpha_{BX}({\rm int})\,=\,1.5,
$$
$$
\alpha_{KX}({\rm int})\,=\,1.
$$

\subsubsection{Comparison with Data}

X-ray count rate data from the PSPC detector on the ROSAT satellite
was provided by Siebert \etal (1997). 
The data applies to the broad soft X-ray band: 0.1-2.4keV and is available
for 266 of the 323 sources in the Drinkwater \etal sample.
We have converted the observed counts and available upper limits
into monochromatic X-ray fluxes at 1keV assuming a
pure power-law
continuum ($f_{\nu_{X}}\propto\nu^{-\alpha_{X}}$) between 0.1-2.4keV.
Since we are interested in total observed fluxes in the
presence of absorption, no absorption corrections have been made.
A mean slope of $\alpha_{X}=1.2$ as found for radio loud LBQS quasars
in the ROSAT band: 0.1-2.4keV was assumed (Schartel \etal 1996). 
These authors also find that $\alpha_{X}$ for radio quasars appears
to become flatter with increasing redshift in the range: $z\sim0-2$,
contrary to previous claims (Canizares \& White, 1989 and references therein).
The value of $\alpha_{X}=1.2$ assumed from Schartel \etal (1996), 
corresponds to our median redshift of 1.
Detector efficiency factors and method for the counts-to-flux conversion
are given in the ROSAT Users' Handbook (1994).

\begin{figure}
\vspace{-1.5in}
\hspace{-0.7cm}
\plotonesmall{1}{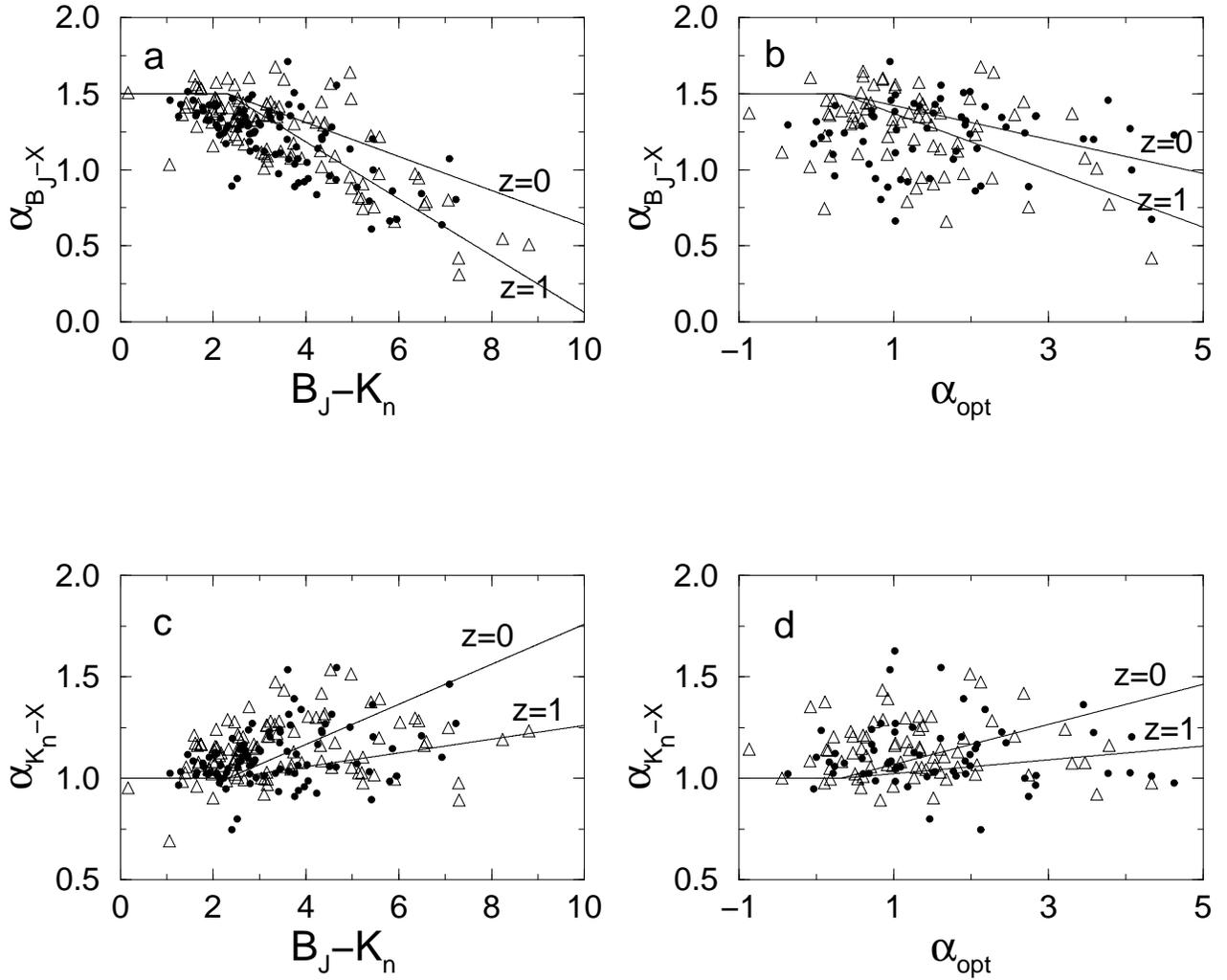}
\caption[Observed and predicted relations between continuum slopes 
involving the soft X-ray band] 
{{\bf a.} $B_{J}$-to-1keV power-law continuum slope
as a function of $B_{J}-K_{n}$ colour and
{\bf b.} $B_{J}$-to-1keV continuum slope as
a function of optical continuum slope.
{\bf c} and {\bf d:} same as {\it a} and {\it b} but for
$K$-to-1keV power-law slope.
Triangles represent lower limits on the
spectral indices and correspond to upper limits
on the measured X-ray flux.
Our galactic gas-dust
model predictions are shown for two absorber redshifts.}
\label{slopesX}
\end{figure}

Using the derived 1keV fluxes and upper limits, we have computed the
observed spectral indices: $\alpha_{KX}$ and $\alpha_{BX}$ for Parkes quasars. 
These are shown as a function of $B-K$ colour 
and optical continuum slope
in Fig.~\ref{slopesX}.
Taking into account the lower limits on spectral indices,
we have computed Kendall's tau rank correlation coefficient in each
case. We find that all correlations are significant at the 5\% level.
Also shown in Fig.~\ref{slopesX} 
are the predictions of our simple dust model
for absorber redshifts: $z_{d}=0$ and 1.  
The observed correlations appear consistent with our model predictions
which assume
a galactic ISM extinction-to-gas column relation and metal 
abundances, and intrinsic slopes
as observed in optically selected quasars.
For $B-K\,<\,(B-K)_{{\rm int}}=2.3$ 
and $\alpha({\rm opt})<\alpha_{int}({\rm opt})=0.3$ 
(ie. the flat portions),
we have assumed 
$\tau_{B}=\tau_{K}=\tau_{X}=0$ so that the observed slopes should be comparable 
to the intrinsic (unabsorbed) values in Eqn.~\ref{intslopes}. 
Indeed, there is very good agreement between
our assumed intrinsic slopes 
$\alpha_{BX}({\rm int})$ and $\alpha_{KX}({\rm int})$, and
those of blue Parkes quasars with $B-K\simlt2.3$. 

The anti-correlations involving 
$\alpha_{BX}$ (Figs.~\ref{slopesX}a and b)
implies that on a relative scale,  
the observed flux in $B_{J}$ must suffer greater absorption than that
at 1keV. In other words, the effective optical depth in $B_{J}$ appears
to be greater than that at 1keV by a factor $\tau_{B}/\tau_{X}\sim3$
(Eqns.~\ref{optdepths}). 
This anticorrelation is also consistent with that claimed by
McDowell \etal (1989). These authors however interpret this trend as 
being due to an intrinsically varying optical-UV
``Big Blue Bump''. 
Our analysis of emission line EWs in section~\ref{ewweak} 
argues against this possibility.

The weaker correlations involving $\alpha_{KX}$
(Figs.~\ref{slopesX}c and d) however, indicates that all of the $B_{J}$
to 1keV continuum flux appears to be absorbed relative to that in $K$.
This result is analogous to the correlation between radio-to-soft
X-ray slope and radio-to-optical slope claimed by Ledden \& O'Dell (1983)
for several optically reddened radio quasars.
They conclude that absorption by associated gas and dust is the
most likely explanation.
Their statistics however are too low from which to draw any firm
conclusion.

The correlations in Fig.~\ref{slopesX} and their consistency 
with our simple dust model implies that the near-IR, optical
and soft-X-ray continuum emission must all be affected 
by the same dust component.
As suggested by previous studies to explain the 
difference in soft X-ray spectral properties of radio loud
and radio quiet quasars,
a significant fraction
of the soft X-ray flux in radio-quasars is believed to be 
due to a synchrotron self-Compton component 
associated with the $\simgt10$kpc radio emission (eg. Ciliegi \etal 1995). 
If this is true, then our hypothesised gas+dust component must also 
extend to such scales. If for instance the ``large scale'' soft X-ray
emission were unaffected
by absorption so that effectively $\tau_{X}\approx0$, 
then we expect both a steeper 
anticorrelation than that shown in Figs.~\ref{slopesX}a and b, and
the opposite behaviour to that seen in Figs.~\ref{slopesX}c and d
(ie. an anticorrelation). 
We find no significant correlation between soft X-ray 
and radio flux
to support a SSC origin for the bulk of the soft X-ray emission.
A strong correlation between optical and soft X-ray flux however exists 
suggesting a common emission mechanism in these two wavebands (eg. as
predicted from standard accretion disk models for AGN).

We conclude from Fig.~\ref{slopesX} that 
if the observed optical reddening in Parkes quasars is due to dust, then
soft X-rays are absorbed at a level consistent 
with this hypothesis.

\subsection{Dust Location: Extrinsic or Intrinsic?}

Since the optical extinction increases rapidly with frequency, while
X-ray absorption decreases even more sharply with frequency 
(eg. Fig.~\ref{Xext}),
one would also expect a similar dependence of these two extinction
measures on absorber redshift in an observer's frame.
Fig.~\ref{Xext} then suggests that one could constrain 
the redshift
of an absorber $z_{d}$ given observed measures of the optical extinction
and X-ray attenuation.
We must note that this method of estimating $z_{d}$
crucially depends on the extinction-to-gas column conversion
and metal abundance model assumed (see section ~\ref{Theory}).

Due to the absence of sufficient soft X-ray spectral data 
for Parkes
quasars, we are unable to estimate the X-ray attenuation directly and
hence estimate absorber redshifts using Fig.~\ref{Xext}.
We can however apply a simple qualitative test to the broadband 
X-ray data to 
decide which of the following
two possibilities most likely applies: Dust
intrinsically associated with the sources, or dust lying in an unrelated 
extrinsic system (or
systems) along the line-of-sight.

\begin{figure}
\vspace{-2.7in}
\plotonesmall{1}{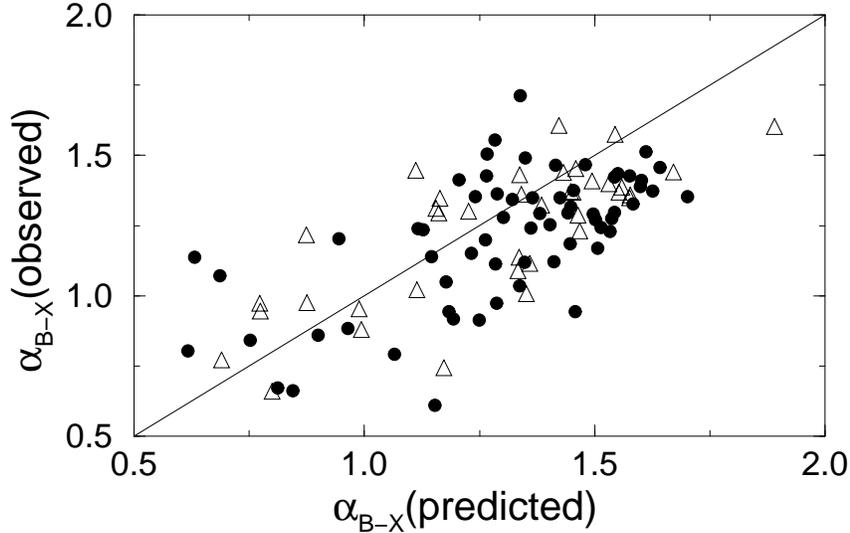}
\vspace{-2.6in}
\caption[Observed versus predicted 4400\AA-to-1keV spectral slope]{Observed 4400\AA-to-1keV spectral slope versus
that predicted using a simple dust model (see Eqn.~\ref{alphaOX})
which assumes the absorbing material towards each source is located
at its measured spectroscopic redshift.
Triangles represent lower limits on the observed slopes.
}
\label{zX}
\end{figure}

We assume the extrinsic 
model as the null hypothesis.
We have tested this by 
comparing the observed optical--to--soft X-ray slopes to those predicted
using our simple dust model (Eqn.~\ref{alphaOX}) assuming the 
dust is located at the measured spectroscopic
redshift of each source.
These predicted slopes are compared with those observed in Fig.~\ref{zX}.
There is a rather strong correlation indicating
that on average, the absorbing material towards most sources is likely to
reside
in their local environment.
Results of a Kendall's tau test shows that 
the hypothesis that the dust is totally extrinsic (ie. that the
observed and predicted slopes are uncorrelated) is rejected at the
95\% confidence level.
The observed scatter about the diagonal (equality) line
may then be attributed to a dispersion in intrinsic spectral slope 
$\alpha_{BX}$(int), and/or physical properties of the gas-dust absorber. 

\subsection{Summary}

In this section, we have shown that there exist
significant correlations between optical
(or near-IR)--to--soft X-ray continuum slope and optical--to--near-IR 
continuum reddening
that are consistent with that expected from a simple dust model 
(Fig.~\ref{slopesX}).
This model assumes an extinction-to-gas column conversion relation
derived empirically 
in the local ISM by previous authors and galactic metal abundances.
Using this model to predict the optical--to--soft X-ray slopes expected
if dust were located at the redshift of each source,
we find that they are in good agreement with those observed.
This indicates that for red Parkes quasars in general, 
the hypothesised absorbing material 
is most likely located in the source environment. 

\section{Implications on Soft X-ray QSO surveys}
\label{impsurv}

Given a measure of the optical extinction, we can 
use the corresponding X-ray attenuation (Fig.~\ref{Xext})
to predict the fraction of QSOs lost from a soft X-ray flux limited
sample as a function of redshift. 
We consider the observed energy bandpass of the ROSAT
detector: 0.1-2.4 keV and  
assume that the absorbing material is intrinsically 
associated with the sources. 

\begin{figure}
\vspace{-2.7in}
\plotonesmall{1}{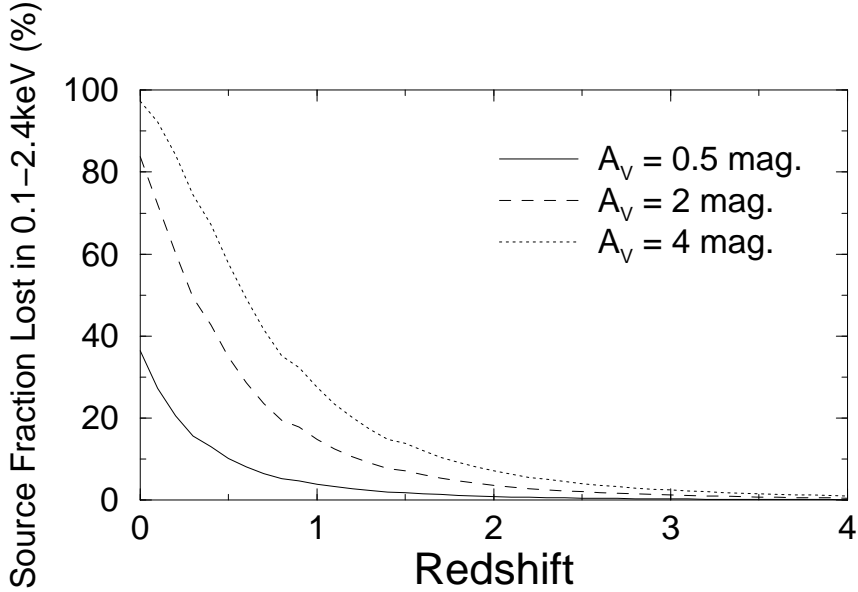}
\vspace{-2.6in}
\caption[QSO fraction lost in soft X-rays due to intrinsic absorption]{Fraction of QSOs lost due to
intrinsic absorption in the soft energy bandpass: 0.1-2.4keV
as a function of redshift (see section~\ref{impsurv}).}
\label{fmissX}
\end{figure}

Since the effective X-ray attenuation decreases rapidly with energy 
(Fig.~\ref{Xext}), we expect the higher redshift sources to suffer
relatively less absorption in a fixed observing 
bandpass than those at low redshift.
Thus, a similar dependence on the fraction of sources lost 
in a flux limited sample as a function of redshift is expected.
To explore this, we assume for simplicity a power-law
cumulative luminosity function (LF), $\Phi\propto L^{-\beta}$. 
We assume a slope
$\beta\simeq2.2$ corresponding to the bright part of the
soft X-ray LF as determined by Page \etal (1996) from
a ROSAT AGN survey.
For a power-law LF, the fraction of 
sources missing at some observed energy $E$ and redshift $z$ 
can be written (see section~\ref{simplepl}):
\begin{equation}
f_{miss}(E,z)\,=\,1\,-\,\exp{\left[-\beta\,\sigma_{(1+z)E}
\,N_{H}(A_{V},z)\right]}.
\label{fraclostX}
\end{equation}
Assuming our galactic ISM gas-dust model to determine $\sigma$ and $N_{H}$
(see section ~\ref{Theory}), 
the fraction lost within the range 0.1-2.4 keV  
is computed by integrating Eqn.~\ref{fraclostX}. 
This fraction is shown as a function of redshift
in Fig.~\ref{fmissX} for three observed extinction measures 
$A_{V}$. 

As expected, the fraction of sources lost due to absorption
decreases rapidly with increasing redshift.
This has severe implications for soft X-ray quasar surveys. 
For an intrinsic absorption comparable  
to the mean value $A_{V}\sim2$ as observed
in Parkes quasars, a soft X-ray 
survey is expected to be seriously incomplete at
the lowest redshifts $z\simlt0.5$.
A consequence of this is that those sources which are heavily
absorbed intrinsically in soft X-rays, 
are more likely to be detected at high redshifts.
We should therefore expect a larger fraction of optically 
reddened sources to be  
found at high redshifts in a soft X-ray flux-limited sample. 

A soft X-ray quasar sample was used by Boyle \& di Matteo (1995)
to investigate whether dust extinction was present at a level
consistent with that claimed by Webster \etal (1995) for Parkes quasars. 
They found that optical extinction was neglible, consistent with
our prediction that `heavily absorbed' low-$z$ sources would not be
selected in soft X-rays.
Their sample contained very few high redshift sources, 
presumably due to limited flux sensitivity, 
and hence their observations cannot be used to test the
hypothesis that most quasars are obscured optically by dust. 
Their conclusion that dust is unlikely to significantly bias quasar
searches in the optical must be viewed with skepticism.
As implied by Fig.~\ref{Xext}, hard X-ray surveys should have very little
bias against absorption by dust. 
Surveys with the Einstein X-ray satellite, whose bandpass extends
to $\sim4.5$keV has indeed revealed significant numbers of highly
reddened AGN in the optical (Kruper \& Canizares, 1989).
These AGN samples also show a relatively large scatter in 
optical-to-hard X-ray continuum slope, consistent 
with possible reddening by dust (eg. Stocke \etal 1991). 

\section{Discussion}
\label{disredm}

In this section, we first discuss the possibility 
of detecting infrared thermal emission from the dust
component in Parkes quasars.
Physical properties of the dust grains involved as directly
implied by the available data, and further 
predictions are then discussed.

\subsection{Dust Mass and Thermal Emission?}

If the observed reddening in Parkes quasars is to be interpreted as due to
extinction by dust, what is the total mass of dust implied?
How does this compare with other determinations of dust mass estimates in AGN?
To get an approximate estimate from
the observed extinction, we assume for simplicity a spherical dust
distribution centered on the central source with uniform dust mass density. 
We assume grain extinction efficiencies for a graphite and silicate
mixture with mean grain size $a\sim0.1\mu$m and
density $\rho\sim2{\rm gm}\,{\rm cm}^{-3}$ characteristic of the
galactic ISM (eg. Goudfrooij \etal 1994).
The dust mass in terms of an optical extinction measure $A_{V}$
according to these assumptions can be derived  
using the method presented in section~\ref{scale} (ie. Eqn.~\ref{to}).
We find that the dust mass scales as:  
\begin{equation}
M_{d}\,\simeq\,2.5\times10^{5}\left(\frac{R}{{\rm kpc}}\right)^{2} 
\left(\frac{a}{0.1{\rm m}}\right)
\left(\frac{\rho_{d}}{2{\rm gr}\,{\rm cm}^{-3}}\right)A_{V}\,\,M_{\odot}. 
\label{dustmass}
\end{equation}

Assuming a $1/\lambda$ type extinction law, we find that the reddest
Parkes quasars (with $E_{B-K}\simgt5$) correspond to a mean rest frame
optical extinction
$\langle A_{V}\rangle\sim2$.
Given this value, we find that if all the dust were distributed on a scale
comparable to the size of the narrow-line emitting region where
typically $R\sim0.1-1$kpc, then dust masses of $\simgt10^{3}M_{\odot}$
are implied assuming the dust properties above. 
On a typical host galaxy scale of $R\sim10$kpc, dust masses of
$\simgt10^{7}M_{\odot}$ are implied.
The normalisation and scaling of the dust mass on spatial extent $R$ 
defined by Eqn.~\ref{dustmass} is remarkably similar to that
derived by Sanders \etal (1989) using thermal dust reprocessing models 
of the infrared emission from a warped disk geometry surrounding an AGN
(see their table 1).
Furthermore, if dust were distributed on a typical 
host galaxy scale, $R\sim10$kpc,
then our dust estimate using Eqn.~\ref{dustmass}
is consistent to within an order of magnitude 
with that derived from the 
thermal emission components found at mm/sub-mm wavelengths   
in a number of high redshift radio-selected quasars by 
McMahon \etal (1994).

Depending on the sizes of dust grains, their location relative
to the central continuum source, and temperatures
of the associated gas, thermal emission from heated dust 
is expected to peak at $\lambda\sim2-10\mu$m for the
hottest grains ($T_{d}\simgt1000$K) and $\lambda\simgt1$mm  
for $T_{d}\simlt100$K - as typically observed in the galaxy. 
A detection of thermal infrared emission
associated with the dust component in Parkes quasars will thus not only 
strengthen the dust hypothesis, but also provide important diagnostics 
on dust location and physical grain properties. 
Presently however, we have very little spectral information at
wavelengths $\lambda>2\mu$m from which to detect thermal dust components.
One would however expect the infrared SEDs of flat spectrum
radio quasars to be strongly dominated by beamed non-thermal emission,
preventing an unambiguous
detection of such components
(eg. Impey \& Neugebauer, 1988). 

We can give an estimate of the flux density expected 
at an observed
wavelength $\lambda\sim0.1$mm if we assume the 
dust 
in Parkes quasars emits thermal radiation 
similar to that of local galactic dust.
Given a value of the dust mass and temperature, the flux density
expected can be computed using Eqn.~\ref{Mdust1} in section~\ref{otherdiag}. 
Since the dust mass as estimated from Eqn.~\ref{dustmass} above is likely
to be an upper limit due to our assumption of a spherical dust
geometry of uniform mass density, our derived flux density will also
be an upper limit.
We assume the dust is distributed to $R\sim10$kpc, which may be
considered an upper limit if dust does not
significantly extend beyond the quasar host galaxy. 
With a dust mass of $M_{d}\simlt10^{8}M_{\odot}$ on such scales 
(as predicted from Eqn.~\ref{dustmass} 
with $\langle A_{V}\rangle\sim2$),  
and assuming a typical galactic dust temperature $T_{d}\sim100$K (Chini
\& Kr\"{u}gel, 1994),
we find $S(0.1{\rm mm})_{dust}\simlt10$mJy.
The SEDs of radio quasars and blazars are remarkably
flat from radio to sub-mm wavelengths (eg. Impey \& Neugebauer, 1988)
and hence in Parkes quasars, 
we expect flux densities at sub-mm wavelengths to be of order 
comparable to the radio flux limit: $S_{2.7{\rm GHz}}\simgt0.5$Jy. 
This implies that the SEDs are likely to be dominated by non-thermal
mechanisms. 

An underlying thermal dust component 
may still however be detectable 
to flux levels $<1$mJy
through angular size measurements using present day sub-mm interferometers. 
Non-thermal models for the far-IR to sub-mm continuum
such as synchrotron emission require brightness temperatures 
$T_{B}=m_{e}c^{2}/k\simgt10^{10}$K which imply angular sizes of order
microarcseconds (eg. O'Dell, 1979).
On the other hand, it can be shown that angular sizes
for thermal models in the low frequency (Rayleigh-Jeans) 
limit predict a lower bound (eg. Sanders \etal 1989):
\begin{equation}
\theta\,\simgt\,0.2\,\left(\frac{\lambda_{o}}{0.1{\rm mm}}\right)^{3/2}
\,\left(\frac{T_{d}}{100{\rm K}}\right)^{-1/2}\,
\left(\frac{f_{\lambda_{o}}}{{\rm Jy}}\right)^{1/2}\,\,{\rm arcsec}.
\label{angsize}
\end{equation} 
Thus, based on such angular size constraints,
one can in principle search whether thermal dust emission 
is present to some flux level which would otherwise
escape detection from a direct spectral study.

\subsection{Physical Dust Properties}

Based on the high identification rate of Parkes sources at optical to
near-IR wavelengths and the strong detections of emission lines in the
optical, the dust responsible for the observed optical reddening
must be associated with an ``optically thin'' or diffuse component.
The observations argue against a component which is significantly
``optically-thick'' so as to completely absorb the optical
continuum emission. 
Optically-thick dust is usually associated with cold dense, molecular
gas clouds, such as those in the outer regions of the galaxy
where the relatively low reddening observed 
is consistent with a high abundance of
large dust grains ($a\simgt1\mu$m) (Mathis, 1990).
Extrapolations from IR-extinction measures have lead
to individual cloud optical depths exceeding $A_{V}\sim10$mag.
Optically thick dusty molecular gas
is expected to be a major
constituent of the geometrically thick tori believed to
surround AGN, strongly absorbing the optical-to-soft X-ray continuum 
and BLR emission when the inclination is such that the nucleus
is hidden from direct view (see Antonucci, 1993 for a review).
Constraints on the extinction optical depth of such material 
come from measurements and non-detections of hard X-rays.
X-ray data on Seyfert 2 galaxies indicate absorbing columns in the range
$10^{23}-10^{25}{\rm cm}^{-2}$ (Mulchaey \etal 1992). 
Using typical torus scale lengths of tens of parsecs
as inferred from observations of IR emission (eg. Pier \& Krolik, 1993), 
these column densities correspond to mean dust extinctions of 
$A_{V}>10^{5}$ mag/kpc.

Thus, it is unlikely that our line of sight to the central continuum source
is intercepted by cold, opaque dusty molecular gas. 
Extinction measures significantly larger than our observed mean
value $\langle A_{V}\rangle\sim2$ would be expected. 
As implied by unified models, our line-of-sight to the
central continuum source in flat spectrum radio quasars is unlikely to
intercept a torus. 
If a galactic-type dusty molecular cloud significantly larger 
than the continuum source was intercepted,
the continuum source would either be 
completely blocked or unaffected at all. 
If however, 
the clouds are considerably smaller than the continuum source, 
the relatively large associated grains 
would nonetheless cause neglible reddening. 

In fact, a search for possible molecular absorption
in a number of red Parkes quasars was 
conducted by
Drinkwater, Combes \& Wiklind (1996). 
Assuming a local ratio of CO to H of $\simeq10^{-4}$ from Dickman (1978), 
these authors calculated upper limits to the molecular absorption
corresponding to columns $N_{H}<10^{16}-10^{19}{\rm cm}^{-2}$.
These measures are 
significantly lower than those expected if the observed reddening
were due to dust similar to that in the galaxy where 
$N_{H}\simgt 10^{21}{\rm cm}^{-2}$ for $\langle A_{V}\rangle\sim2$ is predicted 
(see section ~\ref{Theory}).

As discussed by Laor \& Draine (1993), only ``optically thin'' dust
in AGN can produce ``observable'' reddened continua and emission lines. 
As shown by their grain extinction models (see their Fig.6),
such optically thin dust can only redden
if the grain size distribution, $n(a)\propto a^{-\beta}$, is
dominated by small enough grains.
A model distribution with $\beta\sim3.5$ for a range of grain sizes:
$0.005\mu {\rm m}<a<0.25\mu {\rm m}$
as used by Mathis \etal (1977) to interpret the
extinction law of local diffuse ISM dust is a typical example. 
The models of Laor \& Draine (1993) show that grain distributions 
with $\beta<3$ and $a_{max}\simgt1\mu{\rm m}$ result in an almost
wavelength independent extinction law and hence cannot have a large
reddening effect. 
Thus, the observed reddened continua of Parkes quasars appears to be 
consistent with a grain size distribution
very similar to that
in the local diffuse ISM.

Although no convincing explanation for the apparent grain size distribution 
of diffuse dust in the galactic ISM yet exists, 
there are a number of processes in an AGN environment that can lead
to dust dominated by small grains.
If dust originates from the envelopes of cool evolved stars losing 
mass at a prodigious rate near the center of AGNs
(Scoville \& Norman, 1988),
then recently injected grains will be subject to a high flux 
of hard photons within a few hundred parsecs.   
Grains may be rapidly destroyed by photo-disintegration 
on a timescale significantly shorter than that required 
to form large ($a\sim1\mu$m) grains 
from accretion in the gas phase (eg. Chang \etal 1987).
For conditions typical of the central regions of 
AGN, the timescale to form 1$\mu$m-sized grains is expected to be of order 
$\tau_{form}\sim a/\stackrel{.}a_{acc}\simgt10^{9}$yr (where
$\stackrel{.}a$ is the accretion rate defined by Draine \& Salpeter, 1979a).
For comparison, simulations by Chang \etal (1987) 
predict photo-disintegration timescales of order $\sim10^{7}$yr within
a few kiloparsecs of the central AGN. 
Other mechanisms that may be efficient at maintaining a grain size distribution 
dominated by small grains (or depleted in large grains),
include thermal sputtering by hot gas
and destruction by shocks
either associated with star-forming regions or those induced by 
propagating radio jets 
(Draine \& Salpeter, 1979b).

\subsection{Further Predictions of the Dust Model}
\label{predchem}

The hypothesis that reddening by ``optically-thin'' dust in the environs
of Parkes quasars accounts for their relatively red optical continua
does make a number of other predictions which can be tested by future studies.
The following predictions may 
further strengthen this hypothesis and
provide strong constraints on
physical properties of the absorbing medium:
\\\indent 1. As discussed above, 
IR emission from dust through thermal reprocessing of the central
UV/optical continuum emission is expected to be present at some level.
Since optically-thin dust dominated by small grains is expected to be
maintained at a higher equilibrium temperature than optically-thick type dust
out to distances of a few hundred parsecs (Laor \& Draine, 1993), 
emission in the near-IR ($\lambda\simlt10\mu$m) is more likely. 
Dust further out however, may emit at sub-mm/mm wavelengths. 
A multiwavelength infrared spectroscopic study to search for thermal
dust components is required. 
\\\indent 2. Angular size constraints predicted by thermal emission models 
(see Eqn.~\ref{angsize}) 
can provide a stronger test at detecting an underlying dust component.
These are accessible to the next generation of sub-mm interferometers. 
Such studies can be combined with variability and/or polarimetry
studies to confirm the 
nature of the emission component 
(eg. thermal vs. non-thermal emission). 
Those showing the lowest mean polarisations in the far to near-IR
are expected to exhibit the strongest underlying thermal dust componenents.
It would be encouraging if this was also found to correlate
with optical continuum reddening. 
\\\indent 3. An optical continuum reddened by dust should be accompanied by a 
characteristic spectral curvature into the extreme UV 
($\lambda\simlt2000$\AA$\,$). 
Since ``optically thin'' dust produces an extinction that varies
as $\tau_{\lambda}\propto 1/\lambda$ and the resulting 
observed flux is reduced according to 
$f_{\lambda}\propto\exp{(-\tau_{\lambda})}$,  
a sharp exponential-like turnover is expected as one approaches 
shorter wavelengths.
For an observed mean value of $\langle\tau_{V}\rangle\sim2$, 
and an intrinsic (unabsorbed) optical-UV continuum: 
$f_{\nu}\propto\nu^{-0.3}$, 
the observed flux is expected to be reduced by more than two orders
of magnitude from the optical to a wavelength of $\sim1000\mu$m. 
This can be tested using HST observations.
Observations of 
shorter UV rest wavelengths using high redshift reddened quasars  
can provide a stronger test.
\\\indent 4. Optical continuum reddening by an extended distribution 
of optically-thin dust may produce 
a diffuse optical halo from scattering of radiation into the
line-of-sight. 
This may be detected from high resolution optical imaging using HST.
As discussed by Martin (1970), such a scattering halo is expected
to be more pronounced in soft X-rays of a few keV. 
\\\indent 5. If associated soft X-ray absorption is present,
then the K-absorption edges of heavy metals may also be detected
in high resolution X-ray spectra. 
The clearest X-ray signature of dust would be the detection
of a shift in the K-edge energy by a few eV due to the potential field
of neighbouring atoms in a dust grain as predicted by Martin (1970).  
Our reddest sources however are too weak to extract 
any meaningful high S/N X-ray spectral information in the near future.
\\\indent 6. High resolution near-IR spectroscopy may be used to detect
the characteristic silicate emission features at 9.7 and $18\mu$m
as observed from dust in the local ISM. 
There are also a number of emission features associated with polycyclic
aromatic hydrocarbons (eg. Allamandola \etal 1985). 
Their existence
sensitively depends on physical conditions in the ISM, and
hence can 
provide constraints on gas densities and temperatures.

\section{Conclusions}
\label{concfive}

In this chapter, we have explored two classes of reddening models to 
describe the relatively large dispersion in continuum slopes 
observed in Parkes {\it quasars}.
The first class consists of mechanisms whereby the reddening is
associated with intrinsically `red' emission and the second,
with mechanisms where line-of-sight absorption primarily by dust and gas is
affecting our view to the quasars. 
Using the available data, 
we arrive at the
following main conclusions:
\vspace{0.5cm}
\\{\bf Intrinsically Red Emission:}
\\\indent 1. The $B-K$ colours do not correlate with either the strength of
radio core emission or the equivalent widths of most broad lines.
This is contrary to that expected 
if the colours were purely due to a beamed synchrotron component
dominating the near-IR flux.
\\\indent 2. Some of the reddest Parkes quasars have near-IR--to--optical 
spectral turnovers which 
have to be as extreme as possible as allowed by simple
synchrotron theory. Unrealistic particle energy distributions 
that are nearly monoenergetic in nature are required.
\\\indent 3. The observed redness as a function of emission line EW
argues against a mechanism where the reddening is due to an 
intrinsically varying optical/UV continuum source.
\\\indent 4. The colours are also inconsistent with thermal
emission from hot dust in the `steep' Wien (high) frequency limit
of the blackbody radiation law. 
\vspace{0.5cm}
\\{\bf Extinction by Dust:}
\\\indent 1. Since dust is expected to reside at distances larger than the BLR, 
the continuum and emission lines are expected to be reddened equally.
This explanation is consistent with the observation that the
redness does not correlate with the EWs of broad emission lines. 
\\\indent 2. There appears to be a correlation between optical and $B-K$ 
continuum slope that is consistent with the 
generic $1/\lambda$ type dust extinction law. 
The observation that optical slopes appear redder than those
interpolated between $B_{J}$ and $K_{n}$ is not expected under
the synchrotron model.
\\\indent 3. There is a significant correlation between the Balmer-line
ratio, ${\rm H}\alpha/{\rm H}\beta$, and 
optical continuum slope that is consistent with reddening by dust. 
Due to the reduced sample size in which such lines can be measured, 
this result must {\it not} be taken as conclusive 
evidence that dust provides all of the reddening observed.
\\\indent 4. Correlations between 
continuum reddening and optical(or near-IR)--to--soft X-ray (1keV) slopes 
are observed that are consistent 
with the predictions of a simple (galactic ISM) gas-dust model.
Soft X-rays are absorbed at a level consistent with that
implied by the optical reddening if the latter is due to dust.
\\\indent 5. The observed distribution in optical--to--soft X-ray slopes
is consistent with that predicted by our simple model if the
absorbing material is assumed located at the redshift of each source. 
\\\indent 6. Our analysis of the soft X-ray data in Parkes quasars
suggests that QSO surveys selected at soft X-ray energies
will be strongly biased (like those selected optically) 
against sources with significant amounts of dust in their line-of-sight. 

Some physical properties of the absorbing medium in Parkes quasars directly
implied from the observations are the following: 
\\\indent 1. The amount of dust extinction required to explain the reddened
continua is relatively small, being typically $\langle A_{V}\rangle\simeq2$ 
($E_{B-V}\sim0.5$).
This is considerably smaller than that invoked for obscuring
molecular tori in AGN and, 
is consistent with unified models since our line-of-sight to
the central AGN in 
flat spectrum radio quasars is not expected to intercept a torus. 
\\\indent 2. The observed degree of reddening 
argues in favour of an ``optically-thin''
(possibly hot) diffuse dust component with a grain size distribution dominated
by small grains, similar to that invoked for the local ISM. 

On the basis of these simple observational tests, we conclude
the data on red
Parkes {\it quasars} can be better explained by the dust
extinction model rather than `purely' by an intrinsic emission property.
The reddening in sources exhibiting weak-lined BL-Lac type properties however 
is likely to be purely intrinsic (eg. beamed synchrotron) in origin. 
A beamed synchrotron component 
dominating the near-IR flux in the quasars (presumably 
to a lesser extent than in the BL-Lacs) cannot be ruled out
completely. A combination 
of both absorption mechanisms and intrinsically red 
emission could then be present. 
This will be further investigated using polarisation observations in
Chapter 7.
\chapter{Contribution From the Host Galaxy} 

\vspace{1mm}
\leftskip=4cm
 
{\it ``There are two kinds of light--the glow that
illumines, and the glare that obscures.''}
 
\vspace{1mm}
 
\hfill {\bf --- James Thurber, 1989}
 
\leftskip=0cm

\section{Introduction}

The $B-K$ colours of
normal radio 
galaxies are known
to be quite red, exhibiting a similar dispersion to those observed in
the Parkes sample.
These sources are often associated with giant ellipticals and
their colours 
appear to be uniformly distributed over the range
$3\simlt B_{J}-K\simlt 7$ for redshifts $z\simlt2$
(McCarthy, 1993 and references therein).
It is possible that the host galaxies of Parkes {\it quasars} 
also exhibit similar properties.
We therefore consider whether the
relatively red $B-K$ colours of Parkes quasars
are due to a possible 
contribution from a ``red'' stellar component from their host galaxies.
To determine whether this is true, we need to quantify the contribution
from the host galaxy throughout the optical to near-IR wavelength region.

Determining the host galaxy properties of radio-quasars and BL-Lac-type 
sources are crucial
for studies of AGN evolution and testing unified schemes.
A significant number of BL-Lacs and other core-dominated radio loud AGN 
are surrounded by nebulosities whose optical spectra 
are very similar to those of giant ellipticals (eg. Miller, French \& 
Hawley, 1978; Ulrich, 1988; Taylor \etal 1996). 
A majority of these
host galaxies
have been detected in sources at relatively low redshifts;
$z\simlt0.1$, where the galaxy flux is easily detected. 
Since most quasars are at high redshifts, host galaxy detections 
have been
difficult due to contamination by their strong nuclear emission.
Recent high resolution imaging using HST
has revealed that a majority of high redshift quasars 
also reside in giant ellipticals 
(Hutchings \& Morris, 1995; Bahcall \etal 1995).

The optical spectral energy distributions (SEDs) 
of ellipticals are all similar in shape.
At $\lambda\simgt3000$\AA, they rise steeply up to $\sim4000$\AA$\,$
where a sharp step-like cutoff is observed. This cutoff is often
referred to as the ``4000 Angstrom break'' and is 
caused by the dearth of hot and young, usually O and B-type stars
with time, whose spectra mostly dominate at wavelengths
$\lambda<4000$\AA~ into the UV.
For $\lambda\simlt4000$\AA, 
the existence of a multiplet of heavy metal absorption lines arising from
stellar atmospheres, 
results in a steepening of the break. 
The presence of this feature thus provides a signature
for determining whether 
an underlying elliptical host is contributing to the total light.

In this chapter, we describe an unbiased
method that uses the size of this 
characteristic break in the optical spectra of Parkes quasars 
to quantify the host galaxy contribution.
We use this method to investigate whether the host galaxies of Parkes
quasars can significantly
contribute to the optical and near-IR continua and hence 
cause the spread in $B_{J}-K_{n}$ colours observed.

The plan of this chapter is as follows:
In section~\ref{hubblelaw}, we present a preliminary discussion on the host 
galaxy contribution in Parkes quasars by comparing their 
$K$-band Hubble relation to that observed for radio galaxies. 
In section~\ref{outgc}, we outline our method used to quantify the host galaxy
contribution. A discussion of the spectral data 
and assumed input parameters
is given in section~\ref{data}.
Results are presented in section~\ref{regcp}.
Implications for unified schemes are explored in section~\ref{tuf} and
results discussed in section~\ref{diseven}.
All results are summarised in section~\ref{concsix}.

\section{Preliminary Investigation}
\label{hubblelaw}

As an initial quick test to assess the relative galactic contribution
in the near-IR in Parkes quasars, we compare their $K$-$z$ relation
with that observed for radio galaxies.
We assume that the host galaxy $K$-band luminosities
of Parkes quasars are similar to those of normal radio galaxies. 
In other words, we assume that the host galaxy luminosities
of flat spectrum radio sources and extended powerful radio galaxies
are not significantly biased by selection effects in the radio.

\begin{figure}
\vspace{-3in}
\hspace{-1.5cm}
\plotonesmall{1}{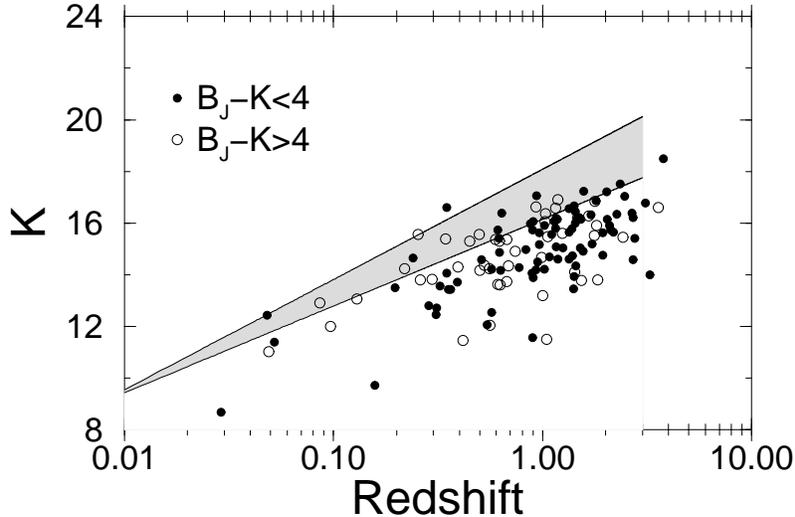}
\vspace{-2.5in}
\caption[Hubble diagram for radio-loud galaxies and Parkes sources]{Observed $K$-$z$ relation
for radio-loud ellipticals given by the shaded region which 
approximately corresponds to the spread in $K$ magnitudes
(McCarthy, 1993).
The open circles represent all Parkes sources with colours
$B_{J}-K>4$
and filled circles are those with $B_{J}-K<4$.}
\label{hubble}
\end{figure}

The $K$-$z$ relation or Hubble diagram
has been one of the most effective means
of assessing the luminosity evolution in radio galaxies
since the $K$-band is mostly dominated by
``old'' passively evolving stellar populations. 
Contamination from active evolutionary effects such as young starbursts 
are minimal.
These wavelengths are also less sensitive to effects of 
a changing rest frame
bandpass with redshift as compared to the optical-UV. 
The shaded region in Fig.~\ref{hubble} shows the observed
range of $K$-band magnitudes for radio galaxies
as a function of redshift as determined from a number of independent
studies (see McCarthy, 1993 and references therein).
At low redshifts ($z\simlt0.1$), the magnitudes span $<0.4$ magnitudes,
while towards $z\sim2$, the spread increases to $\sim2$ magnitudes.
It has been suggested that this increase is
due to the onset of AGN line 
and continuum emissions
that enter the $K$-band at high redshifts (Eales \& Rawlings, 1993).
 
Also plotted in Fig.~\ref{hubble} 
are the observed $K$ magnitudes for Parkes sources 
corresponding to colours $B_{J}-K<4$ (filled 
circles) 
and $B_{J}-K>4$ (open circles).
Overall, 
very little overlap between the radio galaxy and Parkes magnitudes
is observed.
A majority of the Parkes sources
are more than 2 magnitudes
brighter than the mean in the galaxy $K$-$z$ relation.
This cannot be explained by a possible difference in the 
size of aperture used in the photometry to determine $K$ magnitudes
since it can only account for 
at most an uncertainty of $\sim0.2$ mag in $K$ (McCarthy, 1993). 
Thus, if the host galaxies of Parkes sources have the typical
$K$ magnitudes of 
radio galaxies, we estimate from Fig.~\ref{hubble} that 
the galaxy will only contribute on average $\simlt20$\% of the total light
at $z\sim1$. 

There are also a number of sources in Fig.~\ref{hubble}
which appear to lie within the
shaded region of observed radio galaxy magnitudes.   
A majority of these do indeed show extended $K$-band images that
can be interpreted as diffuse emission from a host galaxy.
In fact, $\sim10\%$ of Parkes sources
have extended $K$-band images that can be attributed to a galaxy
component.

We conclude that as a first measure, a majority of Parkes sources 
appear significantly brighter at near-IR wavelengths than the
locus defined by radio galaxies in the $K$-$z$ relation.
Emission from the central AGN is likely to dominate the
near-IR flux and hence $B_{J}-K$ colours observed in Parkes sources.
This conclusion is strongly supported by
the observation that a majority of Parkes sources appear very
compact and bright in $K$.
This is expected, since for quasars in general which comprise
the majority of identifications in the Parkes sample, their
redshifts are high enough that contamination by a red stellar
component would be minimal.

\section{Outline of Method}
\label{outgc}

The galaxy contribution can be estimated using an
algorithm that determines the relative strength of the 
characteristic 4000\AA$\,$ break in the optical spectra of Parkes sources
in an unbiased way. 
No other feature in a generic quasar spectrum will mimic this feature.
In this section, we describe this algorithm.

\subsection{Assumptions}

There are two important assumptions that will be required by our
algorithm. 
First, we assume that the spectral shape defining an underlying quasar
optical continuum is `smooth' and contains no breaks. 
This is justified by our current knowledge of quasar optical spectra.
Our choice for this shape is somewhat arbitrary, and will
be further discussed in section~\ref{method}.  

\begin{figure}
\vspace{-3.1in}
\hspace{0.7cm}
\plotonesmall{1}{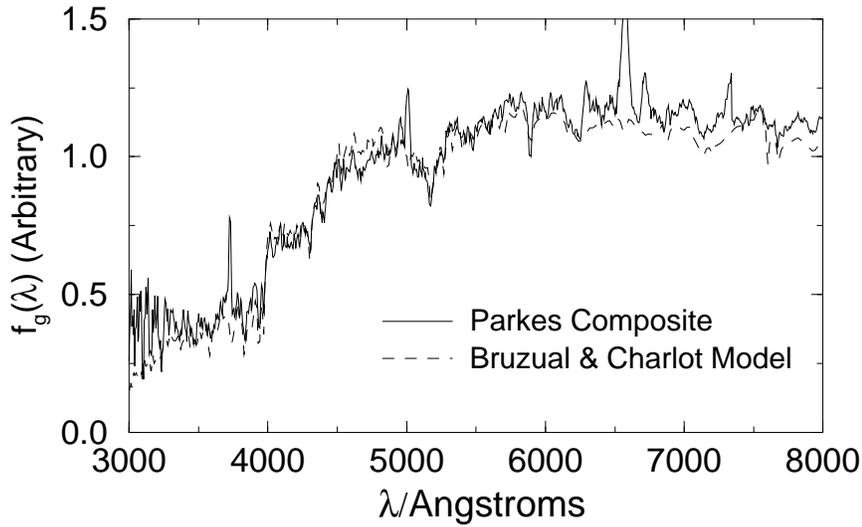}
\vspace{-2.6in}
\caption[Composite spectrum of spatially extended Parkes sources]{Composite optical spectrum of 18 Parkes sources which appear 
spatially extended in $B_{J}$
(solid curve),
and a model spectrum for an `old' 
elliptical SED from Bruzual \& Charlot (1993).}
\label{galspec}
\end{figure}

Second, we need to assume
the input galaxy spectrum which defines the shape of the 4000\AA$\,$
break to be used in our analysis.
We have formed a composite optical spectrum of 18 Parkes sources all of which
appear `spatially extended' in $B_{J}$.
This composite is shown as the solid curve in Fig.~\ref{galspec} and
clearly shows the characteristic 4000\AA$\,$ break.
Superimposed (dashed) is an elliptical galaxy SED predicted from the
stellar population synthesis models of 
Bruzual \& Charlot (1995; see Bruzual \& Charlot, 1993).
This model is for a galaxy of age 8Gyr, and 
assumes a 1Gyr burst described by a Salpeter initial
mass function with no star formation
thereafter. 
In fact, any `old aged' model with long completed star formation
will in general be similar. 
Models assuming `long lived' constant star formation rates
lead to considerably bluer optical-UV continua, characteristic 
of those observed in 
spirals.
The composite and model in Fig.~\ref{galspec}
are in excellent agreement, except for the
emission lines arising from ISM gas in the composite.
In our analysis, we shall assume that this model represents the 
underlying elliptical SED in each of our Parkes sources. 

\begin{figure}
\vspace{-3in}
\hspace{-1.2cm}
\plotonesmall{1}{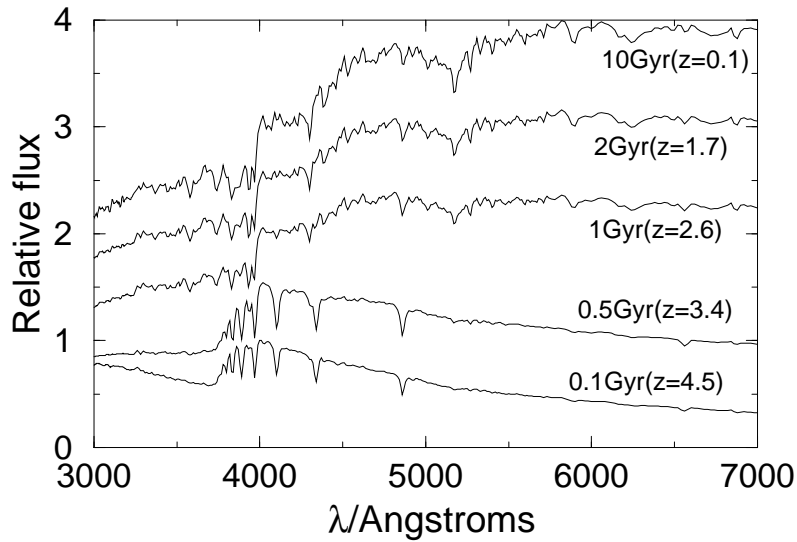}
\vspace{-2.4in}
\caption[Spectral evolution of a 1Gyr burst model]{
Spectral evolution of the 1Gyr burst model of Bruzual \& Charlot (1993).
The age in Gyr and the approximate corresponding
redshift for a formation epoch $z_{f}=5$ with
($q_{0},h_{50}$)=(0.5,1) is shown beside each spectrum.}
\label{specevol}
\end{figure}

For sources with redshifts $z\simgt1$, the $4000$\AA$\,$ break is redshifted 
out of the wavelength range available in the spectra of Parkes quasars 
(see section~\ref{apal}) and thus, cannot be used in our algorithm.
We are therefore restricted to $z<1$.
For simplicity, we assume that the general shape or our model 
elliptical SED is independent of redshift to $z\sim1$.
The colours of radio galaxies observed to $z\sim3$ are consistent
with formation redshifts $4\simlt z_{f}\simlt 20$ 
(Spinrad \& Djorgovski, 1987; Dunlop \etal 1989). 
These formation redshifts are also suggested by models of
galaxy formation in the CDM scenario (eg. White \& Frenk, 1991). 
The spectral synthesis models  
predict that the general form of an elliptical SED shown in 
Fig.~\ref{galspec} can be immediately established following an
almost instantaneous ($\sim1$Gyr) burst of star formation. 
Adopting the 1Gyr burst model of Bruzual \& Charlot (1993) (updated
1995 models), we show in Fig.~\ref{specevol} the spectral evolution as a
function of age since the initial starburst.
Also shown beside each age is the approximate
redshift for a formation epoch of $z_{f}=5$.
Initially, the 4000\AA$\,$ break region is diluted by starburst activity.
After a few gigayears,
the stellar population evolves passively, maintaining 
an almost uniform SED shape (see Bruzual \& Charlot, 1993). 
For formation epochs $z\simgt5$, 
it is apparent that the generic elliptical SED shape of Fig.~\ref{galspec} 
is easily established by $z\sim1$. 

\subsection{The Algorithm}
\label{method}

In general, the total flux at a given wavelength observed in a Parkes source,
$f_{T}(\lambda)$, can be modelled as the sum of light contributed
by the central quasar or AGN, $f_{q}(\lambda)$, and any underlying host
galaxy $f_{g}(\lambda)$ (eg. Fig.~\ref{galspec}). 
We write the relationship 
between these quantities as:
\begin{equation}
f_{q}(\lambda)\,=\,f_{T}(\lambda) - c\,f_{g}(\lambda), 
\label{fq}
\end{equation} 
where $c$ is a scaling factor giving an arbitrary measure of the
amount of galaxy light we wish to determine for a particular source. 

Given $f_{T}(\lambda)$ for a particular source and an
arbitrary galaxy spectrum $f_{g}(\lambda)$ (Fig.~\ref{galspec}), our aim is to 
determine the value of $c$ such that $f_{q}(\lambda)$ looks 
something like a quasar spectrum. 
From our knowledge of quasar optical spectra, 
an obvious choice is to require that $f_{q}(\lambda)$ be ``smooth'' and
contain no breaks. 
Thus, the basis of this algorithm involves subtracting 
an arbitrary amount
of galaxy flux, $c\,f_{g}(\lambda)$ from $f_{T}(\lambda)$ such that  
the resulting spectrum $f_{q}(\lambda)$ appears smooth (see below).
When this is achieved, the fraction of total light at a given wavelength
contributed by the host galaxy can be estimated by normalising:
\begin{equation}
F_{gal}(\lambda)\,=\,\frac{c\,f_{g}(\lambda)}{f_{T}(\lambda)}. 
\label{galfrac}
\end{equation}

In order to implement the above algorithm, we need to define
an acceptable form for the shape of the quasar spectrum $f_{q}(\lambda)$. 
Our only requirement is that this spectrum be smooth and hence our 
choice is somewhat arbitrary.
We choose $f_{q}(\lambda)$ to be a power-law, parameterised by:
\begin{equation}
f_{q}(\lambda)\,=\,f_{q}(\lambda_{max})\left(\frac{\lambda}
{\lambda_{max}}\right)^{\alpha}, 
\label{pl2}
\end{equation} 
where the slope $\alpha$ is determined between two fixed wavelengths,
$\lambda_{min}$ and $\lambda_{max}$ (see below):
\begin{equation}
\alpha\,=\,\frac{\ln\left[f_{q}(\lambda_{min})/f_{q}(\lambda_{max})\right]}
{\ln\left[\lambda_{min}/\lambda_{max}\right]}.
\label{alpha2}
\end{equation} 
Using Eqn.~\ref{fq}, the fluxes defined in Eqn.~\ref{alpha2} can be written: 
\begin{equation}
f_{q}(\lambda_{min})\,=\,f_{T}(\lambda_{min}) - c\,f_{g}(\lambda_{min})
\label{fqminmax}
\end{equation}
$$
f_{q}(\lambda_{max})\,=\,f_{T}(\lambda_{max}) - c\,f_{g}(\lambda_{max}).
$$
For a discussion on how the wavelengths 
$\lambda_{min}$ and $\lambda_{max}$ are chosen 
and the fluxes in Eqn.~\ref{fqminmax} measured, see section~\ref{apal}. 

\begin{figure}
\vspace{-1.2in}
\plotonesmall{0.9}{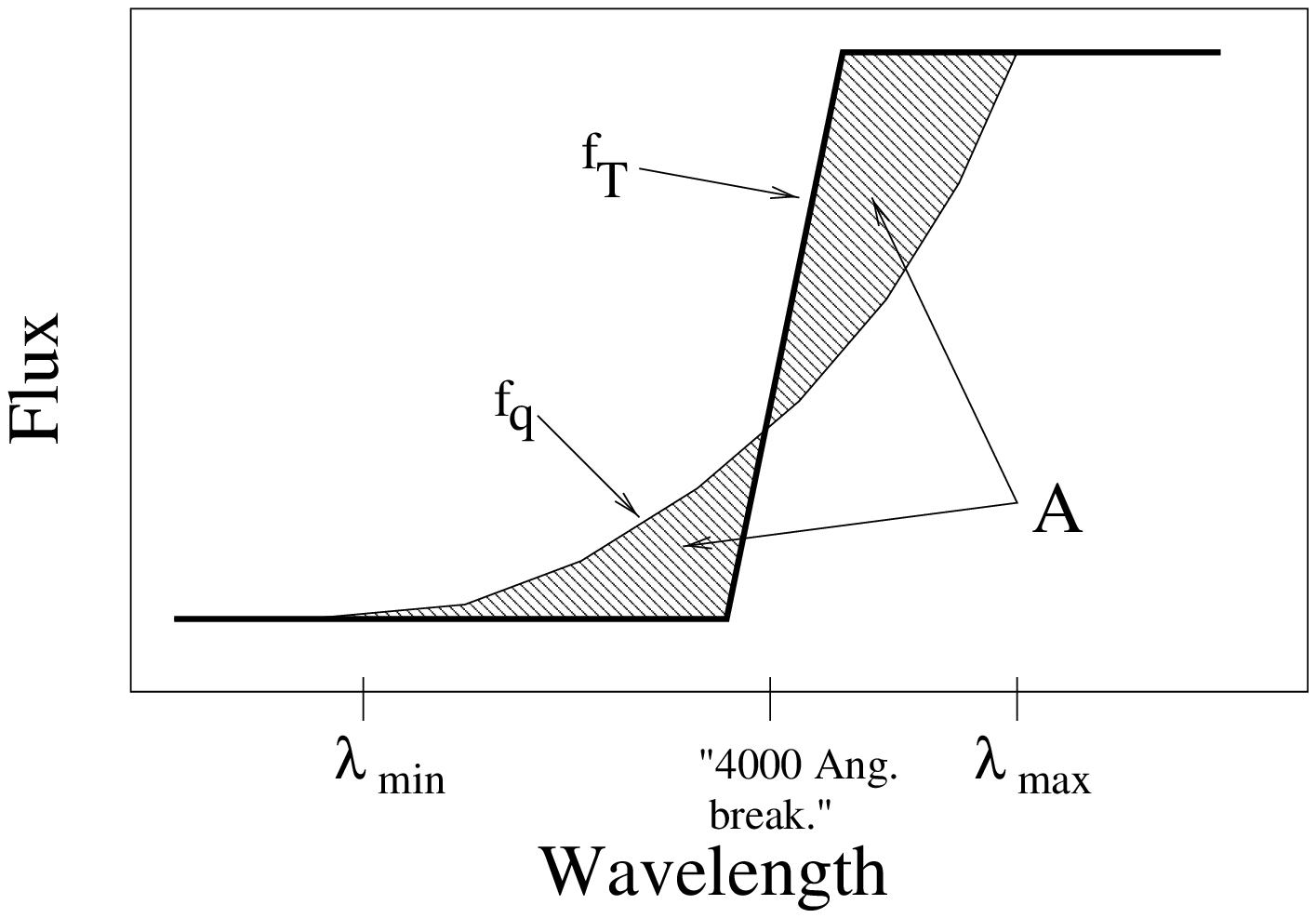}
\caption[Definition of our ``figure of merit'' $A$]{Definition of our ``figure of merit'' $A$.
For a particular source spectrum $f_{T}(\lambda)$ (whose 4000\AA$\,$ break
region is shown exaggerated), we subtract an amount of galaxy
flux, $f_{g}(\lambda)$ (Fig.~\ref{galspec}) until $A$ is a
minimum and a
smooth power-law $f_{q}(\lambda)$ results. 
}
\label{schematic}
\end{figure}

To apply this algorithm in a self consistent way to each of our
optical spectra, we need to define a figure of merit indicating 
the point at which the maximum amount of galaxy spectrum has been 
subtracted and a ``smooth'' quasar spectrum (ie. a power-law) is
achieved.  
Let us first consider the rest frame optical spectrum of a source,
$f_{T}(\lambda)$,  
suspected of containing a 4000\AA$\,$ break. This is illustrated in  
Fig.~\ref{schematic}.
Furthermore, let us consider a ``smooth'' power-law (ie. 
the underlying quasar spectrum $f_{q}(\lambda)$)
between two wavelengths
$\lambda_{min}$ and $\lambda_{max}$ on either side of the 
4000\AA$\,$ break region as shown. 
We define our figure of merit as representing the
area $A$ of the shaded region in Fig.~\ref{schematic}.
As certain amounts of galaxy spectrum,
$c\,f_{g}(\lambda)$ (where $f_{g}(\lambda)$ is given by Fig.~\ref{galspec}) 
are gradually 
subtracted from $f_{T}(\lambda)$, $A$ will decrease
and becomes a minimum when the break disappears.
We can thus determine the value of $c$ when this occurs, allowing us
to estimate the fractional galaxy contribution from Eqn.~\ref{galfrac}.

For a given amount of
subtracted galaxy flux, $c\,f_{g}(\lambda)$,
we can write $A$ in terms of $c$ from Eqn.~\ref{fq} as follows:
\begin{equation}
A\,=\,\int^{\lambda_{max}}_{\lambda_{min}}
{\left| f_{q}(\lambda) - f_{T}(\lambda) + c\,f_{g}(\lambda)\right|}
\,\,d\lambda. 
\label{Aone}
\end{equation}
With $f_{q}(\lambda)$ defined by Eqns.~\ref{pl2},~\ref{alpha2} and 
~\ref{fqminmax}, $A$ can be written:
\begin{equation} 
A\,=\,\int^{\lambda_{max}}_{\lambda_{min}}
\left| \left[f_{T}(\lambda_{max})-c\,f_{g}(\lambda_{max})\right]
\left(\frac{\lambda}{\lambda_{max}}\right)^{\alpha (c,f_{T},f_{g})}
-\,f_{T}(\lambda) + c\,f_{g}(\lambda)\right|\,\,d\lambda,
\label{Atwo}
\end{equation}
where $\alpha(c,f_{T},f_{g})$ is defined by Eqns.~\ref{alpha2} and 
~\ref{fqminmax}. 
Thus, one would only have to minimise $A$ with respect to $c$ in order to
determine the maximal galactic contribution. 
We introduce however,  
an additional factor in Eqn.~\ref{Atwo} whose purpose 
will be to make best use of the 
available data 
and optimise our algorithm.

\subsubsection{Data Optimisation}

Due to systematic effects, each Parkes optical spectrum suffers from a
considerable amount of noise at the range of observed wavelengths.
Although the noise may have some wavelength dependence, we assume it is
constant. 
Given this assumption, we have used an optimal method that gives
more weight to those wavelengths in a particular spectrum where the residual 
$\left| f_{q}(\lambda)-f_{T}(\lambda)\right|$ (see Fig.~\ref{schematic})
is likely to be a maximum.
In other words, the 
galaxy subtraction process will depend most sensitively on the
spectral shape around the 4000\AA$\,$ break region.
Our assumption that the noise is wavelength independent will greatly 
simplify our optimisation method,
since otherwise, 
we would have to simultaneously optimise
those observed spectral regions with
the highest signal-to-noise.

The basis of this method simply involves convolving the integrand in 
Eqn.~\ref{Atwo} with a weighting function $G(\lambda)$ that gives
more weight to regions on either side of the $4000$\AA$\,$ break 
within $\lambda_{min}<\lambda<\lambda_{max}$ (see Fig.~\ref{schematic}).
We choose to define $G(\lambda)$ purely from the galaxy spectrum
$f_{g}(\lambda)$ (dashed curve in Fig.~\ref{galspec}).
This is defined as the difference
(or residual) between a power-law and the galaxy spectrum
within the range
$\lambda_{min}<\lambda<\lambda_{max}$, containing the $4000$\AA$\,$ break:
\begin{equation}
G(\lambda)\,=\,\left|f_{gPL}(\lambda) - f_{g}(\lambda)\right|, 
\label{G}
\end{equation} 
where 
$$
f_{gPL}(\lambda)\,=\,f_{g}(\lambda_{max})\left(
\frac{\lambda}{\lambda_{max}}\right)^{\alpha_{g}}
$$
and
$$
\alpha_{g}\,=\,\frac{\ln\left[f_{g}(\lambda_{min})/f_{g}(\lambda_{max})\right]}
{\ln\left[\lambda_{min}/\lambda_{max}\right]}.
$$
The function $G(\lambda)$ will peak
at wavelengths on either side of the
$4000$\AA$\,$ break. Thus, by convolving $G(\lambda)$ with Eqn.~\ref{Atwo}, 
relatively more weight will be given to spectral data at these 
wavelengths, where our algorithm is most sensitive.

\begin{figure}
\vspace{-3.5in}
\hspace{0.7cm}
\plotonesmall{1}{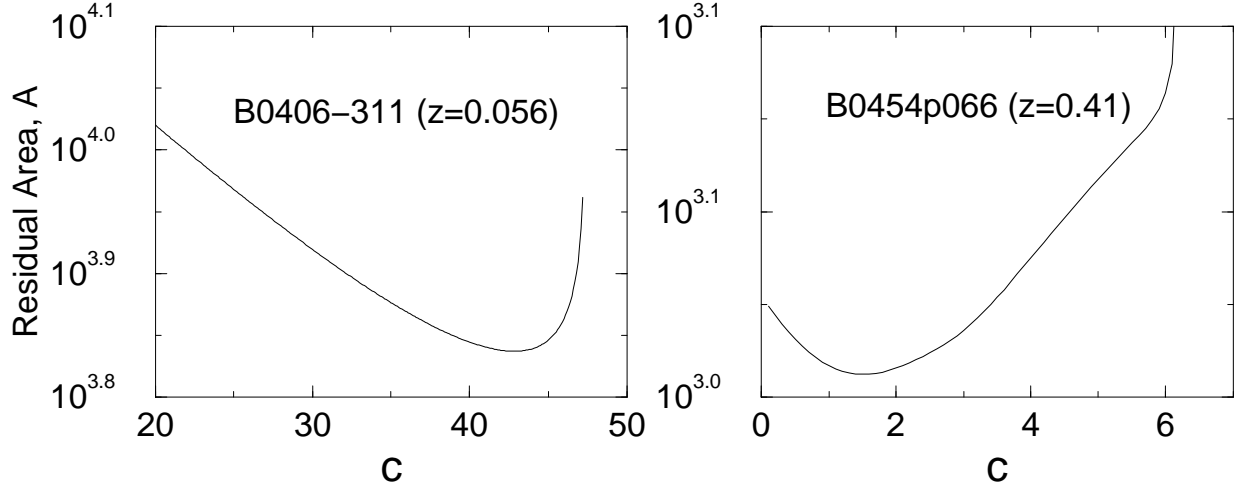}
\vspace{-2.5in}
\caption[Examples of our figure of merit function $A$]{
Examples of our figure of merit function $A$ (defined by
Eqn.~\ref{Athree})
as a
function of the parameter $c$. Source spectra are shown in Fig.~\ref{fits1}.}
\label{mineg}
\end{figure}

Thus, the function to minimise can now be written:
\newpage
\begin{eqnarray}
\nonumber
A\,& = & \,\int^{\lambda_{max}}_{\lambda_{min}}
\put(0,20){\line(0,-1){35}}\left[f_{T}(\lambda_{max})\,-\,c\,f_{g}(\lambda_{max})\right]
\left(\frac{\lambda}{\lambda_{max}}\right)^{\alpha (c,f_{T},f_{g})}\\ 
& &{\rm\hspace{0.5in}}\,-\,f_{T}(\lambda)\, 
+\, c\,f_{g}(\lambda){\rm\hspace{0.05in}}\put(0,20){\line(0,-1){35}}\,\,
\,G(\lambda)\,\,d\lambda,
\label{Athree}
\end{eqnarray}
where $G(\lambda)$ is defined by Eqn.~\ref{G}.
Our aim is to determine the
value of $c$ for a particular source spectrum $f_{T}(\lambda)$ 
such that $A$ is a minimum.                          
In Fig.~\ref{mineg}, we show examples of the figure of merit function $A$
(Eqn.~\ref{Athree}) as a function of the parameter $c$ for two sources
whose spectra are shown in Fig.~\ref{fits1}.
If the weighting factor $G(\lambda)$ in 
Eqn.~\ref{Athree} is neglected, we find that the values of $c$ at which 
$A$ is a minimum decrease by up to $5\%$ in most sources
where the break feature appears relatively weak or absent.
In sources with strong 4000\AA$\,$ breaks, there is no 
significant difference.
Given the value of $c$ that minimises $A$, the fractional 
galaxy contribution at some wavelength can now be computed using 
Eqn.~\ref{galfrac}.

\subsection{Error Determination}
\label{errdes}

In spectra where the galaxy contribution is relatively 
weak, an upper limit on its contribution at some appropriate
level of significance
would be required.
In order to do so, both random and systematic errors need to be investigated.
The magnitude of these two types of errors are estimated and compared in 
section~\ref{sys.vs.ran}. 
Here, we briefly outline the method in their determination.

\subsubsection{Random (Statistical) Errors}

The random error at some confidence level 
in the galaxy contribution is estimated by computing the value of $c$
corresponding to the statistical error in $A$ (Eqn.~\ref{Athree}).
Since Eqn.~\ref{Athree} is actually a discrete sum over wavelength bins
$\lambda_{i}$, from $\lambda_{min}$ to $\lambda_{max}$,
the statistical error is determined by adding the error for each
individual bin in quadrature, so that
\begin{equation}
\sigma(A)\,=\,\sqrt{\sum_{\lambda_{i}=\lambda_{min}}^{\lambda_{max}}
\sigma^{2}_{i}(I)}, 
\label{rand}
\end{equation}
where $\sigma_{i}(I)$ is the error in the integrand $I$ of 
Eqn.~\ref{Athree} for bin $\lambda_{i}$. 
$\sigma_{i}(I)$ will depend only on uncertainties in the measured source fluxes
$f_{T}(\lambda_{max})$ and $f_{T}(\lambda_{min})$ 
in Eqn.~\ref{Athree}.
As shall be discussed in section~\ref{apal}, these fluxes are
estimated by calculating the median continuum flux in
wavelength bins centered on $\lambda_{max}$ and $\lambda_{min}$. 
We estimate the corresponding uncertainties by computing the
rms deviation from the mean spectral flux in these wavelength bins. 

\subsubsection{Systematic Errors}

Our definition of the underlying quasar continuum $f_{q}(\lambda)$,
in each source only requires that it be smooth and contain no breaks.
This implies that the shape of $f_{q}(\lambda)$ is somewhat arbitrary
and thus it is possible that our quantitative measures of the galaxy 
contribution may strongly depend on its assumption in our algorithm.
The uncertainty introduced by this possible 
systematic effect will be investigated. 

All our calculations assume that $f_{q}(\lambda)$
is a power-law (ie. Eqn.~\ref{pl2}), since the continua of optical quasar
spectra are well represented by $f_{\nu}\propto\nu^{-\alpha}$ where
$\alpha\sim0.2-0.3$ (Francis, 1996).
To investigate the effects of assuming a 
different form for $f_{q}(\lambda)$ however,
we also apply our algorithm by assuming for simplicity that
$f_{q}(\lambda)$ is a straight line joining $\lambda_{min}$ and
$\lambda_{max}$ in Fig.~\ref{schematic}.
This is parameterised as follows:
\begin{equation}
f_{q}(\lambda)_{line}\,=\,\frac{\left[f_{q}(\lambda_{max}) - 
f_{q}(\lambda_{min})\right]}{\lambda_{max} - \lambda_{min}}
(\lambda - \lambda_{max})\,+\,f_{q}(\lambda_{max}), 
\label{sl}
\end{equation}
where $f_{q}(\lambda_{min})$ and $f_{q}(\lambda_{max})$ 
are defined by Eqn.~\ref{fqminmax}.
A comparison in the fractional galaxy contributions
resulting from our use of a power-law (Eqn.~\ref{pl2}) 
and a straight line (Eqn.~\ref{sl})
for $f_{q}(\lambda)$ in our algorithm, will enable us to estimate
the magnitude of this systematic effect. 
Results are presented in 
section~\ref{sys.vs.ran}.
Other possible sources of systematic error are also discussed in this
section.

\subsection{Summary}

To summarise, we have presented in this section a method to determine
the relative galaxy contribution in each Parkes source in a robust 
way. 
Our algorithm requires the following two input assumptions:
First, an elliptical optical SED, $f_{g}(\lambda)$, defining 
the shape of the characteristic 4000\AA$\,$ break. This we assume is a model
from Bruzual \& Charlot (1993) (see Fig.~\ref{galspec}). Second,
we require a spectral shape 
defining the underlying quasar spectrum $f_{q}(\lambda)$.
Unless otherwise specified, all our calculations shall assume a power-law
for $f_{q}(\lambda)$.

With the only requirement that $f_{q}(\lambda)$ be smooth and contain
no breaks, the ``suspected'' 4000\AA$\,$ break in each source spectrum
$f_{T}(\lambda)$, is subtracted until the residual between
$f_{q}(\lambda)$ and galaxy subtracted source spectrum is a minimum.
The galaxy contribution is estimated from the amount of galaxy, 
$f_{g}(\lambda)$ subtracted.
To apply this procedure in a self-consistent manner,
we have defined a figure of merit given by Eqn.~\ref{Athree}.
This is minimised with respect to the parameter $c$ from which the
fractional galaxy contribution can be easily computed using Eqn.~\ref{galfrac}. 

\section{Data and Input Parameters}
\label{data}

Out of the 323 sources in the Drinkwater \etal (1997) sample,
we have optical spectra for 194 or 60\% of the sample.
Some of these are from the compilation of Wilkes \etal (1983)
and others are from recent observations on the AAT 
and ANU 2.3m (see Drinkwater \etal 1997).
For these latter observations (65 sources), 
the spectra cover the observed wavelength range:
$3200{\rm \AA}\simlt\lambda\simlt10000{\rm \AA}$ 
with a resolution of $\sim5$\AA$\,$ 
in the blue and $\sim20$\AA$\,$ in the red ($\simgt5200$\AA). 
Other spectra have 
typically a mean resolution
$\sim10$\AA$\,$ and cover the range: 
$3200{\rm \AA}\simlt\lambda\simlt8000{\rm \AA}$.

\subsection{Applying the Algorithm}
\label{apal}

Before applying our algorithm, each Parkes optical spectrum is redshifted
to its rest frame.
We then define the fixed rest frame wavelengths $\lambda_{min}$ and
$\lambda_{max}$ within which the suspected 4000\AA$\,$ break and our
figure of merit (Eqn.~\ref{Athree}) is defined (see Fig.~\ref{schematic}).
To choose these wavelengths, we need to consider the following:

First, the highest redshift sources
will have the rest wavelength $\lambda_{max}$
redshifted out of the observational wavelength range of the spectra.
These sources will not be able to be used in our algorithm.
We therefore need to choose $\lambda_{max}$
such that the number of sources in which our algorithm
can be applied is not significantly reduced.
Second, we need a wavelength range
$\lambda_{min}<\lambda<\lambda_{max}$
that makes `optimal' use of 
the shape of the 4000\AA$\,$ break
region defining our figure of merit $A$ 
(the residual area in Fig.~\ref{schematic}).
In other words, we need to ensure that this region
is unambiguously defined and clearly represented in each source spectrum. 
As a compromise, we assume 
$\lambda_{min}=3500$\AA$\,$ and $\lambda_{max}=5080$\AA$\,$ in every source. 

Relative measures of the fluxes: $f_{T}(\lambda_{min,max})$ and
$f_{g}(\lambda_{min,max})$ in the source and galaxy rest frame spectra
respectively (see Eqn.~\ref{Athree}), are determined as follows.
We first define 
wavelength bins of width $\sim200$\AA$\,$ and
$\sim400$\AA$\,$ centered on $\lambda_{min}$
and $\lambda_{max}$ respectively, and then calculate the median
{\it continuum} flux in each bin.
The wavelength regions defining these bins however may
contain absorption
and emission lines. 
Such lines
are likely to bias
our estimates of the continuum level in these regions.
From the bin widths defined above, the rest wavelength regions
of interest 
are:  
$3400{\rm\AA}<\lambda<3600$\AA$\,$ and $4880{\rm\AA}<\lambda<5280$\AA$\,$. 
From the available source spectra, we find that 
no lines are likely to 
contaminate the short wavelength bin. 
For the long wavelength bin however, we find that the emission line
doublet [OIII]$\lambda\lambda$4959,5007 and a weak absorption 
feature at $\sim5170$\AA$\,$ (possibly from MgI) are strong contaminants.
To avoid significant contamination, our algorithm excludes regions of width 
10\AA$\,$ centered on these lines. 

Given the definitions above, our rest wavelength range required by our
algorithm will therefore 
be $3400{\rm \AA}\simlt\lambda_{rest}\simlt5280{\rm \AA}$.
With a maximum observed wavelength of
$\lambda_{max}({\rm obs})\simeq10000$\AA$\,$
in $\sim30\%$ of the available spectra, we find that only sources
with redshifts $z\simlt0.9$ can be used
in our algorithm.
In total, we have about 53 spectra in which our algorithm can be applied.
 
\section{Results}
\label{regcp}

\subsection{Spectral Fits}

With the input parameters from the previous section, 
Eqn.~\ref{Athree} is minimised numerically
with respect to the parameter $c$ for each spectrum.
Having found the value $c_{min}$ that minimises Eqn.~\ref{Athree},
we can reconstruct the initial source spectrum $f_{T}(\lambda)$
by superimposing the maximum amount of galaxy spectrum, 
$c_{min}f_{g}(\lambda)$ generated by the algorithm, and a power-law
representing the underlying ``smooth'' quasar continuum $f_{q}(\lambda)$. 
From Eqns.~\ref{fq} and~\ref{pl2}, 
these reconstructed model spectra can be represented:
\begin{eqnarray}
\nonumber
f_{T}(\lambda)_{model}\,& = &\,c_{min}f_{g}(\lambda)\,+\,f_{q}(\lambda)\\
&\equiv & c_{min}f_{g}(\lambda)\,+\,\left[f_{T}(\lambda_{max})-c_{min}\,f_{g}(\lambda_{max})\right]
\left(\frac{\lambda}{\lambda_{max}}\right)^{\alpha (c_{min},f_{T},f_{g})}.
\label{fTmodel}
\end{eqnarray}

\begin{figure}
\includegraphics{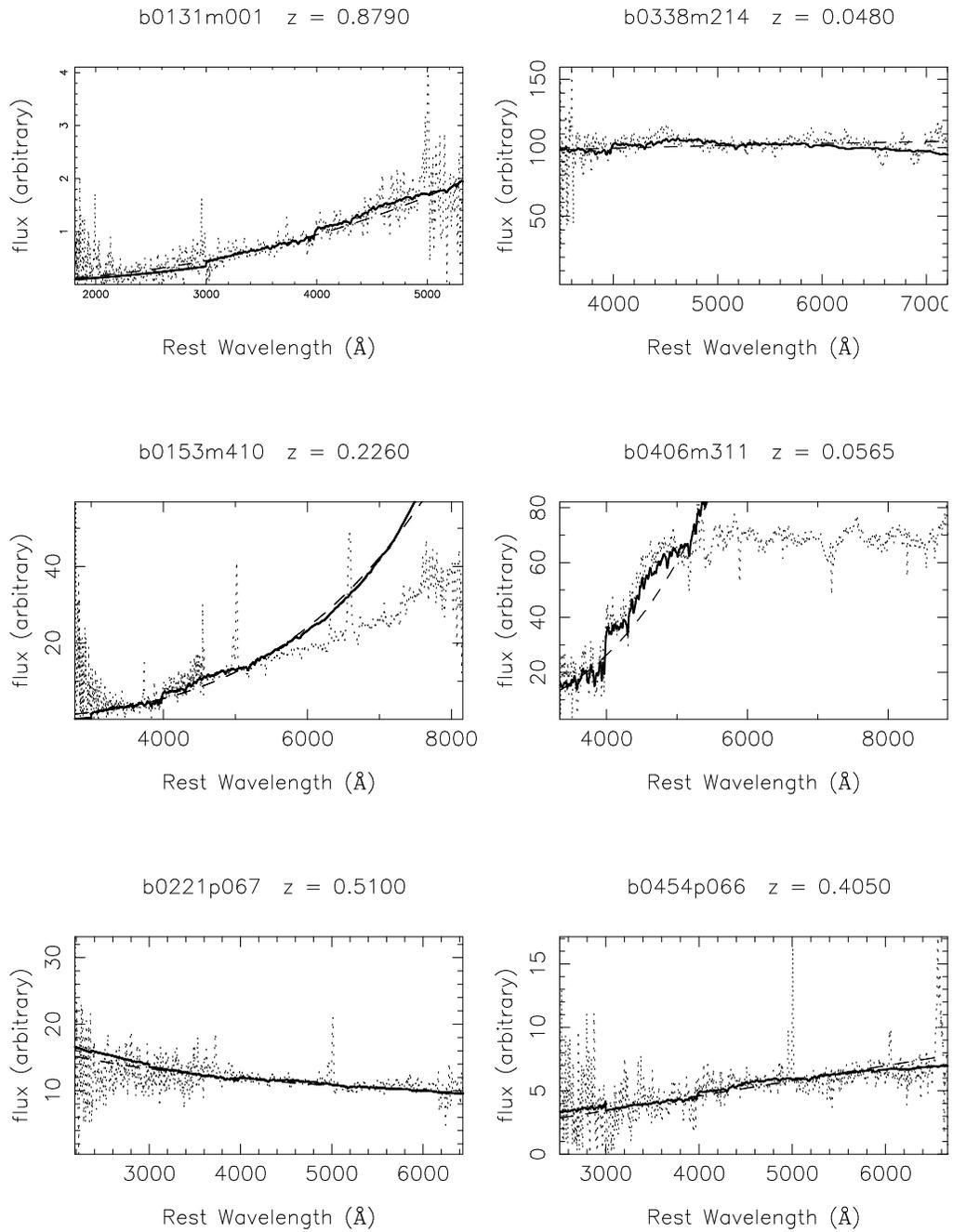} 
\vspace{5.5in}
\caption[Observed and reconstructed model spectra]{Observed rest frame spectra (dotted) for a number of
sources in the Parkes sample. Also shown are
reconstructed model spectra (solid) as defined by Eqn.~\ref{fTmodel} and
power-law fits representing the underlying quasar continuum (dashed).}
\label{fits1}
\end{figure}

A visual comparison between these model and observed spectra in the range
$\lambda_{min}<\lambda<\lambda_{max}$ will allow us to investigate the
accuracy of our algorithm in reproducing the observed spectra
around the 4000\AA$\,$ break region.
Reconstructed spectra $f_{T}(\lambda)_{model}$ are compared with the observed 
spectra, $f_{T}(\lambda)$, for a number of sources in Fig.~\ref{fits1}.
On each spectrum, we also show our power-law fit $f_{q}(\lambda)$.  
As expected, observed spectra showing strong 4000\AA$\,$ breaks with SEDs
similar to that given in Fig.~\ref{galspec} are reproduced very accurately
about the break region. 

In a majority of observed spectra where no significantly strong
breaks are discernible to the human eye however, our algorithm
nevertheless attempts to fit for a break.
Unbiased estimates of the galaxy contribution using the
relative sizes of these breaks, however weak, are presented in
section~\ref{galcont}.

\begin{figure}
\vspace{-4in}
\hspace{-2in}
\plotonesmall{1}{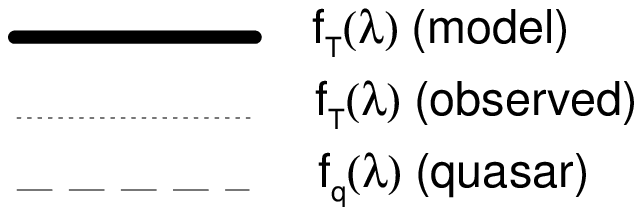}
\vspace{-3.5in}
\end{figure}

\begin{figure}
\includegraphics{fitstwo.ps}
\vspace{5.5in}
\centerline{Fig.~\ref{fits1} continued.}
\end{figure}

\subsection{Systematic vs. Random Errors}
\label{sys.vs.ran}

\begin{figure}
\vspace{-2.5in}
\plotonesmall{1}{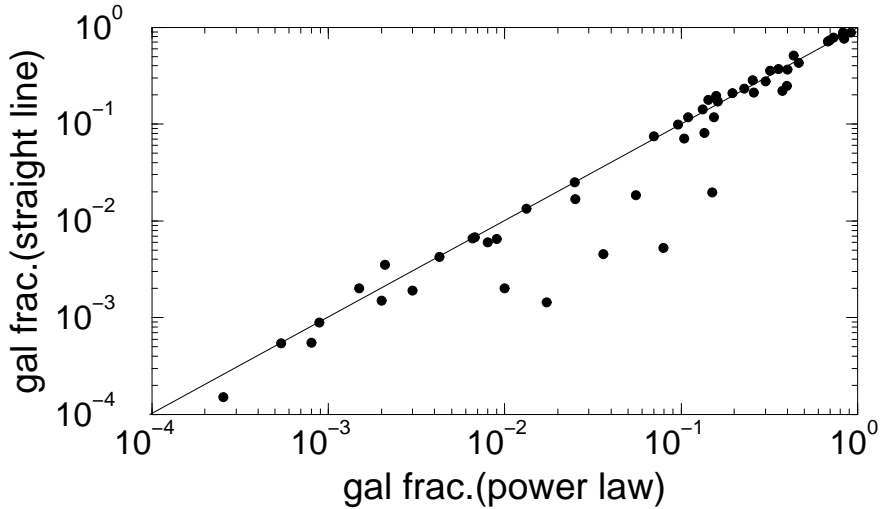}
\vspace{-3in}
\caption[Fractional galaxy contributions for different model assumptions]{Fractional galaxy contribution at 5000\AA$\,$ (rest frame) 
assuming a straight line
for $f_{q}(\lambda)$ (vertical axis) and a power-law (horizontal axis).
The diagonal line is the line of equality. (see section~\ref{sys.vs.ran}).}
\label{sys}
\end{figure}

As discussed in section~\ref{errdes}, 
a possible source of systematic uncertainty
is in our assumption of the shape of the underlying smooth quasar
continuum $f_{q}(\lambda)$.
To explore this, we compare the relative galaxy contribution obtained 
by assuming first, a power-law (PL) (Eqn.~\ref{pl2}) for $f_{q}(\lambda)$
and second, a straight line (L) (Eqn.~\ref{sl}).
Using our algorithm and these two
definitions for $f_{q}(\lambda)$, we have 
computed the fractional galaxy contribution
at 5000\AA$\,$ (rest frame). Results are shown in Fig.~\ref{sys}.

At first glance, estimates for the galaxy fraction using the power-law
and straight line for $f_{q}(\lambda)$ agree very well.
There is relatively little scatter about the diagonal line defining
the equality ${\rm frac}_{L}={\rm frac}_{PL}$, except for
a distinct population with ${\rm frac}_{L}<{\rm frac}_{PL}$.
No distinguishing feature in the optical spectra of this latter
class is immediately apparent.
It is likely that 
a power-law (rather than a straight line) 
within $\lambda_{min}<\lambda<\lambda_{max}$ for these sources
provides a better representation of our figure of merit $A$ 
in Fig.~\ref{schematic}. 
We quantify the systematic error from the rms scatter in the
difference:
$\delta={\rm frac}_{L}-{\rm frac}_{PL}$, which we denote by $\sigma(\delta)$.
For this systematic effect, we therefore estimate a 
$1\sigma$ uncertainty in the galaxy fraction 
at 5000\AA$\,$ of at most 
$\sigma_{frac.}\simeq0.02$.

Another possible source of systematic error is
in our selection of the wavelengths $\lambda_{max}$ and $\lambda_{min}$,
within which the suspected 4000\AA$\,$ break and our figure of merit
(Eqn.~\ref{Athree}) is defined
(see Fig.~\ref{schematic}).
As discussed in section~\ref{apal}, the values 
$\lambda_{min}=3500$\AA$\,$ and $\lambda_{max}=5080$\AA$\,$ 
were chosen as a compromise between: 
first, to maximise the number
of sources in which the
4000\AA$\,$ break remains within the
observed wavelength after redshifting, and second, to 
make optimal use of the break region. 
What are the effects on the galaxy fraction if 
a different wavelength range were assumed?

To explore this, we choose to vary
$\lambda_{max}$ alone. 
Due to the
relatively small wavelength range at
$\lambda<4000$\AA$\,$ available in the galaxy spectrum (Fig.~\ref{galspec}), 
we are not as flexible in
varying $\lambda_{min}$.
We therefore keep $\lambda_{min}$ fixed at 3500\AA.
Assuming the same bin widths (200\AA$\,$ and 400\AA$\,$) 
centered on $\lambda_{min}$ and $\lambda_{max}$,
and the PL definition for $f_{q}(\lambda)$,
we have computed
galaxy fractions at 5000\AA$\,$ with 
$\lambda_{max}=5500$\AA$\,$, and 
$\lambda_{max}=6500$\AA$\,$. 
Combined with our estimates using
$\lambda_{max}=5080$\AA$\,$, we find that changing $\lambda_{max}$ makes 
negligible difference in the galaxy fraction. The fractions differ 
by no more than 1\%.

\begin{figure}
\vspace{-2.7in}
\plotonesmall{1}{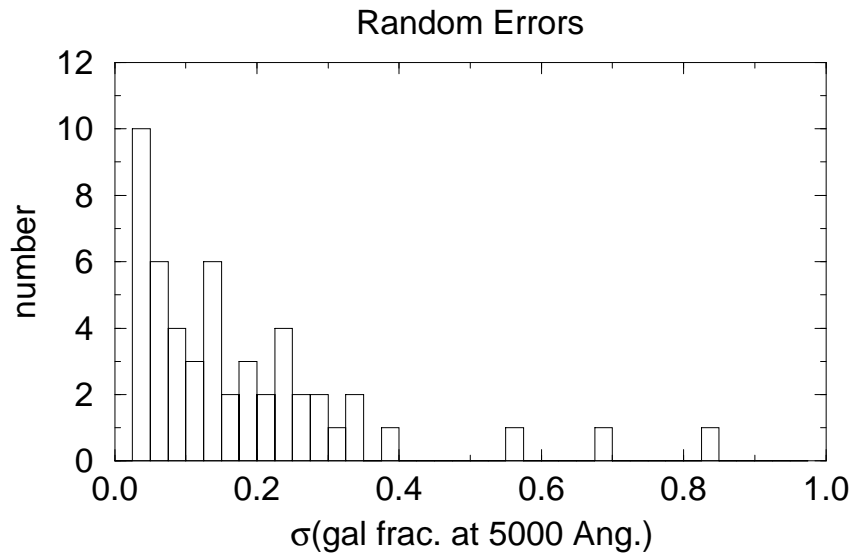}
\vspace{-2.6in}
\caption[Distribution of $1\sigma$ random errors in the fractional galaxy contribution]{Distribution of $1\sigma$ random errors in the 
fractional galaxy contribution at 5000\AA~(rest frame)
for Parkes sources.
(see section~\ref{sys.vs.ran}).}
\label{randfig}
\end{figure}

We now compare these systematic uncertainties with estimates of
the random errors.
Fig.~\ref{randfig} shows 
the distribution of random errors in the fractional galactic contribution
at 5000\AA$\,$ as determined from our algorithm (see section~\ref{errdes}).
The range in random errors is significantly broad, with a majority
of values exceeding our maximum $1\sigma$ systematic 
uncertainty of 0.02 deduced from Fig.~\ref{sys}. 
In the remaining sections, we therefore quote all 
uncertainties in the relative galaxy contribution
as purely statistical, based
on random errors alone.

\subsection{Galactic Contribution to the Observed Optical-Near IR
Continuum.}
\label{galcont}

Using our algorithm, we have computed the fractional galaxy
contribution in the individual bandpasses $B_{J}$ 
$(\lambda\simeq4400{\rm \AA})$ and $K$ $(\lambda\simeq2.2\mu{\rm m})$,
expected in an observer's frame for each source. 
We estimate these using Eqn.~\ref{galfrac} where fluxes are
approximated by computing the median spectral flux at the central wavelength
of each bandpass. 
Since our observed spectra do not extend to near-IR wavelengths,
we estimate the observed spectral flux in $K$ using our observed
$B_{J}-K$ colours and extrapolate from 
the spectral flux corresponding to $B_{J}$. 
Unfortunately, not all of the 53 sources used in
our algorithm have a measured $K$ magnitude and hence $B_{J}-K$ colour.
Of these sources, about 34 have known $K$ magnitudes.

\begin{figure}
\vspace{-1in}
\plotonesmall{1}{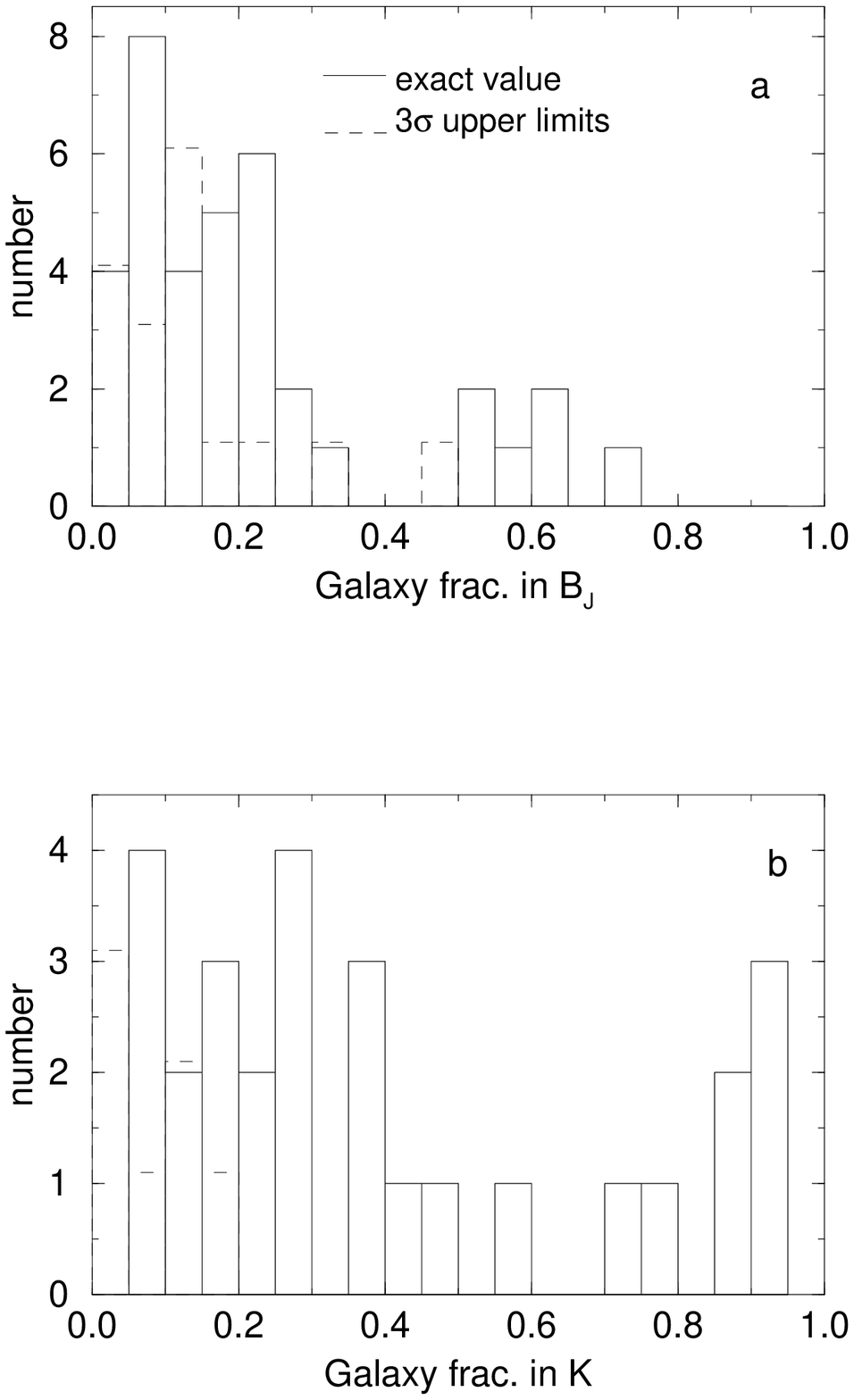}
\caption[Distribution in fractional galaxy contributions in $B_{J}$ and $K_{n}$]{(a) Distribution in fractional galaxy contributions
in $B_{J}$ 
and (b) $K$
for $z<1$.
Dashed portions represent $3\sigma$ upper limits,
(see section~\ref{galcont}).}
\label{fracBK}
\end{figure}

The distributions in galaxy fractions in $B_{J}$ and $K$ are shown in
Figs.~\ref{fracBK}a and b respectively.
Sources with mean galaxy fractions 
$\simlt5\%$ are replaced by their
$3\sigma$ upper limits (dashed histograms). 
The distributions in Figs.~\ref{fracBK}a and b appear very similar, except
however for a greater number of sources
with galaxy contributions $>70\%$ in the $K$-band. 
These are all low redshift sources with strong 4000\AA$\,$ 
breaks in their
spectra. Their light is expected to be dominated
by evolved stellar populations, and hence
strongest in $K$. 
We must also note that the $B_{J}$ magnitudes
used to estimate the $K$ spectral fluxes via 
$B_{J}-K$ colours
are only accurate to
$\sim1$ mag (Drinkwater \etal 1997). 
This photometric uncertainty is likely 
to contribute significant scatter  
in our estimates of the $K$ galaxy fraction in Fig.~\ref{fracBK}b. 

\begin{figure}
\vspace{-2.7in}
\plotonesmall{1}{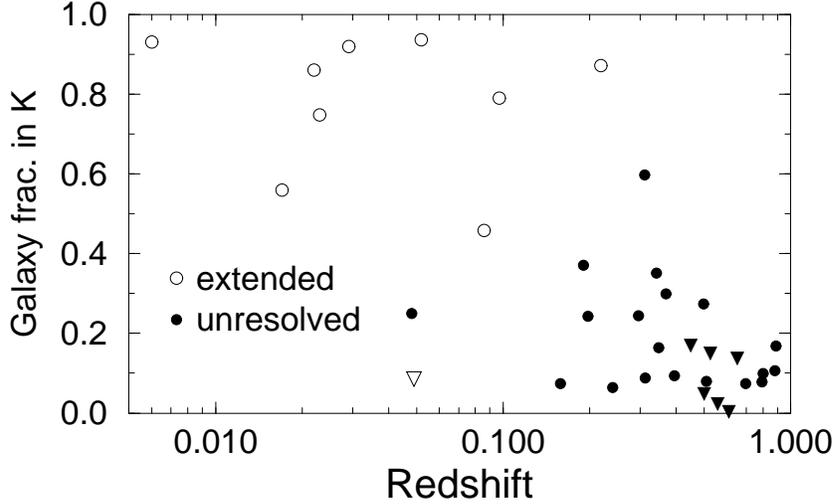}
\vspace{-2.6in}
\caption[Galaxy contribution in resolved and unresolved sources]{Fractional galaxy contribution in $K$ as a function of
redshift for resolved sources
(extended on $K$ and $B_{J}$ images; open symbols) and
unresolved sources (closed symbols). Triangles represent $3\sigma$ upper limits
on the galaxy fraction.}
\label{Kfracvsz}
\end{figure}

We have divided the $K$-band identifications into two populations:
those which show extended (resolved) structure, and those which remain
unresolved.
The galaxy fraction in $K$ 
for these two populations is shown as a function of $z$
in Fig.~\ref{Kfracvsz}.
Open symbols represent resolved sources, and filled symbols, unresolved
sources.
As expected, those sources exhibiting resolved structure are also those
which show large galaxy fractions and are at relatively low redshifts. 

We now investigate whether emission from the host 
galaxies of Parkes quasars can 
significantly contribute to their observed $B_{J}-K$ colours. 
We do this by computing the $B_{J}-K$ colour of the hypothesised
underlying ``quasar'', $(B_{J}-K)_{q}$,  
we would expect if contribution from the host galaxy was absent in each source.
If the observed colours were entirely due to galactic emission, then we
expect the distribution in $(B_{J}-K)_{q}$ to show a relatively small
scatter, i.e. similar to that observed for 
optically selected quasars where typically 
$(B_{J}-K)_{q}\simeq2.5$.

\begin{figure}
\vspace{-1in}
\plotonesmall{1}{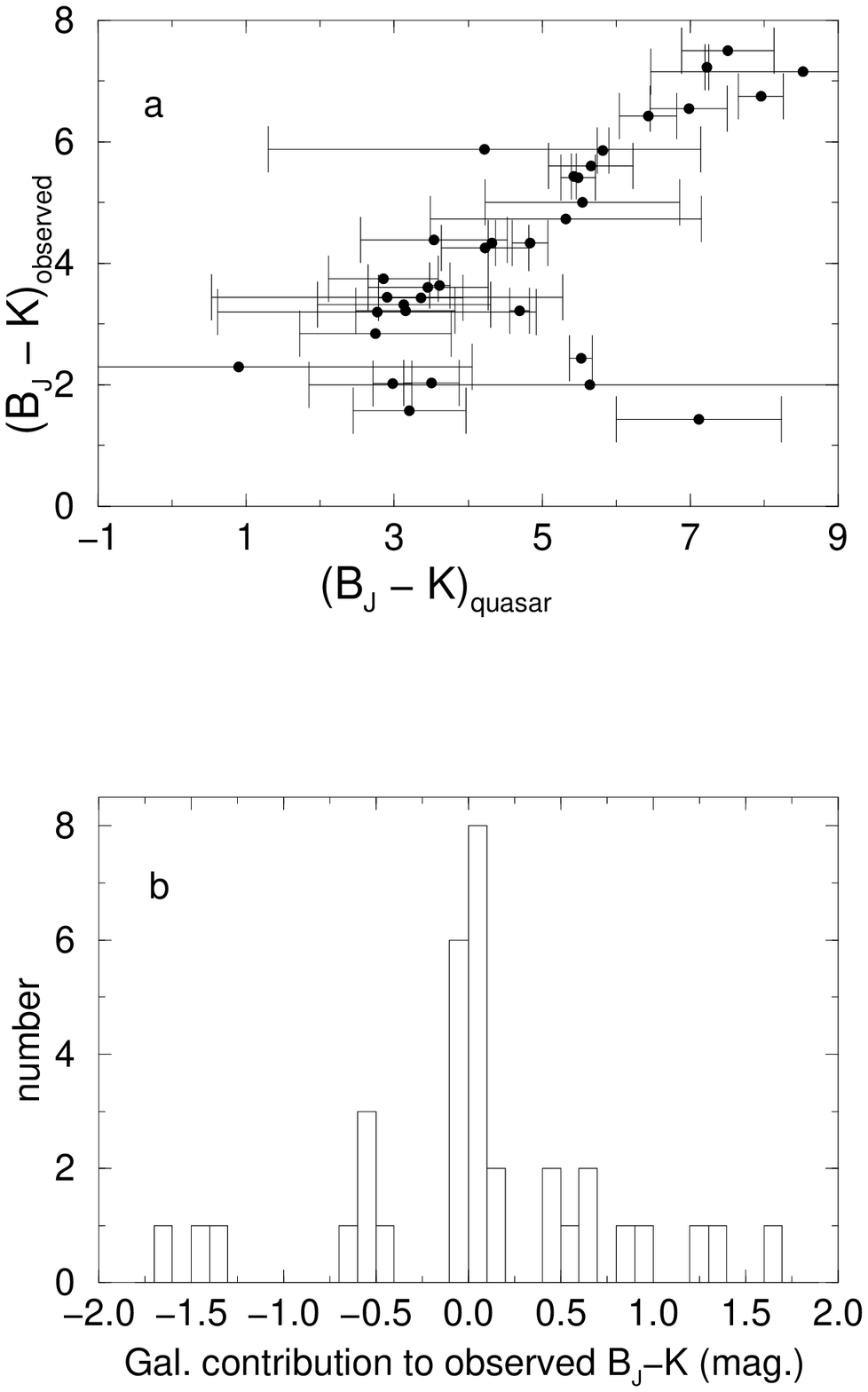}
\caption[Galaxy contribution to observed $B_{J}-K_{n}$ colours]{(a) Observed $B_{J}-K$ colour vs. colour of
the underlying quasar (i.e. resulting galaxy subtracted colour:
$(B_{J}-K)_{q}$). Error bars correspond to $1\sigma$ significance.
(b) Distribution in galaxy contribution to 
observed $B_{J}-K$ colours in magnitudes.} 
\label{BminusKgal}
\end{figure}

The colour of an underlying quasar, $(B_{J}-K)_{q}$, can be written in terms
of the observed colour $(B_{J}-K)_{obs}$ and the galaxy
fractional contributions $F_{gal}(B_{J})$ and $F_{gal}(K)$ as follows: 
\begin{equation}
(B_{J}-K)_{q}\,=\,(B_{J}-K)_{obs}\,+\,2.5\log{\left[\frac{1-F_{gal}(K)}
{1-F_{gal}(B_{J})}\right]}.  
\label{BKq}
\end{equation}
$(B_{J}-K)_{q}$ is plotted against $(B_{J}-K)_{obs}$ in Fig.~\ref{BminusKgal}a. 
As can be seen, the scatter in galaxy subtracted colours,
$(B_{J}-K)_{q}$, remains and is extremely similar to that of
the observed colour distribution. 
We quantify the galaxy contribution
to the observed $B_{J}-K$ colours (in magnitudes)
in Fig.~\ref{BminusKgal}b. 
From Figs.~\ref{BminusKgal}a and b, we conclude that the observed spread
in colours cannot be due to emission from the host galaxies of Parkes
quasars. An independent mechanism
must be involved.

\section{A Test for the Unified Model}
\label{tuf}

Studies of the host galaxy properties of BL-Lacs and radio-quasars
can be used as a test of the unified scheme for radio-loud AGN. 
Motivated by the 
canonical axisymmetric model for AGN,
the basis of this scheme is 
that the appearance of an extragalactic radio source 
is primarily determined
by viewing geometry.
Extended FRI and FRII-type radio galaxies are believed
to represent the parent (misaligned) populations of 
the more compact BL-Lacs and radio quasars
respectively (Urry \& Padovani, 1995
and references therein). 
If classification is purely based on orientation, 
then intrinsic properties such as 
host galaxy luminosity should  
be approximately uniform throughout. 
In this section, we shall test this hypothesis.

Studies have shown that for quasars and 
FRII radio galaxies at redshifts $z\simlt0.3$, 
the situation is not
entirely clear. 
From comparisons of their mean host galaxy luminosities,
some studies have shown that FRII hosts are fainter by $\sim0.5$-1 mag. 
(Smith \& Heckman, 1989), while others have concluded that they
are comparable (eg. Taylor \etal 1996).
The main difficulty in these studies was finding sufficiently large
samples of radio galaxies and quasars matched both in 
radio power and redshift.
There is strong observational evidence however that the low redshift
BL-Lacs reside in giant ellipticals with mean optical
luminosities and de Vaucoleurs $r^{1/4}$ law profiles similar to those in
FRIs (Ulrich, 1989; Stichel \etal 1993). 
Very little is known about the host galaxies
of compact radio sources at higher redshifts.
Using our algorithm however, we can get estimates
of host galaxy $K$ magnitudes for sources up to $z\sim1$. 

We can predict the host galaxy $K$ magnitudes of Parkes sources 
directly from our estimates of the fractional galaxy contribution in the
$K$-band, $F_{gal}(K)$ (see section~\ref{galcont}). 
The host galaxy magnitude in an observer's frame can be written:
\begin{equation}
K_{gal}\,=\,K_{source}\,-\,2.5\log{\left[F_{gal}(K)\right]},
\label{Kgalobs}
\end{equation}
where $K_{source}$ is the observed $K$ magnitude of the source.
As discussed in section~\ref{galcont}, $F_{gal}(K)$ is determined 
from the median spectral flux at $\lambda\simeq2.2\mu$m (Eqn.~\ref{galfrac}), 
which we estimate 
using the observed spectral flux at $\lambda\simeq4400$\AA$\,$
and $B_{J}-K$ colour. 

\begin{figure}
\vspace{-3in}
\plotonesmall{1}{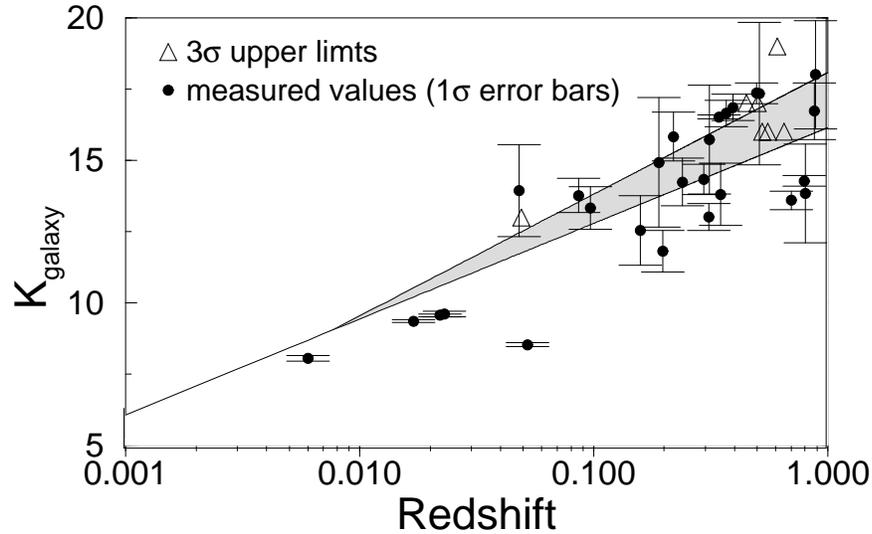}
\vspace{-2.7in}
\caption[Host galaxy $K$ magnitude as a function of redshift]{Host galaxy $K$ magnitude as a function of redshift
for Parkes sources (symbols) and the range observed for radio galaxies
(shaded region; McCarthy, 1993).}
\label{Kgalfig}
\end{figure}

Estimates of $K_{gal}$ as a function of redshift are shown 
in Fig.~\ref{Kgalfig}. 
For comparison, we also show the range
observed for radio galaxies (shaded region) as determined from a number of
independent studies (McCarthy, 1993; see section~\ref{hubblelaw}). 
Within our quoted uncertainties, there appears to be no significant
difference in the mean host galaxy magnitude of 
``compact'' Parkes sources and extended radio galaxies at the
redshifts indicated. 
The compact sources however appear to show a larger scatter in $K_{gal}$
at some redshift.
Since the $K_{gal}$ values were determined from
non-contemporaneous measures of observed $B_{J}$ spectral fluxes and
$B_{J}-K$ colours, this may be attributed to variability in the underlying AGN.
We are unable at present to quantify this uncertainty.
From Fig.~\ref{Kgalfig}, we conclude that the host galaxy 
luminosities of these two classes
of radio source is consistent with that required by the unified model. 

\section{Discussion}
\label{diseven}

Our results of section~\ref{galcont} clearly show that galactic emission
is unlikely to fully explain
the dispersion in $B_{J}-K$ colours observed.
This conclusion is only valid however for sources at redshifts $z\simlt 0.9$.
At higher redshifts, the 4000\AA$\,$
break feature on which our algorithm is based
is redshifted out of our observational wavelength range.
Taking into account our completeness in spectral data (section~\ref{data}), 
only 53 of the 323 
sources in the Drinkwater \etal (1996) sample have been analysed using our
algorithm.

Since we are limited to redshifts $z\simlt 0.9$, it is possible that we are 
biased towards detecting
relatively large galaxy contributions.
Galaxy flux will be more easily detected at the lowest redshifts. 
This indeed may appear to be the case if we compare the fractional
galaxy contribution in $K$ from our algorithm (Fig.~\ref{fracBK}b),
with that predicted from the Hubble diagram (Fig.~\ref{hubble})
containing sources to $z\sim3$.
Comparing the radio galaxy and source $K$ magnitudes, 
the Hubble diagram predicts a ``mean'' fractional galaxy contribution
of $\sim20\%$ for redshifts $z\simlt3$, 
while the mean in Fig.~\ref{fracBK}b is $\sim50\%$.
This discrepancy is expected since galactic flux for sources at $z>0.9$ 
will obviously be relatively weak.

We have three pieces of observational evidence that supports 
a minimal galaxy contribution from the high redshift sources: 
First, all sources with $z\simgt0.5$ appear very 
compact in $K$ (eg. Fig.~\ref{Kfracvsz}), 
and exhibit broad-line equivalent widths
typical of those
observed in quasars in the optical.
Second, the observed $B_{J}-K$ colours correlate with optical-UV 
continuum slope (see section~\ref{optvscol}), implying that 
optical-UV colours are also relatively red and display large scatter. 
The observed 
optical-UV colours of ellipticals at $z\simgt0.9$ however are observed to be  
quite blue (Guiderdoni \& Rocca-Volmerange, 1987 and references therein) 
and hence cannot explain this observation.
Third, significantly high levels of linear polarisation ($\simgt5\%$)
are observed at near-IR wavelengths in a majority of sources 
(see Chapter 7).
This strongly indicates 
that the emission is dominated by a non-thermal mechanism.

\section{Conclusions}
\label{concsix}

In this chapter, we have investigated whether emission from the host galaxies
of Parkes quasars can significantly
contribute to the relatively large spread in $B_{J}-K$ colours observed. 
If the hosts are classical giant ellipticals and their flux strongly 
contributes, then this would be expected since elliptical colours 
are known to be quite red in $B-K$ to $z\sim2$.

We have devised an algorithm that measures the
relative galaxy contribution in each source in an unbiased way using
the characteristic 4000\AA$\,$ break feature of elliptical galaxy SEDs.
The basis of the algorithm involves subtracting a generic elliptical SED  
from each source spectrum until the 4000\AA$\,$ break feature disappears 
and what is left is a ``smooth'' spectrum containing no breaks. 
This ``smooth'' spectrum we refer to as the underlying quasar continuum.
The only requirement by our algorithm is that this remaining 
spectrum be smooth.
The galactic 
contribution, relative to the total light at any wavelength 
is estimated from the amount of galaxy subtracted.

The main conclusions are:
\\\indent 1. For $z\simlt0.9$, (for which the 4000\AA$\,$ feature remains
observable in our spectra),
we find broad and almost bimodal distributions in the 
relative galaxy fraction in $B_{J}$ and $K$.
Most sources
($\simgt70\%$) have galaxy
fractions $<0.3$ at the $3\sigma$ level in both
$B_{J}$ and $K$. 
The remainder have large galaxy contributions and are
predominately low redshift galaxies with strong 4000\AA$\,$ breaks. 
All of these latter sources are spatially extended and resolved 
on $B_{J}$ and $K$-band images.
In particular, there is a clear distinction in the strength of the
4000\AA$\,$ break for resolved and unresolved sources. 
\\\indent 2. Using these estimates, we find that the mean
$K$-band magnitude of the host galaxies of flat spectrum radio quasars
is consistent with that of extended radio galaxies at $z\simlt0.9$. 
This is consistent with the unified model for radio-loud
AGN.
\\\indent 3. By subtracting the galaxy contribution in each 
bandpass from the observed $B_{J}-K$ colours 
of Parkes sources, we find that at the $1\sigma$ confidence level, 
the relatively large spread in colours still 
remains. 
We conclude that in a majority of cases, the 
relatively red colours
must be due to a mechanism
other than that contributed by a ``red'' stellar component.
\chapter{A Search for Near-Infrared Polarisation} 

\vspace{1mm}
\leftskip=4cm
 
{\it ``For the truth of the conclusions of physical science, 
observation is the supreme Court of Appeal.''} 
 
\vspace{1mm}
 
\hfill {\bf --- Sir Arthur Eddington,}

\hfill {\bf\it The Philosophy of Physical Science, 1958}
 
\leftskip=0cm

\section{Introduction}

As explored in previous chapters, 
the two primary models used to explain the
redness observed in Parkes quasars are:
the synchrotron emission model and the dust model.
We can further distinguish between these two 
models using broadband polarimetry.
In the near-infrared,
both the strength and wavelength dependence of linear polarisation
should be different for each of these mechanisms. 

Linear polarisation is
an excellent signature of synchrotron emission (Pacholczyk, 1970).
If the reddening in Parkes quasars 
is due to a red synchrotron component
dominating the near-IR flux, then strong polarisation is expected. 
Observations of the wavelength dependence of polarisation in
a number of flat spectrum radio-quasars shows
that the polarisation generally decreases strongly with decreasing wavelength
from the near-IR to optical (eg. Wills \etal 1988; 1992). 
This can be explained if the synchrotron continuum
is diluted
by unpolarised light from the ``big blue bump'' emission
responsible for photoionizing emission line regions. 

This wavelength dependence has mostly been observed
in the optically brightest
and ``bluest'' radio-quasars
with $B\simlt16$ mag.
Observations by Wills \etal (1992) and
Impey \& Tapia (1990) find that about half of their flat-spectrum
radio sources are polarised in the optical with
$3\%\simlt p_{V}\simlt20\%$. 
Due to the faintness of many of the 
``reddest'' Parkes quasars in the optical however, 
little information on their
optical polarisation properties exists.
Most of these red sources however 
are relatively bright in the
near-IR ($K_{n}\simlt14$) where broadband polarimetry is possible.

As shown in Chapter 5, if the reddest Parkes quasars are to be
explained purely by the synchrotron model, then the spectra require particle
energy distributions that are nearly monoenergetic or step-like in nature.
Such particle distributions make a number of other predictions for the
wavelength dependence of polarisation that can also be tested.

For the dust reddening model, both the strength and 
wavelength dependence of linear polarisation
is expected to be 
different from that in the
synchrotron model. 
If the dust and physical conditions in reddened Parkes quasars is 
similar to that in the galactic ISM, then polarisations
from magnetically aligned dust grains
of order a few percent in $K$ are expected (eg. Whittet, 1992 and references
therein). 
As observed in the Galaxy, the linear polarisation shows a wavelength dependence
that increases with decreasing wavelength primarily in the near-IR.
Thus, the wavelength dependence of
polarisation in the synchrotron and dust model are
expected to be different in the near-IR. 
By measuring the linear polarisation in the near-IR, we may be able to 
unequivocally distinguish between these two models.

The observations presented in this chapter are designed to search for
linear polarisation in the near-infrared in a number of red quasars.
Our aim is to determine whether there are any sources with dust-like 
wavelength dependences in polarisation as well as exploring
the previous modelling of the polarisation dependence in (blue) radio-quasars. 

This chapter is organised as follows:
In sections~\ref{obs} and~\ref{datared}, 
we describe our observations and data reduction.
Results are presented in section~\ref{resdes} and compared with previous 
polarisation studies in the optical where available.
Possible evidence for variability and implications for the 
reddening mechanism are also discussed. 
In section~\ref{orip}, we briefly outline the predictions of 
a number of polarisation mechanisms and compare them with the data.
In section~\ref{twocm}, we analyse our 
results using fits of a two-component model
for the source emission.
This model also includes the possible effects of reddening by dust.
Implications and further predictions 
are discussed in section~\ref{discuss}, and all results are
summarised in section~\ref{concsev}.

\section{The Observations}
\label{obs}

Observations were made on the 4 nights of 27th Feb. - 2nd Mar. 1997
using the infrared imager and spectrometer (IRIS) mounted at the 
Cassegrain focus of the 3.9m Anglo Australian Telescope at Siding Spring.
IRIS was used in its polarimetry mode with the $f/15$ intermediate
field setup (the polarimeter was provided by Prof. J. Hough of the
University of Hertfordshire).
The detector is a hybrid array of $128\times128$ pixel format made
of mercury cadmium telluride. The image scale is $0.61''$ per pixel for the
$f/15$ intermediate field setup and corresponds to a field of view of
$\sim1.2'$. This setup was used to allow for good sampling in the 
moderate seeing conditions typical at Siding Spring. 
Before describing our broadband polarimetry measurements, we first
give a brief review of the method used to 
measure linear polarisation with the IRIS polarimeter.

\subsection{Measuring the Polarisation with IRIS}
\label{IRISobs}

\begin{figure}
\vspace{-0.8in}
\plotonesmall{0.6}{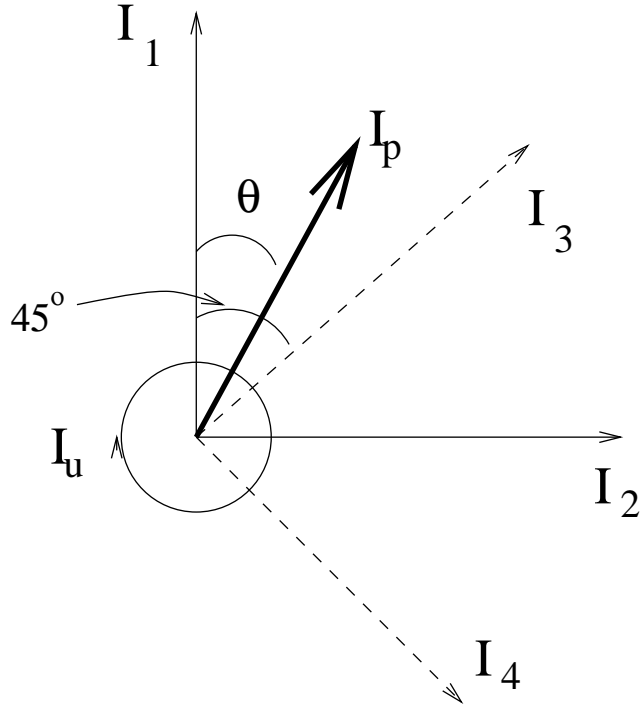}
\caption[Schematic of the components of the polarisation vector]{Schematic representation of the components of the polarisation
vector with intensity $I_{p}$. The sets of orthogonal components (images)
$I_{1},I_{2}$ and $I_{3},I_{4}$ are used to compute the two relevant
Stokes parameters $Q$ and $U$. $I_{u}$ is the unpolarised beam
which contributes equally in each component such that
the total intensity is $I_{TOT}=I_{p}+I_{u}$.}
\label{stokes}
\end{figure}

Our aim is to measure the degree of polarisation. This is achieved
by measuring the Stokes parameters, three of which are relevant for
linear polarisation. One of the parameters, $I$, is the total
intensity of the light, both polarised and unpolarised. 
The other two, normally called $U$ and $Q$, characterise the degree of
polarisation along two orientations, with the plane of polarisation
separated by $45^{\circ}$ (see below).
To measure either $U$ or $Q$ along a given angle on the sky, the intensity of
the light is measured through a polariser aligned along some direction
(giving intensity $I_{1}$) and the intensity 
measured with the polariser at right angles to this (giving $I_{2}$).
One Stokes parameter is then given by (say $U$):
\begin{equation}
U\,=\,\frac{I_{1} - I_{2}}{I_{1} + I_{2}},
\label{stokesU}
\end{equation}
where we have normalised with respect to the total intensity: $I=I_{1}+I_{2}$.
Likewise, the other Stokes parameter ($Q$) is defined analogously, but with
the plane of polarisation rotated by $45^{\circ}$ relative to its initial
orientation. Given orthogonal component intensities $I_{3}$ and $I_{4}$ 
in this new orientation, we have:
\begin{equation}
Q\,=\,\frac{I_{3} - I_{4}}{I_{3} + I_{4}}.
\label{stokesQ}
\end{equation}
See Fig.~\ref{stokes} for a schematic representation. 

Once the Stokes parameters are measured, the fractional polarisation $P$
and its orientation $\theta$ can be found.
Let us assume that
the incident light is the incoherent sum of an unpolarised beam of
intensity $I_{u}$ (which contributes equally to each orthogonal component
image; see Fig.~\ref{stokes}), and a perfectly linearly polarised beam
of intensity $I_{p}$ with the $E$ vector lying at an angle $\theta$
to the instrument's reference angle. It can be shown by simple
trigonometry from Fig.~\ref{stokes} that the Stokes parameters
(Eqns.~\ref{stokesU} and~\ref{stokesQ}) can be written:
\begin{equation}
U\,=\,\frac{I_{p}\cos{2\theta}}{I_{p} + 2I_{u}},
\label{stokestrig}
\end{equation}
$$
Q\,=\,\frac{I_{p}\sin{2\theta}}{I_{p} + 2I_{u}}.
$$
The fraction of the total light in the polarised component (the percentage
polarisation) is then defined:
\begin{equation}
P(\%)\,=\,100\frac{I_{p}}{I_{p} + 2I_{u}}\,\equiv\,100\sqrt{U^{2} + Q^{2}},
\label{ppercent}
\end{equation}
and its orientation is given by:
\begin{equation}
\theta\,=\,0.5\,{\rm Tan}^{-1}{\left(\frac{Q}{U}\right)}.
\label{theta}
\end{equation}

IRIS measures the two orthogonal polarisations (ie. $I_{1}$ and $I_{2}$ 
or $I_{3}$ and $I_{4}$ above) 
simultaneously
using a Wollaston prism made of ${\rm Mg}{\rm F}_{2}$ - a 
birefringent material. 
The structure of the material is such that the velocities of the light ray
will depend upon the polarisation of the radiation, and hence, two 
components polarised in different directions will be refracted by different 
amounts when emerging from the prism.
This produces two images on the same chip, one in the ``ordinary'' light
(ie. that polarised in one direction, say $I_{1}$ above), and the 
other in the ``extraordinary light'' (that polarised in the orthogonal
direction, ie. $I_{2}$). 
The light rays producing these two images are usually referred to as the
`{\it o}-ray' and `{\it e}-ray'.

In principal then, since the ``orthogonal'' intensities
$I_{1},I_{2}$ (Eqn.~\ref{stokesU}) or $I_{3},I_{4}$ (Eqn.~\ref{stokesQ})
can be measured simultaneously, a Stokes parameter ($U$ or $Q$) can
be measured from a single exposure.
In practice however, there will be some instrumental polarisation
originating primarily from inclined planar reflections in the optical
elements.
To take this out, the observing sequence on the IRIS polarimeter consists
by following each exposure with a second one with the plane of polarisation
rotated by $90^{\circ}$. What was the ``ordinary image'' (say $I_{1}$)
is now the ``extraordinary'' image ($I_{2}$), ie. the two will have flipped
places on the chip.  
This will allow any instrumental polarisation to be calibrated out and
shall be discussed in more detail in our reduction procedure in 
section~\ref{dataredins}. 

\begin{figure}
\vspace{-0.5in}
\plotonesmall{0.9}{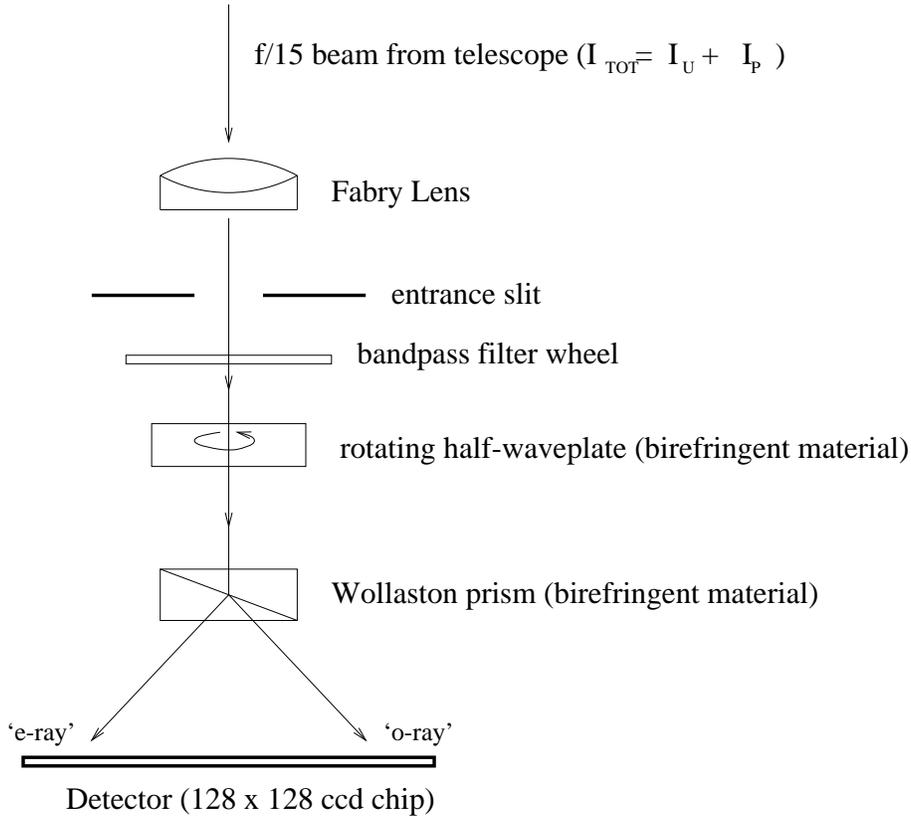}
\caption[Setup of optical components in the IRIS polarimeter]{Basic setup of optical components in the IRIS polarimeter.}
\label{setup}
\end{figure}

The plane of polarisation is
rotated by using an optical element known as a half-waveplate 
above the fixed Wollaston prism.
This device is also made of birefringent material and is rotated until the
{\it o}- and {\it e}-rays travel in the same direction. The emergent rays
will have altered phase delay relative to each other and hence with
the polarisation vector rotated.
A general plan of the major optical components in the IRIS polarimeter
is shown in Fig.~\ref{setup}.

Thus, a measure of the linear polarisation is obtained by taking 4 exposures,
each of which contains two orthogonal polarisations (images) on the same chip.
The first pair of exposures has the plane of polarisation separated by
$90^{\circ}$ and are combined to measure one of the Stokes parameters 
(see section~\ref{dataredins}). The second pair also contains polarised images
with the polarisation vector separated by $90^{\circ}$. This pair however
has the plane of polarisation displaced by $45^{\circ}$ relative to the 
first pair and are combined to measure the other Stokes parameter. 
Once the Stokes parameters are known, the
fractional polarisation and its orientation 
can be computed from Eqns.~\ref{ppercent} and~\ref{theta}. 

\subsection{Observational Strategy}

The seeing during our observing run ranged from $4''$ (FWHM) on the first
night to $0.8''-1.5''$ for the subsequent 3 nights. Observing
conditions were mainly photometric throughout. We obtained
polarimetric images
for 9 sources in total from our sample
of flat spectrum radio sources (Drinkwater \etal 1997).
In order to reasonably model the wavelength dependence of polarisation,
we obtained images in $J$, $H$ and $K_{n}$ for most of these sources.
Follow-up observations showed that three of the sources are 
identified as moderately red quasars (with optical continuum 
slopes $\alpha_{opt}\simgt1$ and colours $B_{J}-K_{n}\simgt4$),
three as `blue' quasars, and three as BL-Lac objects. 
A further discussion on individual source characteristics is given in 
section~\ref{resdes}. 
Contemporaneous measurements in the $J$, $H$ and $K_{n}$ bandpasses
was required to obtain a reliable 
polarisation spectrum, as the polarisation
in such sources may vary substantially with time (eg. Impey \& Tapia, 1990).

If the reddening in Parkes quasars is due to dust, then 
using empirical relations between extinction and polarisation from
dust in the galaxy (see section~\ref{poldust}), we predict for a 
mean observed reddening of $A_{V}\sim2$, a polarisation of order
a few percent in $K_{n}$. 
Due to the possibility of 
such relatively low levels, we initially sought to measure a
$1\%$ polarisation at $5\sigma$ confidence level. 
Since crudely, $\sigma_{p}\sim\sigma_{U}/I\,\sim\sigma_{Q}/I$
(where $I$ is the total intensity), we estimated that a 
$S/N\sim500$ in each of the individual 4 exposures (required for a
polarisation measure-see above) was required. 
However, the accuracy of our measurements is mainly limited by
photon statistics because of the faintness of our sources in the near-infrared.
For a quasar with $K_{n}\sim15$ (or $H,J\sim16$), this would have taken
$\sim3$ hours integration time.
Since the observations are sky limited ($S/N\propto c\sqrt{t}$, where $c$
are the counts and $t$ the exposure time), if the magnitude were
increased by one, the exposure time would be increased by a factor of 6.25. 
The reddest of our sources have typically $J-K_{n}\sim2$ and hence a red
source with $K_{n}\sim15$ implies more than 12 hrs integration in $J$.
Thus, to keep the integration times short and maximise the number of objects
that can be measured, we only observed red sources with $K_{n}<14$.

Our aim of a Poisson uncertainty of 0.2\% in a 1\% polarisation level
is unlikely to be limited by other sources of contaminant polarisation. 
An example may be 
polarisation
from transmission through aligned dust grains in the galaxy. 
All sources however
have galactic latitude $>20^{\circ}$ where 
measurements of the polarisation
towards distant stars suggest $p_{V}\simlt0.2\%$ 
(Markkanen, 1979) and hence $p_{K}\simlt0.1\%$ using the empirical polarisation
law.
Furthermore, the IRIS polarimeter is capable of an accuracy
$\sigma_{p}=\pm 0.01\%$, given at least $10^{8}$ photons.
The analysing efficiency is close to 100\% as demonstrated by inserting
a wire-grid (polariser) calibrator into the beam and as confirmed
using polarised standards.

As discussed above, 4 separate exposures corresponding to different
rotations of the half-waveplate are required to measure the two
relevant Stokes parameters and hence polarisation.
The observing sequence for each source and each band $J$, $H$ and $K_{n}$,
consisted of repeated sets of these 4 exposures with each set
corresponding to a raster position on the sky.
For each consecutive set of four exposures, the raster positions
were displaced by $\sim30''$ east to west on the CCD chip.
Each individual exposure consisted of effectively 60 second integrations
(formed from repeated integration cycles) at which
sufficient photons were accumulated.
The sets of rasters were formed close enough to each other in time
as possible so that 
uncertainties caused by variations in sky brightness and its polarisation
are minimised when the sets are combined to form `sky frames'. This
ensured 
that the final signal-to-noise in the `sky-subtracted' polarimetric
images was maximised (see section~\ref{datared}
for more details).

On the 3 nights at which the seeing was reasonably good (ie. $\sim1''$),
typical ``total'' integration times for our brightest sources
(with $K_{n}\simeq 11-12$ mag) were about 30 min in each band. 
For fainter sources with $K_{n}\simgt13$, total integration times
were of order 60-90 min.
Approximately $\sim120$ min was spent in $J$ for our faintest source
with $K_{n}\sim14.4$.
A log of our observations is shown in Table~\ref{obssum}. Total
numbers of rasters and 
exposures (each of effectively 60 sec. duration for all bands)
are shown.  
A more complete description
of the identification of each source from its optical spectrum is 
given in Table~\ref{proptab}. 
Column 4 in Table~\ref{obssum} 
lists the $K_{n}$ magnitudes as measured from previous runs
by P. Francis and M. Drinkwater (private communication). The
dates of observation for these magnitudes are given in parentheses.

\begin{table}
\vspace{-0.7in}
\leftskip=-0.6in
\begin{tabular}{|c|c|c|c|c|c|c|}
\hline
Source & RA (B1950) & Dec (B1950) & $K_{n}$ (Date) & 
\multicolumn{3}{|c|}{Exposure (min.)}\\ 
\hline
\hline
 & & & &\multicolumn{1}{|c|}{J} & \multicolumn{1}{|c|}{H} & \multicolumn{1}{|c|}{$K_n$} \\
\cline{5-7}
PKS0537$-$441 & 05 37 21.00 & $-$44 06 46.8 & 11.56 (13 Feb 95) & $14\times4$ & 
$14\times4$ & $14\times4$\\ 
\hline
PKS0622$-$441 & 06 22 02.68 & $-$44 11 23.0 & 14.34 (12 Feb 95) & - & - & $10\times4$\\ 
\hline
PKS0829+046 & 08 29 10.89 & +04 39 50.8 & 11.37 (25 Apr 94) & $7\times4$ & $7\times4$ &
$7\times4$\\
\hline
PKS1020$-$103 & 10 20 04.18 & $-$10 22 33.4 & 13.50 (13 Feb 95) & $14\times4$ &
$14\times4$ & $14\times4$\\ 
\hline
PKS1036$-$154 & 10 36 39.48 & $-$15 25 28.1 & 14.37 (3 Apr 94) & $21\times4$ &
$30\times4$ & $21\times4$\\ 
\hline
PKS1101$-$325 & 11 01 08.51 & $-$32 35 06.2 & 13.43 (12 Feb 95) & $14\times4$ & - &
$14\times4$\\ 
\hline
PKS1313$-$333 & 13 13 20.05 & $-$33 23 09.7 & 12.28 (25 Apr 94) & $7\times4$ &
$7\times4$ & $7\times4$\\ 
\hline
PKS1510$-$089 & 15 10 08.90 & $-$08 54 47.6 & 13.44 (13 Aug 95) & $14\times4$ & - &
$14\times4$\\
\hline
PKS1546+027 & 15 46 58.29 & +02 46 06.1 & 11.45 (12 Aug 95) & $14\times4$ & $14\times4$ &
$14\times4$\\ 
\hline
\end{tabular}
\caption[Summary of observations]{Summary of observations.}
\label{obssum}
\end{table}

\section{Data Reduction}
\label{datared}

The initial stages of the CCD reductions were carried out using the FIGARO
package which includes software specifically designed for the IRIS detector.
This comprised mainly of corrections for bias and dark current and
non-linearity between photon counts and output electrons.
The final stages were carried out using the IRAF package. The principal
tasks here were flat-field corrections, bad pixel corrections,
sky subtractions and integrated photometry.
Each of these steps is discussed in turn below.

\vspace{0.6cm}

\noindent
{\it BIAS AND DARK CURRENT:} We first subtracted the readout `bias level'
and `dark current' from each frame. The bias level is the offset value
which a CCD adds to a signal before it converts its output to digital values.
It allows the zero light level to be kept at a positive value, and avoids
feeding negative values to the A/D converter. 
For the IRIS detector, the bias counts (typically 25,000-30,000 counts/pixel)
are sufficiently uniform from pixel to pixel that they can be corrected for
by subtracting the `dark current' alone.
The dark current of a CCD, 
which includes the bias level, is due to an accumulation
of thermally created photons within the silicon multiplex layer that 
register as counts. This also includes stray photons that may be present in 
the IRIS dewar. Dark frames were taken with exposure times
identical to those of the individual data frames with the shutter closed. 
The dark current was estimated to be relatively low, which averaged
$\simlt0.1$cnts/pixel/sec. The noise associated with readout of the CCD
was also considerably low, being $<3$cnts rms over the entire chip. 

\vspace{0.6cm}

\noindent
{\it NON-LINEARITY:} The conversion of the numbers of photons falling onto
the CCD chip into output electron counts is not perfectly linear.
This particularly applies when the number of incident photons is relatively
high or nears saturation (typically 65,000 cnts/pixel for the IRIS chip).
Corrections for non-linearity in the IRIS chip involves computing correction
factors from polynomial fits to the `count vs. photon' relation for each
pixel.
This is built into the FIGARO software. 

\vspace{0.6cm}

\noindent
{\it FLAT FIELDS:} The sensitivity of a CCD chip is generally not uniform.
Differences from pixel-to-pixel most commonly arise from imperfections
introduced during the manufacturing process. 
These include variations in pixel sizes and quantum efficiency changes
across the chip.
To correct for these effects, we divided each data exposure by a similar
exposure taken through the same filter.
The flat field image must be taken on a uniformly emitting source.
Our flat fields were obtained by exposing the chip
on a uniformly illuminated section of the inside of the dome.
These `dome flats' were formed by subtracting flat-fields with the
dome lamp off from those with the lamp on. The reason for this was
so to correct for any 
thermal glow associated with the telescope, the
instrument, and the inside of the dome. 
In addition to the pixel sensitivity variations, the IRIS chip
has a considerable number of intrinsically ``bad'' pixels which were also
corrected for by interpolating over them.
Once the above corrections were applied, the frames were flat to better than
1\% of sky.

\vspace{0.6cm}

\noindent
{\it SKY SUBTRACTION:} In order to correct for uncertainties caused by
temporal variations in sky brightness, polarisation and 
instrumental `glow', and hence increase the final
signal-to-noise ratio in images, a sky frame needs to be formed and subtracted
from the data frames.
As discussed above, raster positions were formed for every consecutive
set of 4 exposures ($\sim4$ min), 
each of which corresponds to a different waveplate position as 
required for a polarisation measure.
To minimise variations in sky brightness, sky frames were formed by taking 
a median of  5 to 7 data frames corresponding to the same waveplate
position from the rasters. 
These sets of sky frames were then 
subtracted from the initial data frames (taken at approximately the
same time as the respective sky frames) corresponding to the
respective waveplate positions. This yielded a signal-to-noise ratio
$\sim5$ times greater than before sky subtraction. 

\vspace{0.6cm}

\noindent
{\it PHOTOMETRY:} Once data frames were sky subtracted, integrated counts
for each of the two images on each frame (ie. corresponding to the 
orthogonal `{\it e}' and
`{\it o}' polarisation states - see section~\ref{IRISobs}) were estimated
using normal aperture photometry.
For the fainter sources, 
the seeing conditions presented the major limitation,
and for these, the 
accuracy of the counts was improved by using
larger apertures on the sky. 
For each source then, we have 4 frames each containing 2 images. These
can now be used to calculate the polarisation and its position angle
as discussed in the following section. 

\subsection{Instrumental Effects and Measuring the Polarisation}
\label{dataredins}

\begin{figure}
\vspace{-0.8in}
\plotonesmall{0.8}{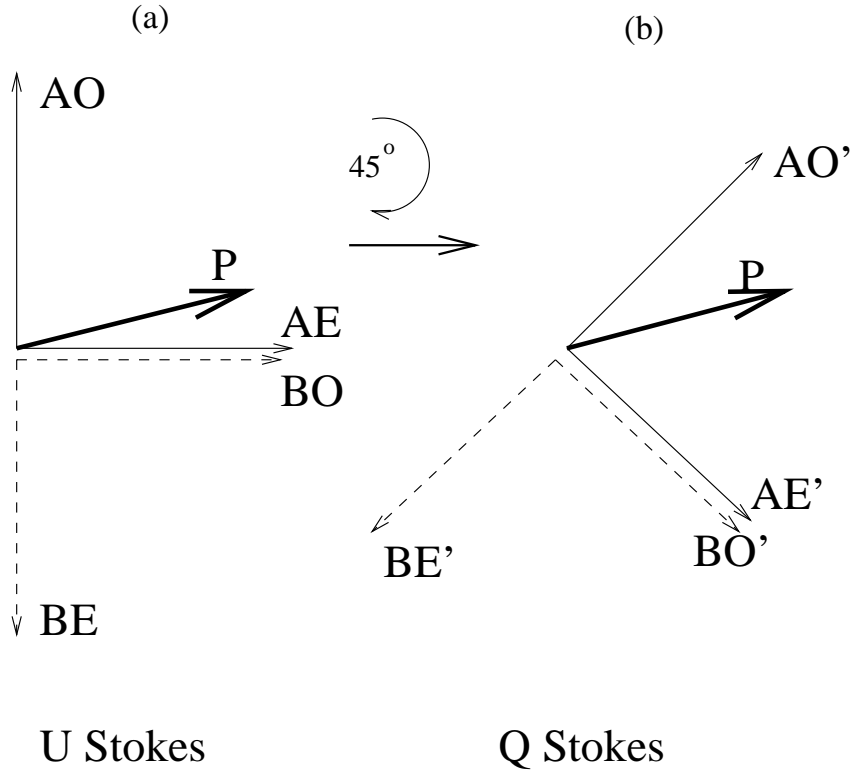}
\caption[Vectorial representation of polarimetric images]{Vectorial representation of the $4\times2$ images which correspond
to components of the polarisation vector {\bf P}.} 
\label{insrot}
\end{figure}

As discussed in section~\ref{IRISobs}, the Stokes parameters $U$ and $Q$
(Eqns.~\ref{stokesU} and~\ref{stokesQ}) contain all the information
needed to compute the fractional linear polarisation.
To compute each Stokes parameter, we have two frames, each taken
with the plane of polarisation rotated by $90^{\circ}$. As discussed, these
two frames are to be used to correct for any possible instrumental
polarisation.
For one of the Stokes parameters (say $U$), 
let us call these the $A$ and $B$ frames. Each frame contains two images,
the `ordinary' ($O$) and the `extraordinary' ($E$) image. 
The same applies for the 
second pair of frames used to compute the $Q$ Stokes parameter, 
whose plane of
polarisation is rotated through $45^{\circ}$ relative to the
first pair. To compute each Stokes parameter then,
we have 4 fluxes, let us call these: $AO$, $AE$, $BO$, and $BE$.
Figure ~\ref{insrot} shows a schematic representation.  

There are two algorithms that we can use for measuring the Stokes
parameters from these 4 fluxes
(J. Bailey; private communication), a `subtraction' algorithm and a `ratio'
algorithm.
The former is likely to give a slightly better signal-to-noise ratio
if conditions are fairly uniform (as it uses arithmetic rather than
geometric means for the latter).

For the first algorithm, the first step is to construct a correction factor
$R_{inst}$ for any instrumental polarisation.
This would show up as a systematic difference between the $O$ and $E$
images. If there was no such effect, $AE$ and $BO$ should be identical,
and $BE$ and $AO$ likewise. This correction factor is given by:
\begin{equation}
R_{inst}\,=\,\frac{AE + BE}{AO + BO}.
\label{Rinst}
\end{equation}
There is a second correction factor we can form to account
for possible variations in transparency
between the $A$ and $B$ frames.
If sky transparency fluctuations are negligible, the sum $AO+AE$ and
$BO+BE$ should be identical. The relevant correction factor can 
thus be written: 
\begin{equation}
R_{see}\,=\,\frac{AO+AE}{BO+BE}. 
\label{Rsee}
\end{equation} 
The Stokes parameter (say $U$; Eqn.~\ref{stokesU}) can be computed 
by averaging over the $A$ and $B$ frames ($I_{1}=(AO+BE)/2$ and
$I_{2}=(AE+BO)/2$), and modifying these by the above correction factors
so that:
\begin{equation}
U\,=\,\frac{R_{inst}AO + R_{see}BE - (AE + R_{inst}R_{see}BO)}
{R_{inst}AO + R_{see}BE + AE + R_{inst}R_{see}BO}.
\label{stokesUnew}
\end{equation}
The same applies for the second set of frames (Fig.~\ref{insrot}b)
to estimate $Q$.
If $R_{inst}$ and $R_{see}$ are stable during the night (or over a given set
of data frames), they can be measured to high precision, increasing
the final signal-to-noise.

The second algorithm uses a ratio technique.
Firstly, a ratio is calculated:
\begin{equation}
r\,=\,\left(\frac{AO}{AE}\frac{BE}{BO}\right)^{1/2},
\label{goem}
\end{equation}
which is the geometric mean of the ratio of image intensities along
two orthogonal polarisations.
Any one of the Stokes parameters (say $U$) is then given by: 
\begin{equation}
U\,=\,\frac{r-1}{r+1}.
\label{Ur}
\end{equation}
This ratio technique normalises out instrumental polarisation effects
as well as atmospheric transparency variations, though it is likely
to give a lower final signal-to-noise ratio.

We have computed the correction factors $R_{inst}$ and $R_{see}$ 
(Eqns.~\ref{Rinst} and~\ref{Rsee}) for each set of 
4 frames in each night and found that they 
were reasonably constant, varying by $\simlt3\%$
around unity for the brighter sources ($K_{n}\simeq11$) and
$\simlt7\%$ for the fainter sources ($K_{n}\simgt13$).
Thus we have used the first algorithm above to compute our Stokes parameters.
For the repeated sets of 4 frames we have for each source 
(in $J$, $H$ and $K_{n}$; see Table~\ref{obssum}), average values of the
correction factors and hence Stokes parameters (Eqn.~\ref{stokesUnew})
were computed.
Statistical errors in Stokes parameters were computed by finding the rms
uncertainty over a given set of frames, $\sigma_{rms}$, so that final
Poisson uncertainties in the mean values can be estimated: 
$\sigma_{p}\simeq\sigma_{rms}/\sqrt{N}$, where $N$ is the
number of (sets of 4 waveplate) observations.
Thus, by taking repeated sets of observations (ie. increasing $N$),
it can be seen that the final signal-to-noise ratio
can be significantly improved.

For comparison, we also computed the Stokes parameters and errors using the
`ratio technique' and found that signal-to-noise ratios were lower 
by up to 5\% than those found with the `subtraction' algorithm. 
Once the Stokes parameters and corresponding errors were known, the fractional
polarisation and its orientation angle was 
computed from Eqns.~\ref{ppercent} and~\ref{theta}. 
This orientation angle was converted into a 
position angle $\theta_{pa}$ on the sky
by using a wire-grid calibrator (or polariser) in the beam path to
calibrate the zero point and confirmed using polarised standards.

During the run, unpolarised standards were also taken so that
integrated photometric magnitudes in $J$, $H$ and $K_{n}$ can also be
determined. 
This was done by summing the counts in the two `orthogonal'
polarisation images on each frame and taking the mean over all frames.
All results are presented in the next section.

\section{Results}
\label{resdes}

\begin{figure}
\vspace{-2in}
\plotonesmall{1}{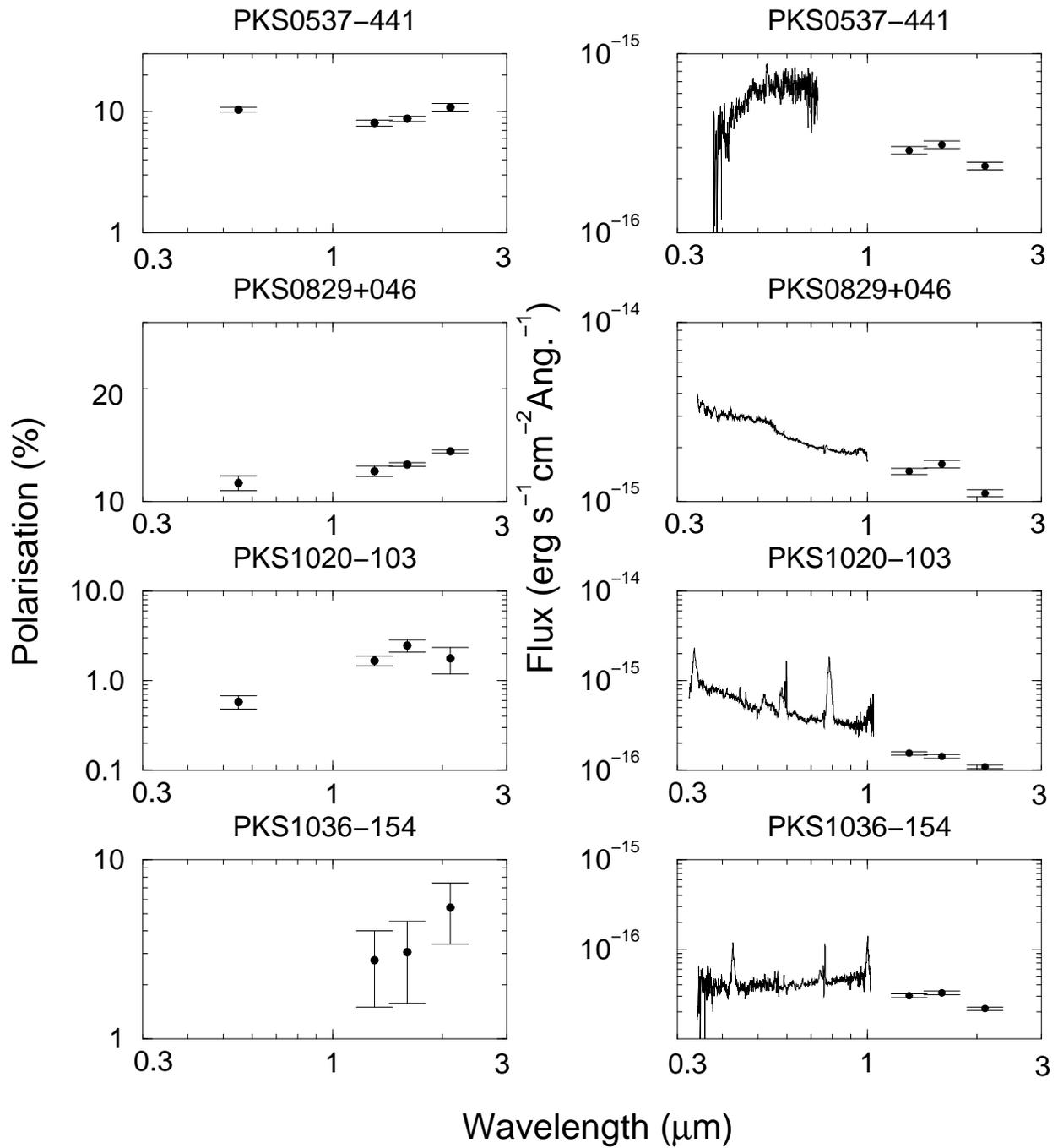}
\vspace{0.3in}
\caption[Observed wavelength dependence of percentage polarisation and
total flux spectra]{
Wavelength dependence of percentage polarisation (left) and total 
flux spectra (right).
Polarisations in $J$, $H$ and $K$ are from this study. 
The total flux spectra include extrapolated fluxes from 
$J$, $H$ and $K$ photometric measurements.} 
\label{spectra}
\end{figure}

A summary of the source properties
is given in Table ~\ref{proptab}. 
Sources have been divided into into 2 classes, classified
according to their optical spectral properties. Optical spectra where
available are shown in Fig.~\ref{spectra} and references are given in
Table~\ref{proptab}.
Two of the BL-Lacs (PKS0537$-$441 \& PKS0829+046) 
were initially selected on the basis of the absence
of emission lines and relatively high levels of polarisation in the optical 
as found by previous authors.
The quasars were all selected on the basis
of strong to intermediate
broad lines.
The sources PKS1313$-$333 and PKS1546+047 were identified as quasars in the
early 1970's (Burbidge \& Strittmatter, 1972) who claimed to detect
lines of intermediate strength. Neither however have
published spectra. To be certain of their classification, 
allowing for any possible variability over this period,
optical spectra were obtained
on the ANU 2.3m by P. Francis in April 1997. 
These observations revealed that the source PKS1313$-$333 now exhibits
BL-Lac characteristics, and hence, is listed under the BL-Lac
category in Table~\ref{proptab}. 

A description of the columns in Table~\ref{proptab}
is as follows:
Columns (1) to (4) list the source name, redshift, 
$B_{J}-K_{n}$ colour (from contemporaneous photometry; see below) 
and optical continuum slope, $\alpha_{opt}$, respectively (where 
$f_{\nu}\propto\nu^{-\alpha_{opt}}$). 
Column (5) gives a brief description of the relative strength of optical
emission lines for each source (see Fig.~\ref{spectra} for optical
spectra) and the relative amount of reddening as indicated by the
optical slope alone.
Balmer decrements (H$\alpha$/H$\beta$) where measurable are also shown.
Spectrum references are given in parentheses.

Our selection of objects for this observing run was based on 
$B_{J}-K_{n}$ colours derived from $B_{J}$ and $K_{n}$ measurements taken
at different epochs.
Contemporaneous photometry was obtained
by P. Francis in April 1997 on the ANU 2.3m.
Significant differences in broadband magnitudes were obtained, resulting
in revised colours for these sources.
From the new spectroscopic and photometric observations,
only three of the quasars (PKS0622$-$441, PKS1020$-$103 and PKS1036$-$154)
appear `moderately reddened' according to optical
continuum slope and $B_{J}-K_{n}$ colour.
None of these sources exhibit large
Balmer decrements indicative of significant reddening by
dust. The Balmer decrements for two of these red quasars however
appear somewhat larger than those in the three `bluer' quasars,
suggesting extinction may be important.

Table~\ref{contemp} shows our 
measured $J$, $H$ and $K_{n}$ magnitudes from this polarisation run.
For comparison, contemporaneous $J$, $H$ and $K_{n}$ measurements
from a later observing run in April 1997 are also shown.
RMS uncertainties based on noise statistics in these magnitudes 
are typically $\simlt\pm0.1$mag.
Combining the two sets of $K_{n}$-band measures in Table~\ref{contemp} 
with those found from previous runs (ie. those listed in Table~\ref{obssum}), 
we list in column (4) the maximum difference in $K_{n}$ magnitude
($K_{max}-K_{min}$) from measurements over these three epochs. 
See section~\ref{varia} for more details.

A summary of our measurements of the percentage linear polarisation
and position angle for $J$, $H$ and $K_{n}$ is given in Table~\ref{poltab}.
Column (2) lists the polarisations and associated rms uncertainties
based on Poisson noise statistics. Column (3) lists the corresponding
position angle in degrees with associated rms uncertainties.
In column (4), we list for comparison optical polarisation measures
(at an effective wavelength $\lambda=0.55\mu$m) where available from the
studies of Impey \& Tapia (1990) and Wills \etal (1992).
In Fig.~\ref{spectra} we show the dependence of our polarisation
measures with wavelength including cases where an optical measure
was available.
On the right of each plot we show the optical spectra together with
near-IR fluxes scaled from the $B$-band spectral flux
using contemporaneous $B$, $J$, $H$ and $K$
magnitudes.
The polarisation position angles ($\theta_{pa}$)
do not show any significant
wavelength dependence from $J$ to
$K_{n}$
for each source.
In most cases, a scatter of $\sim20\%$
about the mean value in $\theta_{pa}$ for the three bands
is observed.

\begin{table}
\vspace{-0.7in}
\begin{center}
\begin{tabular}{|c|c|c|c|p{1.6in}|}
\multicolumn{5}{c}{Quasars}\\ 
\hline
Source & $z$ & $B_{J}-K_{n}$ & $\alpha_{opt}$ & 
\multicolumn{1}{|c|}{Comments}\\
(1) & (2) & (3) & (4) & \multicolumn{1}{|c|}{(5)}\\ 
\hline
\hline
PKS0622$-$441 & 0.688 & 4.25 & 1.24 & 
`moderately red' with intermediate lines (W83)\\ 
\hline
PKS1020$-$103 & 0.196 & 3.92 & 1.01 & `moderately red' with strong lines,
H$\alpha$/H$\beta$=4.4 (D97)\\ 
\hline
PKS1036-154 & 0.525 & 4.91 & 2.27 & `red' with strong lines,
H$\alpha$/H$\beta$=4.6 (D97)\\
\hline
PKS1101$-$325 & 0.355 & 2.38 & 0.01 & `blue' with strong lines,
H$\alpha$/H$\beta$=3.9 (W83)\\
\hline
PKS1510$-$089 & 0.362 & 3.98 & $-0.08$ & `blue' with strong lines,
H$\alpha$/H$\beta$=2.5 (W83)\\
\hline 
PKS1546+027 & 0.415 & 4.42 & 0.13 & `blue' with strong lines, 
H$\alpha$/H$\beta$=2.1 (S97)\\
\hline
\end{tabular}

\vspace{0.5cm}

\begin{tabular}{|c|c|c|c|p{1.6in}|}
\multicolumn{5}{c}{BL-Lacs}\\
\hline
Source & $z$ & $B_{J}-K_{n}$ & $\alpha_{opt}$ &
\multicolumn{1}{|c|}{Comments}\\
\hline
\hline
PKS0537$-$441 & 0.893 & 4.92 & \hspace{3mm}2.68 & very weak lines (W86)\\
\hline
PKS0829+046 & - & 4.48 & \hspace{3mm}1.33 & no lines visible (D97)\\
\hline 
PKS1313$-$333 & 0.901 & 4.44 & \hspace{3mm} 0.40 & 
one weak line present (S97)\\ 
\hline
\end{tabular}
\caption[Properties of sources in polarimetry study]{Source properties; {\small REFERENCES}.---W83: Wilkes \etal (1983);
W86: Wilkes (1986); D97: Drinkwater \etal (1997); S97: spectra provided by
a run on the ANU 2.3m by P. Francis (April 1997).}
\label{proptab}
\end{center}
\end{table}

\begin{table}
\vspace{-1.3in}
\begin{center}
\begin{tabular}{|c|c|c|c||c|c|c|c|}
\multicolumn{8}{c}{Quasars}\\
\hline
Source &
\multicolumn{3}{|c||}{27 Feb.-2 Mar. 1997} &
\multicolumn{3}{|c|}{12-18 Apr. 1997} &
$\Delta K_{n}(max)$\\
(1) & \multicolumn{3}{|c||}{(2)} & \multicolumn{3}{|c|}{(3)} &
(4)\\
\hline
\hline
  &\multicolumn{1}{|c|}{J} & \multicolumn{1}{|c|}{H} & \multicolumn{1}{|c||}{$K_
{n}$}
& \multicolumn{1}{|c|}{J} & \multicolumn{1}{|c|}{H} & \multicolumn{1}{|c|}{$K_{n
}$}
& \\
\cline{2-7}
PKS0622$-$441 & - & - & 14.15 & - & - &  - & 0.2\\
PKS1020$-$103 & 14.96 & 14.27 & 13.33 & 14.85 &
14.11 & 13.27 & 0.23\\
PKS1036$-$154 & 17.37 & 16.44 & 15.79 & 17.23 &
16.32 & 15.63 & 1.42\\
PKS1101$-$325 & 14.34 & - & 13.11 & 14.55 & 13.93 & 13.16 &
0.32 \\
PKS1510$-$089 & 14.55 & - & 12.87 & 15.24 & 14.09 & 13.27 &
0.57 \\
PKS1546+027 & 14.73 & 13.61 & 12.82 & 14.97 &
13.88  & 13.02 & 1.57\\
\hline
\end{tabular}

\vspace{0.5cm}

\begin{tabular}{|c|c|c|c||c|c|c|c|}
\multicolumn{8}{c}{BL-Lacs}\\
\hline
Source &
\multicolumn{3}{|c||}{27 Feb.-2 Mar. 1997} &
\multicolumn{3}{|c|}{12-18 Apr. 1997} &
$\Delta K_{n}(max)$\\
\hline
\hline
  &\multicolumn{1}{|c|}{J} & \multicolumn{1}{|c|}{H} & \multicolumn{1}{|c||}{$K_
{n}$}
& \multicolumn{1}{|c|}{J} & \multicolumn{1}{|c|}{H} & \multicolumn{1}{|c|}{$K_{n
}$}
& \\
\cline{2-7}
PKS0537$-$441 & 15.68 & 14.85 & 14.03 & 14.76 &
13.85 & 13.02 & 2.47\\
PKS0829+046 & 13.41 & 12.61 & 11.84 & 13.45 &
12.52 & 11.80 & 0.47\\
PKS1313$-$333 & 14.62 & 13.78 & 12.95 & 15.46 &
14.36 & 13.69 & 1.41\\
\hline
\end{tabular}
\caption[Contemporaneous photometry of sources in polarimetry study]{Contemporaneous photometry. Combined with
$K_{n}$ magnitudes from earlier runs (Table~\ref{obssum}), 
column (4) lists the
maximum difference in $K_{n}$ magnitude from these three epochs.}
\label{contemp}
\end{center}
\end{table}

\begin{table}
\vspace{-0.8in}
\leftskip=-0.8in
\begin{tabular}{|c|c|c|c||c|c|c|c|}
\multicolumn{8}{c}{Quasars}\\
\hline
Source &
\multicolumn{3}{|c||}{$p\pm\sigma(p)$ (\%)} &
\multicolumn{3}{|c|}{$\theta_{pa}$ (deg)} & 
$p_{V}$ (\%)\\
(1) & \multicolumn{3}{|c||}{(2)} & \multicolumn{3}{|c|}{(3)} &
(4)\\
\hline
\hline
  &\multicolumn{1}{|c|}{J} & \multicolumn{1}{|c|}{H} & \multicolumn{1}{|c||}{$K_{n}$} 
& \multicolumn{1}{|c|}{J} & \multicolumn{1}{|c|}{H} & \multicolumn{1}{|c|}{$K_{n}$}
& \\
\cline{2-7}
PKS0622$-$441 & - & - & $1.4\pm0.9$ & - & - & $3.2^{+1.0}_{-0.9}$ & -\\
PKS1020$-$103 & $1.7\pm0.2$ & $2.5\pm0.4$ & $1.8\pm0.6$ & $42.5^{+7.3}_{-8.3}$ &
$35.0^{+6.2}_{-6.9}$ & $51.8^{+7.1}_{-8.1}$ & 0.58 (W92) \\
PKS1036$-$154 & $2.8\pm1.3$ & $3.1\pm1.5$ & $5.4\pm2.0$ & $56.5^{+6.1}_{-5.2}$ &
$70.6^{+8.8}_{-9.1}$ & $60.9^{+7.9}_{-8.2}$ & - \\
PKS1101$-$325 & $0.2\pm0.3$ & - & $0.3\pm0.3$ & $87^{+14.5}_{-11.1}$ & - 
& $89^{+10.1}_{-9.5}$ &
- \\
PKS1510$-$089 & $3.6\pm0.2$ & - & $2.7\pm0.2$ & $58^{+6.0}_{-5.2}$ & - 
& $61^{+8.1}_{-7.2}$ &
$1.9\pm0.4$ (I90) \\
PKS1546+027 & $6.2\pm0.2$ & $8.5\pm0.2$ & $8.6\pm0.2$ & $5.0^{+1.1}_{-0.8}$ & 
$3.7^{+0.9}_{-0.9}$ & $3.7^{+0.7}_{-0.6}$ & 3.4 (W92)\\
\hline
\end{tabular}

\vspace{0.5cm}

\begin{tabular}{|c|c|c|c||c|c|c|c|}
\multicolumn{8}{c}{BL-Lacs}\\
\hline
Source &
\multicolumn{3}{|c||}{$p\pm\sigma(p)$ (\%)} &
\multicolumn{3}{|c|}{$\theta_{pa}\pm\sigma(\theta_{pa})$ (deg)} &
$p_{V}$ (\%)\\
\hline
\hline
  &\multicolumn{1}{|c|}{J} & \multicolumn{1}{|c|}{H} & \multicolumn{1}{|c||}{$K_{n}$}
& \multicolumn{1}{|c|}{J} & \multicolumn{1}{|c|}{H} & \multicolumn{1}{|c|}{$K_{n}$}
& \\
\cline{2-7}
PKS0537$-$441 & $8.0\pm0.5$ & $8.7\pm0.4$ & $10.8\pm0.8$ & $31.8^{+5.5}_{-5.1}$ &
$34.4^{+6.3}_{-5.1}$ & $28.8^{+4.1}_{-5.2}$ & $10.4\pm0.5$ (I90)\\
PKS0829+046 & $12.0\pm0.4$ & $12.5\pm0.2$ & $13.6\pm0.1$ & 
$67\pm^{+7.1}_{-6.2}$ &
$69.0^{+7.1}_{-7.2}$ & $69.0^{+6.9}_{-7.2}$ &11.2 (W92)\\
PKS1313$-$333 & $12.7\pm0.3$ & $12.9\pm0.5$ & $13.2\pm0.5$ & $3.5^{+0.6}_{-0.5}$ &
$2.0^{+0.3}_{-0.3}$ & $2.6^{+0.4}_{-0.4}$ & - \\
\hline
\end{tabular}
\caption[Polarimetry results]{Polarimetry results; REFERENCES.---I90: Impey \& Tapia (1990);
W92: Wills \etal (1992).}
\label{poltab}
\end{table}

\begin{figure}
\vspace{-2in}
\plotonesmall{1}{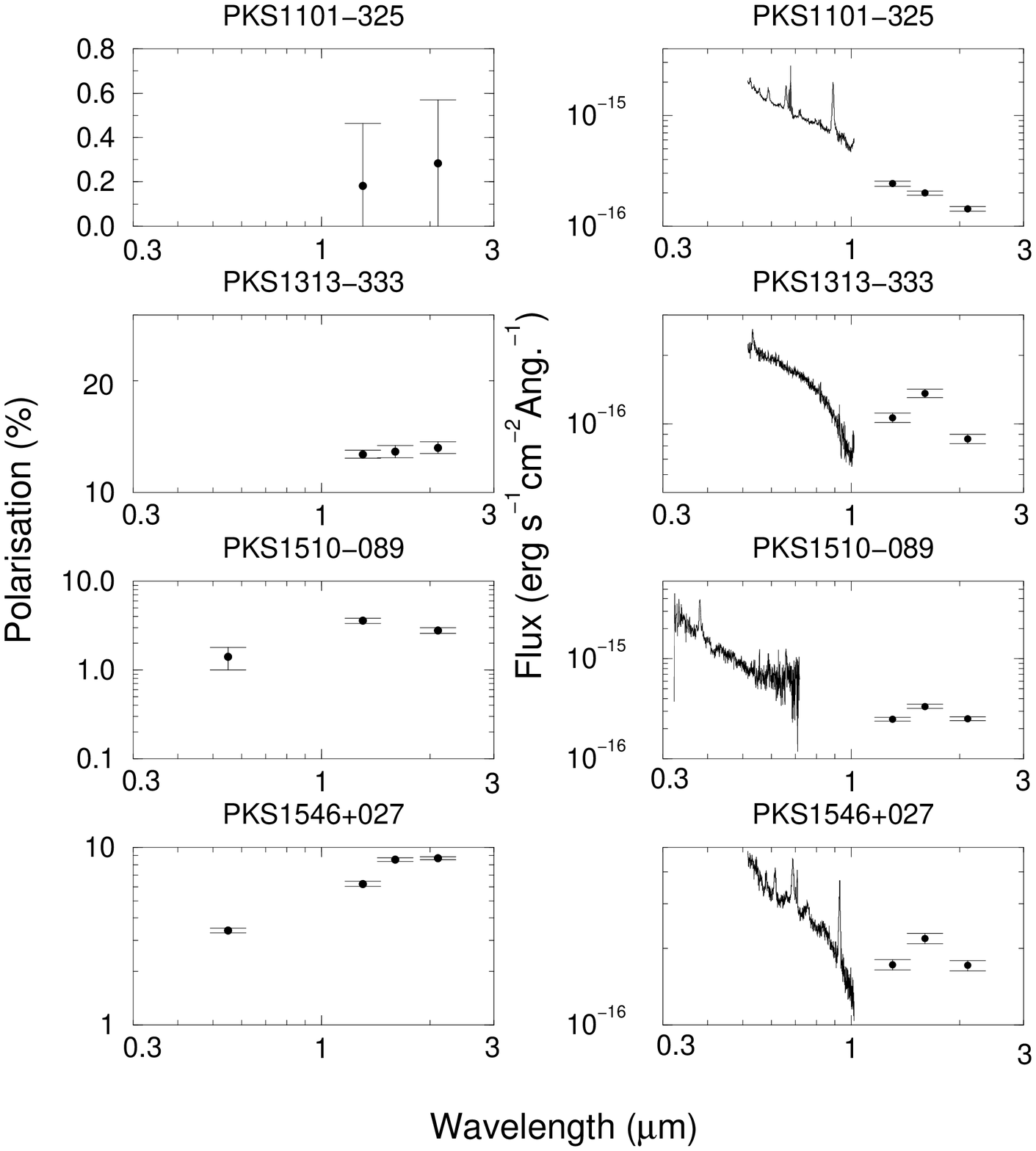}
\vspace{0.5in}
Fig.~\ref{spectra} continued.
\end{figure}

\newpage
\subsection{Polarisation versus Variability and $B-K$ Colour?}
\label{varia}

\begin{figure}
\vspace{-3.5in}
\plotonesmall{1}{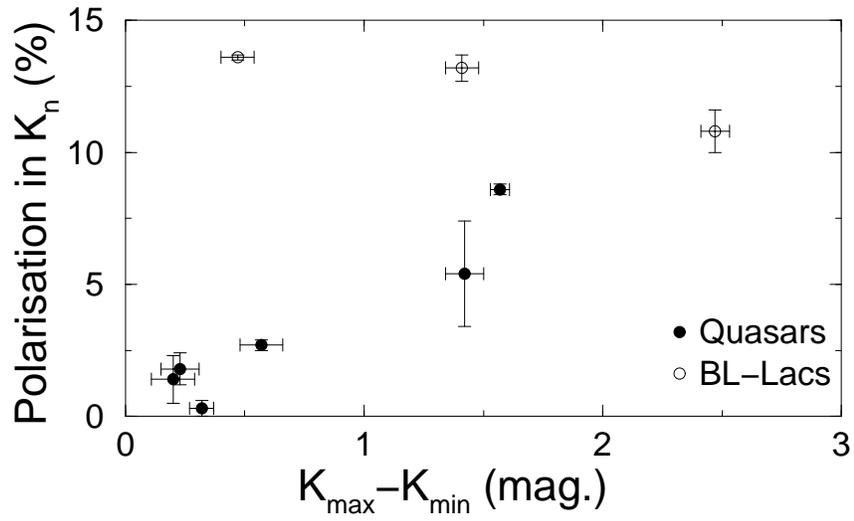}
\vspace{-2.5in}
\caption[$K_{n}$ polarisation versus variability in $K_{n}$ flux]{Polarisation at $K_{n}$ ($\lambda\simeq2.1\mu$m)
as a function of the difference in maximum and minimum $K_{n}$ magnitudes
from measurements over three epochs (see section~\ref{varia}).}
\label{var}
\end{figure}

A comparison of the maximum difference in $K_{n}$ magnitudes measured from the
three epochs
(see column (4) in Table~\ref{contemp}) 
with $K_{n}$ polarimetry measures for our six quasars
shows a correlation between the two.
This is shown in Fig.~\ref{var} (filled  circles) and is
significant at the $99\%$ confidence level.
The three BL-Lacs however (open circles) are all highly polarised
with $p_{K}>10\%$ and two show changes of $>1$ mag
in $K_{n}$-band flux over a timescale of $\sim2$ years.

The significant changes in $K_{n}$ flux for the highly polarised sources
suggests variability in the source emission. 
If so, then a correlation with the degree of polarisation can be explained
if the near-IR is dominated by ``beamed'' synchrotron emission,
possibly from a relativistic jet. This mechanism both enhances
the amplitude and decreases the timescale of flux variations
in an observer's frame.
This is the standard model used to interpret the 
properties of highly variable AGN at optical and radio wavelengths, ie. 
the classical 
blazars and BL-Lac-type objects (eg. K\"{o}nigl, 1981), and is indeed consistent
with observations of our three BL-Lacs in the near-infrared.
Fig.~\ref{var} indicates that there are also sources with 
normal quasar-like properties in the optical but 
BL-Lac-like properties in the near-infrared. 
Relatively high
near-IR polarisation levels in quasars which are `normal' in the
optical were also
observed by Brindle \etal (1986), Smith \etal (1986) and Wills \etal (1988).

\begin{figure}
\vspace{-2.2in}
\plotonesmall{1}{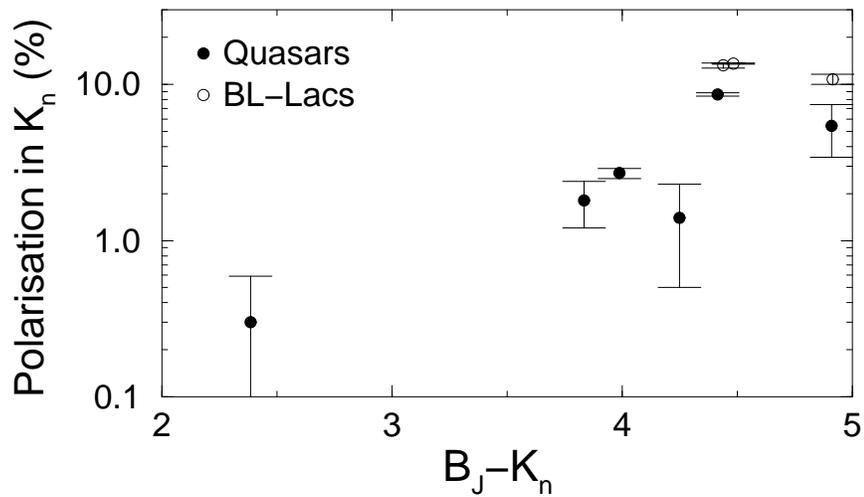}
\vspace{-3in}
\caption[$K_{n}$-band polarisation as a function of $B_{J}-K_{n}$ colour]{$K_{n}$-band polarisation
as a function of $B_{J}-K_{n}$ colour.} 
\label{BKpol}
\end{figure}

The polarisations (and hence changes in $K_{n}$ flux) appear
to have a bimodal distribution with respect to $B_{J}-K$ colour. 
The four bluest quasars (with $B_{J}-K_{n}$\simlt4.4) in our sample 
are also those which exhibit the
lowest polarisations as compared to the redder quasars and BL-Lacs with  
$p_{K}\simgt5\%$. The $K_{n}$ polarisations are shown as a function of
$B_{J}-K_{n}$ colour in Fig.~\ref{BKpol}. The relatively
high near-IR polarisations (and variability) in the red quasars 
suggests that their colours
may be associated with an
intrinsically red synchrotron
component dominating the near-IR flux. 
Implications of this result on models
for the source emission and other reddening
mechanisms will be discussed in section~\ref{discuss} . 

\newpage

\section{Origin of the Polarisation?}
\label{orip}

The reddening models explored in the previous chapters make a number
of specific predictions for both the strength and wavelength 
dependence of polarisation.
In this section, we outline these predictions and compare them with
the data. We shall primarily discuss three possible mechanisms that
can give rise to a wavelength dependent polarisation. 
The first is polarisation by transmission through aligned dust grains.
The second and third mechanisms are implied by the synchrotron 
reddening model. These are respectively: polarisation from a `pure'
synchrotron radiation spectrum and, a synchrotron spectrum
modified by `unpolarised' components which we refer to as the `two
component model'.
These are discussed in turn below.

\subsection{Polarisation from Dust} 
\label{poldust}

If dust is contributing to the reddening 
observed in Parkes quasars, then we need to assume 
a model for the grains and physical conditions in order to explore
the polarisation properties expected from the dust. 
As a first working hypothesis,
we assume that the extinction and polarisation properties
of the grains  
are the same as those derived from observations in the galaxy. 

Transmission of light through aligned dust grains in the interstellar medium
is responsible for the linear polarisation of starlight in our galaxy
and other galaxies (Elvius 1978; Hough \etal 1987; Clayton, Martin \&
Thompson, 1983).
The accepted model is `linear dichroism' or directional extinction
resulting from the presence of non-spherical grains aligned by the
galactic magnetic field.
In the galaxy, the degree of polarisation in the visual passband 
($\lambda\simeq5500$\AA) shows a correlation with the amount of
reddening lying in the range: 
\begin{equation}
P_{V}\,\simeq\,(2.3-9)E_{B-V}\,\,\%\,{\rm mag}^{-1}, 
\label{pvred}
\end{equation}
or in terms of extinction:
\begin{equation}
P_{V}\,\simeq\,(0.6-2.2)A_{V}\,\,\%\,{\rm mag}^{-1}
\label{pvext}
\end{equation}
(Serkowski \etal 1975).
The most efficient polarising medium is one where the grains can be
modelled as `infinite cylinders' (ie. length$\gg$radius) 
with diameters comparable to the wavelength,
and their long axes parallel to each other. From Mie theory computations,
such a model predicts a theoretical upper limit on the polarisation of
\begin{equation}
P_{V}\,\simlt\,13.8A_{V}\,\,\%\,{\rm mag}^{-1}
\label{pvredul}
\end{equation}
(Whittet \etal 1992 and references therein).

\begin{figure}
\vspace{-2.7in}
\plotonesmall{1}{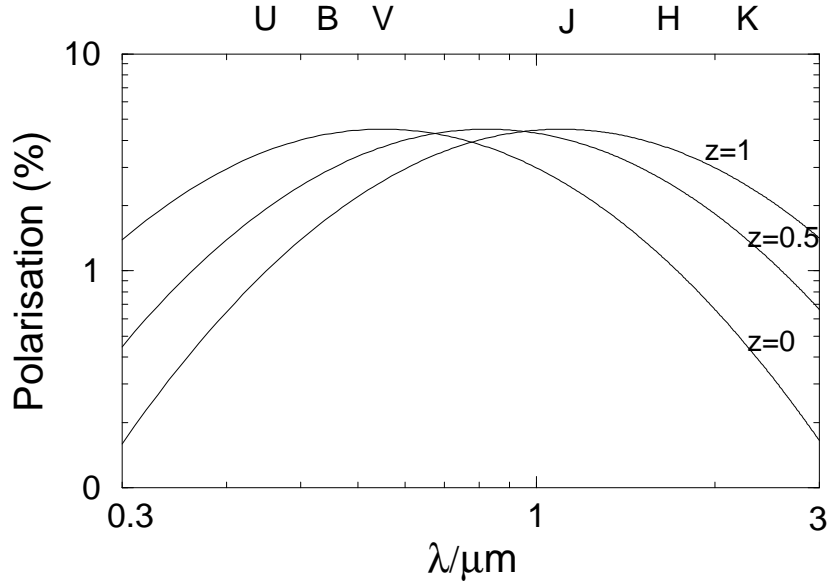}
\vspace{-2.5in}
\caption[Predicted wavelength dependence of polarisation from dust]{
Wavelength dependence of polarisation as observed in diffuse
regions of the galaxy for three redshifts of the dust. 
The polarisation levels are for optimum grain alignment,
deduced from the polarisation-reddening relation (Eqn.~\ref{pvext})
with $A_{V}=2$ - a mean value inferred from the reddest Parkes quasars. 
}
\label{dusteg}
\end{figure}

The wavelength dependence of linear polarisation from stars in the galaxy
from the $K$ to $U$ passband has been described by the following
empirical formula (Wilking \etal 1982; Whittet \etal 1992):
\begin{equation} 
p(\lambda)\,=\,p_{max}\exp{\left[-C\ln^{2}{\left(\frac{\lambda_{max}(1+z)}{\lambda}
\right)}\right]}, 
\label{emp}
\end{equation}
where $p_{max}$ is the maximum polarisation occurring at $\lambda_{max}$
and $C$ an inverse measure of the width of the polarisation curve.
Observations show that $C$ and $\lambda_{max}$ are linearly related:
\begin{equation}
C\,=\,-0.1+1.86\lambda_{max}.
\label{clmax}
\end{equation}
As discussed below, $\lambda_{max}$ is related to the average size
of the polarising grains and lies in the range:
\begin{equation}
0.4\mu{\rm m}\simlt\lambda_{max}\simlt0.8\mu{\rm m}, 
\label{lmaxrange}
\end{equation}
with a mean value $0.55\mu$m for observations in the diffuse ISM. 
We have included the redshift factor $(1+z)$ in Eqn.~\ref{emp} 
to account for the
decrease in rest frame wavelength with source redshift.
With $\lambda_{max}=0.55\mu$m, and adopting the upper bound
$p_{max}=9E_{B-V}\%$ with a mean reddening $E_{B-V}\simeq0.5$ 
($A_{V}\simeq2$), as inferred from the reddest Parkes quasars, Fig.~\ref{dusteg}
shows Eqn.~\ref{emp} for three redshifts.

The polarisation is seen to increase monotonically from $K$ to $V$-band 
wavelengths at the lowest redshifts, 
and this dependence remains strongest in the near-IR for
dusty sources up to moderate redshifts; $z\simlt 1$. 
Within our interested wavelength range, ie. $J$ to $K$, the
polarisation is predicted to decrease by as much as a factor of 3. 
The fact that this characteristic decrease in $p$ with $\lambda$
only occurs at the longest wavelengths ($\lambda>0.55\mu$m), is the
crucial distinguishing feature of this model.

\subsubsection{Comparison with Observations}

As discussed above (see Table~\ref{proptab}), we have three quasars in
our sample whose optical continuum slopes appear moderately redder
(with $\alpha_{opt}\simgt1$) than those of the three bluer quasars
having $\alpha_{opt}\simlt0.1$. 
Although the Balmer decrements available for two of the `red' quasars 
do not indicate significant amounts of dust
along the line-of-sight, they are nonetheless consistent with their
optical depths (see below) deduced from the slopes if we assume
the intrinsic lower limit $H\alpha/H\beta\simeq4$ (as 
observed in the `bluest' radio-quasars where 
$H\alpha/H\beta=4.8\pm0.9$; see section~\ref{Balmpks}).
These Balmer decrements are also on average larger than those of the
bluer quasars in Table~\ref{proptab}. 

If the `moderate' optical 
reddening in these sources were due to dust similar to that 
in diffuse regions of the galaxy, 
then what polarisation levels should we expect in the near-infrared? 
Assuming an `unreddened' optical continuum slope of 
$\alpha_{opt}\simeq0.3$ ($f_{\nu}\propto\nu^{-\alpha}$), which is a 
mean value observed for optically selected quasars (Francis, 1996),
we can find approximate values for the extinctions, $A_{V}$ for our 
three quasars classified as moderately reddened in Table~\ref{proptab}.
From the optical continuum slopes, we find that in the {\it rest frame},
$A_{V}\simeq0.6,\,0.7$ and 1.4 mag respectively for the sources:
PKS0622$-$441, PKS1020$-$103 and PKS1036$-$154. 
Using the upper bound in the interstellar polarisation-reddening relation
(Eqn.~\ref{pvext}) to represent $p_{max}$
in Eqn.~\ref{emp}, we predict from Eqn.~\ref{emp} with 
$\lambda_{max}=0.55\mu$m and $z=0.5$ that
\begin{equation}
P_{K}\,\simlt\,0.8A_{V}\,\,\%\,{\rm mag}^{-1}.
\label{pk}
\end{equation}
For these sources specifically, 
upper limits in the $K$-polarisations 
predicted for the above extinction measures
in an {\it observer's} frame are respectively: 
$P_{K}\simlt$0.7\%, 0.5\%, and 1.3\%.
As compared to the values observed: $P_{K}=1.4\pm0.9\%$, 
$1.8\pm0.6\%$ and $5.4\pm2.0\%$,
in only one case does the observed polarisation 
marginally agree within observational
errors with the maximum value predicted for galactic-dust. 
The reddening observed in the other two sources is incompatible with
their observed polarisation levels if due to dust like that in the local 
diffuse ISM. 

Considerably greater grain alignment would be required for the
observed polarisations in these sources 
to be due to transmission through aligned dust grains.
The polarisation levels in $K$ are however consistent with the
theoretical upper limit (Eqn.~\ref{pvredul}) for perfectly aligned grains 
modelled as `ideal' infinite cylinders. If such grains are assumed
to have diameter:  
$a\sim\lambda_{max}=0.55\mu$m and dielectric
properties like those of in the galactic ISM, 
then the predicted wavelength dependence
of polarisation (Greenberg, 1968)
can be approximated using the galactic law (Eqn.~\ref{emp}). 
Using Eqn.~\ref{emp} we find that this
model predicts:
\begin{equation}
P_{K}\,\simlt\,5A_{V}\,\,\%\,{\rm mag}^{-1}.
\label{pktheory}
\end{equation}
Table 7.5 summarises the predictions from Eqns.~\ref{pk}
and~\ref{pktheory} for these three quasars.
As one may expect, if the dust were located near the quasar central engine,
the magnetic fields are likely to be strong, giving optimum grain
alignment and hence high polarisation levels. The assumption of
grain shapes in the form of `infinite cylinders' on which 
Eqn.~\ref{pktheory} is based however is an
oversimplification and implausible.
It is unlikely that transmission through aligned dust grains
can fully explain the observed polarisation levels.

\begin{table}
\vspace{-0.3in}
\begin{center}
\begin{tabular}{|l|c|c|c|c|}
\hline
Source & $A_{V}$ (mag) & $P_{K}$({\it obs}) (\%)& $P_{K}$({\it gal})& 
$P_{K}$({\it theory})\\ 
\hline
\hline
PKS0622$-$441 & 0.6 & $1.4\pm0.9$ & $<0.7$ & $<3$\\
PKS1020$-$103 & 0.7 & $1.8\pm0.6$ & $<0.5$ & $<3.5$\\
PKS1036$-$154 & 1.4 & $5.4\pm2.0$ & $<1.3$ & $<7$\\
\hline
\end{tabular}
\caption[Observed and dust model predictions for polarisation in 
three `moderately reddened'
quasars]{Comparisons of $K$-polarisations in three `moderately reddened'
quasars, $P_{K}$({\it obs}),
with two dust model predictions: maximum value from measurements
of local-galactic dust, 
$P_{K}$({\it gal}) (Eqn.~\ref{pk}), and theoretical upper limit for 
perfectly aligned `ideal' infinite cylinders, 
$P_{K}$({\it theory}) (Eqn.~\ref{pktheory}).} 
\end{center}
\label{dustsimp}
\end{table}

It is immediately evident that in two of our moderately reddened quasars 
(PKS1020$-$103 and PKS1036$-$154) for 
which a polarisation spectrum could be deduced (Fig.~\ref{spectra}), the
wavelength dependence of polarisation cannot be fitted by the
galactic-dust polarisation curve (Fig.~\ref{dusteg}) in the near-infrared. 
If an underlying dust-like polarisation spectrum were present
and superimposed on a polarisation spectrum from an independent
mechanism (say synchrotron-see next section), then the 
red quasars in our sample should systematically show a flatter polarisation
curve on average than the bluer quasars. 
A comparison of the average polarisation slopes ($dP/d\lambda$)
in these two `red' quasars with the three `blue' quasars however, shows that
there is no significant difference.

If line-of-sight dust is contributing to the optical reddening
observed in our  
three `moderately reddened' quasars,
then the polarisation data does not imply a total
failure for this model.
The polarisation
observed from transmission through aligned dust grains in the galaxy
is known to critically depend on a number of factors. 
The observed polarisation crucially depends on the
orientation of the aligned grains and hence magnetic field relative 
to the observer. Since on average the magnetic field in the galaxy
is parallel to the galactic plane (Heiles, 1987), 
grains will be aligned predominately in the plane of the line-of-sight,
giving us an 
almost perfect geometry with which to observe 
interstellar polarisation through the galactic plane. 
Furthermore, theoretical models show that the grain alignment
sensitively depends on the size, temperature and composition of the grains
(Mathis \etal 1986), as well as physical conditions of the 
associated gas component and nature of the radiation field 
(Whittet, 1992 and references therein). 
The presence of random or turbulent small scale magnetic fields
along the line-of-sight can 
also significantly destroy uniformity
in grain alignment on large scales.
There is no doubt that the physical conditions
in powerful radio-quasars 
are considerably different from those in the galaxy, and hence
the presence of dust may not necessarily give a polarisation spectrum
similar to that observed in the galactic ISM.

Mie theory computations show that the wavelength $\lambda_{max}$, corresponding
to the maximum polarisation $p_{max}$ in Eqn.~\ref{emp} is related to the
average size of the polarising grains (Greenberg, 1968).
Dielectric cylinders of radius $a$ and refractive index $n$ are shown to
polarise most efficiently when $2\pi a(n-1)/\lambda\sim1$ and thus,
$\lambda_{max}\sim2\pi a(n-1)$. 
Thus, the value $\lambda_{max}\simeq0.55\mu$m
as observed in the diffuse galactic ISM requires that
$\langle a\rangle\simeq0.15\mu$m for n=1.6 (typical of silicates).
This is consistent with that predicted from the
grain size distribution used to model the extinction curve.
If the observed decrease in polarisation towards short wavelengths in our
two red quasars were due to dust, then $\lambda_{max}$ would have to be
at wavelengths $>2\mu$m in the rest frame. 
This is so that the region with $dP/d\lambda>0$ in the dust polarisation curve
(Fig.~\ref{dusteg}) falls in the $J$, $H$ and $K$ passbands.
A wavelength of $\lambda_{max}>2\mu$m would correspond to effective
grain sizes $\sim3.5$ times larger than those in the diffuse galactic ISM
(for the same composition and geometry).
Grain sizes this large however will cause very little optical-UV
continuum reddening, contrary to that observed. 
Using the empirical relation, 
$R_{V}=A_{V}/E(B-V)\simeq5.6\lambda_{max}$ for diffuse dust in the galaxy
(Whittet \& van Breda, 1978), we find that
$\lambda_{max}>2\mu$m requires $R_{V}>10$ or $A_{B}/A_{V}<1.1$.
The optical continuum reddening in Parkes quasars implies 
$A_{B}/A_{V}\simgt1.3$, consistent with a $1/\lambda$ or galactic-type
extinction law. Thus, such large grains cannot be responsible for
both the reddening and polarisation observed. 

On the basis of this very small sample of `reddened' quasars,
we conclude that absorption by aligned grains cannot explain their
observed polarisation.
If the observed reddening is to be attributed to dust, then the grains
and/or physical conditions must be quite different from those found
in the interstellar medium of our and other nearby galaxies.
An alternative polarisation mechanism is required.

\subsection{Polarisation from the Synchrotron Mechanism}
\label{synchmod}

\subsubsection{`Pure' Synchrotron Emission} 

Polarisation is a well known property of the synchrotron process, being
primarily due to the presence of ordered magnetic fields.
Polarised emission is a characteristic exhibited by blazars and most
flat radio spectrum core dominated sources, in most cases extending
over more than five decades
in frequency.
Large polarisations ($>10\%$)
in the optical continua of a significant number of core-dominant
radio quasars are known to exist
(eg. Impey \& Tapia, 1990 and references therein), 
and provide strong evidence for a synchrotron component.

The percentage polarisation observed
is dependent on a number of factors: firstly, on the
magnetic field geometry, where it is greatest in the case when magnetic
fields are highly aligned, secondly, on the spectral
index of the emission spectrum
and thirdly on the effects of
depolarisation due to the presence of ionized plasma in front
of the synchrotron source (eg. Faraday rotation).
The effects of depolarisation 
by  Faraday rotation
are greatest at long wavelengths (mm to radio)
where synchrotron self-absorption effects are important, while they are
minimal at high frequencies (ie. optical to near-IR; 
Bj\"{o}rnsson \& Blumenthal, 1982). 

Bjornsson \& Blumenthal (1982) have shown that the fractional
polarisation at some frequency
$\nu$ expected from a synchrotron emitting plasma can be written
\begin{equation}
P(\nu)\,=\,{\cal F}(\nu)\left[\frac{1+\alpha(\nu)}
{\frac{5}{3}+\alpha(\nu)}\right].
\label{pures}
\end{equation}
${\cal F}(\nu)$ is a factor describing the magnetic field geometry. For
perfectly aligned fields, ${\cal F}(\nu)=1$, while as magnetic
fields become more disordered, ${\cal F}(\nu)\rightarrow 0$.
$\alpha(\nu)$ is the spectral index of the observed emission at the
frequency $\nu$, where $f_{\nu}\propto\nu^{-\alpha(\nu)}$ and hence
\begin{equation}
\alpha(\nu)\,\equiv\,-\frac{d\ln f_{\nu}}{d\ln\,\nu}.
\label{specindex}
\end{equation}
Thus, if the spectral index is independent of frequency as is the case
if the particle energy distribution is a power law,
and if ${\cal F}(\nu)$ is also constant, then the
polarisation is frequency independent.
The polarisation will be frequency dependent
only for a synchrotron radiation spectrum which exhibits some
degree of curvature (ie. where $\frac{d\alpha(\nu)}{d\ln{\nu}}\neq 0$).

\begin{figure}
\vspace{-2.8in}
\plotonesmall{1}{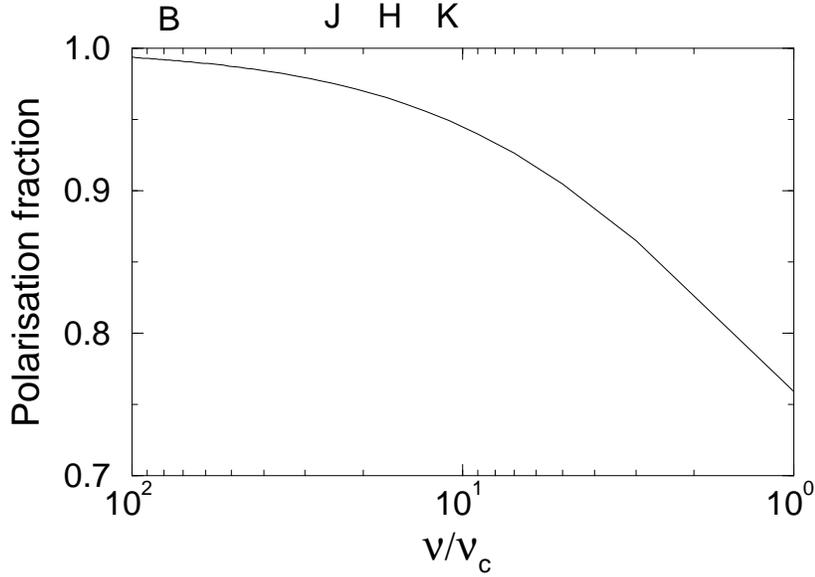}
\vspace{-2.5in}
\caption[Predicted polarisation in the monoenergetic synchrotron model]
{Fractional polarisation as a function of the
scaled frequency $\nu/\nu_{c}$ (where $\nu_{c}$ is
defined by Eqn.~\ref{nuce}) for a synchrotron spectrum exhibiting the maximum possible
spectral curvature
(eg. a monoenergetic synchrotron spectrum; see section~\ref{monofits}).
Typically, $\nu_{c}\sim10^{13}$Hz and for this value, the $K$, $H$, $J$ and
$B$ bands are also shown.} 
\label{pure}
\end{figure}

As discussed in section~\ref{monofits} of Chapter 5, a majority of the reddest
Parkes sources (with say $B_{J}-K\simgt5$) show
concave-`downward' emission spectra (where $\frac{d\alpha(\nu)}{d\ln{\nu}}>0$)
that are as extreme as can possibly be as predicted by 
conventional synchrotron models.
Some of the sources require particle energy distributions
that have to be nearly monoenergetic or ``step-like'' in nature.
As discussed, the observed spectral curvature for such distributions
will be the maximum possible via the synchrotron mechanism
(ie. the single particle radiation spectrum, Eqn.~\ref{singlep}).
Thus, as predicted from Eqn.~\ref{pures}, 
this spectrum will lead to the {\it strongest}
possible polarisation dependence with frequency.
Combining Eqns.~\ref{singlep},~\ref{pures} 
(with ${\cal F}(\nu)=1$) and~\ref{specindex},
Fig.~\ref{pure} shows this dependence in terms of the scaled frequency
$\nu/\nu_{c}$, where $\nu_{c}$ is given by Eqn.~\ref{nuce}.
Typically, $\nu_{c}\sim10^{13}$Hz - ie. the turnover frequency $\nu_{T}$
discussed in section~\ref{monofits}.

As shown in Fig.~\ref{pure}, 
if the reddest Parkes quasars are due to a pure
synchrotron radiation spectrum 
from a monoenergetic or step-like particle distribution,
the polarisation is expected to increase with frequency. 
The frequency dependence will be strongest at the lowest frequencies.
This characteristic dependence (with $dP/d\lambda<0$) has been observed
in the optical-to-near-IR of a number of BL-Lac type objects by Brindle \etal
(1986). 
Polarisations as high as those shown in Fig.~\ref{pure} 
have never been observed,
and thus imply that either unpolarised components must be present 
(see next section) 
or that the emission may arise from regions without perfectly
aligned magnetic fields.

The observed wavelength dependence of polarisation in our 
moderately reddened
quasars (Fig.~\ref{spectra}) is opposite to that predicted from a 
`pure' synchrotron radiation spectrum exhibiting the maximum 
(`concave downward') spectral
curvature (Fig.~\ref{pure}).
To achieve polarisation spectra in this model that are compatible with the
observations would require concave-`upward' total flux spectra 
(ie. $\frac{d\alpha(\nu)}{d\ln{\nu}}<0$; see Eqn.~\ref{pure}). 
This however would be contrary to the observed high frequency spectral shapes
of our sources and 
blazar-like AGN in general (eg. see Fig.~\ref{multi}). 
We conclude that this mechanism {\it alone} cannot be used to explain the
the polarisation and hence
reddening observed. In the next section, we discuss a general model that
may be more realistic of the emission mechanism in radio-quasars.

\subsubsection{A Simple model with Two Spectral Components}

Smith \etal (1986) and Wills \etal (1992) showed that the observed decrease
in polarisation from near-IR to optical wavelengths in most core dominated
radio quasars can be modelled using the sum of two spectral components.
These components are as follows:

The first component is the `big blue bump' (BBB) feature characteristic of the
optical-UV continuum spectra of quasars.
This is a broad feature with $\alpha\sim0$ ($f_{\nu}\propto\nu^{-\alpha}$)  
lying between $\sim1\mu$m and 1000\AA$\,$ and is usually attributed to
thermal emission from an accretion disk
or optically thin free-free emission from hot gas (Barvainis, 1993 and
references therein). 
This component is only weakly polarised, $P_{V}<1\%$, 
and is responsible for photoionizing emission line regions.
The second component is a power-law `BL-Lac' type component 
with $\alpha\sim1-2$ which extends from the near-IR to optical-UV and is
attributed to synchrotron emission.
This component is assumed to have to have the same origin as that 
associated with the highly polarised and variable continua of BL-Lac
objects.
In most observations of BL-Lacs, the fractional polarisation of this
almost pure synchrotron component is approximately independent of
wavelength (eg. see sources in Fig.~\ref{spectra}).
If the synchrotron emission is relatively bright (eg. relativistically
beamed), then contributions from unpolarised components in the near-IR
such as emission from the host galaxy, emission lines and dust emission
will be unimportant. 
By summing the unpolarised flat BBB component at short wavelengths with this
polarised power-law synchrotron component, a decrease in polarisation
towards shorter wavelengths is expected as the polarised power-law
becomes increasingly diluted by the BBB. 

If we denote the flux in the synchrotron component at some wavelength as
$F_{s}(\lambda)$ with intrinsic wavelength independent polarisation $p_{s}$,
and that in the BBB component as $F_{B}(\lambda)$, then the fractional
polarisation can be written:
\begin{equation}
p(\lambda)\,=\,\frac{p_{s}F_{S}(\lambda)}{F_{S}(\lambda) + F_{B}(\lambda)}.  
\label{poldep}
\end{equation}
If we represent $F_{S}(\lambda)$ and $F_{B}(\lambda)$ in the rest frame as
power-laws:
\begin{equation}
F_{S}(\lambda)\,=\,F_{S}(\lambda_{o})\left(\frac{\lambda}
{\lambda_{o}}\right)^{\alpha_{ _{S}}-2}\,\,{\rm and {\hspace{3mm}}}\,\, 
F_{B}(\lambda)\,=\,F_{B}(\lambda_{o})\left(\frac{\lambda}
{\lambda_{o}}\right)^{\alpha_{ _{B}}-2},
\label{pldep}
\end{equation}
where $\alpha_{ _{S}}$ and $\alpha_{ _{B}}$ are power-law slopes
($f_{\nu}\propto\nu^{-\alpha}$), then the observed polarisation
wavelength dependence can be written:
\begin{equation}
p(\lambda)\,=\,\frac{p_{s}}{1 + f_{BS}(\lambda_{o})\left(\frac{\lambda}
{\lambda_{o}}\right)^{\alpha_{ _{B}}-\alpha_{ _{S}}}
(1+z)^{\alpha_{ _{S}}-\alpha_{ _{B}}}}. 
\label{poldep2}
\end{equation}
The redshift factor accounts for the $K$-correction and $f_{BS}(\lambda_{o})$
represents the relative amount of BBB flux to synchrotron flux at $\lambda_{o}$,
ie, $f_{BS}(\lambda_{o})\equiv F_{B}(\lambda_{o})/F_{S}(\lambda_{o})$.

An increase in the degree of relativistic beaming of the synchrotron
component will cause the $p(\lambda)$ curve to approach the 
intrinsic polarisation $p_{s}$ of the synchrotron spectrum $F_{S}(\lambda)$.
Any beaming factors will be contained in the normalisation factor
$f_{BS}(\lambda_{o})$ in Eqn.~\ref{poldep2}. If $\delta$ is the Doppler 
factor, then $f_{BS}\propto\delta^{-(2+\alpha_{ _{S}})}$ for a
continuous jet and hence, $p(\lambda)\rightarrow p_{s}$ as 
$\delta\rightarrow\infty$.  
The total flux spectrum,
$F_{TOT}(\lambda)=F_{S}(\lambda) + F_{B}(\lambda)$, 
will obviously steepen towards $F_{S}(\lambda)$ with slope
$\alpha_{ _{S}}$. 

\begin{figure}
\vspace{-2.8in}
\plotonesmall{1}{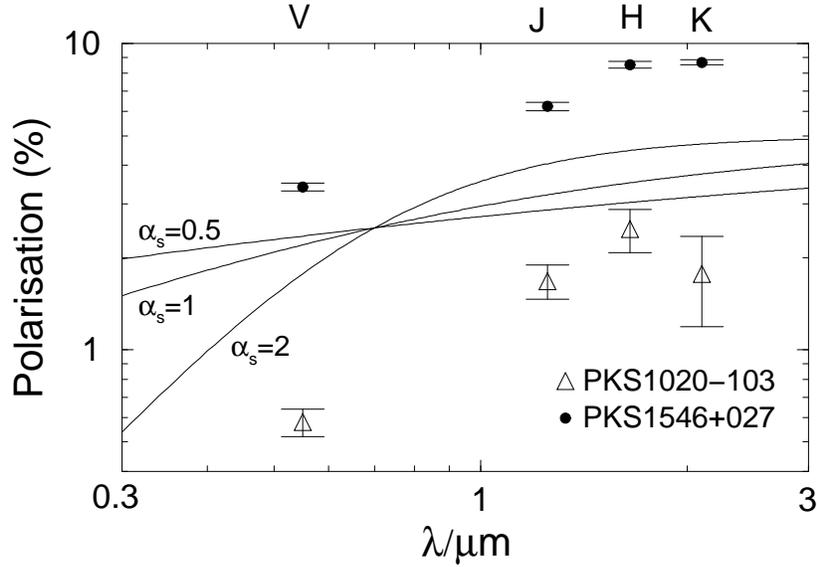}
\vspace{-2.5in}
\caption[Polarisation vs. wavelength for a simple two-component model]{Polarisation as a function of wavelength for a simple two
component model (solid lines) comprised of a flat
($\alpha=0$), unpolarised BBB component and a power-law
synchrotron component with intrinsic polarisation, $p_{s}=6\%$. 
Models are shown for three values of the synchrotron power-law slope
$\alpha_{ _{S}}$. The $J$, $H$, $K$ polarisations shown for the two
quasars are from the present run, and those in the $V$-band are from
Impey \& Tapia (1990).} 
\label{psyncheg}
\end{figure}

As a simple illustration, we assume the synchrotron `BL-Lac' component
has an intrinsic wavelength independent polarisation, $p_{s}=6\%$, and the
`flat' BBB component has $\alpha_{B}=0$. 
We assume these two components contribute equal flux densities at a rest
wavelength $\lambda_{o}=7000$\AA$\,$, ie. $f_{BS}(7000{\rm\AA})=1$. 
We take $z=0.5$.
Fig.~\ref{psyncheg} shows $p(\lambda)$ for three values of the synchrotron
power-law slope: $\alpha_{B}=0.5$, 1, 2. 
For comparison, the wavelength dependence of polarisation observed in two of
our quasars from Fig.~\ref{spectra} having measured optical ($V$-band) 
polarisations from previous studies are also shown.
PKS1546+027 is classified as a `blue' quasar with $\alpha_{opt}\sim0.1$,
and PKS1020$-$103 as a `moderately red' quasar with $\alpha_{opt}\sim1$. 
With the normalisation ($p_{s}$) being a free variable, a model with
synchrotron slope $\alpha_{ _{S}}\sim2$ provides a good fit to these sources. 

This simple model can explain the general wavelength dependence
of polarisation observed in a majority of our sources in Fig.~\ref{spectra}.
According to this model, the synchrotron component is expected to be 
dominating the optical to near-IR continua of the BL-Lacs (giving almost
wavelength independent polarisation) and the near-IR continua of the quasars. 
This is particularly important for the quasars 
exhibiting relatively redder colours ($B_{J}-K_{n}\simgt4$), since they
also appear to be those most highly
polarised in the near-IR (see Fig.~\ref{BKpol}).
Furthermore, if the observed significant changes in $K$-band flux over a
timescale of $\sim2$yrs (see Fig.~\ref{var}) can be attributed to variability, 
then this is consistent with a synchrotron dominated (possibly beamed)
continuum in the near-IR.

We must note that this two component model also makes a specific
prediction for the general shape of the total flux spectrum 
and the equivalent widths of emission lines observed.  
The additional effects of reddening
by dust is likely modify the total
flux spectra.
Thus, by combining our polarimetry measurements with 
observations of the total flux spectra,
we can constrain both the reddening mechanism and
relative contribution of the components in the above model. 
More realistic and detailed model fits to each source are
presented in the following section.

\section{The Two Component Model: Fits to Observations}
\label{twocm}

The simple two component model used to explain the wavelength dependence of
polarisation must also satisfy two observational constraints:
first, the total continuum flux spectra and second, the equivalent widths
(EW) of emission lines observed.
If the synchrotron component does not photoionize emission line regions
as expected if it were beamed into a narrow relativistic jet in the 
rest frame of the emission line gas, then the magnitude of this
component must be such that it does not significantly reduce the EWs
of emission lines observed.
Thus, the EWs of optical-UV emission lines can provide a constraint on the
relative magnitude of the synchrotron component allowed.

As presented in Chapter 5, there is considerable evidence suggesting
that the optical reddening observed in some Parkes quasars may be due to dust.
If dust were reddening the UV-to-optical continua,
then the shapes of the total flux spectra
observed are likely to be different than those predicted assuming 
a superposition of a synchrotron and a BBB component alone.
Thus, a comparison of the total flux spectrum observed with that predicted 
from the two component model
(which is made to satisfy both the polarisation and EW data), 
will enable us to determine whether dust is
required to explain the reddening observed. 

Assuming the simple two component model is an adequate representation
of the optical-to-near-IR emission from flat spectrum radio-quasars,
we present in this section a more generalised model that includes
the effects of reddening by dust.
We shall fit this model to our data in Fig.~\ref{spectra} 
according to three observational
constraints: first, the wavelength dependence of polarisation, 
second, the total optical-to-near-IR flux spectra and third, the
EWs of emission lines.
Our aim is to explore the relative contribution of the BBB and synchrotron
components in each of the sources, and any possible reddening contributed 
by dust. 

\subsection{Model Assumptions} 
\label{modass}

We shall initially assume that the presence of 
dust reddens both the (polarised) synchrotron and
(unpolarised) BBB component equally. 
Since dust in the local quasar environment can exist at distances 
from $\sim1$pc (to avoid sublimation) to host galaxy or NLR (kiloparsec) 
scales from the central AGN, 
this assumption requires the
emission from 
both components 
to arise from relatively small scales. 
This can be easily justified for the BBB component since 
the emission has been observed to vary on time
scales from weeks to a few months, implying scales 
$r\simlt0.1$pc (Barvainis, 1993 and references therein). 

Evidence that the optical-to-near-IR synchrotron component also originates 
on a relatively small scale is supported indirectly from 
observational and theoretical studies of flat spectrum radio-quasars and 
BL-Lacs. 
Wills \etal
(1992) found that
a significant correlation exists between the luminosity of the optical 
polarised component and radio core luminosity. 
Since VLBI observations have shown that in most cases, 
the compact radio cores 
appear unresolved within parsec to sub-parsec scales, 
this suggests that the
(unresolved) optical-to-near-IR
synchrotron emission also originates from the same inner region of the
quasar, perhaps at the base of a relativistic jet.
A similar relation has been observed in BL-Lacs by Impey \& Neugebauer (1988).
Furthermore, the small scale nature of high frequency synchrotron 
emission is also supported on various theoretical grounds. 
Such emission requires relatively high electron energies whose lifetimes
are very short, and hence, must be produced in a small volume (where
magnetic fields are likely to be more aligned).
This assumption implies that the observed polarisation (both its
strength and wavelength dependence) will be independent of dust optical
optical depth. 

According to this assumption,   
the total observed flux spectrum with the presence of
a uniform dust screen in front of the emission regions for these
components can be written:
\begin{equation}
F_{TOT}(\lambda)\,=\,\left[F_{S}(\lambda) + F_{B}(\lambda)\right]\,
\exp{(-\tau_{\lambda})},
\label{Ftotdust}
\end{equation}
where $F_{S}$ and $F_{B}$ are the flux in the synchrotron and BBB
components respectively, and 
$\tau_{\lambda}$ is the dust extinction optical depth. We assume this follows
the simple $1/\lambda$ dependence:
\begin{equation}
\tau_{\lambda}\,=\,\tau_{B}\left(\frac{\lambda}{\lambda_{B}}\right)^{-1},
\label{tautottwo}
\end{equation}
where $\tau_{B}$ is the rest frame optical depth at $\lambda_{B}=4400$\AA$\,$
and is a free model parameter.

A second assumption is that the emission from the synchrotron
component does not significantly photoionize
emission line regions. 
If the optical-to-near-IR synchrotron emission is beamed
along the 
line-of-sight
(as suggested by correlations with the 
radio core emission discussed above), then
this is somewhat expected. 
Simple relativistic jet models predict that the effective solid angle 
subtended by the beaming cone in an observer's frame is given by 
$\Omega\sim4\pi\gamma^{-2}$, where $\gamma$ is the Lorentz factor. 
Statistical studies in the radio find that
typically $\gamma>5$ (Cohen \& Vermeulen, 1992; Padovani \& Urry, 1992)  
and may be higher in the optical (Impey \& Tapia, 1990).
This corresponds to a jet covering factor: $\Omega/4\pi<4\%$ and hence it
is unlikely that a majority of the emission line gas can 
be photoionized by a beamed synchrotron component.
We assume such a component only contributes 
to the continuum emission and hence will affect the EWs of
emission lines observed.
This is consistent with observations of the weak-lined BL-Lac type sources or
radio loud AGN with very low line EWs.
These sources also appear to be strongly polarised, highly variable,
and exhibit relatively high measures of core-to-extended radio flux 
(eg. Impey \& Tapia, 1990; Wills \etal 1992).
Furthermore, as suggested by such observations, 
the flux in the synchrotron component is assumed to be independent
of that in the photoionizing unpolarised BBB component.

For our two-component model of the previous section, the observed EW of
an emission line with wavelength $\lambda$ can be written:
\begin{equation}
W_{\lambda}\,=\,\frac{(1+z)W_{o}
F_{B}(\lambda)}{F_{B}(\lambda) + F_{S}(\lambda)}, 
\label{EW}
\end{equation}
where $W_{o}$ represents the intrinsic rest frame EW width in a `normal'
quasar with no synchrotron component present.
In our modelling of the observed EWs,  
we shall assume the mean values and standard deviations 
in $W_{o}$ for the various lines
as found in optically selected quasars (eg. Francis \etal 1991).
Amongst all five quasars (with an available polarisation spectrum), 
only three broad lines of 
measurable strength are observed: 
MgII(2798\AA), H$\beta$(4861\AA) and  H$\alpha$(6563\AA). 
The intrinsic EWs ($W_{o}$) 
assumed for MgII and H$\beta$ are respectively:
$50\pm20$\AA$\,$ and $58\pm23$\AA$\,$ (Francis \etal 1991).
Due to redshift limitations, the statistics for the
H$\alpha$ line are generally quite poor.
We can however approximate
its intrinsic EW from $W(H\beta)$ using the mean 
observed Balmer ratio $I(H\beta)/I(H\alpha)\sim4.5$ 
(see section~\ref{Balmpks}), and 
assuming a power-law continuum with $\alpha=0.3$.
We find that typically, $W(H\alpha)\sim7W(H\beta)=
406\pm160$\AA, which agrees with the observed distribution in Parkes quasars.
Only `broad' emission lines have been used in order to avoid any possible
effects of anisotropic dust obscuration on the EWs. These effects may be
more important on emission lines produced at larger distances from the central 
AGN (eg. in the NLR, see Fig.~\ref{EWdata} in Chapter 5).

Both the synchrotron and BBB components are assumed to be
represented by power-laws as defined by Eqn.~\ref{pldep}. 
Instead of assuming a purely flat spectrum for the BBB component
as used in Fig.~\ref{psyncheg}, we assume the spectral slope to be fixed at the 
mean value:
$\alpha_{ {B}}=0.3$, as observed in optically selected quasars 
(Francis, 1996). 
This value agrees well with a composite spectrum for
optically selected quasars extending into the near-IR as derived by 
Holman (1997). 

In summary, Eqns.~\ref{poldep}, ~\ref{Ftotdust} and~\ref{EW} 
are the three model equations used to fit for
the following respective
observational constraints for each source:
the polarisation spectrum, the total flux spectrum and emission line EWs. 
Together with 
Eqns.~\ref{pldep} and~\ref{tautottwo}, 
the model is defined by four parameters: the power-law slope of the
synchrotron component $\alpha_{ _{S}}$, 
the intrinsic (wavelength independent) 
polarisation of the synchrotron component $p_{s}$,
the ratio of synchrotron flux to
BBB flux at a fixed fiducial wavelength, $f_{BS}(\lambda_{o})$, and,
the $B$-band dust optical depth $\tau_{B}$.
All these parameters refer to the source {\it rest} frame and all
relevant K-corrections will be accounted for.

\subsection{Model Fits and Results}
\label{resfit}

The best fit values for the model parameters 
($\alpha_{ _{S}},p_{s},f_{BS}(\lambda_{o}),
\tau_{B}$) for each source 
are determined as follows: 
\\\indent 1. We first combine two of the 
observational constraints: the fractional polarisation and total flux
spectra, and find the best fit parameters by 
minimising $\chi^{2}$. Only polarisation measures in the near-IR are used, as
those in the $V$-band shown in Fig.~\ref{spectra} are from the literature and
not contemporaneous. 
The total flux spectra used were determined
from contemporaneous photometric measurements in
$B$, $V$, $R$, $I$, $J$, $H$ and $K$ for all sources taken by
P. Francis in April 1997. 
\\\indent 2. Given this set of parameters, we then predict the
intrinsic underlying EWs ($W_{o}$) expected if no synchrotron
component were present. This is done by inverting Eqn.~\ref{EW},
where the observed EWs ($W_{\lambda}$) for MgII, H$\beta$ and  H$\alpha$ 
are determined from the optical spectra in 
Fig.~\ref{spectra}.
\\\indent 3. If the predicted intrinsic EWs from step 2 
lie outside the assumed 1$\sigma$ ranges
for $W_{o}$ discussed above, the parameters are re-adjusted
until 
the predicted $W_{o}$'s lie within the assumed ranges and step 1 is repeated. 

In summary, our fitting procedure consists of fixing the model parameters
according to the fractional polarisation and total flux spectra with the 
constraint that the predicted intrinsic EWs lie within the 
assumed ranges for $W_{o}$ as seen in `normal' optically selected quasars.  

The observed emission line EWs, $W_{\lambda}$,
in our five quasars are not in general
lower than the assumed intrinsic values
$W_{o}$ discussed above.
All of them lie within one standard deviation of the mean 
intrinsic values expected.
This indicates that the
`non-photoionizing' polarised
synchrotron component must not significantly contribute in the
optical-UV region.
Most of the observed EWs require that the synchrotron component, $F_{S}$, 
not contribute more than 30\% to the total flux 
at rest wavelengths $<6000$\AA$\,$ in order 
to be compatible with the assumed ranges in $W_{o}$.

\begin{figure}
\vspace{-2in}
\plotonesmall{0.9}{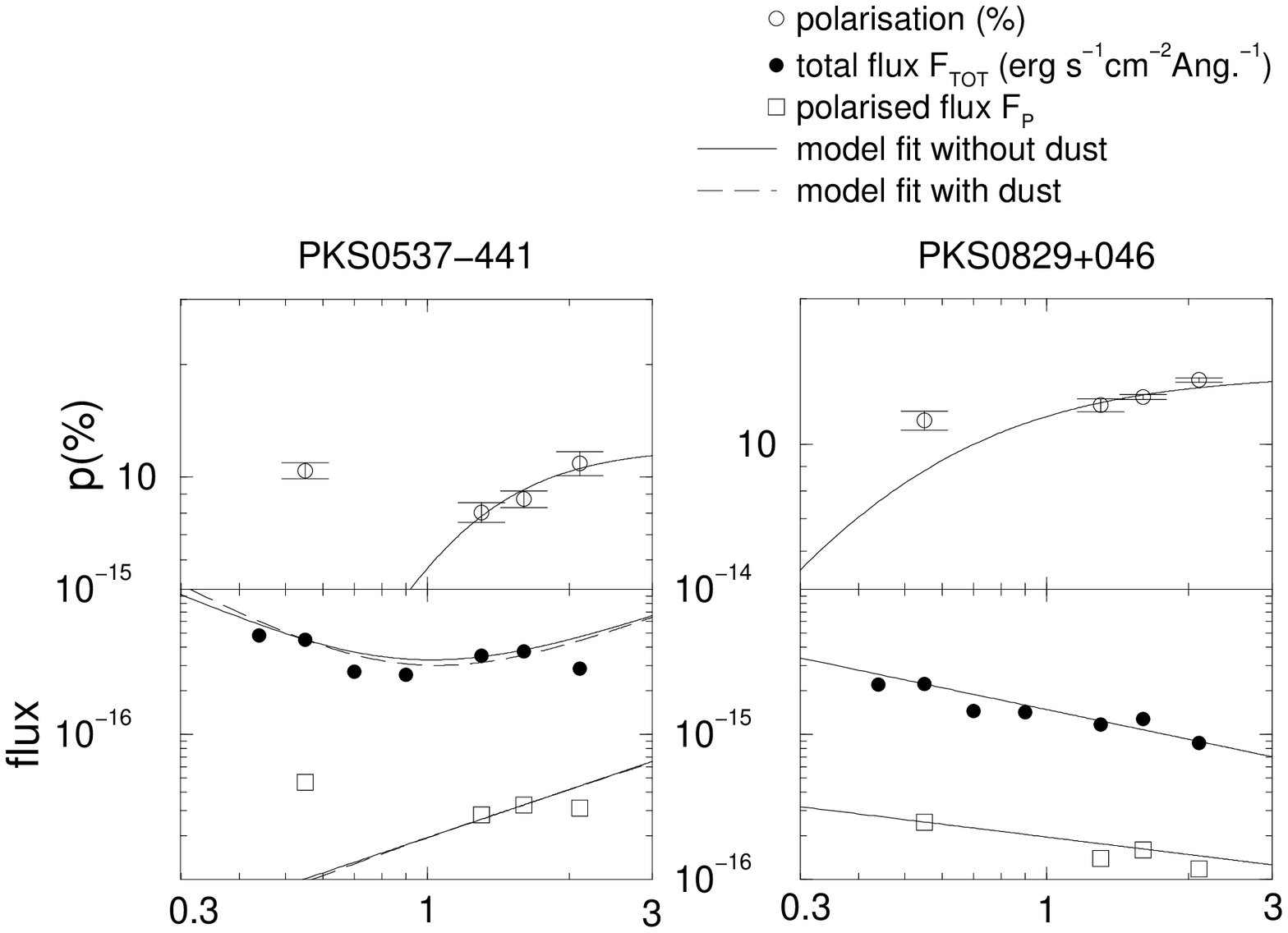}
\vspace{-3.7in}
\plotonesmall{0.9}{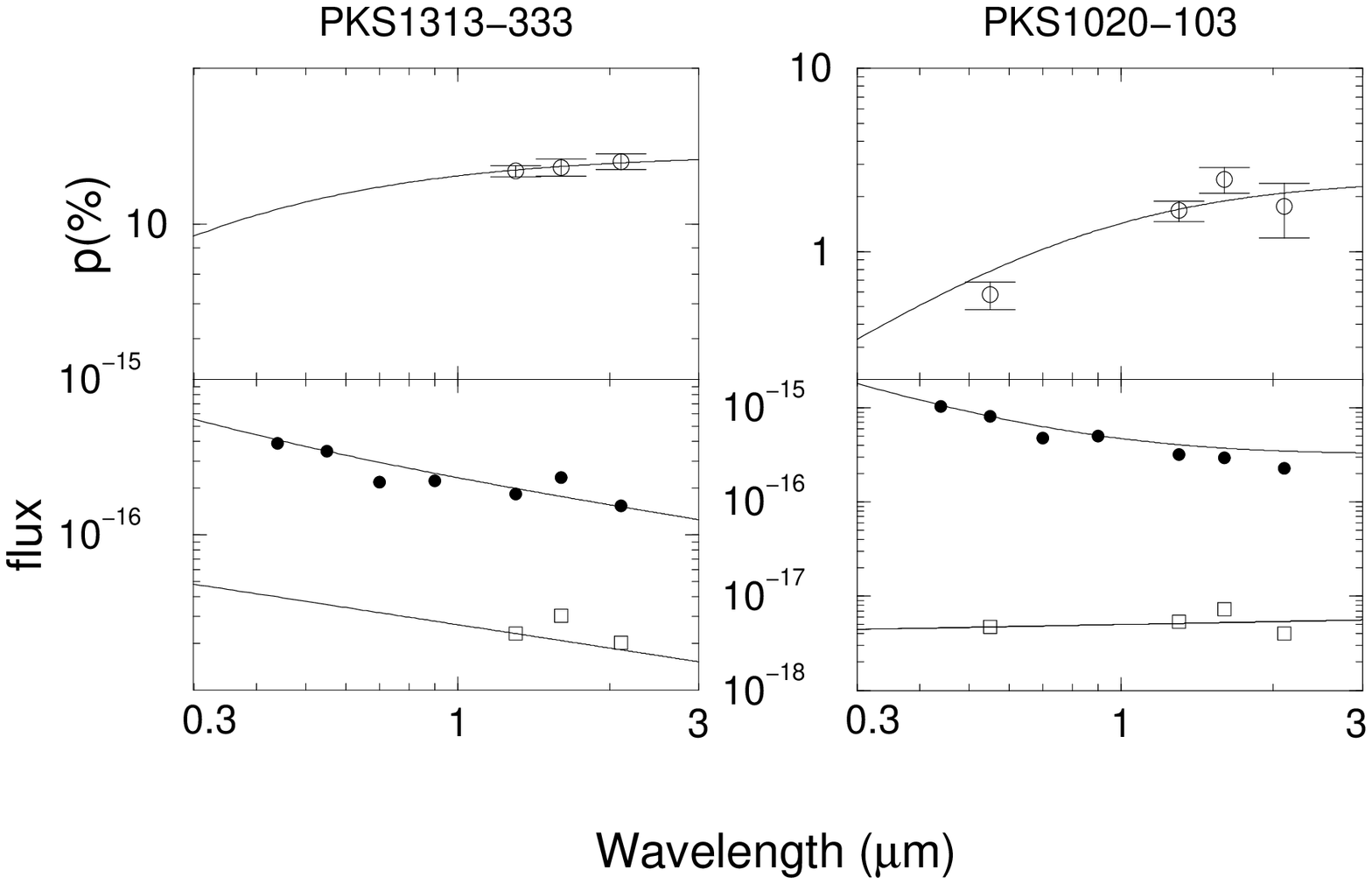}
\vspace{-1in}
\caption[Model fits to polarisation and total flux spectra]
{Model fits to polarisation and total flux spectra
for each source. Top panels show polarisation spectra, $p(\lambda)$
(open circles),
and bottom panels, total flux spectra, $F_{TOT}(\lambda)$
(in ${\rm erg s}^{-1}{\rm cm}^{-2}{\rm\AA}^{-1}$,
filled circles), together with polarised flux spectra
($=p(\lambda)F_{TOT}$, squares)
on observed wavelength scales. Model fits for sources
which require appreciable
dust optical depths are shown by dashed curves.
Where available, the $V$-band polarisations (at $\lambda\simeq0.55\mu$m) are
not contemporaneous with those in the near-IR and are {\it not} included in
the model fits.}
\label{fits2}
\end{figure}

Model fits to each of the eight sources were performed by
fitting for two sets of parameters. The first set of parameters
neglects dust and sets
$\tau_{B}=0$, and the second set assumes $\tau_{B}\ne0$. 
Results of the best fit model parameters
for $\tau_{B}=0$ and $\tau_{B}\ne0$ are listed in Table~\ref{fittab}.
Our model fits to the three observational constraints 
are compared with 
the data in Fig.~\ref{fits2}.
Cases consistent with appreciable dust optical depths $\tau_{B}$ are also shown.
The top part of each panel in Fig.~\ref{fits2}
shows the wavelength dependence of the
degree of polarisation, $p(\lambda)$. The bottom portion shows the total
flux density spectrum, $F_{TOT}(\lambda)$ (filled circles) 
and the polarised flux spectrum, 
$F_{P}(\lambda)=p(\lambda)F_{TOT}(\lambda)$ (squares).

For the three BL-Lac objects (PKS0537$-$441, PKS0829+046 and PKS1313\\ 
$-$333) classified according to optical spectral properties,
one would naively 
expect their emission to be dominated by the synchrotron spectrum 
($F_{TOT}(\lambda)\equiv F_{S}(\lambda)$) as predicted by the
beaming model. 
If this were true, then the intrinsic polarisation 
$p_{s}$ in these sources should equal $p(\lambda)$ and 
be close to wavelength independent as initially assumed.
An examination of the BL-Lac polarisation 
spectra however (Fig.~\ref{fits2}) shows
that $dp(\lambda)/d\lambda<0$ in the near-IR
(note that the optical polarisation measures
are not contemporaneous
with those in the near-IR).
Thus, if indeed $F_{TOT}\equiv F_{S}$ in these sources, then  
$p_{s}$ must decrease towards short wavelengths.
This is contrary to previous modelling where $p_{s}=$constant 
or dependences with $dp_{s}/d\lambda>0$
was seen to fit the BL-Lac data quite well (eg. Brindle \etal 1986). 
It is possible that our BL-Lacs are a special class
where in fact $p_{s}$=constant and their polarisation spectra
are modified by a BBB component.
We must also note that the optical spectra used to identify the
BL-Lacs are not contemporaneous with the near-IR polarisation data,
and hence the
synchrotron component may have varied significantly.

Given the polarised flux spectrum,
$F_{P}(\lambda)$ (squares in Fig.~\ref{fits2}), the actual
synchrotron spectrum is related to this
by $F_{S}(\lambda)=F_{P}(\lambda)/p_{s}$, where $p_{s}$ is the
intrinsic polarisation of the synchrotron spectrum.
If $p_{s}$ were wavelength independent in most of
our sources as initially assumed, then the
shape of $F_{P}(\lambda)$ should represent $F_{S}(\lambda)$ well in most cases.
If $p_{s}$ however decreases towards short wavelengths
as implied by the BL-Lacs, then the true form 
for $F_{S}(\lambda)$ will decline less steeply than $F_{P}(\lambda)$ at
short wavelengths.

In only one of the quasars, PKS1036$-$154, does the model which 
incorporates dust provide a better fit to its total flux spectrum.  
For this source, the model without dust predicts an optical spectrum with
a characteristic upward curvature towards short wavelengths, where the
BBB with $f_{\nu}\propto\nu^{-0.3}$ dominates.
Such a model can explain the observed wavelength dependence
of polarisation and
emission
line EWs, but fails to satisfy the total flux spectrum observed.
Reddening
by dust of the total flux spectrum
(with extinction law defined by Eqn.~\ref{tautottwo})
in this source is required.
This is consistent with its relatively red optical continuum slope of
$\alpha_{opt}\simeq 2.3$ as
compared to $\alpha_{opt}\simlt0.1$ for the blue quasars. 
Its Balmer decrement value of $H\alpha/H\beta\sim4.6$ is
also consistent with the required value $\tau_{B}\sim0.6$ if we
assume an intrinsic Balmer ratio $H\alpha/H\beta=4$ (a lower limit
observed in the `bluest' radio-quasars; see section~\ref{Balmpks}).
The data for the BL-Lac PKS0537$-$441 is also 
consistent with a moderate extinction ($\tau_{B}\sim0.2$).
This source however can also be equally
explained by a model with no dust.
As expected, the blue quasars are all consistent with the no dust model,
or require very small optical depth measures 
($\tau_{B}<0.01$) as shown in Table~\ref{fittab}. 
The same applies to the three BL-Lacs.
The introduction of dust in these sources does not improve the model
fits as seen in the $\chi^{2}$ values for the `no-dust' and `dust' cases
in Table~\ref{fittab}.

\begin{table}
\vspace{-0.5in}
\begin{center}
\begin{tabular}{|c|c|c|c|c||c|c|c|}
\multicolumn{8}{c}{Quasars}\\
\hline
Source &
\multicolumn{4}{|c||}{$\tau_{B}=0$} &
\multicolumn{3}{|c|}{$\tau_{B}\neq0$}\\
\hline
\hline
  &\multicolumn{1}{|c|}{$\alpha_{ _{S}}$}
& \multicolumn{1}{|c|}{$f_{BS}(5500{\rm\AA})$}
& \multicolumn{1}{|c|}{$p_{s}$(\%)}
& \multicolumn{1}{|c||}{$\chi^{2}$}
& \multicolumn{1}{|c|}{$\alpha_{ _{S}}$}
& \multicolumn{1}{|c|}{$\tau_{B}$}
& \multicolumn{1}{|c|}{$\chi^{2}$}\\
\cline{2-8}
PKS1020$-$103 & 2.2 & 1.60 & 2.5 & 0.56 & 2.2 & $<0.01$ & 0.56\\
\hline
PKS1036$-$154 & 3.2 & 3.00 & 5.6 & 2.10 & 3.1 & 0.66 & 1.45\\
\hline
PKS1101$-$325 & 0.1 & 0.22 & 0.3 & 0.13 & 0.1 & $<0.01$ & 0.13\\
\hline
PKS1510$-$089 & 1.9 & 0.94 & 4.0 & 1.07 & 1.7 & $<0.01$ & 1.07\\
\hline
PKS1546+027 & 2.6 & 0.68 & 8.5 & 2.63 & 2.3 & $<0.01$ & 2.63\\
\hline
\end{tabular}

\vspace{0.5cm}

\begin{tabular}{|c|c|c|c|c||c|c|c|}
\multicolumn{8}{c}{BL-Lacs}\\
\hline
Source &
\multicolumn{4}{|c||}{$\tau_{B}=0$} &
\multicolumn{3}{|c|}{$\tau_{B}\neq0$}\\
\hline
\hline
  &\multicolumn{1}{|c|}{$\alpha_{ _{S}}$}
& \multicolumn{1}{|c|}{$f_{BS}(5500{\rm\AA})$}
& \multicolumn{1}{|c|}{$p_{s}$(\%)}
& \multicolumn{1}{|c||}{$\chi^{2}$}
& \multicolumn{1}{|c|}{$\alpha_{ _{S}}$}
& \multicolumn{1}{|c|}{$\tau_{B}$}
& \multicolumn{1}{|c|}{$\chi^{2}$}\\
\cline{2-8}
PKS0537$-$441 & 3.2 & 0.52 & 11.0 & 1.08 & 3.0 & 0.20 & 1.08\\
\hline
PKS0829+046 & 1.9 & 0.49 & 14.0 & 0.44 & 1.9 & $<0.01$ & 0.44\\
\hline
PKS1313$-$333 & 1.5 & 0.10 & 13.7 & 0.11 & 1.5 & $<0.01$ & 0.11\\
\hline
\end{tabular}
\caption[Parameter fits for the `two-component model']{Model parameter fits for $\tau_{B}=0$ and $\tau_{B}\neq0$.
Values of $p_{s}$ and $f_{BS}$ for the case $\tau_{B}\neq0$
are very similar as those corresponding to $\tau_{B}=0$ and are
not shown.}
\label{fittab}
\end{center}
\end{table}

The three observational constraints on the red quasar,
PKS1036$-$154, are also
consistent with our assumption that the
dust reddens
both the (polarised) synchrotron and (unpolarised) BBB component equally.
Both the `dust' and `no dust' model 
can explain the observed polarisation spectrum
and emission line EWs of this source. 
If the dust used to explain $F_{TOT}(\lambda)$ 
for instance were assumed to only affect the BBB component,
$F_{B}(\lambda)$, 
then a considerably flatter dependence in 
$p(\lambda)$ than that observed would result. 
By assuming such a model, 
we find that an 
optical depth $\tau_{B}\simlt0.1$ is required to explain the 
observed $p(\lambda)$ dependence in PKS1036$-$154. 
Such a value however would violate the 
total flux spectrum observed. 
In order to satisfy both $p(\lambda)$ and $F_{TOT}(\lambda)$, the
dust must affect both components equally.

\begin{figure}
\vspace{-2in}
\plotonesmall{0.9}{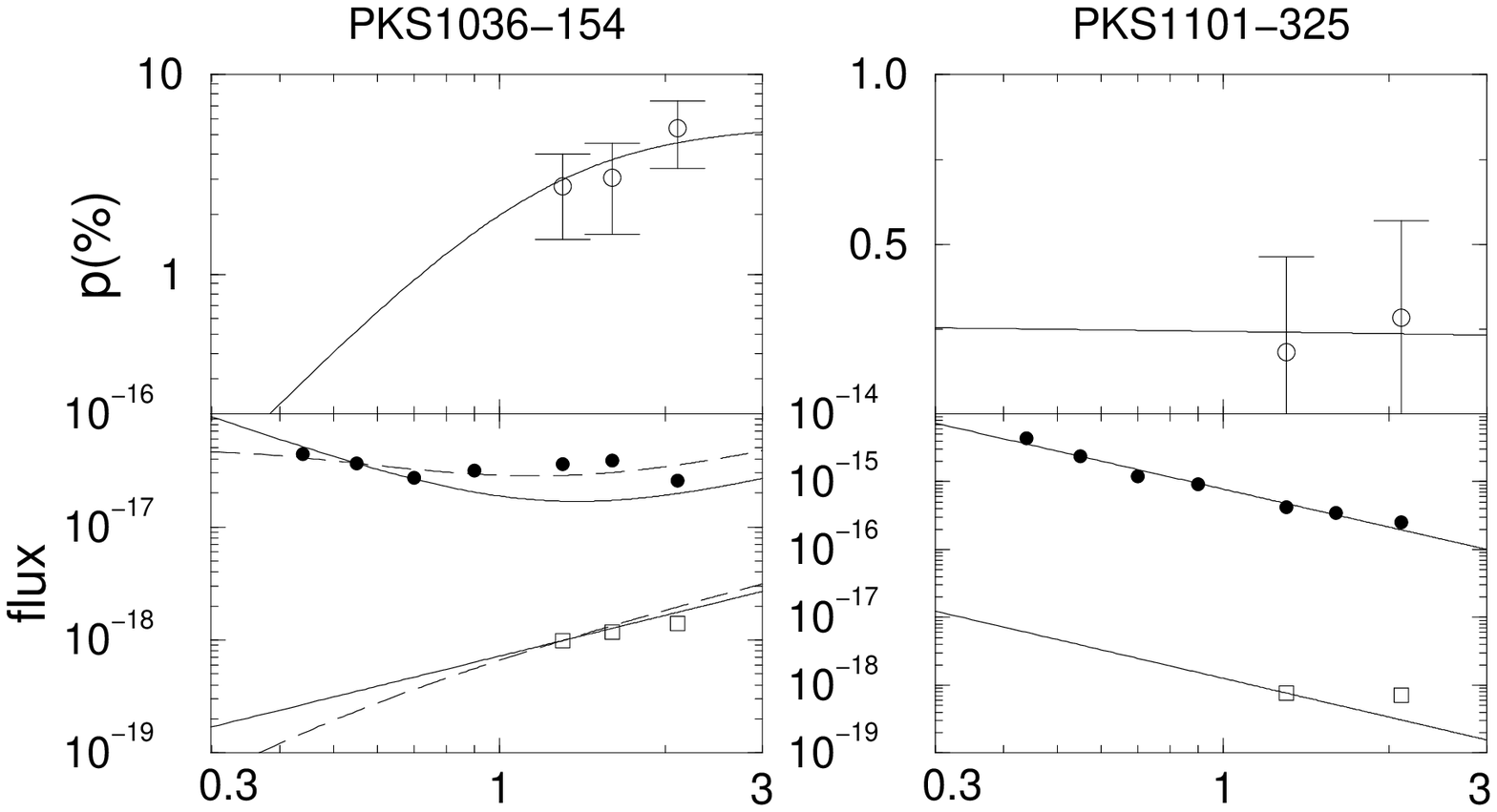}
\vspace{-3.7in}
\plotonesmall{0.9}{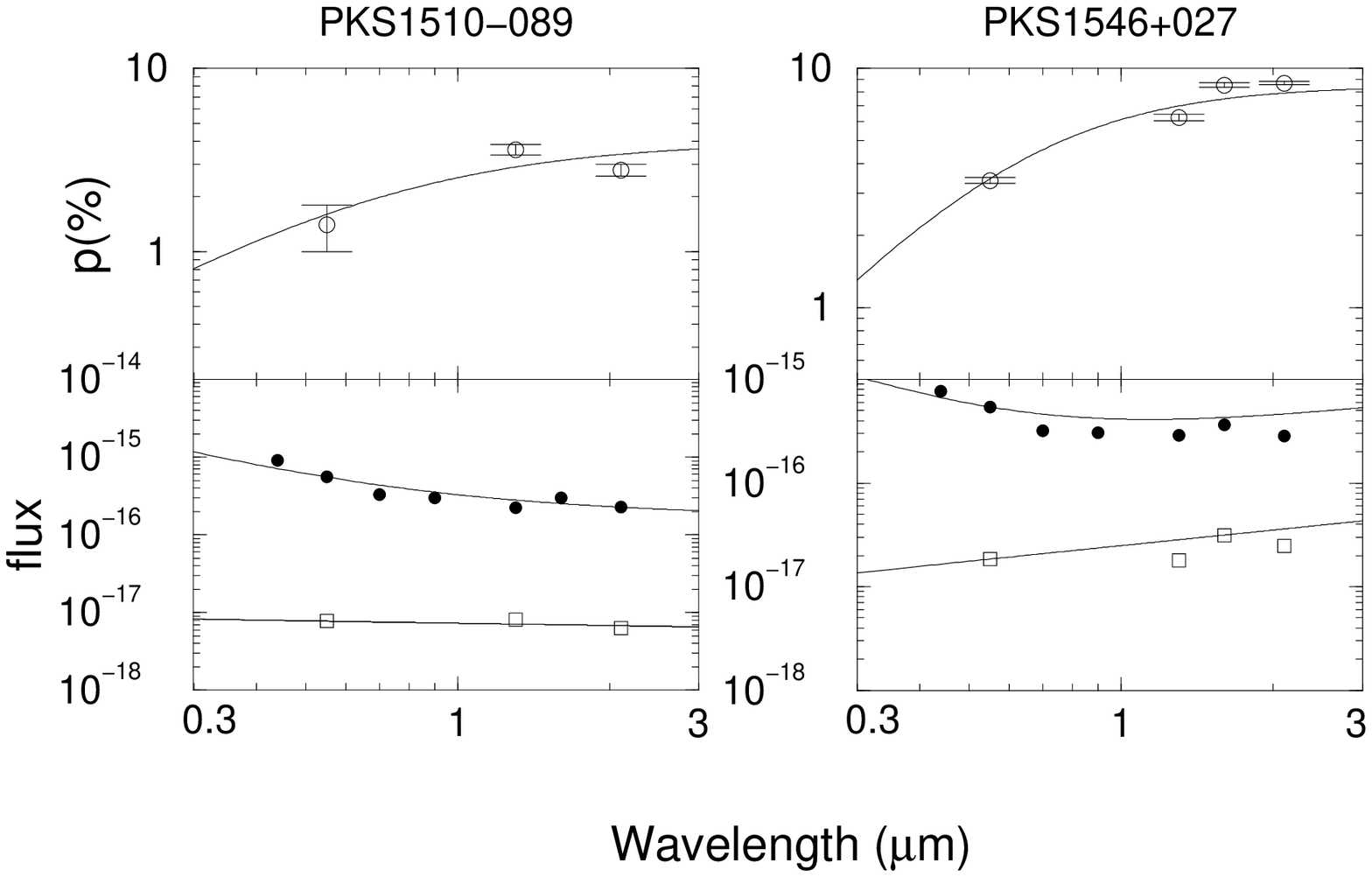}
\vspace{-1in}
Fig.~\ref{fits2} continued.
\end{figure}

\subsection{Summary}

In conclusion, our results appear consistent with those found by
previous authors for flat-spectrum radio quasars.  
The predicted underlying synchrotron continua are extremely steep in
all cases; most are well fitted by $F_{S}\propto\nu^{-1.5}$ to 
$F_{S}\propto\nu^{-3}$. 
Subtracting this synchrotron component from the total flux spectra
in the `blue' quasars, we are left with an unpolarised spectrum that
agrees well in shape with those seen in optically selected (radio quiet)
quasars with a prominent BBB signature.
Due to the low statistics of sufficiently reddened
quasars in this study, we are
unable to make any strong conclusions on the nature of the
reddening mechanism.
The only `red' quasar in our sample does however indicate that its
optical reddening
cannot be modelled as the sum of a `pure' synchrotron component and a
BBB component alone.
This simple model can explain the polarisation wavelength dependence and
emission line EWs,
but reddening by dust
is required to explain its total flux spectrum.

\section{Discussion}
\label{discuss}

Another polarisation mechanism which we have not explored in our
analysis is that due to scattering of light in anisotropic geometries.
Theoretical models 
show that for dust grains with sizes very small compared with the
wavelength ($2\pi a/\lambda\ll 1$), the polarisation from dust `absorption'
becomes vanishingly small (Greenberg, 1968).
The polarisation produced from scattered light in this regime however can
approach 100\% and will generally be wavelength independent.
The scattering cross-section or optical depth will behave
as $\tau_{s}\propto\lambda^{-4}$ (ie. the Rayleigh scattering regime).
Polarisation by scattering from very small dust grains or electrons
has been used by Wills \etal (1992) to account for the
high polarisation level ($\sim8\%$) and its wavelength dependence
observed in an IRAS selected quasar.
In this source, the scattering is believed to be from material above and
below a dusty torus viewed almost edge on.
According to the unified model for radio loud AGN, such a scattering geometry
is not expected in flat-spectrum radio quasars. The fact that all of our 
sources exhibit relatively high measures of the ratio of compact 
(on 10 mas scales) to
extended emission at 2.3GHz (eg. Duncan \etal 1993) 
also argues against such a geometry. 

If the polarisation in our sources
were due to scattering from small grains (or electrons)
rather than synchrotron, high polarisation 
levels can be achieved. The characteristic blue spectrum
(ie. where $\tau_{s}\propto\lambda^{-4}$) and wavelength independent 
nature of polarisation from scattering however, would be
inconsistent with both the wavelength dependence of polarisation and continuum
slopes 
observed. 
Dilution by an unpolarised big 
blue bump component in the optical-UV is likely 
to further violate the polarisation wavelength dependence and slopes observed. 
This is due to the `flatness' of the big blue bump in the optical-UV and the 
prediction that the scattered flux will steeply rise towards the blue
with intrinsic wavelength independent
polarisation.

According to our relatively high desired signal-to-noise ratio, the
major limitation of this polarisation study is the requirement that most of
the sources be sufficiently bright in the near-IR ($K\simlt14$).
This limitation may have lead to a selection bias in that those sources which
are relatively bright may also be those most strongly
polarised.
The nine sources in this study may thus be biased in favour of those
where a synchrotron (possibly beamed) continuum dominates in the near-IR.
This is indeed indicated by the significant correlation observed
between $B_{J}-K_{n}$ colour and $K_{n}$ polarisation as shown in 
Fig.~\ref{BKpol}.
The strongest polarised sources, namely the three BL-Lacs with 
$p_{K}\sim10\%$ are also those brightest in $K_{n}$.
Furthermore, our initial selection of `red' quasars
for this polarisation study was based on 
$B_{J}-K_{n}$ colour alone. This implies that we are {\it not} totally unbiased
towards sources whose colours are 
intrinsically reddened by synchrotron emission
dominating in the near-IR, rather than reddening by dust in the optical.
This is consistent with the relatively blue optical-UV continua
and moderately low Balmer decrements observed in most of our sources.

Most of the Parkes quasars with reddened optical-UV continua and
large Balmer decrements
suspected of being reddened by dust are those faintest in $K$  
(with $K\simgt16$). Polarimetry measurements will thus be exceedingly difficult
for our desired levels of accuracy. 
Thus, future polarisation studies with the next generation of near-IR
detectors shall be aimed at targeting these `faint' optically reddened
quasars. 
Given a sufficiently large statistical sample of reddened quasars, one
can then determine whether such sources systematically show a wavelength
dependence in polarisation significantly different from that in the 
bluer quasars. 
More specifically, if the general observed increase in polarisation with
wavelength in radio quasars can be explained by a simple BBB plus synchrotron
component model, then the presence of an underlying dust
polarisation spectrum (eg. Fig.~\ref{dusteg}) in 
the redder quasars should lead to a `flatter'
$p(\lambda)$ curve on average for these sources.
A sufficiently large sample however is required to draw any 
reasonable conclusion on the nature of the reddening mechanism 
on the basis of near-IR polarimetry alone.

\subsection{Further Predictions}

If dust in the local quasar environment is contributing to the 
reddening observed in Parkes quasars, then there are two predictions
that can be tested using future polarimetric observations
of the fainter, optically reddened sources. 
The first prediction is based on the rapid time variability 
in flux and polarisation often
observed at optical-to-near-IR wavelengths in BL-Lacs and 
highly polarised radio quasars (eg. Impey \& Neugebauer, 1988).  
If this is due to outbursts in the synchrotron component
(enhanced by Doppler boosting), then the presence of an extended dust
distribution on a scale larger (or cospatial) with the synchrotron emission
region is expected to modify the degree of variability in the
polarised component somewhat. 
Scattering of synchrotron photons in an extended optically-thin dust
distribution on scales of a few hundred parsecs, 
is expected to significantly increase variability timescales, or
average-out flux variations altogether.  
This will strongly depend on the geometry of the dust and its location
relative to the synchrotron emitting region.
As discussed in section~\ref{modass} however, it is likely that 
the high frequency synchrotron emission is produced on relatively small scales
which has a greater chance of being embedded or intercepted by dust. 
Thus, if our line-of-sight to the near-IR synchrotron emission region
is intercepted by dust, which is significantly spatially extended,
then such sources should show a degree of variability
in their polarised flux significantly different on average than
bluer `unreddened' radio quasars. 

Our second prediction is that if the hypothesised dust component in Parkes
quasars acts as an efficient polarising medium (either through scattering
or absorption), then the broad emission lines are expected to exhibit
some degree of polarisation (eg. Antonucci \& Miller, 1985). 
Any significant emission line polarisations cannot be caused by the
synchrotron mechanism.
Synchrotron emission is only expected to contribute to the continuum 
polarisation and thus any polarisation in the emission line flux must be
attributed to an alternative mechanism. 
Thus,
this test avoids polarisations from mechanisms other than those expected
directly from dust. 
Such a test however will require accurate spectropolarimetric measurements which
at present will be extremely difficult (even in the near-IR)
for the faintest and `reddest' Parkes quasars. 

\section{Conclusions}
\label{concsev}

The initial aim of this investigation was to use near-IR linear polarimetry
as a further test of the dust hypothesis used to explain the redness
observed in Parkes quasars. 
The aim was to determine whether any of the reddened quasars show
wavelength dependences in polarisation in the near-IR similar to that
observed from absorption by aligned dust grains in the Galaxy.
Due to the faintness of many of the reddest quasars however, our
observations were mostly limited to `blue' and `moderately reddened'
quasars.
Subsequent spectroscopic and contemporaneous
photometric studies of our nine sources
revealed that three were identified with BL-Lacs and six with strong-lined
quasars. 
Three of these quasars have `blue' optical continua (where
$\alpha_{opt}\simlt0.1$) and moderately low Balmer decrements.
The other three appear to be `moderately reddened' with
$\alpha_{opt}\simgt1$ and two of which exhibit somewhat larger
Balmer decrement measures.
Furthermore, only two of these moderately reddened quasars have 
broadband polarimetry
measurements in all $J$, $H$ and $K_{n}$ bandpasses. 
Our main observational results and conclusions are the following: 
\\\indent 1. All sources have been detected to be significantly polarised in the
$J$, $H$ and $K_{n}$ bands at greater than $2\sigma$ confidence.
As expected, the three BL-Lacs are the most strongly polarised with 
$K_{n}$ band polarisation $p_{K}\sim10\%$. 
\\\indent 2. A comparison of $K_{n}$ photometric measurements from three
different epochs for each source shows that the scatter in $K_{n}$ magnitude
is significantly correlated with $p_{K}$. 
If these flux variations are due to variability in the source emission,
then this suggests that the near-IR may be dominated by a variable
(possibly beamed) synchrotron component. 
\\\indent 3. There also exists a significant correlation between 
$p_{K}$ and
$B_{J}-K_{n}$ colour. 
Together with the previous conclusion, this suggests the reddening in 
$B_{J}-K_{n}$ colour can be explained by an intrinsically `red'
synchrotron component dominating the near-IR flux.
We believe this is due to a selection effect imposed by the 
relatively bright flux limit in the near-IR required to achieve reliable
polarisation measures. 
\\\indent 4. For the five quasars in which a broadband polarisation spectrum 
in the near-IR can be deduced, the 
polarisation level is observed to decrease towards 
shorter wavelengths.
The level of polarisation decreases by more than 40\% from $K_{n}$ to
$H$ in most cases. The three spectroscopically identified 
BL-Lacs also exhibit a `contemporaneous' polarisation curve in the
near-IR with a similar wavelength dependence.
If these BL-Lacs are in fact dominated by a (beamed) synchrotron
continuum, then the intrinsic polarisation of this component must also
decrease towards shorter wavelengths.
This is contrary to previous modelling of BL-Lacs where $dp/d\lambda\leq0$ 
was often implied by the data. 
\\\indent 5. Both the strength and wavelength dependence of polarisation in
our two moderately reddened quasars cannot be explained by transmission through 
aligned
dust grains like that observed in diffuse regions of the galaxy. 
This mechanism is characterised by $dp/d\lambda<0$, opposite to that observed. 
Polarisation by scattering from small grains or electrons is also
inconsistent with the polarisation wavelength dependence.
A larger sample of reddened quasars is needed to test such models.
\\\indent 6. By combining our 
polarisation spectra with observational constraints
on emission line EWs and total flux spectra from contemporaneous photometric
measurements, the data can be well explained using a simple
two component model.
This consists of an unpolarised, flat-spectrum `Big Blue Bump' (with
$f_{\nu}\propto\nu^{-0.3}$), and a steep-spectrum synchrotron component 
($f_{\nu}\propto\nu^{-1.5}$ to $\nu^{-3}$) with wavelength independent
intrinsic polarisation.
Our results are in general consistent with previous modelling of
flat-spectrum radio quasars.
\\\indent 7. Only one of our optically `reddened' quasars, 
PKS1036$-$154, has a total flux spectrum that cannot be explained 
using the two component model alone, while 
its polarisation and line EW data can be. 
Equal reddening of both the BBB and synchrotron component by dust with 
$\tau_{B}\simeq0.7$ 
is required to satisfy the total flux spectrum.

Although we are unable to make any strong conclusions on the nature of the
reddening mechanism in Parkes quasars on the basis of these few sources alone,
we have shown that near-IR polarimetry can provide a powerful diagnostic
for constraining various reddening models.
Given that the polarisation properties
of mostly `blue' radio quasars studied up to date
are moderately well known, future polarimetric
studies shall now be aimed towards the reddest (and faintest) quasars
in the Parkes sample.
\chapter{Conclusions and Future Prospects} 

\vspace{1mm}
\leftskip=4cm
 
{\it ``Upon a slight conjecture I have ventured on a dangerous journey,
and I already behold the foothills of new lands. Those who have
the courage to continue the search will set foot upon them.''}
 
\vspace{1mm}
 
\hfill {\bf --- Immanuel Kant,}

\leftskip=56mm
\noindent
{\it Allgemeine Naturgeschichte und Theorie des\\ 
Himmels, 1755\\}
 
\leftskip=0cm

\section{Conclusions}

This thesis has explored two issues: the problem of incompleteness in
optical quasar surveys due to obscuration by dust, and the interpretation
of the reddening observed in a complete sample of radio quasars 
(the `Parkes sample'). Part I investigated the observational consequences
of an intervening cosmological dust component and uses the 
Parkes sample to constrain the amount of dust involved. 
Part II examined the continuum properties of Parkes quasars in the
framework of a number of absorption and emission mechanisms to assess the
importance of extinction by dust.
The key results are as follows:

\subsection{{\it Part I:} Cosmologically Distributed Dust}

\begin{itemize}

\item
Dust is more effective at obscuring background sources when diffuse or 
extended. Obscuration of background sources in the optical will be
most effective when $\tau_{B}<1/\beta$, where $\beta$ is the slope
of their cumulative luminosity function. 
This criterion is satisfied for quasars observed 
through diffusely distributed dust on scales typical of rich galaxy
clusters. The presence of an undetected `diffuse' IGM component,
even of moderately low density, will severely affect optical observations
of the background universe at $z\simgt2$. 

\item
Obscuration due to dust in galaxies randomly distributed along the 
line-of-sight is strongly model dependent, but can still be significant with
minimal assumptions.
Models assuming no evolution in dust content, and dust galaxy properties
at the lower limits of current observations predict at least 10\% of
quasars at $z>3$ to be under-represented in existing optical surveys. 
Optically-selected quasars are strongly biased against reddening by
line-of-sight dust, and are expected to exhibit only a few percent of the
total reddening to any redshift.

\item
We have explored the implications on Mg{\small II} and C{\small IV} 
absorption-line statistics
using our intervening dusty galaxy model. A local calibration
for the relation between absorption-line equivalent width and dust 
reddening was assumed.
We find that up to 40\% of lines
with equivalent widths $\simgt0.5$\AA$\,$ (the typical median)
may have been missed
in metal absorption-line surveys by dust in 
systems obscuring background QSOs. 

\item
The reddening observed in Parkes quasars is
independent of redshift and cannot be fully explained by a non-evolving
distribution of dusty galaxies along the line-of-sight. Motivated by this
observation, we have simulated the effects of galactic obscuration
assuming evolution in the dust content. 
For evolutionary parameters consistent with previous studies of the evolution
of heavy element QSO absorption-line systems and 
cosmic chemical evolution models,
the mean optical depth to redshifts $z>1$ is smaller by at least a 
factor of 3 compared to non-evolving models. 
Obscuration by dust is not as severe as found in previous studies 
if effects of evolution are accounted for. 

\item
Dust models which allow for even moderately low amounts of evolution do
not have a significant effect on existing optical studies of QSO evolution. 
Correcting for dust obscuration, evolving dust models
predict the `true' (intrinsic) space density of bright quasars to
decrease beyond $z\sim2.5$, as observed, contrary to 
previous non-evolving models where a continuous monotonic
increase was predicted.
Nonetheless, the enhancement in true QSO numbers over that observed may 
still be significant in evolving dust models. 

\item
We have constrained this generalised evolutionary model using the 
Parkes
quasar data.  
The `flat' redshift dependence of the mean reddening observed is well
explained with evolutionary parameters consistent with other studies. 
We require a sky covering factor in dusty absorbers to some redshift that 
scales as $\sim(1+z)^{0.15}$, as compared to $\sim(1+z)^{1.5}$ for
non-evolving models in a $\Omega=1$ ($\Lambda=0$) universe. 
The fraction of sources reddened at any redshift however cannot be explained
using the dust properties of present-day galaxies. More than 50\% of the
sample is reddened with $\tau_{B}\simgt1$ to $z\sim0.5$. 
This requires a covering factor 
in dusty absorbers more than two
orders of magnitude greater than that predicted assuming the 
average properties of local galaxies.  
The amount of dust required is inconsistent with theoretical
constraints derived for the local mass density in metals, 
observed limits on galactic
dust-to-gas ratios, and the far-IR background radiation. 
The reddening cannot be {\it totally} due to dust located
in extrinsic systems that may happen to lie in the line-of-sight
of otherwise `normal' blue quasars. 

\end{itemize}

\subsection{{\it Part II:} The Reddening Mechanism in Parkes Quasars}

\begin{itemize}

\item
We have explored the relatively large dispersion in continuum slopes
of Parkes sources in the framework of three independent 
reddening models: `intrinsically red' AGN emission models, the 
dust extinction model, and the `host galaxy' red emission model. 
`Intrinsically red' synchrotron emission, that may be characteristic  
of flat spectrum radio-quasars was the primary mechanism of the
class of red AGN emission models explored.
A comparison of the data
with simple models shows that the reddening in sources
exhibiting {\it quasar-like} properties
can be better explained
by dust extinction rather than `purely' by an intrinsic
emission property. 

\item
The reddening cannot be explained by emission from a red stellar component
in the host galaxy as motivated by the characteristic red colours of
ellipticals associated with radio galaxies. 
To explore this, a new spectral fitting method was developed which measures
the relative galaxy contribution in an AGN using the size
of the 4000\AA$\,$ break feature characteristic of `old' elliptical SEDs.

\item
The following evidence suggests that extinction by dust 
contributes to the observed reddening:
spectral curvatures from the near-IR to UV (rest frame) that 
are consistent with the generic $1/\lambda$ extinction law,
Balmer decrements that correlate with reddening, 
and soft X-ray absorption at a level 
consistent with the optical reddening expected from
metal abundances and dust-to-gas ratio characteristic of the galactic ISM.

\item
Our analysis of the soft X-ray data in Parkes quasars
suggests that QSO surveys selected at soft X-ray energies
are expected to be strongly biased (like those selected optically) 
against sources with significant amounts of dust in their line-of-sight. 

\item
The amount of dust extinction required is
typically $\langle A_{V}\rangle\simeq2$. 
This is considerably smaller than that invoked for obscuring
molecular tori in AGN and
is consistent with unified models since our line-of-sight to
the central AGN in 
flat spectrum radio-quasars is not expected to intercept a torus. 
The observed degree of reddening 
argues in favour of an `optically-thin'
diffuse dust component with grain size distribution dominated
by small grains, similar to that invoked for the local ISM. 

\item
The reddening in sources exhibiting weak-lined BL-Lac type properties however
is likely to be purely intrinsic (eg. beamed synchrotron) in origin.
A beamed synchrotron component
dominating the near-IR flux in the {\it quasars} (presumably
to a lesser extent than in the BL-Lacs) cannot be ruled out
completely.
A combination of both dust absorption and intrinsically red emission
could be present.

\item
A near-infrared polarisation study of mostly `blue' and `moderately
reddened' quasars shows that significant linear polarisation levels
($\simgt5\%$) are present, suggesting that the near-IR may 
indeed be dominated by a synchrotron
(possibly beamed) component. 
This is supported by correlations of polarisation level
with $B-K$ colour and variations in $K$-flux on monthly timescales. 
We must emphasize that this finding may be due to a selection effect
imposed by the 
relatively bright flux limit in the near-IR required to achieve reliable
polarisation measures. 

\item
The polarisation level is observed to decrease by more than 40\%
from $K$(2.1$\mu$m)
to $H$(1.3$\mu$m) in the quasars. 
Both the strength and wavelength dependence of polarisation
observed in our `moderately reddened' quasars cannot be explained by
transmission through, or scattering off aligned dust grains.
The data are consistent with previous studies of flat spectrum
radio-quasars 
using
a two component model: 
an unpolarised `big blue bump' (BBB), and a steep spectrum, synchrotron 
component dominating in the near-IR. 

\item

Only one of our optically `reddened' quasars, 
PKS1036$-$154, has a total flux spectrum that cannot be explained 
using this simple two component model alone, while its polarisation
data can be.
Equal reddening of both the BBB and synchrotron component by dust with 
$\tau_{B}\simeq0.7$
is required to satisfy its 
total flux spectrum. 
The modelling of this source shows that near-IR polarimetry
can provide a powerful diagnostic for constraining various reddening
mechanisms and source geometries in radio-quasars. 

\end{itemize}

\newpage
\section{Further Work and Future Prospects}

This section outlines some avenues for further research drawn from
my work on Parkes quasars. 
Observational prospects and 
theoretical issues centered on the 
dust extinction hypothesis are primarily discussed. 

\subsection{Intervening Cosmological Dust} 

Radio quasars provide a random sample of sight-lines to the
high redshift universe with no bias against extinction by dust. 
Such sources can therefore be used to assess incompleteness in studies
which use bright optical quasars to detect absorption lines
and gravitational lens systems and measure the evolution of
gas and dust with redshift. 
Previous studies of the relative abundances of heavy elements in QSO 
absorption-line systems
suggest that dust is present at a level which roughly scales in proportion to 
metallicity 
(see section~\ref{hzdev}).
A survey for heavy element absorption in Parkes quasars
will thus provide the strongest constraint on the contribution of 
an extrinsic, cosmologically distributed dust component to the 
observed continuum reddening. 

The detection of a sufficiently large number of absorption systems
unassociated with the quasars will be of enormous potential value 
in studying the global evolution of 
metallicity and
star formation with cosmic time in an `unbiased' way. 
Detection of absorption lines however, particularly those associated with 
heavy metals is rather difficult. Optical spectra of
high signal-to-noise ratio and 
resolutions typically less than a few Angstroms are required to
detect even the strongest lines (eg. the resonance doublets 
C{\small IV}$\lambda\lambda$1548, 1551 and 
Mg{\small II}$\lambda\lambda$2795, 2802). 
Due to the faintness of many of the reddest Parkes quasars in the optical
($B_{J}\simgt22$), 
this calls for the high sensitivity and resolution 
capabilities of {\it HST}.

\subsection{Dust Geometry and Location in the Quasar Environment} 

The possibility that dust resides in the local environs of
flat-spectrum
radio quasars
raises
questions about its configuration, kinematic and chemical properties. 
These issues can be addressed using observational diagnostics
from which most of our existing knowledge of the location and geometry of 
circumnuclear gas and dust in AGN has been derived.
Observational tests to constrain possible chemical properties were 
discussed in section~\ref{predchem}.
Here we shall present a number of tests that may constrain the dust
geometry and/or location.

The near-infrared, optical and soft X-ray
data on Parkes quasars 
implies the following observational constraints:
the dust must lie outside the
broad line region (consistent with the minimum sublimation radius), 
be distributed as an `optically thin' diffuse medium with 
$\langle A_{V}\rangle\sim2$,  
and be dominated by small grains ($a\simlt0.01\mu$m). 
Since more than 50\% of the sample is reddened 
(with $A_{V}\simgt2$), the material is likely to have a large covering 
factor.

\begin{figure}
\vspace{-0.8in}
\plotonesmall{1.1}{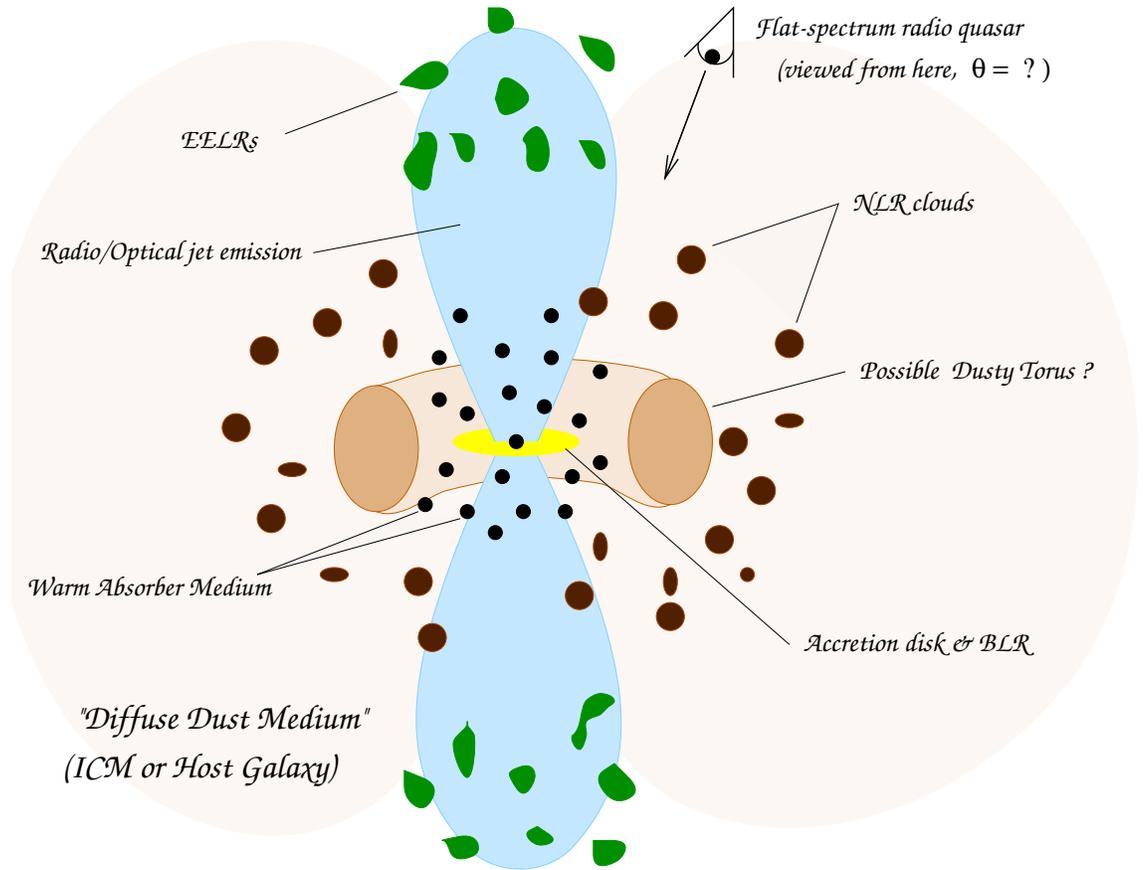}
\vspace{0.3in}
\caption[Schematic showing possible locations of the dusty medium]{Schematic showing possible locations of the
dusty medium in Parkes quasars. According to
simple unified models for radio-loud AGN, these sources are believed to
be observed more-or-less in the
direction shown.}
\label{geomfs}
\end{figure}

With these constraints, the most likely locations (or geometries)
of the dust are:
(i) the `Warm X-ray Absorber' (or warm scattering medium) 
(ii) the narrow line region (either in clouds or the intercloud medium),
(iii) the host galaxy, (iv) in a hydromagnetic, 
centrifugally disc-driven wind (K\"{o}nigl \& Kartje, 1994), or 
(v) in extended emission line regions (EELRs). 
A schematic diagram illustrating some of these regions 
is shown in Fig.~\ref{geomfs}. 
Some of these possibilities can be explored using the following
observational tests:

\subsubsection{\large\it A Warm Absorber?}

A prominent feature in the X-ray spectrum of many type-1 
AGN and BL-Lacs is absorption
by partially ionized, optically thin material
(Pan, Stewart \& Pounds, 1990; Madejski \etal 1991).
This is also believed to represent 
the scattering medium implied by polarisation studies of Seyfert-2 galaxies. 
The high ionization state of this `warm' absorber suggests it to be 
photoionized material of temperature $T\simgt10^{4}$K situated within 
$\sim10$pc of the primary continuum source (Reynolds \& Fabian, 1995).
This feature is usually identified with K-absorption edges 
at (rest energies) 0.7-1keV due to O{\small VII} 
and O{\small VIII} and excess emission below 
$\sim0.7$keV due to a reduction in absorber opacity. 
K-shell absorption edges from highly ionized iron are also sometimes seen.

In a recent study of type-1 AGN, Reynolds (1996) 
found that sources displaying significant optical reddening also
displayed the strongest OVII edges.
This suggests the existence of dusty warm photoionized plasma.
Dusty warm absorbers were also claimed by Brandt \etal (1996) and 
Komossa \& Fink (1997).
Their origin are explained 
as due to a radiatively driven wind from
a molecular torus. 
Due to the tenuous and `photoionization dominated' state of the warm 
highly ionized gas, 
these authors show
that dust can easily exist, avoiding 
destruction against sputtering for a wide range of physical conditions. 

Motivated by these results, one can also conduct an 
an X-ray spectral study to
search for signatures of  
warm absorbers in reddened Parkes quasars. 
Due to their faintness in the 
{\it ROSAT} energy band however, 
this may be difficult with existing instruments such as 
{\it Ginga} and {\it ASCA}.  
The high signal-to-noise spectra required may be achieved 
with the {\it AXAF} and {\it XMM} satellites.

\subsubsection{\large\it The Narrow Line Region (NLR)}

There is considerable evidence suggesting 
that the (0.1-1kpc scale ) NLR of AGN contains dust
(eg. Wills \etal 1993). Netzer \& Laor (1993) have shown that
if dust absorption is taken into account, the narrow line emission
implies covering factors $\simgt30\%$ for the NLR gas. 
Photoionization modelling and line intensity ratios suggest low to
moderately low optical depths ($A_{V}\sim0.1-1$) to the central AGN. 
Emission line ratios provide the most powerful
diagnostic for detecting dust in the NLR. 
Draine \& Bahcall (1981) present an exhaustive list of line pairs 
to measure dust extinction in AGN.
If lines of different ionization state and/or velocity 
width can be used as a distance indicator from the
photoionizing continuum source, then line ratios can also 
be used to constrain the spatial dust distribution.  

The profiles of narrow emission lines can also provide a useful diagnostic.
The narrow line spectra of Seyfert galaxies, and to a lesser extent
high redshift quasars often exhibit lines showing some degree of asymmetry.
Both `redward' (excess red flux) and `blueward' asymmetries are found 
(eg. Boroson \& Green, 1992).
These are interpreted as due to a combination of net radial motion of the
clouds and some extended source of opacity such as dust. 
The observed asymmetries imply modest extinctions $A_{V}\sim1$. 
Correlations with other reddening indicators may also
constrain the dust distribution. 

Our existing optical spectra are not of 
sufficient high signal-to-noise and resolution to detect any
narrow lines favourable for the above diagnostics.
A high resolution spectral study of just
a few reddened Parkes quasars will provide vital clues as to whether
reddening by NLR dust is important.

\subsubsection{\large\it Extended Emission Line Regions (EELRs)}

Many radio-loud AGN show extended line emission on scales from
1kpc to hundreds of kpc, often cospatial with radio and optical jets
(Baum \& Heckman, 1989). 
The line emitting gas is distributed rather
inhomogeneously and is likely to be an extension of the smaller scale
NLR.
Its coincidence with the radio morphology 
and larger than usual velocity widths
suggests that shock heating by non-thermal radio plasma
also plays a r\^{o}le. 
A recent study
of the EELRs of a number of nearby radio galaxies has
found evidence for dust mixed in with the gas 
(Villar-Mart\'{\i}n \& Binette, 1996). 
This was inferred from the relative faintness and absence of 
(dust sensitive) calcium lines,
suggesting severe depletion onto grains. 

If flat spectrum radio-quasars are interpreted as 
sources with radio jets pointing more-or-less along the line-of-sight,
then it is possible that our line-of-sight to the central AGN
intercepts a large fraction of 
dusty EELR gas. 
A large number of Parkes quasars show resolved radio 
structure on arcsecond scales
to millijansky flux levels. 
An obvious test would then be to look for spatially extended
line emission on scales coincident with radio morphologies. 
Optically thin dust cospatial with this emission would produce
a diffuse halo from scattering of line and continuum radiation
into the line-of-sight. 
This project would be possible with the 
recent emergence of high quality Fabry-Perot etalons for 
high resolution imaging spectroscopy. 
The capabilities of {\it HST} however would be necessary for 
narrow line imaging of higher
redshift quasars. 

\subsection{Some Theoretical Considerations}

The presence of dust physically associated with 
flat-spectrum radio quasars also
raises a number of theoretical issues:
What is its origin? Is it external (eg. from a cooling flow),
or does it originate from the ISM of the host galaxy? 
What are its implications for chemical evolution 
of the associated gas and hence
star formation history? 
How does it interact with the radiation field and plasma in
jets and associated shocks?
Does it provide an efficient cooling mechanism and
a means of transporting angular momentum from the central engine?
These latter issues could be explored through simulations of 
(relativistic) jets in a dusty medium. 

Is radiation pressure from the primary continuum source
effective in dispersing grains to large scales?
This will depend on whether grains are strongly coupled
to magnetic fields or gas, or concentrated
in gas clouds so that 
the inward gravitational force of the  
central black hole will
dominate over radiation pressure. 
For gravity to dominate, 
grains may have to be embedded in massive and
hence optically-thick clouds to avoid ejection from the nuclear region.
This would be inconsistent with the requirement 
of an `optically-thin' medium in Parkes quasars. 
If these sources are indeed enshrouded by a
dust `cocoon' on large enough scales, then
the observed SEDs of jets 
from optical to soft X-ray energies are likely to be modified.
Current models of the 
high frequency SEDs of BL-Lacs would thus need to be revised.

\subsection{Reddened {\it Radio-Quiet} quasars?}

On the basis of a dichotomy in the
ratio of optical to radio emission, 
over 80\% of optically-selected QSOs are classified as radio-`quiet'.
If the diversity in continuum slopes in Parkes quasars is not
an intrinsic property of the sources but due to 
reddening by dust, then it is quite plausible
that radio-quiets are also
reddened by comparable amounts of dust. 
The narrow range in continuum slopes observed for optically-selected QSOs
suggests that existing bright optical surveys may be 
severely incomplete.
The observed spread in continuum slopes of Parkes quasars predicts
an incompleteness of least 50\% in 
a typical QSO survey
of magnitude limit $B_{J}\sim19$. 

How can a reddened population of radio-quiet quasars be detected?
Since the extinction 
implied by reddening in Parkes quasars is 
relatively `low', this population need {\it not} 
comprise the ultraluminous IRAS galaxies (and type-2 AGN) 
believed to contain buried, heavily obscured quasars. 
Surveys at wavelengths where no or little bias against
dust obscuration would be optimal.
Optical identifications of {\it hard} ($\simgt10$keV) X-ray selected samples 
which are known to contain a high fraction of extragalactic objects  
and quasars would be promising.
However, the recent emergence of large format, 
high quality CCDs and multi-object
fibre spectrographs such as
{\it 2dF} on the {\it AAT} will provide
the necessary technology to search for faint 
(possibly reddened) quasars. 

\newpage
\section{Summary}

This thesis has studied the problem of incompleteness in optical
quasar surveys due to obscuration by dust.
This was primarily explored using optical and near-infrared observations of 
a large complete sample of radio-selected quasars.
I have developed simple models and
novel techniques
to explain the properties of these sources 
in the framework of a number
of absorption and emission mechanisms.
Previously unknown correlations between spectral properties 
were predicted,
and some identified observationally. 
These I hope will provide
powerful diagnostics for constraining quasar models in future.
Amongst the mechanisms believed to 
operate in radio-loud quasars,
the data strongly suggests that 
a substantial number are
reddened optically by several magnitudes of 
dust extinction, mostly in their environs.
To determine whether radio-quiet quasars have similar amounts of
dust in their line-of-sight 
will require much observational work.
This study has substantial room for expansion and should further improve
our knowledge of the nature and distribution of quasars.
\appendix

\chapter{References}

\begin{center}
{\large ABBREVIATIONS}
\end{center}

\begin{tabbing}
\hspace{1.7cm}{\it A\&A\hspace{2.5cm}}\={\it Astronomy \& Astrophysics}\\ 
\hspace{1.7cm}{\it A\&ASuppl}  \>{\it Astronomy \& Astrophysics Supplements}\\
\hspace{1.7cm}{\it A\&ARv}   \>{\it Astronomy \& Astrophysics Review}\\
\hspace{1.7cm}{\it AJ}       \>{\it Astronomical Journal}\\
\hspace{1.7cm}{\it ApJ}      \>{\it Astrophysical Journal}\\
\hspace{1.7cm}{\it Ap\&SS}   \>{\it Astrophysics \& Space Science}\\
\hspace{1.7cm}{\it ApJS} \>{\it Astrophysical Journal Supplements}\\
\hspace{1.7cm}{\it ARA\&A} \>{\it Annual Review of Astronomy and Astrophysics}\\
\hspace{1.7cm}{\it BAAS} \>{\it Bulletin of the American Astronomical Society}\\
\hspace{1.7cm}{\it Fund. Cos. Phys.} \>{\it Fundamentals of Cosmic Physics}\\
\hspace{1.7cm}{\it Lick Obs. Bull.}\>{\it Lick Observatory Bulletin}\\
\hspace{1.7cm}{\it Lund Obs. Not.}\>{\it Lund Observatory Notices}\\ 
\hspace{1.7cm}{\it MNRAS}\>{\it Monthly Notices of the 
Royal Astronomical Society}\\
\hspace{1.7cm}{\it PASA}\>{\it Publications of the Astronomical Society
of Australia}\\
\hspace{1.7cm}{\it PASP}\>{\it Publications of the Astronomical Society
of the Pacific}\\
\hspace{1.7cm}{\it P\&SS} \>{\it Physics \& Space Science}\\
\hspace{1.7cm}{\it Phys. Lett. B} \>{\it Physics Letters B}\\
\hspace{1.7cm}{\it Phys. Rep.} \>{\it Physics Reports}\\
\hspace{1.7cm}{\it Rep. Prog. Phys.} \>{\it Reports \& Progress in Physics}\\
\hspace{1.7cm}{\it Rev. Mod. Phys.} \>{\it Reviews of Modern Physics}
\end{tabbing}

\newpage
 
\def\paper#1#2#3#4#5{\hangindent=3pc \hangafter=1#1, #2. {\it #3}, {\bf #4},
#5.}
 
{\parindent=0pt

\paper{Allamandola,L.J., Tielans,A.G.G.M. \& Barker,J.R.}{1985}{ApJ}{290}{25}

\paper{Allen,S.W.}{1995}{MNRAS}{276}{947}

\paper{Angel,J.R.P. \& Stockman,H.S.}{1980}{ARA\&A}{18}{321}

\paper{Antonucci,R. \& Miller,J.S.}{1985}{ApJ}{297}{621}

\paper{Antonucci,R.}{1993}{MNRAS}{31}{473}

\paper{Awaki,H., Koyama,K., Kunieda,H. \& Tawara,Y.}{1990}{Nature}
{346}{544}

\paper{Babul,A. \& Rees,M.J.}{1992}{MNRAS}{255}{346}

\paper{Bahcall,J.N., Kirhakos,S., Saxe,D.H. \& Schneider,D.P.}
{1995}{ApJ}{479}{642}

\paper{Ballard,K.R., Mead,A.R.G., Brand,P.W.J.L., Hough,J.H.,
Bailey,J.A. \& Brindle,C.}{1988}{in BL Lac Objects}{\rm Proceedings: Como,
Italy}{p.181} 

\paper{Bartelmann,M. \& Schneider,P.}{1993}{A\&A}{268}{1}

\paper{Bartelmann,M., Schneider,P. \& Hasinger,G.}{1994}{A\&A}{290}{399}

\paper{Barvainis,R.}{1987}{ApJ}{320}{537}

\paper{Barvainis,R.}{1993}{ApJ}{412}{513}

\paper{Baum,S.A. \& Heckman,T.M.}{1989}{ApJ}{336}{702}

\paper{Beichman,C.A., Pravdo,S.H., Neugebauer,G., Soifer,B.T., Matthews,K.
\& Wooten,A.A.}{1981}{ApJ}{247}{780} 

\paper{Bergeron,J. \& Boisse,P.}{1991}{A\&A}{243}{344}

\paper{Bj\"{o}rnsson,C. \& Blumenthal,G.R.}{1982}
{ApJ}{259}{805}

\paper{Blain,A.W. \& Longair,M.S.}{1993a}{MNRAS}{264}{509}
 
\paper{Blain,A.W. \& Longair,M.S.}{1993b}{MNRAS}{265}{L21}

\paper{Blandford,R.D. \& Rees,M.J.}{1978}{in Pittsburgh Conf. on BL-Lac Objects}
{\rm ed. A.M. Wolf (Pittsburgh: Univ. Pittsburgh, Dept. of Physics and Astronomy)}
{p.328}

\paper{Blandford,R.D. \& Kochanek,C.S.}{1987}{in Dark Matter in the 
Universe}{\rm eds. J.Bahcall,T.Piran \& S.Weinberg}{p.133}

\paper{Blandford,R.D. \& Eichler,D.}{1987}{Phys. Rep.}{154}{1}

\paper{Blandford,R.D. \& K\"{o}nigl,A.}{1979}{ApJ}{232}{34}

\paper{Blumenthal,G.R. \& Gould,R.J.}{1970}{Rev. Mod. Phys.}{42}{237}

\paper{Bogart,R.S. \& Wagoner,R.V.}{1973}{ApJ}{181}{609}

\paper{Bohlin,R.C., Savage,B.D. \& Drake,J.F.}{1978}{ApJ}{224}{291}

\paper{Boroson,T.A. \& Green,R.F.}{1992}{ApJS}{80}{109}

\paper{Bouchet,P., Lequeux,J., Maurice,E., Pr\'{e}vot,L. \&
Pr\'{e}vot-Burnichon,M.L.}{1985}{A\&A}{149}{330}

\paper{Boyle,B.J., Fong,R. \& Shanks,T.}{1988}{MNRAS}{231}{897}

\paper{Boyle,B.J., Shanks,T. \& Peterson,B.A.}{1988}{MNRAS}
{235}{935}

\paper{Boyle,B.J., Fong,R., Shanks,T. \& Peterson,B.A.}{1990}
{MNRAS}{243}{1}

\paper{Boyle,B.J. \& di Matteo,T.}{1995}{MNRAS}{277}{L63}

\paper{Brandt,W.N., Fabian,A.C. \& Pounds,K.A.}{1996}{MNRAS}{278}{326}

\paper{Bregman,J.N., Lebofsky,M.J., Aller,M.F., Rieke,G.H., Aller,H.D., 
Hodge,P.E., Glassgold,A.E. \& Huggins,P.J.}{1981}{Nature}{293}{714}

\paper{Bregman,J.N. \etal}{1986}{ApJ}{301}{708}

\paper{Brindle,C., Hough,J.H., Bailey,J.A., Axon,D.J. \& Hyland,A.R.}{1986}
{MNRAS}{221}{739}

\paper{Bruzual,G. \& Charlot,S.}{1993}{ApJ}{405}{538}

\paper{Burbidge,E.M. \& Strittmatter,P.A.}{1972}{ApJ}{174}{L57}

\paper{Byun,Y.I.}{1993}{PASP}{105}{993}

\paper{Calzetti,D., Kinney,A.L. \& Storchi-Bergmann,
T.}{1994}{ApJ}{429}{582}

\paper{Canizares,C.R. \& White,J.L.}{1989}{ApJ}{339}{27}

\paper{Chang,C.A., Schiano,A.V.R. \& Wolfe,A.M.}{1987}{ApJ}{322}{180}

\paper{Cheney,J.B. \& Rowan-Robinson,M.}{1981}{MNRAS}{195}{831}

\paper{Chini,R., Kr\"{u}gel,E., Kreysa,E. \& Gemund,H.P.}{1989}{A\&A}{216}{L5} 

\paper{Chini,R. \& Kr\"{u}gel,E.}{1994}{A\&A}{288}{L33}

\paper{Ciliegi,P., Elvis,M. Wilkes,B.J. Boyle,B.J, McMahon,R.G.
\& Maccacaro,T.}{1995}{MNRAS}{277}{1463}

\paper{Clayton,G.C., Martin,P.G. \& Thompson,I.}{1983}{ApJ}{265}{194}

\paper{Cotton,W.D.}{1983}{ApJ}{271}{51}

\paper{Cohen,M.H. \& Vermeulen,R.C.}{1992}{in Proc. Conf. on 
Extragalctic Radio Sources--From Beams to Jets}{\rm ed. E. Feigelson}
{Paris:Inst. d'Astrophys}

\paper{Crampton,D. (Editor)}{1991}{in the Space Distribution of Quasars}
{\rm ASP Conf. series}{vol.21}

\paper{Cruz-Gonzalez,I. \& Huchra,J.P.}{1984}{AJ}{89}{441}

\paper{De Zotti,G., Franceschini,A., Mazzei,P., Toffolatti,L. \&
Danese,L.}{1995}{P\&SS}{43}{1439}

\paper{Dickman,R.L.}{1978}{ApJS}{37}{407}

\paper{Disney,M., Davies,J. \& Phillipps.S.}{1989}{MNRAS}{239}{939}

\paper{Djorgovski,S., Thompson,D. \& Smith,J.D.}{1993}{in First light in the
Universe: Stars or QSOs? p.67}{\rm ed. B. Rocca-Volmerange, M. Dennefeld, B. Guiderdoni
\& J. Tran Thanh Van}{Gif-sur-Yvette: Editions Frontieres}

\paper{Dopita,M.A.}{1995}{Ap\&SS}{233}{215} 

\paper{Draine,B.T. \& Salpeter,E.E.}{1979a}{ApJ}{231}{77}

\paper{Draine,B.T. \& Salpeter,E.E.}{1979b}{ApJ}{231}{438}

\paper{Draine,B.T.}{1981}{ApJ}{245}{880}

\paper{Draine,B.T. \& Bahcall,J.N.}{1981}{ApJ}{250}{579}

\paper{Draine,B.T. \& Lee,H.M.}{1984}{ApJ}{285}{89}

\paper{Draine,B.T.}{1985}{ApJS}{57}{587}

\paper{Drinkwater,M.J., Webster,R.L. \& Thomas,P.A.}{1993}
{AJ}{106}{848}

\paper{Drinkwater,M.J., Webster,R.L., Francis,P.J., Condon,J.J., Ellison,S.L.,
Jauncey,D.L., Lovell,J., Peterson,B.A. \& Savage,A.}{1997}{MNRAS}{284}{85}

\paper{Drinkwater,M.J., Combes,F. \& Wiklind,T.}{1996}{A\&A}{312}{771}

\paper{Drury,L.O.}{1983}{Rep. Prog. Phys.}{46}{973}

\paper{Duncan,R.A., White,G.L., Wark,R., Reynolds,J.E., Jauncey,D.L.,
Norris,R.P., Taafe,L. \& Savage,A.}{1993}{PASA}{10}{310}

\paper{Dunlop,J.S., Peacock,J.A., Savage,A., Lilly,S.J., Heasley,J.N.
\& Simon,A.J.B.}{1989}{MNRAS}{238}{1171}

\paper{Dunlop,J.S., Guiderdoni,B., Rocca-Volmerange,B., Peacock,J.A.
\& Longair,M.S}{1989}{MNRAS}{240}{257}

\paper{Dunlop,J.S. \& Peacock,J.A.}{1990}{MNRAS}{247}{19}

\paper{Dwek,E., Rephaeli,Y. \& Mather,J.C.}{1990}{ApJ}{350}{104}

\paper{Eales,S.A. \& Rawlings,S.}{1993}{ApJ}{411}{67}

\paper{Edelson,R.A., Malkan,M.A. \& Rieke,G.H.}{1986}{ApJ}{321}{233}

\paper{Edelson,R.A. \& Malkan,M.A.}{1987}{ApJ}{323}{516}

\paper{Edmunds, M.G.}{1990}{MNRAS}{246}{678}

\paper{Efstathiou,G., Ellis,R.S. \& Peterson,B.A.} 
{1988}{MNRAS}{232}{431}

\paper{Elvis,M., Fiore,F., Wilkes,B. \& McDowell,J.}{1994}{ApJ}{422}{60}

\paper{Elvis,M., Wilkes,B., McDowell,J., Green,R.F., Bechtold,J.,
Willner,S.P., Oey,M.S., Polomski,E. \& Cutri,R.}{1994}{ApJS}{95}{1} 

\paper{Elvius,A.}{1978}{Ap\&SS}{55}{49}

\paper{Fabian,A.C., Nulsen, P.E.J. \& Canizares,C.R.}{1991}{A\&ARv}{2}{191}

\paper{Fall,S.M., Pei,Y.C. \& McMahon, R.G.}{1989}{ApJ}{341}{L5}

\paper{Fall,S.M. \& Pei,Y.C.}{1989}{ApJ}{337}{7}

\paper{Fall,S.M. \& Pei,Y.C.}{1993}{ApJ}{402}{479}

\paper{Fall,S.M. \& Pei,Y.C.}{1995}{in QSO Absorption Lines}{\rm
ed: G.Meylan}{Berlin: Springer-Verlag}

\paper{Ferland,G.J. \& Netzer,H.}{1983}{ApJ}{264}{105}

\paper{Fireman,E.L.}{1974}{ApJ}{187}{57}

\paper{Fitzpatrik,E.L. \& Massa,D.}{1990}{ApJS}{72}{163}

\paper{Franceschini, A., Mazzei P. \& De Zotti G.}{1994}{ApJ}{427}{140}

\paper{Francis,P.J., Hewett,P.C., Foltz,C.B., Chaffee,F.H.,
Weymann,R.J. \& Morris,S.L.}{1991}{ApJ}{373}{465}

\paper{Francis,P.J.}{1996}{PASA}{13}{212} 

Francis,P.J., Webster,R.L., Drinkwater,M.J., Masci,F.J. \& Peterson,B.A., 1997. 
\hspace*{12mm}{\it MNRAS}, in press.

\paper{Freeman, K.C.}{1970}{ApJ}{160}{811}

\paper{Fritz,K.D.}{1989a}{A\&A}{214}{14}

\paper{Fritz,K.D.}{1989b}{ApJ}{347}{692}

\paper{Giovanelli,R., Haynes,M.P., Salzer,J.J., Wegner,G., Da Costa,L.N.
\& Freudling,W.}{1994}{AJ}{107}{2036}

\paper{Glazebrook,K., Peacock,J.A., Miller,L. \& Collins,C.A.}{1995}{MNRAS}
{275}{169}

\paper{Goudfrooij,P. \& de Jong,T.}{1995}{A\&A}{298}{784} 

\paper{Goudfrooij,P., de Jong,T., Hansen,L. \& N\o rgaard-Nielsen,H.U.}
{1994a}{MNRAS}{271}{833}

\paper{Goudfrooij,P., Hansen,L., Jorgensen,H.E. \& N\o rgaard-Nielsen,H.U.}
{1994b}{A\&ASuppl}{105}{341}

\paper{Green,P.J., Schartel,N., Anderson,S.F., Hewett,P.C., Foltz,C.B.,
Brinkmann,W., Fink,H. Tr\"{u}mper,J. \& Margon,B.}{1995}{ApJ}{450}{51}

\paper{Greenberg,J.M.}{1968}{in Nebulae \& Interstellar Matter,
Stars and Stellar systems Vol. VII}{\rm eds. B.M. Middlehurst \& L.H. Aller,}
{University of Chicago Press. p.221}

\paper{Greenberg,J.M.}{1971}{A\&A}{12}{240}

\paper{Guiderdoni,B. \& Rocca-Volmerange,B.}{1987}{A\&A}{186}{1}

\paper{Hartwick,F.D.A. \& Schade,D.}{1990}{ARA\&A}{28}{437}

\paper{Katz,N., Quinn,T., Bertschinger,E. \& Gelb,J.M.}{1994}{MNRAS}
{270}{L71}

\paper{Heavens,A.F. \& Meisenheimer,K.}{1987}{MNRAS}{225}{335}

\paper{Heiles,C.}{1987}{in Interstellar Processes}{\rm eds. D. Hollenbach,
H. Thronson}{Reidel,Dordrecht, p.171}

\paper{Heisler,J. \& Ostriker,J.P.}{1988}{ApJ}{332}{543}

\paper{Hewett,P.C., Foltz,C.B \& Chaffee,F.H.}{1995}{ApJ}{109}{1498}

\paper{Hill,G.J., Wynn-Williams,C.G. \& Becklin,E.E.}{1987}{ApJ}{316}{L11}

\paper{Hines,D.C. \& Wills,B.J.}{1992}{BAAS}{24}{1184}

\paper{Holman,B.A.}{1997}{MSc Thesis}{\rm The University of Melbourne}{Ch.2}

\paper{Holmberg,E.}{1958}{Lund Obs. Not.}{2}{136}

\paper{Hough,J.H., Bailey,J.A., Rouse,M.F. \& Whittet,D.C.B.}{1987}{MNRAS}
{227}{1}

\paper{Hoyle,F. \& Wickramasinghe,N.C.}{1991}{The Theory
of Cosmic Grains}{\rm ed. R.L.F. Boyd et al.}{Kluwer}

\paper{Hu,E.M., Cowie,L.L. \& Wang,Z.}{1985}{ApJS}{59}{447}

\paper{Hu,E.M.}{1992}{ApJ}{391}{608}

\paper{Hutchings,J.B. \& Morris,S.L.}{1995}{AJ}{109}{1541} 

\paper{Impey,C.D. \& Neugebauer,G.}{1988}{AJ}{95}{307}

\paper{Impey,C.D. \& Tapia,S.}{1990}{ApJ}{354}{124}

\paper{Jansen,R.A., Knappen,J.H., Beckman,J.E., Peletier,R.F.
\& Hes,R.}{1994}{MNRAS}{270}{373}

\paper{Jones,T.W., Rudnick,L., Owen,F.N., Puschell,J.J., Ennis,D.J.
\& Werner,M.W.}{1981}{ApJ}{243}{97}

\paper{Kellerman,K.I.}{1966}{ApJ}{146}{621}

\paper{Kellerman,K.I. \& Pauliny-Toth,I.I.K.}{1969}{ApJ}{155}{L71}

\paper{Kellerman,K.I. \& Pauliny-Toth,I.I.K.}{1981}{ARA\&A}{19}{373}

\paper{Kirk,J.G. \& Schneider,P.}{1987}{ApJ}{322}{256}

\paper{Knapen,J.H., Hes,R., Beckman,J.E. \& Peletier,R.F.}{1991}{A\&A}{241}{42}

\paper{Kollgaard,R.I., Feigelson,E.D., Laurent-Muehleisen,S.A.,
Spinrad,H., Dey,A. \& Brinkmann,W.}{1995}{ApJ}{449}{61}

\paper{Komossa,S. \& Fink,H.}{1997}{A\&A}{\rm in press}{astro-ph/9707003}

\paper{K\"{o}nigl,A.}{1981}{ApJ}{243}{700}

\paper{Kormendy,J.}{1992}{in Testing the AGN Paradigm}{\rm ed. S.S.Holt,
S.G.Neff, C.M.Urry}{College Park,MD,p.23} 

\paper{Kriss,G.A. \& Canizares,C.R.}{1985}{ApJ}{297}{177}

\paper{Kruper,J.S. \& Canizares,C.R.}{1989}{ApJ}{343}{66}

\paper{La Franca,F., Franceschini,A. Cristiani,S. \& Vio,R.}{1995}{A\&A}
{299}{19}

\paper{Landau,R., Golisch,B., Jones,T.J.,
Jones,T.W., Pedelty,J., Rudnick,L.,
Sitko,M.L., Kenney,J., Roellig,T.,
Salonen,E., Urpo,S., Schmidt,G.,
Neugebauer,G., Mathews,K., Elias,J.H.,
Impey,C., Clegg,P., Harris,S.}{1986}{ApJ}{308}{78}

\paper{Lanzetta,K.M.}{1991}{ApJ}{375}{1}

\paper{Lanzetta,K.M., Wolfe,A.M. \& Turnshek,D.A.}{1995}{ApJ}{440}{435}

\paper{Laor,A. \& Draine,B.T.}{1993}{ApJ}{402}{441}

\paper{Ledden,J.E. \& O'Dell,S.L.}{1983}{ApJ}{270}{434}

\paper{Ledden,J.E. \& O'Dell,S.L.}{1985}{ApJ}{298}{630}

\paper{Lees,J.F., Knapp,G.R., Rupen,M.P. \& Phillips,T.G.}{1991}{ApJ}
{379}{177}

\paper{Lilly,S.J. \& Cowie,L.L}{1987}{In Infrared Astronomy with
Arrays}{\rm eds. C.G. Wynn-Williams \& E.E.Becklin}{p.473}

\paper{Loeb,A.}{1993}{ApJ}{403}{542}

\paper{Longair,M.S.}{1993}{in The Physics of Background Radiation}
{\rm Springer Verlag; SAAS-FEE lecture No.23}{p.83}

\paper{Lonsdale,C.J., Smith,H.E. \& Lonsdale,C.J}{1995}{ApJ}{438}{632}

\paper{Loveday,J., Peterson,B.A., Efstathiou,G. \& Maddox,S.J.}{1992}
{ApJ}{390}{338}

\paper{Low,F.J., Cutri,R.M., Kleinmann,S.G. \& Huchra,J.P.}{1989}
{ApJ}{340}{1}

\paper{Lynden-Bell,D.}{1969}{Nature}{223}{690}

\paper{MacAlpine,G.M.}{1985}{in Astrophysics of Active Galaxies
and Quasi-Stellar Objects}{}{ed. J.S. Miller (Mill Valley:Univ. 
Science books), p.259}

\paper{Madejski,G.M., Mushotzky,R.F., Weaver,K.A. \& Arnaud,K.A.}
{1991}{ApJ}{370}{198}

\paper{Margolis,S.H. \& Schramm,D.N.}{1977}{ApJ}{214}{339}

\paper{Markkanen,T.}{1979}{A\&A}{74}{201}

\paper{Martin,P.G.}{1970}{MNRAS}{149}{221}

\paper{Martin,P.G.}{1978}{Cosmic Dust}{\rm Oxford}{Clarendon Press}

\paper{Masci,F.J. \& Webster,R.L.}{1995}{PASA}{12}{146}

\paper{Mather,J.C. \etal}{1990}{ApJ}{354}{L37}

\paper{Mathis,J.S., Rumpl,W. \& Nordsiek,K.H.}{1977}{ApJ}{217}{425}

\paper{Mathis,J.S.}{1986}{ApJ}{308}{281}

\paper{Mathis,J.S.}{1990}{ARA\&A}{28}{37}

\paper{Mathur,S., Wilkes,B., Elvis,M. \& Fiore,F.}{1994}{BAAS}{184}{1706}

\paper{McAlary,C.W. \& Rieke,G.H.}{1988}{ApJ}{331}{1}

\paper{McCarthy,P.J.}{1993}{ARA\&A}{31}{639}  

\paper{McDowell,J.C.}{1986}{MNRAS}{223}{763}

\paper{McDowell,J.C., Elvis,M., Wilkes,B.J., Willner,S.P.,
Oey,M.S., Polomski,E., Bechtold,J. \& Green,R.F.}{1989}{ApJ}{345}{L13}

\paper{McKee,C.F. \& Petrosian, V.}{1974}{ApJ}{189}{17}

\paper{McKee,C.F., Hollenbach,D.J., Seab,C.B. \& Tielens,A.G.G.M.}
{1987}{ApJ}{318}{674}

\paper{McMahon,R.G., Omont,A., Bergeron,J., Kreysa,E. \& Haslam,C.G.T.}
{1994}{MNRAS}{267}{L9}

\paper{Meisenheimer,K., R\"{o}ser,H.J., Hiltner,P., Yates,M.G.,
Longair,M.S., Chini,R. \& Perley,R.A.}{1989}{A\&A}{219}{63}

\paper{Meyer,D.M., Welty,D.E. \& York,D.G.}{1989}{ApJ}{343}{L37}

\paper{Meyer,D.M. \& Roth,K.C.}{1990}{ApJ}{363}{57}

\paper{M{\o}ller,P. \& Jacobsen,P.}{1990}{A\&A}{228}{299} 

\paper{Moore,R.L. \& Stockman,H.S.}{1984}{ApJ}{279}{465}

\paper{Morris,S.L. \etal}{1991}{AJ}{102}{1627}

\paper{Morrison,R. \& McCammon,D.}{1983}{ApJ}{270}{119}

\paper{Miller,J.S., French,H.B. \& Hawley,S.A.}
{1978}{in Pittsburgh Conf. on BL-Lac Objects}
{\rm ed. A.M. Wolf (Pittsburgh: Univ. Pittsburgh Dept. of Physics and Astronomy)}
{p.176}

\paper{Mulchaey,J.S., Mushotzky,R.F. \& Weaver,K.A.}{1992}{ApJ}{390}{L69}

\paper{Nath,B.B. \& Trentham,N.}{1997}{MNRAS}{\rm in press}{astro-ph/9707177}

\paper{Netzer,H.}{1990}{in Active Galactic Nuclei, Saas-Fee Advanced
Course 20}{\rm Springer}{Berlin}

\paper{Netzer,H. \& Laor,A.}{1993}{ApJ}{404}{L51}

\paper{Neugebauer,G., Oke,J.B., Becklin,E.E. \& Matthews,K.}{1979}
{ApJ}{230}{79}

\paper{O'Dell, S.L.}{1979}{in Active Galactic Nuclei, eds: Hazard, C.
\& Mitton, S.}{\rm 1979}{p.95}

\paper{O'Dell,S.L., Scott,A. \& Stein,W.A.}{1987}{ApJ}{313}{164}

\paper{Oey,M.S. \& Kennicutt Jr,R.C.}{1993}{ApJ}{411}{137}

\paper{Olive,K.A., Schramm,D.N., Steigman,G. \& Walker,T.P.}{1990}
{Phys.Lett.B}{236}{454}

\paper{Orr,M.J.L. \& Browne,I.W.A.}{1982}{MNRAS}{200}{1067}

\paper{Osterbrock,D.E.}{1989}{in Astrophysics of Gaseous Nebulae and
Active Galactic Nuclei}{\rm University Science Books, Mill Valley 
California}{p.326} 

\paper{Ostriker,J.P. \& Heisler,J.}{1984}{ApJ}{278}{1}

\paper{Ostriker,J.P., Vogeley,M.S. \& York,D.G.}{1990}{ApJ}{364}{405}

\paper{Pacholczyk,A.G.}{1970}{in Radio Astrophysics}{\rm San Francisco}
{Freeman}

\paper{Padovani,P. \& Urry,C.M.}{1992}{ApJ}{387}{449}

\paper{Page,M.J., Carrera,F.J., Hasinger,G., Mason,K.O.,
McMahon,R.G., Mittaz,J.P.D., Barcons,X., Carballo,R.,
Gonzalez-Serrano,I. \& Perez-Fournon,I.}{1996}{MNRAS}{281}{579} 

\paper{Pan,H.C., Stewart,G.C. \& Pounds,K.A.}{1990}{MNRAS}{242}{177}

\paper{Peacock,J.A. \& Wall,J.V.}{1981}{MNRAS}{194}{331}

\paper{Peebles, P.J.E.}{1989}{in The Epoch of Galaxy Formation}
{\rm ed. C.S. Frenk \etal (Dordrecht: Kluwer)}{p.1}

\paper{Pei,Y.C., Fall,S.M. \& Bechtold,J.}{1991}{ApJ}{378}{6}

\paper{Pei,Y.C.}{1992}{ApJ}{395}{130}

\paper{Pei,Y.C. \& Fall,S.M.}{1995}{ApJ}{454}{69}

\paper{Peletier,R.F., Valentijn,E.A., Moorwood,A.F.M., 
Freudling,W., Knapen,J.H.
\& Beckman,J.E.}{1995}{A\&A}{300}{L1}

\paper{Pettini,M., Boksenberg,A. \& Hunstead,R.W.} 
{1990}{ApJ}{348}{48}

\paper{Pettini,M., Smith,L.J., Hunstead,R.W. \& King,D.L.} 
{1994}{ApJ}{426}{79}

\paper{Pettini,M., King,D.L., Smith,L.J. \& Hunstead,R.W.} 
{1997}{ApJ}{478}{536}

\paper{Phillipps,S.}{1986}{ApJ}{25}{L19}

\paper{Pier,E.A. \& Krolik,J.H.}{1992}{ApJ}{401}{99}

\paper{Pier,E.A. \& Krolik,J.H.}{1993}{ApJ}{418}{673}

\paper{Press,W.H., Teukolsky,S.A., Vetterling,W.T., Flannery,B.P.}{1994}
{Numerical Recipes in C}{\rm 2nd edition, Cambridge University Press}{p.547}

\paper{Puchnarewicz,E.M., Mason,K.0., Romero-Colmenero,E.,
Carrera,F.J., Hasinger,G. McMahon,R., Mittaz,J.P.D., Page,M.J.
\& Carballo,R.}{1996}{MNRAS}{281}{1243}

\paper{Rao,S. \& Briggs,F.H.}{1993}{ApJ}{419}{515} 

\paper{Readhead,A.C.S.}{1994}{ApJ}{426}{51}

\paper{Rees,M.J.}{1990}{Science}{247}{817} 

\paper{Reynolds,C.S. \& Fabian,A.C.}{1995}{MNRAS}{273}{1167}

\paper{Reynolds,C.S.}{1996}{MNRAS}{286}{513}

\paper{Rieke,M. \& Lebofsky,M.J.}{1979}{ARA\&A}{17}{477}

\paper{Rieke,G.H., Lebofsky,M.J. \& Kinman,T.D.}{1979}{ApJ}{232}{L151}

\paper{Rieke,G.H.}{1982}{In Extragalactic radio sources; (Proceedings,
Albuquerque Symposium)}{\rm Dordrecht}{p.239} 

\paper{Rieke,G.H., Lebofsky,M.J. \& Wi\'{s}niewski,W.Z.}{1982}{ApJ}{263}{73}

\paper{Rieke,M. \& Lebofsky,M.J.}{1985}{ApJ}{288}{618}

\paper{Rix,H.W. \& Rieke,M.}{1993}{ApJ}{418}{123}

\paper{Rocca-Volmerange,B. \& Guiderdoni,B.}{1987}{A\&ASuppl}{74}{185}

\paper{Rodrigues-Williams,L.L. \& Hogan,C.J.}{1994}{AJ}{107}{451}

\paper{Rodrigues-Williams,L.L. \& Hawkins,C.J.}{1995}
{\rm Proc}
{\rm of the 5th Annual Astrophysics Conference}{Maryland}

\paper{Romani,R.W. \& Maoz,D.}{1992}{ApJ}{386}{36}

\paper{ROSAT Users' Handbook}{1994}{\rm Edited by U.G. Briel \etal}
{\rm Max-Planck-Institut f\"{u}r 
extraterrestriche Physik}
{http://heasarc.gsfc.nasa.gov/docs
/rosat/ruh/handbook/handbook.html}  

\paper{Rudnicki,K.}{1986}{in Proc. Internat. of Physics ``Enrico Fermi'',
Summer School}{86}{480}

\paper{Sandage,A.}{1972}{ApJ}{178}{25}

\paper{Sanders,D.B., soifer,B.T., Elias,J.H., Neugebauer,G. \& Matthews,K.}
{1988}{ApJ}{328}{L35} 

\paper{Sanders,D.B., Phinney,E.S.,Neugebauer,G.,Soifer,B.T. \& Matthews,K.}
{1989}{ApJ}{347}{29}

\paper{Sarazin,C.L.}{1986}{Rev. Mod. Phys.}{58}{1}

\paper{Sargent,W.L.W., Boksenberg,A. \& Steidel,C.C.}{1988}{ApJS} 
{68}{539}

\paper{Saunders,W., Rowan-Robinson,M., Lawrence,A., Efstathiou,G.,
Kaiser,N., Ellis,R.S. \& Frenk,C.S.}{1990}{MNRAS}{242}{318}

\paper{Savage,B.D. \& Mathis,J.S.}{1979}{ARA\&A}{17}{73}

\paper{Savaglio,S., D'Odorico,S., \& M\"{o}ller,P.}{1994}{A\&A}{281}{331}

\paper{Schartel,N., Green,P.J.,
Anderson,S.F.,
Hewett,P.C., Foltz,C.B.,
Margon,B.,
Brinkmann,W., Fink,H. \&
Tr\"{u}mper,J.}{1996}{MNRAS}{283}{1015}  

\paper{Schmidt,M. \& Green,R.F.}{1983}{ApJ}{269}{352}

\paper{Schmidt,M. Schneider,D.P. \& Gunn,J.E.}{1988}{in Optical
Surveys for Quasars}{\rm (ed. P.S. Osmer, A.S. Porter, R.F. Green, 
C.B. Foltz}{Provo: Brigham Young Univ. Press), p.87} 

\paper{Scoville,N. \& Norman,C.}{1988}{ApJ}{332}{163}

\paper{Searle,L.}{1971}{ApJ}{168}{327}

\paper{Seaton,M.J.}{1979}{MNRAS}{187}{73}

\paper{Seitz,S. \& Schneider,P.}{1995}{A\&A}{302}{9}

\paper{Serjeant,S. \& Rawlings,S.}{1996}{Nature}{379}{304}

\paper{Serkowski,K., Mathewson,D.E. \& Ford,V.}{1975}{ApJ}{196}{261}

\paper{Shuder,J.M.}{1981}{ApJ}{244}{12}

Siebert,J. \etal 1997. {\it A\&A}, in preparation. 

\paper{Sitko,M.L. \& Zhu,Y.}{1991}{ApJ}{369}{106}

\paper{Smith,E.P. \& Heckman,T.M.}{1989}{ApJ}{341}{658}

\paper{Smith,P.S., Balonek,T.J., Heckert,P.A. \& Elston,R.}{1986}{ApJ}
{305}{484} 

\paper{Soifer,B.T., Houck,J.R. \& Neugebauer,G.}{1986}{ARA\&A}{25}{187} 

\paper{Songaila,A. \& Cowie,L.L.}{1996}{AJ}{112}{335} 

\paper{Songaila,A.}{1997}{ApJ}{\rm in press}{astro-ph/9709046}

\paper{Spinrad,H. \& Smith,H.E.}{1975}{ApJ}{201}{275}

\paper{Spinrad,H. \& Djorgovski,S.}{1987}{in Proc. IAU Symp. No. 126}
{\rm ed: G.Burbidge}{p.129}

\paper{Stebbins,J., Huffer,C.H. \& Whitford,A.E.}{1934}
{Publ. Washburn Obs.}{15}{part V}

\paper{Stickel,M., Fried,J.W. \& K\"{u}hr,H.}{1993}{A\&ASuppl}{98}{393} 

\paper{Stocke,J.T., Morris,S.L., Gioia,I.M., Maccacaro,T.
\& Schild,R.}{1991}{ApJS}{76}{813}

\paper{Stockman,H.S., Moore,R.L. \& Angel,J.R.P.}{1984}{ApJ}{279}{485}

\paper{Takeda,H., Nulsen,P.E.J. \& Fabian,A.C.}{1984}{MNRAS}{208}{261}

\paper{Taylor,G.L., Dunlop,J.S., Hughes,D.H. \& Robson,E.I.}{1996}{MNRAS}
{283}{930}

\paper{Ter\"{a}sranta,H. \& Valtaoja,E.}{1994}{A\&A}{283}{51}

\paper{Thomas,P.A. \& Webster,R.L.}{1990}{ApJ}{349}{437}

\paper{Tinsley,B.M.}{1976}{ApJ}{208}{797}

\paper{Tinsley,B.M.}{1980}{Fund. Cos. Phys.}{5}{287}

\paper{Trumpler,R.J.}{1930}{Lick Obs. Bull.}{114}{154}

\paper{Tytler,D., Boksenberg,A., Sargent,W.L.W. \& Kunth,D.}{1987}
{ApJS}{64}{667}

\paper{Ulrich,M.}{1988}{in Conference on BL Lac Objects (Como, Italy)}
{\rm eds: L.Maraschi, T.Maccacaro, M.Ulrich}{p.45}

\paper{Urry,C.M. \& Padovani,P.}{1995}{PASP}{107}{803}

\paper{Valentijn,E.A.}{1990}{Nature}{346}{153}

\paper{Valentijn,E.A.}{1994}{MNRAS}{206}{614}

\paper{V\'{e}ron-Cetty,M.P. \& V\'{e}ron}{1989}{A Catalogue of Quasars
and Active Nuclei}{\rm 4th Edition}{Munich: ESO} 

\paper{Vila-Costas,M.B. \& Edmunds,M.}{1992}{MNRAS}{259}{121}

\paper{Villar-Mart\'{\i}n,M. \& Binette,L.}{1997}{A\&A}{317}{350}

\paper{Visnovsky,K.L., Impey,C.D., Foltz,C.B., Hewitt,P.C., Weymann,R.J. 
\& Morris, S.L.}{1992}{ApJ}{391}{560}

\paper{Wall,J.V. \& Peacock,J.A.}{1985}{MNRAS}{216}{173} 

\paper{Wang,B.}{1991a}{ApJ}{374}{456}
 
\paper{Wang,B.}{1991b}{ApJ}{383}{L37}

\paper{Warren,S.J. Hewett,P.C. \& Osmer,P.S.}{1994}{ApJ}{421}{412} 

\paper{Webster,R.L., Francis,P.J., Peterson,B.A., Drinkwater,M.J.
\& Masci,F.J.}{1995}{Nature}{375}{469}

\paper{Weinberg,S.}{1972}{Gravitation and Cosmology: Principles and
Applications of the General Theory of Relativity}{\rm New York}{\rm
Wiley}

\paper{Wheeler,J.C., Snedin, C. \& Truran,J.W.}{1989}{ARA\&A}{27}{279}

\paper{White,S.D.M. \& Frenk,C.S.}{1991}{ApJ}{379}{52}

\paper{Whittet,D.C.B. \& van Breda,I.G.}{1978}{A\&A}{66}{57} 

\paper{Whittet,D.C.B.}{1992}{Dust in the Galactic Environment}
{\rm eds. R. Tayler, R. White}{Cambridge University Press} 

\paper{Whittet,D.C.B., Martin,P.G., Hough,J.H., Rouse,M.F., Bailey,J.A.
\& Axon,D.J.}{1992}{ApJ}{386}{562}

\paper{Wilkes,B.J., Wright,A.E., Jauncey,D.L. \& Peterson,B.A.}{1983}
{PASA}{5}{2}

\paper{Wilkes,B.J.}{1986}{MNRAS}{218}{331} 

\paper{Wilkes,B.J. \& Elvis,M.}{1987}{ApJ}{323}{243}

\paper{Wilkes,B.J., Elvis,M., Fiore,F., McDowell,J.C., Tananbaum, H.
\& Lawrence,A.}{1992}{ApJ}{393}{L1}

\paper{Wilkes,B.J., Tananbaum,H., Worrall,D.M., Avni,Y., Oey,M.S.
\& Flanagan,J.}{1994}{ApJS}{92}{53} 

\paper{Wilking,B.A., Lebofsky,M.J. \& Rieke,G.H.}{1982}{AJ}{87}{695}

\paper{Wills,B.J.}{1988}{in BL-Lac Objects, Proc. Como, Italy}
{\rm eds. L. Maraschi, T. Maccacaro}{p.114} 

\paper{Wills,B.J., Wills,D., Evans,N.J.I., Natta,A., Thompson,K.L.,
Breger,M. \& Sitko,M.L.}{1992}{ApJ}{400}{96}

\paper{Wills,B.J., Wills,D., Breger,M., Antonucci,R.R.J. \& Barvainis,R.}
{1992}{ApJ}{398}{454}

\paper{Wills,B.J., Netzer,H., Brotherton,M.S., Mingsheng Han, Wills,D.
Baldwin,J.A., Ferland,G.J. \& Browne,I.W.A.}{1993}{ApJ}{410}{534}

\paper{Wise,M.W., O'Connell,R.W., Bregman,J.N. \& Roberts,M.S.}{1993}
{ApJ}{405}{94} 

\paper{Wolfe,A.M., Turnshek,D.A., Smith,H.E. \& Cohen,R.D.}{1986}
{ApJS}{61}{249}

\paper{Womble,D.S., Sargent,W.L.W. \& Lyons,R.S.}{1996}{in Cold Gas
at High Redshift}{\rm eds. M. Bremer, H. Rottgering, P. van der Werf,
C. Carilli}{(Kluwer)}

\paper{Wright,A.E., Ables,J.G. \& Allen,D.A.}{1983}{MNRAS}{205}{793} 

\paper{Wright,E.L.}{1986}{ApJ}{311}{156}

\paper{Wright,E.L.}{1990}{ApJ}{353}{411}

\paper{Xu,C. \& Buat,V.}{1995}{A\&A}{293}{65}

\paper{Yee,H.K. \& Oke,J.B.}{1978}{ApJ}{226}{753}

\paper{Zaritsky,D.}{1994}{AJ}{108}{1619}

\paper{Zaritsky,D., Kennicutt,R.C., Jr. \& Huchra,J.P.}{1994}{ApJ}
{420}{87}
}

\chapter{Cosmology}

\vspace{1mm}
\leftskip=4cm
 
{\it ``Everything should be made as simple as possible, but not simpler.''} 
 
\vspace{1mm}
 
\hfill {\bf --- Albert Einstein, 1879-1955}

\vspace{7mm}
 
\leftskip=0cm

\noindent
In this thesis, we have assumed the Friedmann-Lema\^{\i}tre cosmological
model.
This is based on a spatially 
homogeneous and isotropic universe with $\Lambda=0$. 
All calculations use a deceleration parameter $q_{0}=0.5$ ($\Omega=1$) 
and a Hubble parameter of $H_{0}=50\,\rm km\,s^{-1}\,Mpc^{-1}$,
unless otherwise stated. In cases where $h_{50}$ and $h_{100}$ are used, these 
are defined via $H_{0}=50h_{50}$ and 
$H_{0}=100h_{100}\,\rm km\,s^{-1}\,Mpc^{-1}$ respectively.

Below are some relations used throughout this thesis to transform 
between apparent and physical quantities of interest. For a brief 
derivation, see Weinberg (1972). 

The apparent angular size of a source, $\theta''$, in terms of 
its physical diameter $r$ is given by:
\begin{equation}
\theta\,=\,\frac{(2.06\times10^{5})(1+z)^{2}\,r}{D_{L}}{\rm\hspace{5mm}
arcsec},
\end{equation}
where $D_{L}$ is the luminosity distance defined as 
\begin{equation}
D_{L}\,=\,\frac{c}{H_{0}} q_{0}^{-2}\left[1 - q_{0} + q_{0}z + (q_{0} -
1)(1+2q_{0}z)^{1/2}\right].
\label{btwo}
\end{equation}
$\frac{c}{H_{0}}$ is the
Hubble distance ($\equiv 6.05\times10^{9}\,h_{50}^{-1}$pc).

To transform luminosity
functions into properties that are directly observable such as
magnitude and redshift distributions,
we need a specification of the evolutionary and $k$-correction for 
the sources under consideration.
The evolutionary correction is a quantity defining the physical
evolution in source luminosity as a function of redshift. 
The $k$-correction is due to sampling of the rest frame SED
at different portions through a fixed observational bandpass as it
is redshifted. Both of these are defined in magnitudes.

Absolute and apparent magnitudes are related by
\begin{equation}
M\,=\,m - 5\log d_{L} - 43.88 + 5\log h_{50} - e(z) - k(z)
\label{bthree}
\end{equation}
where $e(z)$ and $k(z)$ are respectively the evolutionary and $k$-correction
corresponding to the bandpass in which $m$ is measured 
as defined above.
In Eqn.~\ref{bthree}, $d_{L}$ is the luminosity distance in units of $c/H_{0}$
(ie.
$d_{L}\equiv (H_{0}/c)D_{L}$ with $D_{L}$ from Eqn.~\ref{btwo}).

The number counts of sources as a function of apparent magnitude is
an integral of the luminosity function 
$\phi$ over all redshifts.
In general, the number of sources per unit apparent magnitude per
square degree of sky can be written
\begin{equation}
N(m)\,=\,\int^{z_{max}}_{z_{min}}dz\,\left(\frac{dV}{dz}\right)\phi(M,z),
\end{equation}
where $dV/dz$ is the comoving volume element per square
degree of sky and $m$ is the apparent magnitude corresponding to the
absolute magnitude $M$.
In the standard Friedmann cosmology, the comoving volume
element per square degree of sky is
\begin{equation}
\frac{dV}{dz}\,=\,6.5\times10^{7}h_{50}^{-3}(1+z)^{3}(1+2q_{0}z)^{-1/2}d_{L}^{2}
\,\,{\rm\,deg^{-2}Mpc^{3}}.
\end{equation}
 
The redshift distribution, or number of 
sources per unit redshift, per square degree of sky
contributing to some apparent magnitude interval $m_{min}<m<m_{max}$ is
given by 
\begin{equation}
n(z)\,=\,\frac{dV}{dz}\int^{M(m_{max})}_{M(m_{min})}dM\,
\phi(M,z).
\end{equation}

\chapter{Derivations in Evolutionary Dust Model} 

Here we derive exact expressions for the mean and variance in 
total optical depth in galactic dust as a function of redshift in
our evolutionary dust model of Chapter 4. 
The scaling of these quantities with model parameters
was defined by Eqns.~\ref{tauscale} and~\ref{varev}.
The galaxies are modelled as exponential dusty disks, randomly
inclined to the line-of-sight. 

We first derive the average number of galaxies intercepted by a light ray
emitted from some redshift $z$ (ie. Eqn.~\ref{nzev}).
Given a `proper' number density of galaxies at some redshift
$n_{g}(z)$, with each galaxy having an effective cross-sectional area
$\mu\,\sigma$ as viewed by an observer
($\mu$ is a random inclination factor, where
$\mu=\cos\,\theta$ and $\theta$ is the angle
between the sky plane and the plane of a galactic disk),
then the average number of intersections of a light ray along some path
length $ds$ will be given by 
\begin{equation}
dN\,=\,n_{g}(z)\,\mu\sigma\,ds.
\label{dn}
\end{equation}
In an expanding universe we have $n_{g}=n_{0}(1+z)^3$, where $n_{0}$ is
a local comoving number density and is assumed to be constant. Units of proper
length
and redshift are related by
\begin{equation}
{ds\over dz}\,=\,\left({c\over
H_{0}}\right)\,{1\over(1+z)^{2}(1+2q_{0}z)^{1/2}}
\label{ds}
\end{equation}
(Weinberg, 1972). The effective cross-section projected towards an
observer
for a randomly inclined disk is found by averaging over the random
inclination factor
$\mu$, where $\mu$ is randomly distributed between 0 and 1 and
integrating
over the exponential profile assumed for each disk with scale radius
$r_{0}(z)$ (see Eqns.~\ref{expr} and~\ref{rozev}).
The product $\mu\sigma$ in Eqn.~\ref{dn} is thus replaced by
\begin{equation}
\int_{0}^{1}\mu\,d\mu\,\int_{0}^{\infty}e^{-r/r_{0}(z)}\,
2\pi r\,dr\,=\,\pi\,{r_{0}}^2(1+z)^{2\delta}.
\label{eq}
\end{equation}
Thus from Eqn.~\ref{dn}, the average number of intersections up to some
redshift $z$ is given by
$$
\bar{N}(z)\,=\,\int_{0}^{z}\mu\sigma\,n_{g}(z')\,\left({ds\over
dz'}\right)\,dz'
$$
\begin{equation}
=\,n_{0}\pi r_{0}^{2}\left({c\over
H_{0}}\right)\int_{0}^{z}{(1+z')^{1+2\delta
}\over (1+2q_{0}z')^{1/2}}\,dz'.
\label{n}
\end{equation}
With $\tau_{g}$ being defined by $n_{0}\pi r_{0}^{2}\left({c\over
H_{0}}\right)$,
this directly leads to Eqn.~\ref{nzev} for $q_{0}=0.5$.

The mean optical depth $\bar{\tau}$ is derived by a similar argument.
If $\tau_{0}(z)$ is the optical depth observed through a face on galaxy
at
some redshift $z$ (Eqn.~\ref{tboz}), then a galactic disk inclined by some
factor 
$\mu$ will
have its optical depth increased to $\tau_{0}(z)/\mu$. Multiplying this
quantity
by Eqn.~\ref{dn}, the extinction suffered by a light ray along a path
length $ds$
is given by
\begin{equation}
d\tau\,=\,n_{g}(z)\,\sigma\,\tau_{0}(z)\,ds.
\label{dt}
\end{equation}
Thus the mean optical depth to some redshift $z$ can be calculated from
\begin{equation}
\bar{\tau}(z)\,=\,\int_{0}^{z}\sigma n_{g}(z')\tau_{0}(z')\left({ds\over
dz'}\right
)\,dz'.
\label{mt}
\end{equation}
Given $n_{g}(z)$, $\left({ds\over dz}\right)$ and $\sigma$ (from the
integral over $r$ in Eqn.~\ref{eq})
above,
and $\tau_{0}(z')$ from Eqn.~\ref{tboz}, the mean optical depth follows the
general form
\begin{equation}
\bar{\tau}(z)\,=\,2\tau_{g}\tau_{B}\int_{0}^{z}{(1+z')^{2.5+2\delta
}\over (1+2q_{0}z')^{1/2}}\,\left[1 -
{\ln(1+z')\over\ln(1+z_{dust})}\right]\,dz'.
\label{mtz}
\end{equation}
Similarly, the variance in the optical depth distribution is defined as 
the following
\begin{equation}
\sigma_{\tau}^{2}(z)\,=\,\langle\tau^2\rangle -
{\langle\tau\rangle}^{2}\,=\,\int_{0}^{z}\sigma
n_{g}(z'){\tau_{0}}^{2}(z')\left({ds\over dz'}\right)\,dz'.
\label{vart}
\end{equation}
In terms of our model dependent parameters, this becomes
\begin{equation}
\sigma_{\tau}^{2}(z)\,=\,2\tau_{g}{\tau_{B}}^{2}\int_{0}^{z}{(1+z')^{4+2\delta
}\over (1+2q_{0}z')^{1/2}}\,{\left[1
-{\ln(1+z')\over\ln(1+z_{dust})}\right]}^{2}\,dz'.
\label{vartz}
\end{equation}

\begin{titlepage}
 
\vspace*{50mm}
\leftskip=1.5cm
\noindent
{\large\it  That it will never come again,\\
Is what makes life so sweet.\\}
 
\vspace{8mm}
 
\hspace{1in}\bf{--- Emily Dickinson, Poem Number 1741}
 
\end{titlepage}
 
\end{document}